\documentclass[12pt]{article}
\usepackage{a4wide}
\usepackage{latexsym}
\usepackage{amsmath}
\usepackage{amsfonts}
\usepackage{amscd}
\usepackage{amsthm}
\usepackage{shuffle}
\usepackage{cite}
\usepackage{graphicx}
\usepackage{float}
\usepackage{axodraw}
\usepackage{BOONDOX-calo}

\usepackage{pslatex}
\usepackage[latin1,utf8]{inputenc}
\usepackage[OT2,T1]{fontenc}

\usepackage{ifthen}

\newcommand{\bq}{\begin{eqnarray}}
\newcommand{\eq}{\end{eqnarray}}
\newcommand{\eps}{\varepsilon}
\newcommand{\bs}{\begin{small}}
\newcommand{\es}{\end{small}}
\newcounter{exercise}
\newtheorem{algorithm}{Algorithm}

\begin{document}

\thispagestyle{empty}

\begin{flushright}
  MITP/16-109
% \\ version of \today
\end{flushright}

\vspace{1.5cm}

\begin{center}
  {\Large\bf Tales of 1001 Gluons\\
  }
  \vspace{1cm}
  {\large Stefan Weinzierl\\
\vspace{2mm}
      {\small \em PRISMA Cluster of Excellence, Institut f{\"u}r Physik, }\\
      {\small \em Johannes Gutenberg-Universit{\"a}t Mainz,}\\
      {\small \em D - 55099 Mainz, Germany}\\
  } 
\end{center}

\vspace{2cm}

% abstract -------------------------------------------------------------------------
\begin{abstract}\noindent
  {
This report is centred around tree-level scattering amplitudes in pure Yang-Mills theories,
the most prominent example is given by the tree-level gluon amplitudes of QCD.
I will discuss several ways of computing these amplitudes, illustrating in this way 
recent developments in perturbative quantum field theory.
Topics covered in this review include 
colour decomposition, 
spinor and twistor methods, 
off- and on-shell recursion,
MHV amplitudes and MHV expansion,
the Grassmannian and the amplituhedron,
the scattering equations and the CHY representation.
At the end of this report there will be an outlook on the relation between pure Yang-Mills amplitudes
and scattering amplitudes in perturbative quantum gravity.
   }
\end{abstract}

\vspace*{\fill}

% main text ------------------------------------------------------------------------
\newpage

\tableofcontents

% ----------------------------------------------------------------------------------
\newpage
\section{Introduction}
\label{sect:intro}

Each student of physics has encountered the harmonic oscillator during her or his studies, usually more than once.
The harmonic oscillator is a simple system, familiar to the students, and can be used as a pedagogical example
to introduce new concepts -- just think about the introduction of raising and lowering operators for the
quantum mechanical harmonic oscillator.

Another system encountered by any physics student is the hydrogen atom. 
Again, methods to solve this system apply to a wider context.
For example, positronium and charmonium are close cousins of the hydrogen atom.
 
Within the context of quantum field theory the 
computation of tree-level scattering amplitudes in pure Yang-Mills theories
constitutes another example of physical problems every student should know about:
The tree-level scattering amplitudes in pure Yang-Mills theories can be computed in several way
and each possibility illuminates certain mathematical structures 
underlying quantum field theories.

In this report we will take the tree-level scattering amplitudes in pure Yang-Mills theories
as our reference object and explore in a pedagogical way the mathematical structures associated to it.
A prominent special case are tree-level gluon amplitudes of QCD, corresponding to the choice $G=\mathrm{SU}(3)$
as gauge group.
Originally, the study of these amplitudes was motivated by finding efficient methods to compute multi-parton
amplitudes, where the multiplicity ranges from $4$ (corresponding to a $2 \rightarrow 2$ scattering process)
to roughly $10$ (corresponding to a $2 \rightarrow 8$ scattering process).
This is relevant for hadron collider experiments.
In particular at the LHC gluon-initiated processes constitute the bulk of scattering events.
Although we focus on tree-level amplitudes in this review, many of the techniques discussed in this report
have an extension towards higher orders in perturbation theory.
This is useful for precision calculations.
Finally, let me mention that the study of scattering amplitudes has developed in recent years in a field of its own,
revealing hidden structures of quantum field theory which are not manifest in the conventional Lagrangian formulation.

There are other excellent review articles and lecture notes, covering some of the topics presented here, 
for example \cite{Mangano:1990by,Dixon:1996wi,Elvang:2013cua,Henn:2014yza}.
Furthermore the content of this report is not meant to cover the field completely.
More on the contrary, by focusing solely on the tree-level amplitudes in Yang-Mills theory
we try to explain the basic principles and leave advanced topics to the current research literature.
Not included in this report are for example the extension of pure Yang-Mills theory to QCD, i.e.
the inclusion of fermions in the fundamental representation of the gauge group, massless or massive.
Furthermore, not included is the important field of loop calculations, this field alone would fill another
report.
Neither covered in detail are supersymmetric extensions of Yang-Mills theory, the most popular example
of the latter would be ${\mathcal N}=4$ supersymmetric Yang-Mills theory.

We will start with a review of the textbook method based on Feynman diagrams in section~(\ref{sect:textbook}).
Feynman diagrams allows us in principle to compute a tree-level amplitude for any number $n$ of external particles.
However, in practice this method is highly inefficient.

In section~(\ref{sect:efficiency}) we will discuss efficient methods for the computation of
tree-level scattering amplitudes in Yang-Mills theory.
We will learn about colour decomposition, the spinor helicity method and off-shell recurrence relations.
The combination of these three tools constitutes one of the best (probably the best) methods to compute 
tree-level scattering amplitudes numerically.
In section~(\ref{sect:efficiency}) we treat in addition the topics 
of colour-kinematics duality and twistors.
These two topics are not directly targeted at efficiency.
However, colour-kinematics duality is closely related to colour decomposition and 
twistors are closely related to the
spinor helicity method. Therefore these topics are best discussed together.

Section~(\ref{sect:mhv_amplitudes}) is centred around maximally helicity violating amplitudes.
We present the Parke-Taylor formul{\ae} and show that an arbitrary tree-level Yang-Mills amplitude
may be computed from an effective scalar theory with an infinite tower of vertices, where
the new interaction vertices are given by the Parke-Taylor formul{\ae}.

In section~(\ref{sect:BCFW_recursion}) we present the on-shell recursion relations.
On-shell recursion relations are one of the best (probably the best) methods to compute 
tree-level scattering amplitudes analytically.
In addition, they are a powerful tool to prove the correctness of new representations
of tree-level scattering amplitudes in Yang-Mills theory.

Section~(\ref{sect:grassmannian}) is devoted to a geometric interpretation of 
a scattering amplitude as the volume of a generalised polytope (the amplituhedron) in a certain
auxiliary space.
We discuss Grassmannians and the representation of scattering amplitude in this formalism.

In section~(\ref{sect:chy_representation}) we will learn about the Cachazo-He-Yuan (CHY) representation.
Again, we look at some auxiliary space 
and associate to each external particle a variable $z_j \in {\mathbb C}{\mathbb P}^1$.
Of particular interest are the values of the $z_j$'s, which are solutions of the
so-called scattering equations. We will present the scattering equations and show that the
scattering amplitude can be expressed as a global residue at the zeros of the scattering equations.

In section~(\ref{sect:gravity}) we explore tree-level amplitudes in perturbative quantum gravity.
This may at first sight seem to be disconnected from the rest of this review, but we will see 
that there is a close relation between scattering amplitudes in Yang-Mills theory and gravity.
The link is either provided by colour-kinematics duality, the CHY representation or the Kawai-Lewellen-Tye (KLT) relations.
The latter are introduced in this section.

Finally, section~(\ref{sect:outlook}) gives an outlook.

Let me add a word on the notation used in this review: 
Although it is desirable to have a unique notation throughout these notes, with
the variety of topics treated in this course we face the challenge that one topic is best presented in one notation,
while another topic is more clearly exposed in another notation.
An example is given by a central quantity of this report, the tree-level primitive helicity amplitudes.
These amplitudes depend on a set of external momenta $p = (p_1,...,p_n)$, a description of the spin states
of all external particles, either given through a set of helicities $(\lambda_1,...,\lambda_n)$ or through
a list of polarisation vectors $\eps=(\eps_1,...,\eps_n)$. Finally, these amplitudes depend as well on an
external cyclic order, which may be specified by a permutation $\sigma \in S_n$.
Depending on the context we denote these amplitudes either by
\bq
 A_n^{(0)}\left(p_{\sigma_1}^{\lambda_{\sigma_1}}, ..., p_{\sigma_n}^{\lambda_{\sigma_n}} \right)
 & \mbox{or by} &
 A_n^{(0)}\left( \sigma, p, \eps \right).
\eq
A second example is given by the notation used for Weyl spinors.
Also here one finds in the literature various notations, either with dotted/undotted indices $p_{\dot{A}}$, $p_A$,
as bra/ket-spinors $\langle p+|$, $| p+ \rangle$ or in a notation with square/angle brackets $[p|$, $|p\rangle$.
Also here we introduce all commonly used notations, giving the reader a dictionary to translate between
the various notations.

This review grew out of lectures given at the Saalburg summer school 2016.
As such it was natural to include several exercises in the main text,
with solutions to all exercises provided in the appendix.
Students reading this report are encouraged to do these exercises.
The structure of exercises with solutions in the appendix is kept in this report for the following reason:
Many of the exercises provide detailed proofs of statements made in the main text.
Leaving them out is not an option, as it will remove essential information for understanding the main concepts.
Putting these detailed derivations into the appendix has the advantage of not disrupting the argumentation line in the main
text.

% ----------------------------------------------------------------------------------
\newpage
\section{The textbook method}
\label{sect:textbook}

We start our exploration of Yang-Mills theory with a short review of topics which can be found in modern textbooks.
The presentation will be brief and concise, assuming that the reader is already 
familiar with the major part of the material presented in this section.
Any gaps can be closed by consulting a textbook on quantum field theory.
There are many excellent textbooks on quantum field theory, as an example let me just mention the books
by Peskin and Schroeder \cite{Peskin}, Srednicki \cite{Srednicki:2007qs} 
or Schwartz \cite{Schwartz}.
Topics covered in this section include the Lagrangian description of Yang-Mills theory, gauge-fixing
and the computation of amplitudes through Feynman diagrams.

\subsection{The Lagrangian of Yang-Mills theory}

We start with the definition of Yang-Mills theory \cite{Yang:1954ek} in the notation usually used in physics textbooks.
This is almost a synonym for ``a notation with a lot of indices''.
Our convention for the Minkowski metric is 
\bq
 g_{\mu\nu}& = & \mathrm{diag}(1,-1,-1,-1).
\eq
Let $G$ be a Lie group, $\mathfrak{g}$ its Lie algebra and $T^a$ the generators of the Lie algebra
where the index $a$ takes values from $1$ to $\dim G$.
We use the conventions
\bq
\label{convention_Lie_algebra}
 \left[ T^a, T^b \right] = i f^{abc} T^c,
 & &
 \mathrm{Tr}\left( T^a T^b \right) = \frac{1}{2} \delta^{ab}.
\eq
Yang-Mills theory consists of a gauge field $A_\mu^a(x)$.
The action is given as usual by the integral over the Lagrange density
\bq
 S & = & \int d^4x \; {\mathcal L}_{\mathrm{YM}}.
\eq
The {\bf Lagrange density for the gauge field} reads
\bq
\label{Lagrangian_YM}
{\mathcal L}_{\mathrm{YM}} & = & 
 - \frac{1}{4} F^{a}_{\mu\nu} F^{a \mu\nu},
\eq
where repeated indices are summed over and the {\bf field strength} is given by
\bq
F^{a}_{\mu\nu} & = & \partial_{\mu} A^{a}_{\nu} - \partial_{\nu} A^{a}_{\mu}
 + g f^{abc} A^{b}_{\mu} A^{c}_{\nu}.
\eq
The coupling of Yang-Mills theory is denoted by $g$.
The Lagrange density is invariant under the {\bf local transformations}
\bq
 T^a A^a_\mu(x)   & \rightarrow & 
  U(x) \left( T^a A^a_\mu(x) + \frac{i}{g} \partial_\mu \right) U^\dagger(x),
\eq
with
\bq
 U(x) = \exp\left(-i T^a \theta_a(x) \right).
\eq
For later purpose let us define the {\bf covariant derivative in the fundamental representation} by
\bq
 D_\mu & = & \partial_\mu - i g T^a A^a_\mu,
\eq
where we suppressed the indices $i$ and $j$ in $T^a_{ij}$ and the Kronecker symbol $\delta_{ij}$ accompanying $\partial_\mu$.
The {\bf covariant derivative in the adjoint representation} is given by
\bq
\label{def_covariant_derivative_adjoint}
 D^{ab}_\mu  & = & \delta^{ab} \partial_\mu - g f^{abc} A^c_\mu.
\eq

\subsection{Yang-Mills theory in terms of differential forms}

Let us now phrase Yang-Mills theory in the language of differential geometry.
Readers not familiar with differential forms, fibre bundles or the Hodge $\ast$-operator may either
skip this paragraph or look up the basic concepts and definitions in a textbook.
The books by Nakahara \cite{Nakahara} or Isham \cite{Isham:1999qu} 
can be recommended for this purpose.
The rest of this report will not depend on this sub-section.
However, since we are interested in mathematical structures associated to Yang-Mills theory,
the geometric fibre bundle description should not be omitted in this review.

We consider a {\bf principal bundle} $P(M,G)$ with a {\bf connection one-form} $\omega$.
The base space $M$ is the flat Minkowski space,
the fibre consists of the gauge group $G$.
A connection one-form $\omega$ is a Lie-algebra valued one-form on the total space $P$, satisfying two
requirements. 
We may identify a vector $Y$ in the Lie algebra $\mathfrak{g}$ of the Lie group $G$ 
with a tangent vector $Y$ at a point $p$ in the total space $P$.
With this isomorphism the first requirement reads
\bq
\label{connection_projection}
 \omega_p\left(Y\right) & = & Y
\eq
and states that the connection one-form $\omega$ is a projection on the vertical sub-space at $p$.
There is a right action of an element $g \in G$ on a point $p \in P$.
Within a local trivialisation a point $p$ in the total space is given by $p=(x,g')$ and the right action
by $g$ is given by $p g = (x, g' g)$. The right action by $g$ maps the point $p$ to another point $p g$
on the same fibre.
The second requirement for the connection one-form $\omega$ relates the evaluation of $\omega$ on tangent
vectors at different points $p$ and $p g$ on the same fibre:
\bq
\label{connection_adjoint}
 \omega_{p g}\left( R_{g \ast} Y \right)
 & = & 
 g^{-1} \omega_p\left(Y\right) g,
\eq
where $R_{g \ast} Y$ denotes the push-forward of the tangent vector $Y$ at the point $p$ to the point $p g$
by the right action of $g$.

Let $\sigma : M \rightarrow P$ be a (global) section and denote by
$A$ the pull-back of $\omega$ to $M$:
\bq
 A & = & \sigma^\ast \omega.
\eq
In this report we are not interested in topologically non-trivial configurations, therefore
a trivial fibre bundle and a global section are fine for us.
The concepts of this paragraph are easily extended to the general case by introducing an atlas of coordinate patches
and local sections glued together in the appropriate way.
We use the notation
\bq
A & = & A_\mu \; dx^\mu = \frac{g}{i} T^a A^a_\mu \; dx^\mu,
 \;\;\;\;\;\;
 A_\mu = \frac{g}{i} T^a A^a_\mu,
\eq
where $A^a_\mu$ is the gauge field already encountered in the previous paragraph.
Let us now consider two sections $\sigma_1$ and $\sigma_2$, with associated pull-backs $A_1$ and $A_2$.
We can always relate the two sections $\sigma_1$ and $\sigma_2$ by
\bq
\label{relation_sections}
 \sigma_1(x) & = & \sigma_2(x) U(x),
\eq
where $U(x)$ is a $x$-dependent element of the Lie group $G$.
Then we obtain for the pull-backs $A_1$ and $A_2$ of the connection one-form the relation
\bq
\label{relation_pull_backs}
 A_2 & = & U A_1 U^\dagger + U d U^\dagger.
\eq
This is nothing else than a gauge transformation.
\\
\\
\bs
{\it {\bf Exercise \theexercise}: 
Derive eq.~(\ref{relation_pull_backs}) from eq.~(\ref{relation_sections}).
\\
\\
Hint: Recall that the action of the pull-back $A_2$ on a tangent vector is defined as the 
action of the original form $\omega$ on the push-forward of the tangent vector.
Recall further that a tangent vector at a point $x$ can be given as a tangent vector to a curve
through $x$. It is sufficient to show that the actions of $A_2$ and $U A_1 U^\dagger + U d U^\dagger$
on an arbitrary tangent vector give the same result.
In order to prove the claim you will need in addition 
the defining relations for the connection one-form $\omega$,
given in eq.~(\ref{connection_projection}) and eq.~(\ref{connection_adjoint}).
\stepcounter{exercise}
}
\es
\\
\\
The connection one-form defines the {\bf covariant derivative}
\bq
 D_A & = & d + A
 =
 d + A_\mu dx^\mu 
 =
 d + \left( \frac{g}{i} \right) T^a A^a_\mu dx^\mu
 =
 d - i g T^a A^a_\mu dx^\mu.
\eq
If $A$ is a $\mathfrak{g}$-valued $p$-form and 
$B$ is a $\mathfrak{g}$-valued $q$-form, the commutator of the two is defined by
\bq
 \left[ A, B \right] & = & A \wedge B - (-1)^{p q} B \wedge A,
\eq
the factor $(-1)^{p q}$ takes into account that we have to permute $p$ differentials $dx^{\mu_i}$ from $A$ past $q$ differentials
$dx^{\nu_j}$ from $B$.
If $A$ and $B$ are both one-forms we have
\bq
 \left[ A, B \right] & = & A \wedge B + B \wedge A
 =
 \left[ A_\mu, B_\nu \right] dx^\mu \wedge dx^\nu.
\eq
In particular
\bq
 A \wedge A & = &
 \frac{1}{2} \left[ A, A \right] = 
 \frac{1}{2} \left[ A_\mu, A_\nu \right] dx^\mu \wedge dx^\nu.
\eq
We define the {\bf curvature two-form} of the fibre bundle by
\bq
 F & = & 
   D_A A = dA + A \wedge A
 =
  dA + \frac{1}{2} \left[ A, A \right].
\eq
This definition is in close analogy with the definition of the Riemannian curvature tensor.
Also the Riemannian curvature tensor can be calculated through the covariant derivative
of the affine connection.
We have
\bq
 d A 
 & = & d \left( A_\nu dx^\nu \right) = \partial_\mu A_\nu dx^\mu \wedge dx^\nu 
 =
 \frac{1}{2} \left( \partial_\mu A_\nu - \partial_\nu A_\mu \right) dx^\mu \wedge dx^\nu,
 \nonumber \\
 A \wedge A 
 & = & 
 A_\mu dx^\mu \wedge A_\nu dx^\nu = \frac{1}{2} \left( A_\mu A_\nu - A_\nu A_\mu \right) dx^\mu \wedge dx^\nu 
 =
 \frac{1}{2} \left[ A_\mu, A_\nu \right] dx^\mu \wedge dx^\nu,
\eq
and therefore
\bq
 F 
 & = & 
 \frac{1}{2} \left( \partial_\mu A_\nu - \partial_\nu A_\mu + \left[ A_\mu, A_\nu \right] \right) dx^\mu \wedge dx^\nu
 =
 \frac{1}{2} F_{\mu\nu} dx^\mu \wedge dx^\nu.
\eq
With the notation
\bq
 F_{\mu\nu} & = & \frac{g}{i} T^a F^a_{\mu\nu}
\eq
we have
\bq
F & = & 
  \frac{1}{2} F_{\mu\nu} dx^\mu \wedge dx^\nu
  =
  \frac{1}{2} \frac{g}{i} T^a F^a_{\mu\nu} dx^\mu \wedge dx^\nu,
\eq
where
\bq
 F^a_{\mu\nu}
 & = & 
\partial_\mu A^a_\nu - \partial_\nu A^a_\mu + g f^{abc} A^b_\mu A^c_\nu,
\eq
in agreement with the previous notation.
\\
\\
\bs
{\it {\bf Exercise \theexercise}: 
Under a gauge transformation the pull-back $A$ of the connection one-form transforms as
\bq
 A \rightarrow & U A U^\dagger + U d U^\dagger.
\eq
Show that the curvature two-form transforms as
\bq
 F & \rightarrow &
 U F U^\dagger.
\eq
\stepcounter{exercise}
}
\es
\\
\\
In this context we also introduce the {\bf dual field strength}
\bq
\label{def_dual_field_strength}
\tilde{F}_{\mu\nu} \; = \; \frac{1}{2} \eps_{\mu\nu\rho\sigma} F^{\rho\sigma},
 & &
F_{\mu\nu} \; = \; -\frac{1}{2} \eps_{\mu\nu\rho\sigma} \tilde{F}^{\rho\sigma},
\eq
or equivalently, the dual field strength two-form
\bq
 \ast F 
 & = &
 \ast \left( \frac{1}{2} F_{\mu\nu} dx^\mu \wedge dx^\nu \right) 
 =
 \frac{1}{4} \eps_{\mu\nu\rho\sigma} F^{\mu\nu} dx^\rho \wedge dx^\sigma 
 =
 \frac{1}{2} \tilde{F}_{\mu\nu} dx^\mu \wedge dx^\nu.
\eq
In Minkowski space the Hodge $\ast$-operator applied to a two-form satisfies
\bq
 \ast \ast F & = & - F,
\eq
therefore the eigenvalues are $\pm i$.
We call a two-form {\bf self dual}, respectively {\bf anti-self dual}, if
\bq
 \mbox{self dual} & : & F = i \ast F,
 \nonumber \\
 \mbox{anti-self dual} & : & F = -i \ast F.
\eq
Equivalently we have in the notation with indices
\bq
 \mbox{self dual} & : & F_{\mu\nu} = i \tilde{F}_{\mu\nu},
 \nonumber \\
 \mbox{anti-self dual} & : & F_{\mu\nu} = -i \tilde{F}_{\mu\nu}.
\eq
Then the action can be written as
\bq
S & = & - \frac{1}{4} \int d^4x \; F^a_{\mu\nu} F^{a\;\mu\nu}
 = 
 \frac{1}{2g^2} \int d^4x \; \mathrm{Tr} \; F_{\mu\nu} F^{\mu\nu}
 =
 \frac{1}{g^2} \int \; \mathrm{Tr} \; F \wedge \ast F.
\eq
With the Hodge inner product of two $p$-forms $\eta_1$ and $\eta_2$ 
\bq
 \left( \eta_1, \eta_2 \right) & = & \int \eta_1 \wedge \ast \eta_2
\eq
the action can be written as
\bq
S & = & \frac{1}{g^2} \; \mathrm{Tr} \; \left( F, F \right).
\eq
Let us summarise: The gauge potential corresponds to the connection of the fibre bundle, the field strength 
to the curvature of the fibre bundle.

\subsection{Gauge fixing}
\label{sect_gauge_fixing}

Consider the path integral
\bq
 \int {\mathcal D}A^a_\mu(x) \; \exp \left( i \int d^4x \; {\mathcal L}_{\mathrm{YM}} \right).
\eq
The path integral is over all possible gauge field configurations, including ones which are just related
by a gauge transformation.
We will encounter similar situations later on in this report and we will therefore discuss
this issue in more detail.
Gauge-equivalent configuration describe the same physics and it is sufficient to count them only ones.
Technically this is done as follows:
Let us denote a gauge transformation by
\bq
 U(x) = \exp\left(-i T^b \theta_b(x) \right).
\eq
The gauge transformation is therefore completely specified by the functions $\theta_b(x)$.
We denote by $A^a_\mu(x,\theta_b)$ the gauge field configuration obtained from $A^a_\mu(x)$ through the gauge
transformation $U(x)$:
\bq
 T^a A^a_\mu(x,\theta_b)
  & = & 
  U(x) \left( T^a A^a_\mu(x) + \frac{i}{g} \partial_\mu \right) U^\dagger(x),
\eq
$A^a_\mu(x,\theta_b)$ and $A^a_\mu(x)$ are therefore {\bf gauge-equivalent configurations}.
From all gauge-equivalent configurations we are going to pick the one, which satisfies for a given
$G^\mu$ and $B^a(x)$ the equation
\bq
G^\mu A^a_\mu(x,\theta_b) & = & B^a(x).
\eq
Let us assume that this equation gives a unique solution $\theta_b$ for a given $A^a_\mu$.
(This is not necessarily always fulfilled, cases where a unique solution may not exist are known as the Gribov
ambiguity \cite{Gribov:1977wm}.)
We insert the identity \cite{Faddeev:1967fc}
\bq
\label{Faddeev_Popov_identity}
 \int \prod\limits_b {\mathcal D} \theta_b(x) \;\; 
 \delta^n \left( G^\mu A^a_\mu(x,\theta_b(x)) - B^a(x) \right) \; \mbox{det}\;M_G & = & 1
\eq
where
\bq
\left( M_G(x,y) \right)_{ab} & = & \frac{\delta G^\mu A^a_\mu(x,\theta_c(x))}{\delta \theta_b(y)}
\eq
and $n$ is the number of generators of the Lie algebra $\mathfrak{g}$.
The functional derivative is defined by
\bq
 \frac{\delta}{\delta \theta_\beta(y)} F\left(\theta_\alpha(x) \right)
 & = &
 \lim\limits_{\eps \rightarrow 0} \frac{1}{\eps}
 \left[ F\left(\theta_\alpha(x) + \eps \delta_{\alpha\beta}\delta^4(x-y) \right) -F\left(\theta_\alpha(x) \right) \right].
\eq
\bs
{\it {\bf Exercise \theexercise}: 
Prove the equivalent of eq.~(\ref{Faddeev_Popov_identity}) in the finite dimensional case.
Let $\vec{\alpha}=(\alpha_1,...,\alpha_n)$ be a $n$-dimensional vector and let 
\bq
 g_i & = & g_i(\alpha_1,..,\alpha_n), \;\;\; i=1,...,n,
\eq
be $n$ functions of the $n$ variables $\alpha_j$. Show that
\bq
 \int \left( \prod\limits_{j=1}^n d\alpha_j \right)
  \left( \prod\limits_{i=1}^n \delta\left(g_i(\alpha_1,...,\alpha_n)\right) \right)
 \det\left( \frac{\partial g_i}{\partial \alpha_j} \right)
 & = & 1.
\eq
\stepcounter{exercise}
}
\es
\\
From the finite gauge transformation 
\bq
 \left( T^a A^a_\mu(x) \right)'
  & = & 
  U(x) \left( T^a A^a_\mu(x) + \frac{i}{g} \partial_\mu \right) U^\dagger(x),
\eq
we obtain the infinitesimal version
\bq
 \left.A^a_\mu\right.'
 & = &
 A^a_\mu - f^{abc} A^b_\mu \theta^c - \frac{1}{g} \partial_\mu \theta^a.
\eq
Therefore:
\bq
 \frac{\delta}{\delta \theta^b(y)} G^\mu \left.A_\mu^a\right.'
 & = & 
 G^\mu \left[ f^{abc} A^c_\mu \delta^4(x-y) - \frac{1}{g} \delta^{ab} \partial_\mu \delta^4(x-y) \right]
 \;\; = \;\;
 - \frac{1}{g} G^\mu D^{ab}_\mu \delta^4(x-y), \;\;
\eq
where $D^{ab}_\mu$ denotes the covariant derivative in the adjoint representation, defined in eq.~(\ref{def_covariant_derivative_adjoint}).
We now consider 
\bq
 Z[0] & = & \int {\mathcal D}A^a_\mu(x) \; \exp \left[ i \int d^4x \; {\mathcal L}_{\mathrm{YM}}(A^a_\mu(x)) \right]
\eq
and insert eq.~(\ref{Faddeev_Popov_identity}).
Using the gauge invariance of the action and of the measure ${\mathcal D}A^a_\mu(x)$ one arrives at
\bq
 Z[0] 
 & = &
 \left( \int \prod\limits_b {\mathcal D} \theta_b(x) \right)
 \\
 & & 
 \times 
 \int {\mathcal D}A^a_\mu(x) \;  \mbox{det}\;M_G(A^a_\mu(x))
 \delta^n \left( G^\mu A^a_\mu(x) - B^a(x) \right) 
 \exp \left[ i \int d^4x \; {\mathcal L}_{\mathrm{YM}}(A^a_\mu(x)) \right].
 \nonumber
\eq
The integral over all gauge-transformations
\bq
 \int \prod\limits_b {\mathcal D} \theta_b(x)
\eq
is just an irrelevant prefactor, which may be neglected.
We then obtain
\bq
 Z[0] & = &
 \int {\mathcal D}A^a_\mu(x) \;  \mbox{det}\;M_G(A^a_\mu(x))
 \delta^n \left( G^\mu A^a_\mu(x) - B^a(x) \right) 
 \exp \left[ i \int d^4x \; {\mathcal L}_{\mathrm{YM}}(A^a_\mu(x)) \right].
\eq
This functional still depends on $B^a(x)$. As we are not interested in any particular choice of
$B^a(x)$, 
we average over $B^a(x)$ with weight
\bq
\exp \left( - \frac{i}{2 \xi} \int d^4x \; B^a(x) B_a(x) \right)
\eq
and obtain
\bq
\lefteqn{
 \int {\mathcal D}B^a(x) \; Z[0] \exp \left( - \frac{i}{2 \xi} \int d^4x \; B^a(x) B_a(x) \right)
 = } & &
 \nonumber \\
 & = &
 \int {\mathcal D}A^a_\mu(x) \;  \mbox{det}\;M_G
 \exp \left[ i \int d^4x \; {\mathcal L}_{\mathrm{YM}}(A^a_\mu(x)) 
            - \frac{1}{2 \xi} \left( G^\mu A^a_\mu(x) \right) \left( G^\nu A_{\nu\;a}(x) \right) \right].
\eq
The determinant may be exponentiated with the help of Grassmann-valued ghost fields $\bar{c}^a$
and $c^b$.
Ignoring a factor $1/g$ one obtains:
\bq
\mbox{det}\;M_G & = & 
 \int {\mathcal D}c^b(x) \;  {\mathcal D} \bar{c}^a(x) \; 
 \exp \left( i \int d^4x \; \bar{c}^a(x) \left( - G^\mu D^{ab}_\mu \right) c^b(x) \right).
\eq
Up to now we did not specify the quantity $G^\mu$. Different choices for $G^\mu$ correspond
to different gauges.
A popular gauge is Lorenz gauge, corresponding to
\bq
 G^\mu & = & \partial^\mu.
\eq
In Lorenz gauge the {\bf gauge-fixing term} reads
\bq
 {\mathcal L}_{\mathrm{GF}}
 & = &
 - \frac{1}{2 \xi} \left( \partial^\mu A^a_\mu \right) \left( \partial^\nu A^a_{\nu} \right),
\eq
the {\bf Faddeev-Popov term} reads
\bq
 {\mathcal L}_{\mathrm{FP}}
 & = &
 - \bar{c}^a \partial^\mu D^{ab}_\mu c^b.
\eq
The Faddeev-Popov term contributes to loop amplitudes. In this report we are mainly concerned with tree
amplitudes. For tree amplitudes we will only need the gauge-fixing term.

\subsection{Feynman rules}

With the inclusion of the gauge-fixing term and the Faddeev-Popov term we now
have the effective Lagrange density
\bq
 {\mathcal L}
 & = &
 {\mathcal L}_{\mathrm{YM}}
 +
 {\mathcal L}_{\mathrm{GF}}
 +
 {\mathcal L}_{\mathrm{FP}}.
\eq
From the Lagrangian one obtains the Feynman rules.
This is explained in detail in many textbooks of quantum field theory and we present here only the
final cooking recipe.
We first order the terms in the Lagrangian according to the number of fields they involve.
From the terms bilinear in the fields one obtains the propagators, while the terms with three or more
fields give rise to vertices.
Note that a ``normal'' Lagrangian does not contain terms with just one or zero fields.
Furthermore we always assume within perturbation theory that all fields fall off rapidly enough at infinity.
Therefore we can use partial integration and ignore boundary terms.
For the gauge-fixed Yang-Mills Lagrangian one obtains terms with two, three and four gauge fields
plus the Faddeev-Popov term.
\bq
\label{Lagrangian_QCD_expanded}
{\mathcal L} & = & 
 \frac{1}{2} A^{a}_{\mu}(x) \left[ \partial_\rho \partial^\rho g^{\mu\nu} \delta^{ab}
                  - \left( 1 - \frac{1}{\xi} \right) \partial^\mu \partial^\nu \delta^{ab} \right] A^{b}_{\nu}(x)
 \nonumber \\
 & &
 - g f^{abc} \left( \partial_\mu A^{a}_{\nu}(x) \right) A^{b \mu}(x) A^{c \nu}(x)
 - \frac{1}{4} g^2 f^{eab} f^{ecd} A^{a}_{\mu}(x) A^{b}_{\nu}(x) A^{c \mu}(x) A^{d \nu}(x) 
 \nonumber \\
 & &
 + {\mathcal L}_{\mathrm{FP}}.
\eq
Let us first consider the terms bilinear in the fields, giving rise to the propagators.
A generic term for real boson fields $\phi_i$ has the form
\bq
{\mathcal L}_{\mathrm{bilinear}}(x) 
 & = &
 \frac{1}{2} \phi_i(x) P_{ij}(x) \phi_j(x),
\eq
where $P$ is a real symmetric operator that may contain derivatives and must have an inverse. 
Define the inverse of $P$ by
\bq
 \sum\limits_j P_{ij}(x) P_{jk}^{-1}(x-y) & = & \delta_{ik} \delta^4(x-y),
\eq
and its Fourier transform by
\bq
 P_{ij}^{-1}(x) & = & \int \frac{d^4 p}{(2 \pi)^4} e^{-i p \cdot x} \tilde{P}_{ij}^{-1}(p).
\eq
Then the {\bf propagator} is given by
\bq
 \Delta_F(p)_{ij} 
 & = & 
 i \tilde{P}_{ij}^{-1}(p).
\eq
Let's see how this works out for the gluon propagator:
The first line of eq.~(\ref{Lagrangian_QCD_expanded}) gives the terms bilinear in the gluon fields.
This defines an operator
\bq
 P^{\mu\nu\;ab}(x) & = & \partial_\rho \partial^\rho g^{\mu\nu} \delta^{ab}
                  - \left( 1 - \frac{1}{\xi} \right) \partial^\mu \partial^\nu \delta^{ab}.
\eq
For the propagator we are interested in the inverse of this operator
\bq
\label{gluon_prop_inverse_x_space}
 P^{\mu\sigma\;ac}(x) \left( P^{-1} \right)_{\sigma\nu}^{cb}(x-y) & = & g^\mu_{\;\;\nu} \delta^{ab} \delta^4(x-y).
\eq
Working in momentum space we are more specifically interested in the Fourier transform of the inverse
of this operator:
\bq
\label{gluon_prop_Fourier_trafo}
 \left( P^{-1} \right)_{\mu\nu}^{ab}(x) & = & 
  \int \frac{d^4 p}{(2 \pi)^4} e^{-i p \cdot x} \left( \tilde{P}^{-1} \right)_{\mu\nu}^{ab}(p).
\eq
The Feynman rule for the propagator is then given by $(\tilde{P}^{-1})_{\mu\nu}^{ab}(p)$ times the imaginary unit.
For the gluon propagator one finds the Feynman rule
\bq
\label{gluon_propagator}
 \begin{picture}(100,20)(0,5)
 \Gluon(20,10)(70,10){-5}{5}
 \Text(15,12)[r]{\footnotesize $\mu, a$}
 \Text(75,12)[l]{\footnotesize $\nu, b$}
\end{picture} 
 & = & 
  \frac{i}{p^2} \left( - g_{\mu\nu} + \left( 1 -\xi \right) \frac{p_\mu p_\nu}{p^2} \right) \delta^{ab}.
\eq
\bs
{\it {\bf Exercise \theexercise}: 
Derive eq.~(\ref{gluon_propagator}) from eq.~(\ref{gluon_prop_inverse_x_space}) 
and eq.~(\ref{gluon_prop_Fourier_trafo}).
\\
\\
Hint: It is simpler to work directly in momentum space, using the Fourier representation of
$\delta^4(x-y)$.
\stepcounter{exercise}
}
\es
\\
\\
Let us now consider a generic interaction term with $n\ge 3$ fields.
We may write this term as
\bq
 {\mathcal L}_{\mathrm{int}}(x) & = & 
 O_{i_1 ... i_n}\left(\partial_1,...,\partial_n \right)
 \phi_{i_1}(x) ... \phi_{i_n}(x),
\eq
with the notation that $\partial_j$ acts only on the $j$-th field $\phi_{i_j}(x)$.
For each field we have the Fourier transform
\bq
 \phi_i(x) = 
 \int \frac{d^Dp}{(2\pi)^D} \; e^{-i p x} \; \tilde{\phi}_i(p),
 & &
 \tilde{\phi}_i(p) =
 \int d^Dx \; e^{i p x} \; \phi_i(x),
\eq
where $p$ denotes an in-coming momentum.
We thus have
\bq
 {\mathcal L}_{\mathrm{int}}(x) & = & 
 \int
 \frac{d^Dp_1}{(2\pi)^D} ... \frac{d^Dp_n}{(2\pi)^D}
 e^{-i \left(p_1+...+p_n\right) x}
 O_{i_1 ... i_n}\left(-i p_1,...,-i p_n \right)
 \tilde{\phi}_i\left(p_1\right) ... \tilde{\phi}_i\left(p_n\right).
\eq
Changing to outgoing momenta we replace $p_j$ by $-p_j$.
The {\bf vertex} is then given by
\bq
 V & = & 
 i \sum\limits_{\mathrm{permutations}} 
   (-1)^{P_F} O_{i_1 ... i_n}\left(i p_1,...,i p_n \right),
\eq
where the momenta are taken to flow outward.
The summation is over all permutations of indices and momenta of identical particles. 
In the case of identical fermions there is in addition a minus sign for every odd permutation 
of the fermions, indicated by $(-1)^{P_F}$.

Let us also work out an example here.
We look as an example at the first term in the second line of
eq.~(\ref{Lagrangian_QCD_expanded}):
\bq
 {\mathcal L}_{ggg} & = &
 - g f^{abc} \left( \partial_\mu A^{a}_{\nu}(x) \right) A^{b \mu}(x) A^{c \nu}(x).
\eq
This term contains three gluon fields and will give rise to the three-gluon vertex.
We may rewrite this term as
\bq
 {\mathcal L}_{ggg} 
 & = &
 - g f^{abc} g^{\mu\rho}  \partial^\nu_1 \; A^a_\mu(x) A^b_\nu(x) A^c_\rho(x).
\eq
Thus
\bq
 O^{a b c, \mu \nu \rho}\left(\partial_1,\partial_2,\partial_3\right)
 \;\; = \;\;
 - g f^{abc} g^{\mu\rho}  \partial^\nu_1,
 & &
 O^{a b c, \mu \nu \rho}\left(i p_1, i p_2, i p_3 \right)
 \;\; = \;\;
 - g f^{abc} g^{\mu\rho}  i p^\nu_1.
 \;\;
\eq
The Feynman rule for the vertex is given by the sum over all permutations of identical particles of the
function $O^{a b c, \mu \nu \rho}(i p_1, i p_2, i p_3 )$ multiplied by the imaginary unit $i$. 
For the case at hand, we have three identical gluons and we have to sum over $3!=6$ permutations.
One finds
\bq
 V_{ggg} & = &
 i
 \sum\limits_{\mathrm{permutations}}
 \left( - g f^{abc} g^{\mu\rho}  i p_1^\nu \right)
  \nonumber \\
 & = &
 - g f^{abc}
 \left[
          g^{\mu\nu} \left( p_1^\rho - p_2^\rho \right)
        + g^{\nu\rho} \left( p_2^\mu - p_3^\mu \right)
        + g^{\rho\mu} \left( p_3^\nu - p_1^\nu \right)
 \right].
\eq
Note that we have momentum conservation at each vertex, for the three-gluon vertex this implies
\bq
 p_1 + p_2 + p_3 & = & 0.
\eq
In a similar way one obtains the Feynman rules for the four-gluon vertex and the ghost-antighost-gluon
vertex.

Let us summarise the Feynman rules for the propagators and the vertices of Yang-Mills theory:
The gluon propagator (in Feynman gauge, corresponding to $\xi=1$) 
and the ghost propagator are given by
\bq
 \begin{picture}(100,20)(0,5)
 \Gluon(20,10)(70,10){-5}{5}
 \Text(15,12)[r]{\footnotesize $\mu, a$}
 \Text(75,12)[l]{\footnotesize $\nu, b$}
\end{picture} 
& = &
 \frac{-ig^{\mu\nu}}{p^2} \delta^{ab},
 \nonumber \\
\begin{picture}(100,20)(0,5)
 \DashArrowLine(70,10)(20,10){3}
 \Text(15,10)[rb]{\footnotesize $a$}
 \Text(75,10)[lb]{\footnotesize $b$}
\end{picture} 
 & = &
 \frac{i}{p^2} \delta^{ab}.
\eq
The Feynman rules for the vertices are
\bq
\label{Feynman_rules_vertices}
\begin{picture}(100,35)(0,55)
\Vertex(50,50){2}
\Gluon(50,50)(50,80){3}{4}
\Gluon(50,50)(76,35){3}{4}
\Gluon(50,50)(24,35){3}{4}
\LongArrow(56,70)(56,80)
\LongArrow(67,47)(76,42)
\LongArrow(33,47)(24,42)
\Text(60,80)[lt]{$p_{1}$,$\mu$,$a$}
\Text(78,35)[lc]{$p_{2}$,$\nu$,$b$}
\Text(22,35)[rc]{$p_{3}$,$\rho$,$c$}
\end{picture}
 & = &
 g \left( i f^{abc} \right)
 i
 \left[
          g^{\mu\nu} \left( p_1^\rho - p_2^\rho \right)
        + g^{\nu\rho} \left( p_2^\mu - p_3^\mu \right)
        + g^{\rho\mu} \left( p_3^\nu - p_1^\nu \right)
   \right],
 \nonumber \\
\begin{picture}(100,75)(0,50)
\Vertex(50,50){2}
\Gluon(50,50)(71,71){3}{4}
\Gluon(50,50)(71,29){3}{4}
\Gluon(50,50)(29,29){3}{4}
\Gluon(50,50)(29,71){3}{4}
\Text(72,72)[lb]{$\mu$,$a$}
\Text(72,28)[lt]{$\nu$,$b$}
\Text(28,28)[rt]{$\rho$,$c$}
\Text(28,72)[rb]{$\sigma$,$d$}
\end{picture}
 & = &
   i g^2 \left[
                \left( i f^{abe} \right) \left( i f^{ecd} \right) \left( g^{\mu\rho} g^{\nu\sigma} - g^{\nu\rho} g^{\mu\sigma} \right)
 \right. \nonumber \\
 & & \left.
              + \left( i f^{bce} \right) \left( i f^{ead} \right) \left( g^{\nu\mu} g^{\rho\sigma} - g^{\rho\mu} g^{\nu\sigma} \right)
 \right. \nonumber \\
 & & \left.
              + \left( i f^{cae} \right) \left( i f^{ebd} \right) \left( g^{\rho\nu} g^{\mu\sigma} - g^{\mu\nu} g^{\rho\sigma} \right)
  \right],
 \nonumber
\eq
\bq
\begin{picture}(100,35)(0,55)
\Vertex(50,50){2}
\Gluon(50,50)(80,50){3}{4}
\DashArrowLine(50,50)(29,71){3}
\DashArrowLine(29,29)(50,50){3}
\LongArrow(36,59)(29,66)
\Text(28,71)[rb]{$p_1$,$a$}
\Text(82,50)[lc]{$\mu$,$b$}
\Text(28,29)[rt]{$c$}
\end{picture}
 & = &
 i g \left( i f^{abc} \right) p_1^{\mu}.
 \\
 \nonumber \\ 
 \nonumber
\eq
In order to compute tree-level amplitudes will still need a rule for external particles.
A gluon has two spin states, which we may take as the two helicity states $+$ and $-$.
These are described by {\bf polarisation vectors} $\eps^\pm_\mu(p)$, where $p$ is the momentum of the gluon.
The polarisation vectors satisfy
\bq
\label{tranverse_pol_vectors}
 \eps_{\mu}^{\pm}(p) \; p^\mu & = & 0,
\eq
and 
\bq
\label{orthogonality_pol_vectors}
 \eps^+ \cdot \left( \eps^+ \right)^\ast 
 =  
 \eps^- \cdot \left( \eps^- \right)^\ast 
 = -1,  
 & &
 \eps^+ \cdot \left( \eps^- \right)^\ast = 0.
\eq
The Feynman rule for an external gluon is simply to include a polarisation vector $\eps^\lambda_\mu(p)$.

These are all the Feynman rules we will need to compute tree-level amplitudes in Yang-Mills theory.
For completeness let me mention that there are a few additional rules, relevant to amplitudes
with loops and/or fermions. These additional rules are:
\begin{itemize}
\item There is an integration
\bq
 \int \frac{d^4p}{(2\pi)^4}
\eq
for each internal momentum not constrained by momentum conservation.
Such an integration is called a ``loop integration'' 
and the number of independent loop integrations in a diagram
is called the loop number of the diagram.

\item A factor $(-1)$ for each closed fermion loop.

\item Each diagram is multiplied by a factor $1/S$, where $S$ is the order of the permutation group
of the internal lines and vertices leaving the diagram unchanged when the external lines are fixed.
\end{itemize}

\subsection{Amplitudes and cross sections}

Scattering amplitudes with $n$ external particles may be calculated within perturbation theory, assuming that the
coupling $g$ is a small parameter.
We write
\bq
\label{basic_perturbative_expansion}
 {\mathcal A}_n & = & {\mathcal A}_n^{(0)} + {\mathcal A}_n^{(1)} + {\mathcal A}_n^{(2)} + {\mathcal A}_n^{(3)} + ...,
\eq
where ${\mathcal A}_n^{(l)}$ contains $(n-2+2l)$ factors of $g$.
In this expansion, ${\mathcal A}_n^{(l)}$ is an amplitude with $n$ external particles and $l$ loops.
The recipe for the computation of ${\mathcal A}_n^{(l)}$ based on Feynman diagrams
is as follows:
\begin{algorithm}
Calculation of scattering amplitudes from Feynman diagrams.
\begin{enumerate}
\item Draw all Feynman diagrams for the given number of external particles $n$ and the given number of loops $l$. 
\item Translate each graph into a mathematical formula with the help of the Feynman rules.
\item The amplitude ${\mathcal A}_n^{(l)}$ is then given as the sum of all these terms.
\end{enumerate}
\end{algorithm}
Tree-level amplitudes are amplitudes with no loops and are denoted by ${\mathcal A}_n^{(0)}$.
They give the leading contribution to the full amplitude.
The computation of tree-level amplitudes involves only basic mathematical operations:
Addition, multiplication, contraction of indices, etc..
The above algorithm allows therefore in principle for any $n$ the computation of the corresponding 
tree-level amplitude.
The situation is different for loop amplitudes ${\mathcal A}_n^{(l)}$ (with $l\ge 1$).
Here, the Feynman rules involve an integration over each internal momentum not constrained
by momentum conservation.
This constitutes an additional challenge, which we will not touch in this report.
The interested reader is referred to textbooks \cite{Smirnov:2006ry}
and review articles \cite{Weinzierl:2006qs,Weinzierl:2010ps,Duhr:2014woa,Henn:2014qga}.
Let me stress that the calculation of loop amplitudes is important for precision predictions
in particle physics.

Let us consider an example for the calculation of a tree-level amplitude.
To compute the tree-level four-gluon amplitude ${\mathcal A}_4^{(0)}$ 
\begin{figure}
\begin{center}
\includegraphics[scale=0.8]{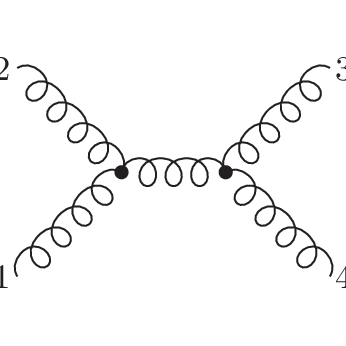}
\hspace*{8mm}
\includegraphics[scale=0.8]{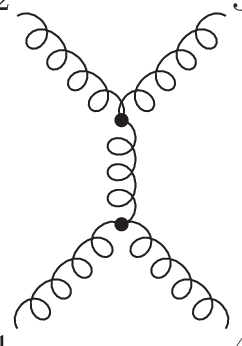}
\hspace*{8mm}
\includegraphics[scale=0.8]{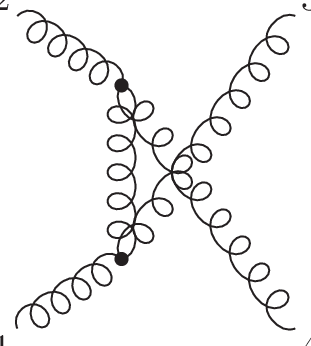}
\hspace*{8mm}
\includegraphics[scale=0.8]{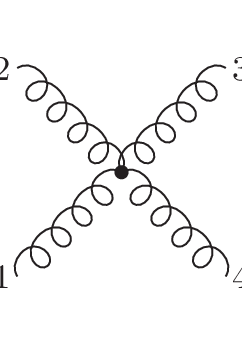}
\caption{\label{fig1} The four Feynman diagrams contributing to the tree-level four-gluon amplitude ${\mathcal A}_4^{(0)}$.}
\end{center}
\end{figure}
we have to evaluate the four Feynman diagrams shown in fig.~(\ref{fig1}).
\\
\\
\bs
{\it {\bf Exercise \theexercise}: 
Compute the amplitude ${\mathcal A}_4^{(0)}$ from the four diagrams shown in fig.~(\ref{fig1}).
Assume that all momenta are outgoing.
The result will involve scalar products $2 p_i \cdot p_j$, $2 p_i \cdot \eps_j$ and $2 \eps_i \cdot \eps_j$.
For a $2 \rightarrow 2$ process (more precisely for a $0 \rightarrow 4$ process, since we take all momenta to be outgoing), the Mandelstam
variables are defined by
\bq
 s \;\; = \;\; \left( p_1 + p_2 \right)^2,
 \;\;\;
 t \;\; = \;\; \left( p_2 + p_3 \right)^2,
 \;\;\;
 u \;\; = \;\; \left( p_1 + p_3 \right)^2.
\eq
The four momenta $p_1$, $p_2$, $p_3$ and $p_4$ are on-shell, $p_i^2=0$ for $i=1,...,4$, and satisfy
momentum conservation. Derive the Mandelstam relation
\bq 
 s + t + u & = & 0.
\eq
This relation allows to eliminate in the result for ${\mathcal A}_4^{(0)}$ one variable, say $u$.
Furthermore we have from eq.~(\ref{tranverse_pol_vectors}) the relation $2 p_i \cdot \eps_i = 0$.
Combined with momentum conservation we may eliminate several scalar products
$2 p_i \cdot \eps_j$, such that for a given $j$ we only have 
$2 p_{j-1} \cdot \eps_j$ and $2 p_{j+1} \cdot \eps_j$, where the indices $(j-1)$ and $(j+1)$ are understood modulo $4$.
You might want to use a computer algebra system to carry out the calculations.
The open-source computer algebra systems FORM \cite{Vermaseren:2000nd} and GiNaC \cite{Bauer:2000cp}
have their roots in particle physics and were originally invented for calculations of this type.
\stepcounter{exercise}
}
\es
\\
\\
The scattering amplitude enters directly the calculation of a physical observable.
Let us first consider the scattering process of two incoming elementary uncoloured spinless
particles with four-momenta $p_a'$ and $p_b'$ 
and $(n-2)$ outgoing particles with four-momenta $p_1$ to $p_{n-2}$.
Let us further assume that we are interested in an observable $O\left(p_1,...,p_{n-2}\right)$ 
which depends on the momenta of the outgoing particles.
In general the observable depends on the experimental set-up 
and 
can be an arbitrary complicated function of the four-momenta. 
In the simplest case this function is just a constant equal to one,
corresponding to the situation where we count every event with $(n-2)$ particles in the final state.
In more realistic situations one takes for example 
into account that it is not possible to detect particles close to the beam pipe. 
The function $O$ would then be zero in these regions of phase space.
Furthermore any experiment has a finite resolution.
Therefore it will not be possible to detect particles which are very soft
or which are very close in angle to other particles.
We will therefore sum over the number of final state particles.
In order to obtain finite results within perturbation theory we have to require that 
in the case where one or more particles become unresolved the value of the observable $O$
has a continuous limit agreeing with the value of the observable for a configuration
where the unresolved particles have been merged into ``hard'' (or resolved) pseudo-particles.
Observables having this property are called {\bf infrared-safe observables}. 
The expectation value for the observable $O$ is given by
\bq
\label{observable_master_electron_positron}
\langle O \rangle
 & = &
 \frac{1}{2 (p_a'+p_b')^2}
 \sum\limits_n
             \int d\phi_{n-2}
             O\left(p_1,...,p_{n-2}\right)
             \left| {\mathcal A}_n \right|^2,
\eq
where $1/2/(p_a'+p_b')^2$ is a normalisation factor taking into account the incoming flux.
The phase space measure is given by
\bq
\label{def_phasespacemeasure}
d\phi_n & = &
 \frac{1}{n! }
 \prod\limits_{i=1}^n \frac{d^{3}p_i}{(2 \pi)^{3} 2 E_i} 
 \left(2 \pi \right)^4 \delta^4\left(p_a'+p_b'-\sum\limits_{i=1}^n p_i\right).
\;\;\;\;\;\;\;
\eq
The quantity
$E_i$ is the energy of particle $i$, for massless particles we have
\bq
 E_i 
 \;\; = \;\;
 \sqrt{\vec{p}_i^2}
 \;\; = \;\;
 p_i^0.
\eq
We see that the expectation value of $O$ is given by the phase space integral over the observable, weighted
by the norm squared of the scattering amplitude.
As the integrand can be a rather complicated function, the phase space integral is usually performed numerically by
Monte Carlo integration.

Let us now look towards a more realistic theory relevant to LHC physics.
As an example consider QCD consisting of quarks and gluons. 
Quarks and gluons are collectively called partons.
There are a few modifications to eq.~(\ref{observable_master_electron_positron}). The master formula reads now
\bq
\label{observable_master_hadron_hadron}
\lefteqn{
\langle O \rangle = 
 \sum\limits_{f_a,f_b}
 \int dx_a f_{f_a}(x_a) \int dx_b f_{f_b}(x_b) 
} & & 
 \\
 & & 
             \frac{1}{2 \hat{s} n_s(a) n_s(b) n_c(a) n_c(b)}
 \sum\limits_n
             \int d\phi_{n-2}
             O\left(p_1,...,p_{n-2}\right)
             \sum\limits_{\mathrm{spins,colour}} 
             \left| {\mathcal A}_n \right|^2.
 \nonumber
\eq
The partons have internal degrees of freedom, given by the spin and the colour of the partons.
In squaring the amplitude we sum over these degrees of freedom. 
For the particles in the initial state we would like to average over these degrees of freedom.
This is done by dividing by the factors $n_s(i)$ and $n_c(i)$, giving the number of spin degrees of
freedom ($2$ for quarks and gluons)
and the number of colour degrees of freedom ($3$ for quarks, $8$ for gluons).
The second modification is due to the fact that the particles brought into collision are not partons, but 
composite particles like protons. At high energies the elementary 
constituents of the protons interact and we have to include
a function $f_{f_a}(x_a)$ giving us the probability of finding a parton of flavour $f_a$ 
with momentum fraction $x_a$ of the original
proton momentum inside the proton.
If the momenta of the incoming protons are $P_a'$ and $P_b'$, then the momenta
of the two incoming partons are given by
\bq
 p_a' 
 \;\; = \;\;
 x_a P_a',
 & &
 p_b' 
 \;\; = \;\;
 x_b P_b'.
\eq
$\hat{s}$ is the centre-of-mass energy squared of the two partons entering the hard interaction.
Neglecting particle masses we have
\bq
 \hat{s}
 \;\; = \;\;
 \left(p_a' + p_b'\right)^2
 \;\; = \;\;
 x_a x_b \left(P_a' + P_b'\right)^2.
\eq
In addition there is a small change in eq.~(\ref{def_phasespacemeasure}). The quantity $(n!)$ is replaced by
$(\prod n_j!)$, where $n_j$ is the number of times a parton of type $j$ occurs in the final state.

It is very convenient to calculate the amplitude with the convention that all particles are out-going.
To this aim we set
\bq
 p_{n-1} \;\; = \;\; - p_a',
 & &
 p_{n} \;\; = \;\; - p_b'
\eq
and calculate the amplitude for the momentum configuration
\bq
 \left\{ p_1, ..., p_{n-2}, p_{n-1}, p_{n} \right\}.
\eq
Momentum conservation reads
\bq
 p_1 + ... + p_{n-2} + p_{n-1} + p_{n} & = & 0.
\eq
Note that the momenta $p_{n-1}$ and $p_{n}$ have negative energy components.

In eq.~(\ref{observable_master_hadron_hadron}) the square of the amplitude, summed over spins
and colours, enters.
In order to perform the spin sum, we may use for each gluon the formula
\bq
\label{spin_sum_gluon}
\sum\limits_{\lambda \in \{+,-\} } \left( \eps_\mu^\lambda(p) \right)^\ast \; \eps_\nu^\lambda(p)
 & = & 
 - g_{\mu\nu} + \frac{p_\mu n_\nu + n_\mu p_\nu}{p \cdot n} - n^2 \frac{p_\mu p_\nu}{(p \cdot n)^2}.
\eq 
Here $n^\mu$ is an arbitrary four vector. The dependence on $n^\mu$ cancels in gauge-invariant
quantities. 
Efficient methods for the colour summation are discussed in section~(\ref{sect:colour}).

At LHC energies the incoming flux of gluons is rather high and the tree-level gluon amplitudes
discussed in this review will give a significant contribution to the jet events observed 
at the LHC experiments.

\subsection{Singular limits of scattering amplitudes}

Let us take a small digression and discuss soft and collinear limits
of tree-level gluon amplitudes.
The scattering amplitudes show a universal factorisation behaviour in these limits.
We will discuss single unresolved limits, corresponding to either one particle becoming soft or
two particles becoming collinear.
In this limit, the amplitude ${\mathcal A}_{n+1}^{(0)}$ will not contribute to an $(n-1)$-particle observable,
but to an $(n-2)$-particle observable.
This gives a next-to-leading order contribution to an $(n-2)$-particle observable.
Double unresolved limits and unresolved limits with even more unresolved partons contribute
to higher orders in perturbation theory, as well as unresolved limits of loop amplitudes.

To make contact with the literature \cite{Bassetto:1984ik,Catani:1997vz}
let us first introduce the colour charge operators ${\bf T}_j$.
The $n$-gluon amplitude ${\mathcal A}_{n}$ has $n$ colour indices $a_j$, one for each gluon $j$.
To make this explicit, let us write the amplitude temporarily as ${\mathcal A}_{n}^{a_1,...,a_n}$.
The colour charge operator ${\bf T}_j$ acts on the gluon $j$ with colour index $a_j$ as
\bq
 \left( {\bf T}_j \right)^{b_1 a_j b_2} \; {\mathcal A}_n^{a_1,...,a_j,...,a_n} 
 & = & 
 \left( i f^{b_1 a_j b_2} \right) \; {\mathcal A}_n^{a_1,...,a_j,...,a_n}.
\eq
In the sequel we will drop again the colour indices and simply write
\bq
 {\bf T}_j {\mathcal A}_n
 & = & 
 \left( {\bf T}_j \right)^{b_1 a_j b_2} \; {\mathcal A}_n^{a_1,...,a_j,...,a_n}.
\eq
Colour conservation implies the relation
\bq
 \sum\limits_{j=1}^n 
 {\bf T}_j{\mathcal A}_n
 & = & 
 0. 
\eq
Let us fist consider the {\bf soft limit}.
We consider the case where particle $j$ becomes soft.
In the soft limit we parametrise the momentum of the soft parton $p_j$ as
\bq
 p_j & = & \lambda q
\eq
and consider contributions to $|{\mathcal A}_{n+1}^{(0)}|^2$ of the order ${\mathcal O}(\lambda^{-2})$.
Contributions to $|{\mathcal A}_{n+1}^{(0)}|^2$ which are less singular than $\lambda^{-2}$ are integrable in the soft limit.
In the soft limit a Born amplitude ${\mathcal A}_{n+1}^{(0)}$ with $(n+1)$ partons behaves as
\bq
\label{soft_limit}
 \lim\limits_{p_j \rightarrow 0} {\mathcal A}_{n+1}^{(0)}
 & = & 
 g \eps_\mu(p_j) {\bf J}^\mu
 {\mathcal A}_{n}^{(0)}.
\eq
The {\bf eikonal current} is given by
\bq
 {\bf J}^\mu & = & \sum\limits_{i \neq j} {\bf T}_i \frac{p_i^\mu}{p_i \cdot p_j}.
\eq
The sum is over the remaining $n$ hard momenta $p_i$.

Let us now consider the {\bf collinear limit}. 
We consider a splitting $\tilde{i} \rightarrow i + j$, where all particles are in the final state.
(Formul{\ae} for a collinear splitting involving initial-state particles can be worked out in a similar
way.) 
In the collinear limit we parametrise the momenta of the two collinear final-state partons $i$ and $j$
as
\bq
\label{parametrisation_collinear}
 p_i & = & z p_{\tilde{i}} + k_\perp - \frac{k_\perp^2}{z} \frac{n}{2 p_{\tilde{i}} \cdot n},
 \nonumber \\
 p_j & = & (1-z)  p_{\tilde{i}} - k_\perp - \frac{k_\perp^2}{1-z} \frac{n}{2 p_{\tilde{i}} \cdot n }.
\eq
Here $n$ is a massless four-vector and the transverse component $k_\perp$ satisfies
$2p_{\tilde{i}} \cdot k_\perp = 2n \cdot k_\perp =0$.
The four-vectors $p_{\tilde{i}}$, $p_i$ and $p_j$ are on-shell:
\bq
 p_{\tilde{i}}^2 
 \;\; = \;\;
 p_i^2 
 \;\; = \;\;
 p_j^2 
 \;\; = \;\;
 0.
\eq
In the collinear limit we have to consider contributions to $|{\mathcal A}_{n+1}^{(0)}|^2$ 
of the order ${\mathcal O}(k_\perp^{-2})$.
In this limit the Born amplitude factorises according to
\bq
\label{collinear_limit}
\lefteqn{
 \lim\limits_{p_i || p_j}
 {\mathcal A}_{n+1}^{(0)}\left(...,p_i,\lambda_i,...,p_j,\lambda_j,...\right) 
 = 
} & &
 \nonumber \\
 & & 
 g 
 \sum\limits_{\lambda_{\tilde{i}}} 
 \; 
 \mbox{Split}(p_{\tilde{i}},p_i,p_j,\lambda_{\tilde{i}},\lambda_i,\lambda_j) 
 \; 
 {\bf T}_{\tilde{i}} 
 \; 
 {\mathcal A}_{n}^{(0)}\left(...,p_{\tilde{i}},\lambda_{\tilde{i}},...\right),
\eq
where the sum is over all polarisations of the intermediate particle.
The variables $\lambda_i$ and $\lambda_j$ denote the polarisations of the particles $i$ and $j$, respectively.
The {\bf splitting function} $\mbox{Split}$ is given by
\bq
\label{def_split}
 \mbox{Split}\left(p_{\tilde{i}},p_i,p_j,\lambda_{\tilde{i}},\lambda_i,\lambda_j\right) 
 & = &
 \frac{2}{2 p_i \cdot p_j} \left[
    \eps^{\lambda_i}(p_i) \cdot \eps^{\lambda_j}(p_j) \; p_i \cdot \left. \eps^{\lambda_{\tilde{i}}}(p_{\tilde{i}}) \right.^\ast
 \right. \\
 & & \left.
  + \eps^{\lambda_j}(p_j) \cdot \left. \eps^{\lambda_{\tilde{i}}}(p_{\tilde{i}}) \right.^\ast \; p_j \cdot \eps^{\lambda_i}(p_i)
  - \eps^{\lambda_i}(p_i) \cdot \left. \eps^{\lambda_{\tilde{i}}}(p_{\tilde{i}}) \right.^\ast \; p_i \cdot \eps^{\lambda_j}(p_j)
 \right]. 
 \nonumber 
\eq
The factorisation formul{\ae} for the soft limit and the collinear limit in eq.~(\ref{soft_limit}) 
and eq.~(\ref{collinear_limit}) are very important for higher-order calculations.
The phase space integration over these infrared regions diverges.
It is therefore necessary to regulate this integration first.
This is usually done by dimensional regularisation \cite{'tHooft:1972fi,Bollini:1972ui,Cicuta:1972jf}.
Within dimensional regularisation one replaces four-dimensional space-time with a $D$-dimensional space-time.
Phase-space integrations and loop integrations are performed consistently in $D$ dimensions.
Setting $D=4-2\eps$ one is interested in the behaviour of the result of these integrations
as $D$ approaches $4$ (or equivalently as $\eps$ approaches zero).
Within dimensional regularisation the original divergences 
will show up as poles in $1/\eps$.
Infrared divergences from the integration over the unresolved phase space
cancel with infrared divergences from the loop integrations for infrared-safe observables.
This is the content of the Kinoshita-Lee-Nauenberg theorem \cite{Kinoshita:1962ur,Lee:1964is}.
Ultraviolet divergences from the loop integrations are removed by renormalisation.

In this report we will not discuss higher-order calculations in perturbation theory.
We included the discussion of soft and collinear limits for another reason:
Suppose we have a new way or a new conjecture for the computation of the tree-level Yang-Mills amplitudes.
If this conjecture is true, it has to give the same result as a calculation based on Feynman diagrams 
(or any other known method for the computation of these amplitudes).
In particular, it has to satisfy the soft and collinear limits 
in eq.~(\ref{soft_limit}) and eq.~(\ref{collinear_limit}).
Checking the correct soft and collinear limits is often simpler than a full proof that a new method
computes the amplitude correctly.
This allows us either to eliminate false conjectures quickly
or gives us further evidence, that a conjecture might be correct.
Of course, the correctness of the soft and collinear limits alone does not constitute a complete proof 
that two methods for the computation of an amplitude agree.

In this context it is also useful to consider the following limit: 
Consider an amplitude ${\mathcal A}_{n}^{(0)}$ with $n$ external particles $(1,2,...,n)$.
Let $I_m = \{i_1,i_2,...,i_m\} \subset \{1,2,...,n\}$ be a subset with $3 \le m \le n-3$.
Denote by $J_{n-m}=\{j_1,...,j_{n-m}\}$ the remaining $(n-m)$ indices not belonging to 
the index set $I_m$.
Let us denote
\bq
 p & = & p_{i_1} + p_{i_2} + ... + p_{i_m},
\eq
and consider the limit $p^2 \rightarrow 0$.
This limit corresponds to the case where an internal propagator of the amplitude goes on-shell.
In this limit the amplitude factorises as
\bq
\label{factorisation_pole}
\lefteqn{
 \lim\limits_{p^2 \rightarrow 0}
 {\mathcal A}_{n}^{(0)}\left(p_1,\lambda_1,a_1,...,p_n,\lambda_n,a_n\right)
 = }
 & & \\
 & &
 \sum\limits_{\lambda}
 {\mathcal A}_{m+1}^{(0)}\left(p_{i_1},\lambda_{i_1},a_{i_1},...,p_{i_n},\lambda_{i_n},a_{i_n},-p,-\lambda,a\right)
 \frac{i \delta^{ab}}{p^2}
 \nonumber \\
 & &
 \times
 {\mathcal A}_{n-m+1}^{(0)}\left(p,\lambda,b,p_{j_1},\lambda_{j_1},a_{j_1},...,p_{j_{n-m}},\lambda_{j_{n-m}},a_{j_{n-m}}\right).
 \nonumber 
\eq
We may view the collinear limit as a special case of eq.~(\ref{factorisation_pole}), 
corresponding to $m=2$ or $m=n-2$,
where the appearance of the three-particle amplitude ${\mathcal A}_{3}^{(0)}$
softens the $1/p^2$-behaviour.

% ----------------------------------------------------------------------------------
\newpage
\section{Efficiency improvements}
\label{sect:efficiency}

In the previous section we presented a method to compute tree-level Yang-Mills amplitudes based on 
Feynman diagrams.
Note that for tree-level amplitudes this algorithm involves only algebraic operations
(contraction of indices, multiplication, summation, ...).
Therefore Feynman diagrams allow us to compute for any $n$ the corresponding tree-level Yang-Mills amplitude
${\mathcal A}_n^{(0)}$ within a finite number of steps.
We might be tempted to call the computation of tree-level Yang-Mills amplitudes a solved problem.
However, the situation is not as simple.
Already for moderate values of $n$, the algorithm based on Feynman diagrams is highly inefficient
and will produce large intermediate expressions.
This can already be inferred from the rapid growth of the
number of Feynman diagrams contributing to the amplitude ${\mathcal A}_n^{(0)}$,
shown for the first few values of $n$ in table~(\ref{table1}).
Thus it is practically impossible to compute with the help of Feynman diagrams
the tree-level amplitude ${\mathcal A}_{1001}^{(0)}$ for $1001$ gluons.

The deficiency of the algorithm based on Feynman diagrams is analogue 
to the deficiency of the simplest algorithm to test 
if an integer $N$ is prime. Consider the following algorithm:
\begin{itemize}
\item For $2 \le j \le \sqrt{N}$ check if $j$ divides $N$.
\item If such a $j$ is found, $N$ is not prime.
\item Otherwise $N$ is prime.
\end{itemize}
This algorithm works in principle, but for large $N$ it does not work in practice.

Now scattering amplitudes with $1001$ gluons are not really needed in particle physics phenomenology, 
but amplitudes with -- say -- up to roughly $10$ gluons are useful to describe physics related to 
the LHC experiments.
A driving force for the study of scattering amplitudes has been the quest to find efficient methods 
for the computation of these amplitudes.
We will now discuss colour decomposition, the spinor-helicity method
and off-shell recurrence relations.
These efficiency improvements offer the opportunity to discuss some formal topics:
The discussion of colour decomposition is followed by an exposition of colour-kinematics duality,
the introduction of the spinor-helicity method by a discussion of twistors.

\subsection{Colour decomposition}
\label{sect:colour}

Let us consider colour first.
The most important examples for the Lie group $G$ of Yang-Mills theory
are the special unitary groups $\mathrm{SU}(N)$ and the unitary groups $\mathrm{U}(N)$.
The unitary group $\mathrm{U}(N)$ consists of all complex $N \times N$-matrices $U$
with
\bq
 U U^\dagger & = & {\bf 1},
\eq
where ${\bf 1}$ denotes the unit matrix.
The unitary group $\mathrm{U}(N)$ is parametrised by $N^2$ real variables.
\begin{table}
\begin{center}
\bq
\begin{array}{l|rrrrrrr}
 n                    & 4 & 5  & 6   & 7    & 8     & 9      & 10 \\
 \hline
\mbox{diagrams}      & 4 & 25 & 220 & 2485 & 34300 & 559405 & 10525900 \\
\end{array}
\eq
\caption{\label{table1} The number of diagrams contributing to the amplitude ${\mathcal A}_n^{(0)}$.}
\end{center}
\end{table}
This is also the number of generators of the associated Lie algebra.

The special unitary group $\mathrm{SU}(N)$ consists of all complex $N \times N$-matrices $U$
with
\bq
 U U^\dagger & = & {\bf 1}
 \;\;\; \mbox{and} \;\;\;
 \det \, U \;\; = \;\; 1.
\eq
The special unitary group $\mathrm{SU}(N)$ is parametrised by $N^2-1$ real variables
and has as many generators.

In both cases we denote the generators by $T^a$ with $a\in\{1,...,N^2-1\}$ for $\mathrm{SU}(N)$
and $a \in \{1,...,N\}$ for $\mathrm{U}(N)$.
We take the generators to be hermitian matrices.
We recall from eq.~(\ref{convention_Lie_algebra}) our conventions for the generators:
\bq
 \left[ T^a, T^b \right] = i f^{abc} T^c,
 & &
 \mathrm{Tr}\left( T^a T^b \right) = \frac{1}{2} \delta^{ab}.
\eq
A useful formula is the Fierz identity.
For $\mathrm{SU}(N)$ the Fierz identity reads
\bq
\label{Fierz_SU_N}
 T^a_{ij} T^a_{kl} & = &  \frac{1}{2} \left( \delta_{il} \delta_{jk}
                         - \frac{1}{N} \delta_{ij} \delta_{kl} \right),
\eq
where a sum over $a$ is understood.
The proof is rather simple:
The matrices $T^a$ with $a=1,...,N^2-1$
and the unit matrix form a basis of the $N \times N$ hermitian matrices, therefore
any hermitian matrix $A$ can be written as
\bq
 A & = & c_0 {\bf 1} + c_a T^a.
\eq
The constants $c_0$ and $c_a$ are determined using the normalisation condition and the fact that the
$T^a$ are traceless:
\bq
 c_0 \;\; = \;\; \frac{1}{N} \mathrm{Tr}\left(A\right),
 & &
 c_a \;\; = \;\; 2 \mathrm{Tr}\left(T^a A\right).
\eq
Therefore
\bq
A_{lk} \left( 2 T^a_{ij} T^a_{kl} + \frac{1}{N} \delta_{ij} \delta_{kl} - \delta_{il} \delta_{jk} \right) 
 & = & 0.
\eq
This has to hold for an arbitrary hermitian matrix $A$, therefore the Fierz identity follows.

Let $X$ and $Y$ be arbitrary matrix products of the generators $T^a$.
An immediate corollary of the Fierz identity are the following formul{\ae} involving traces:
\bq
\label{traces_SU_N}
 \mathrm{Tr}\left( T^a X \right) \mathrm{Tr}\left( T^a Y \right)
 & = & 
 \frac{1}{2} 
 \left[ \mathrm{Tr}\left( X Y \right) - \frac{1}{N} \mathrm{Tr}\left( X \right) \mathrm{Tr}\left( Y \right)
 \right],
 \nonumber \\
 \mathrm{Tr}\left( T^a X T^a Y \right)
 & = & 
 \frac{1}{2} 
 \left[ \mathrm{Tr}\left( X \right) \mathrm{Tr}\left( Y \right) - \frac{1}{N} \mathrm{Tr}\left( X Y \right) 
 \right].
\eq
For $\mathrm{U}(N)$ the Fierz identity reads
\bq
\label{Fierz_U_N}
 T^a_{ij} T^a_{kl} & = &  \frac{1}{2} \delta_{il} \delta_{jk},
\eq
and the analogue of eq.~(\ref{traces_SU_N}) reads
\bq
\label{traces_U_N}
 \mathrm{Tr}\left( T^a X \right) \mathrm{Tr}\left( T^a Y \right)
 & = & 
 \frac{1}{2} 
 \mathrm{Tr}\left( X Y \right),
 \nonumber \\
 \mathrm{Tr}\left( T^a X T^a Y \right)
 & = & 
 \frac{1}{2} 
 \mathrm{Tr}\left( X \right) \mathrm{Tr}\left( Y \right).
\eq
\bs
{\it {\bf Exercise \theexercise}: 
Compute the traces $\mathrm{Tr}\left( T^a T^b T^b T^a \right)$ and $\mathrm{Tr}\left( T^a T^b T^a T^b \right)$
for $\mathrm{U}(N)$ and $\mathrm{SU}(N)$.
\stepcounter{exercise}
}
\es
\\
\\
There is another relation, which is quite useful.
From 
\bq
[ T^a, T^b ] & = & i f^{abc} T^c
\eq
one derives by multiplying with $T^d$ and taking the trace:
\bq
\label{adjoint_to_fundamental}
 i f^{abc} & = & 2 \left[ \mathrm{Tr}\left(T^a T^b T^c\right) - \mathrm{Tr}\left(T^b T^a T^c\right) \right] 
\eq
This yields an expression of the structure constants $f^{abc}$ 
in terms of the matrices of the fundamental representation.
Eq.~(\ref{adjoint_to_fundamental})
together with eq.~(\ref{traces_SU_N}) (for an $\mathrm{SU}(N)$ gauge theory)
or with eq.~(\ref{traces_U_N}) (for an $\mathrm{U}(N)$ gauge theory)
can be used to convert any colour factors into traces of matrices in the fundamental representation
of the gauge group.
Let us consider the following diagram:
\begin{center}
\begin{picture}(100,100)(0,0)
\Vertex(35,50){2}
\Vertex(65,50){2}
\Gluon(5,20)(35,50){4}{4}
\Gluon(5,80)(35,50){4}{4}
\Gluon(35,50)(65,50){4}{3}
\Gluon(65,50)(95,20){4}{4}
\Gluon(65,50)(95,80){4}{4}
\Text(3,20)[r]{$1$}
\Text(3,80)[r]{$2$}
\Text(97,80)[l]{$3$}
\Text(97,20)[l]{$4$}
\end{picture}
\end{center}
The colour part of this diagram is
\bq
 i f^{a_1 a_2 b} i f^{b a_3 a_4}
 & = &
 4 
 \left[ \mathrm{Tr}\left(T^{a_1} T^{a_2} T^b\right) - \mathrm{Tr}\left(T^{a_2} T^{a_1} T^b\right) \right]
 \left[ \mathrm{Tr}\left(T^{a_3} T^{a_4} T^b\right) - \mathrm{Tr}\left(T^{a_4} T^{a_3} T^b\right) \right]
 \nonumber \\
 & = &
   2 \; \mathrm{Tr}\left(T^{a_1} T^{a_2} T^{a_3} T^{a_4}\right)
 - 2 \; \mathrm{Tr}\left(T^{a_1} T^{a_2} T^{a_4} T^{a_3}\right)
 - 2 \; \mathrm{Tr}\left(T^{a_2} T^{a_1} T^{a_3} T^{a_4}\right)
 \nonumber \\
 & &
 + 2 \; \mathrm{Tr}\left(T^{a_2} T^{a_1} T^{a_4} T^{a_3}\right).
\eq
Note in this calculation we obtain the same result, independently if we use
eq.~(\ref{traces_SU_N}) or eq.~(\ref{traces_U_N}).
The terms proportional to $-1/N$ drop out. This is a general feature of {\bf pure} Yang-Mills theories.
The $-1/N$-terms in the Fierz identity of $\mathrm{SU}(N)$ subtract out the trace part of $\mathrm{U}(N)$,
generated by the unit matrix.
One says that the $-1/N$-terms correspond to the exchange of an $\mathrm{\bf U}{\bf (1)}${\bf-gluon}.
The unit matrix commutes with any other matrix, therefore $f^{abc}=0$, whenever one index refers
to the $\mathrm{U}(1)$-gluon and the $\mathrm{U}(1)$-gluon does not couple to any other gauge boson.
However, the $\mathrm{U}(1)$-gluon couples to quarks. In theories with fermions in the fundamental representation
of the gauge group the proper Fierz identities have to be used.

We may now repeat the exercise for any Feynman diagram contributing to an amplitude
${\mathcal A}_n^{(0)}(g_1,g_2,...,g_n)$ and collect all terms proportional 
to a specific trace $\mathrm{Tr}( T^{a_{\sigma(1)}} ... T^{a_{\sigma(n)}})$.
We thus arrive at the {\bf colour decomposition} of the 
tree-level Yang-Mills amplitudes \cite{Cvitanovic:1980bu,Berends:1987cv,Mangano:1987xk,Kosower:1987ic,Bern:1990ux,DelDuca:1999rs,Maltoni:2002mq}:
\bq
\label{colour_decomposition}
{\mathcal A}_{n}^{(0)}\left(g_1,g_2,...,g_n\right) 
 & = & g^{n-2} \sum\limits_{\sigma \in S_{n}/Z_{n}} 
 2 \; \mathrm{Tr} \left( T^{a_{\sigma(1)}} ... T^{a_{\sigma(n)}} \right)
 \;\;
 A_{n}^{(0)}\left( g_{\sigma(1)}, ..., g_{\sigma(n)} \right),
\eq
where the sum is over all non-cyclic permutations of $\{1,2,...,n\}$.
The quantities $A_{n}^{(0)}( g_{\sigma(1)}, ..., \linebreak g_{\sigma(n)} )$ accompanying
the colour factor $2 \; \mathrm{Tr}( T^{a_{\sigma(1)}} ... T^{a_{\sigma(n)}})$
are called {\bf partial amplitudes}.
Partial amplitudes are gauge-invariant.
Closely related are {\bf primitive amplitudes}, 
which for tree-level Yang-Mills amplitudes are calculated 
from planar diagrams with a fixed cyclic ordering of the external legs
and cyclic-ordered Feynman rules \cite{Bern:1990ux,Bern:1994fz,Ellis:2008qc,Ellis:2011cr,Ita:2011ar,Badger:2012pg,Reuschle:2013qna,Schuster:2013aya}. 
Primitive amplitudes are gauge invariant as well.
For tree-level Yang-Mills amplitudes the notions of partial amplitudes and primitive amplitudes
coincide.
However, this is no longer true if one considers amplitudes with quarks and/or amplitudes with loops.
The most important features of a primitive amplitude are gauge invariance and a fixed cyclic ordering of the external legs.
(For amplitudes with quarks and/or loops there will be some additional requirements, which are not relevant here.)
Partial amplitudes are defined as the kinematic coefficients of the independent colour
structures.
Partial amplitudes are also gauge invariant, but not necessarily cyclic ordered.
The leading contributions in an $1/N$-expansion (with $N$ being the number of colours)
are usually cyclic ordered, the sub-leading parts are in general not.

Let us now give the {\bf cyclic-ordered Feynman rules}.
The gluon propagator in Feynman gauge is given by
\bq
 \begin{picture}(100,20)(0,5)
 \Gluon(20,10)(70,10){-5}{5}
 \Text(15,12)[r]{\footnotesize $\mu$}
 \Text(75,12)[l]{\footnotesize $\nu$}
\end{picture} 
& = &
 \frac{-ig^{\mu\nu}}{p^2},
\eq
the cyclic-ordered Feynman rules for the three-gluon and the four-gluon vertices are
\bq
\label{cyclic_ordered_Feynman_rules}
\begin{picture}(100,35)(0,50)
\Vertex(50,50){2}
\Gluon(50,50)(50,80){3}{4}
\Gluon(50,50)(76,35){3}{4}
\Gluon(50,50)(24,35){3}{4}
\LongArrow(56,70)(56,80)
\LongArrow(67,47)(76,42)
\LongArrow(33,47)(24,42)
\Text(60,80)[lt]{$p_1^{\mu_1}$}
\Text(78,35)[lc]{$p_2^{\mu_2}$}
\Text(22,35)[rc]{$p_3^{\mu_3}$}
\end{picture}
 & = &
 i \left[ g^{\mu_1\mu_2} \left( p_1^{\mu_3} - p_2^{\mu_3} \right)
         +g^{\mu_2\mu_3} \left( p_2^{\mu_1} - p_3^{\mu_1} \right)
         +g^{\mu_3\mu_1} \left( p_3^{\mu_2} - p_1^{\mu_2} \right)
   \right],
 \nonumber \\
 \nonumber \\
 \nonumber \\
\begin{picture}(100,35)(0,50)
\Vertex(50,50){2}
\Gluon(50,50)(71,71){3}{4}
\Gluon(50,50)(71,29){3}{4}
\Gluon(50,50)(29,29){3}{4}
\Gluon(50,50)(29,71){3}{4}
\Text(72,72)[lb]{\small $\mu_1$}
\Text(72,28)[lt]{\small $\mu_2$}
\Text(28,28)[rt]{\small $\mu_3$}
\Text(28,72)[rb]{\small $\mu_4$}
\end{picture}
 & = &
  i \left[
        2 g^{\mu_1\mu_3} g^{\mu_2\mu_4} - g^{\mu_1\mu_2} g^{\mu_3\mu_4} 
                                        - g^{\mu_1\mu_4} g^{\mu_2\mu_3}
 \right].
 \nonumber \\
 \nonumber \\
\eq
The cyclic-ordered Feynman rules are obtained from the standard Feynman rules 
by extracting from each formula the coupling constant and by taking the coefficient of the cyclic-ordered colour part.
\begin{figure}
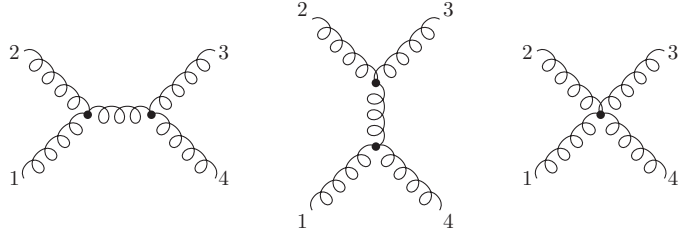

\begin{center}
\includegraphics[scale=0.8]{fig1}
\hspace*{8mm}
\includegraphics[scale=0.8]{fig2}
\hspace*{8mm}
\includegraphics[scale=0.8]{fig4}
\caption{\label{fig2} The three cyclic-ordered Feynman diagrams contributing to the tree-level four-gluon primitive amplitude $A_4^{(0)}(1,2,3,4)$.}
\end{center}
\end{figure}
Note that the Feynman rule for the cyclic-ordered four-gluon vertex is considerably simpler
than the Feynman rule for the full four-gluon vertex.

The primitive amplitude $A_{n}^{(0)}( g_1, g_2, ..., g_n )$ -- or $A_{n}^{(0)}( 1, 2, ..., n)$ for short --
with the external order $(1,2,...,n)$
is calculated from planar diagrams with this external order.
There are fewer diagrams contributing to $A_{n}^{(0)}$ than to ${\mathcal A}_{n}^{(0)}$.
For the case of four gluons there are three cyclic-ordered Feynman diagrams contributing to
the primitive amplitude $A_{4}^{(0)}(1,2,3,4)$.
These are shown in fig.~(\ref{fig2}).
For the next few values of $n$ we list the number of cyclic-ordered diagrams in table~(\ref{table2})
\begin{table}
\begin{center}
\bq
\begin{array}{l|rrrrrrr}
 n                    & 4 & 5  & 6   & 7    & 8     & 9      & 10 \\
 \hline
\mbox{unordered}      & 4 & 25 & 220 & 2485 & 34300 & 559405 & 10525900 \\
\mbox{cyclic ordered} & 3 & 10 & 38  & 154  & 654   & 2871   & 12925 \\
\end{array}
\eq
\caption{\label{table2} The number of diagrams contributing to the full amplitude ${\mathcal A}_n^{(0)}$
and to the cyclic-ordered primitive amplitude $A_n^{(0)}$.}
\end{center}
\end{table}
and compare them with the number of Feynman diagrams contributing to the full amplitude 
${\mathcal A}_{n}^{(0)}$.
The number of diagrams contributing to the cyclic-ordered primitive amplitude $A_n^{(0)}$
is significantly smaller than the number of Feynman diagrams contributing to the full amplitude 
${\mathcal A}_{n}^{(0)}$.

The colour decomposition in eq.~(\ref{colour_decomposition})
allows us to perform the summation over the colour degrees of freedom when calculating
$|{\mathcal A}_{n}^{(0)}|^2$. We have
\bq
 \sum\limits_{\mathrm{colour}}
 \left|{\mathcal A}_{n}^{(0)}\right|^2
 & = &
 g^{2n-4} 
 \sum\limits_{\sigma \in S_{n}/Z_{n}} 
 \sum\limits_{\pi \in S_{n}/Z_{n}} 
 4 \; \mathrm{Tr} \left( T^{a_{\sigma(1)}} ... T^{a_{\sigma(n)}} \right)
 \; \mathrm{Tr} \left( T^{a_{\pi(n)}} ... T^{a_{\pi(1)}} \right)
 \nonumber \\
 & &
 A_{n}^{(0)}\left( \sigma(1), ..., \sigma(n) \right)
 \;\;
 A_{n}^{(0)}\left( \pi(1), ..., \pi(n) \right)^\ast.
\eq
The traces can be evaluated with the help of the formul{\ae} in
eq.~(\ref{traces_SU_N}) or eq.~(\ref{traces_U_N}).

Let us now consider the primitive amplitude $A_n^{(0)}$ for a fixed set of external momenta,
denoted by $p=(p_1,...,p_n)$ and a fixed set of external polarisations,
denoted by $\eps=(\eps_1^{\lambda_1},...,\eps_n^{\lambda_n})$
with $\lambda_j \in \{+,-\}$.
In addition, the primitive amplitude will depend on the cyclic order of the external legs,
specified by $\sigma=(\sigma_1,...,\sigma_n)$.
An obvious question related to the colour decomposition is: 
How many independent primitive amplitudes are there for $n$ external particles?
For a fixed set of external momenta and a fixed set of polarisations
the primitive amplitudes are distinguished by the permutation specifying the order of the external particles.
For $n$ external particles there are $n!$ permutations and therefore $n!$ different orders.
However, there are relations among primitive amplitudes with different external order.
The first set of relations is rather trivial and given by {\bf cyclic invariance}:
\bq
 A_n^{(0)}\left(1,2,...,n\right) & = & A_n^{(0)}\left(2,...,n,1\right)
\eq
Cyclic invariance is the statement that only the external cyclic order matters, 
not the point, where we start to read off the order.
Cyclic invariance reduces the number of independent primitive amplitudes to $(n-1)!$.

The first non-trivial relations are the {\bf Kleiss-Kuijf relations} \cite{Kleiss:1988ne}.
Let
\bq
 \vec{\alpha} = \left( \alpha_1,  \alpha_2, ..., \alpha_j \right),
 & & 
 \vec{\beta} = \left( \beta_1, \beta_2, ..., \beta_{n-2-j} \right)
\eq
be two ordered sequences of numbers, such that
\bq
 \{1\} \cup \{ \alpha_1, ..., \alpha_j  \} \cup \{ \beta_1, ..., \beta_{n-2-j} \} \cup \{ n \}
 & = &
 \{ 1,...,n \}.
\eq
We further set $\vec{\beta}^T = ( \beta_{n-2-j}, ..., \beta_2, \beta_1)$.
\begin{figure}
\begin{center}
\includegraphics[scale=1.0]{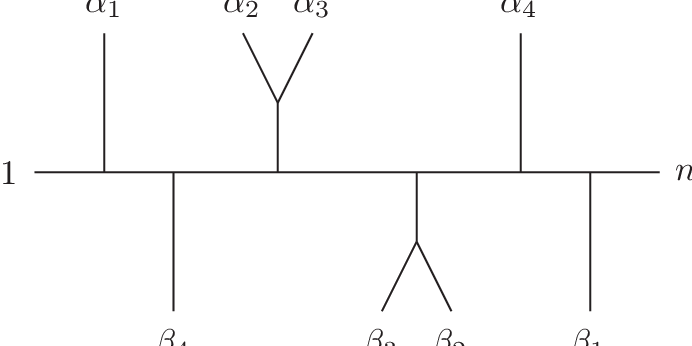}
\caption{\label{figure_Kleiss_Kuijf}
Illustration of the Kleiss-Kuijf relation: 
Only currents consisting of particles from $\alpha_1$, ..., $\alpha_j$
or $\beta_1$, ..., $\beta_{n-2-j}$ couple to the line from $1$ to $n$.
No mixed currents (involving particles from $\alpha_1$, ..., $\alpha_j$
and $\beta_1$, ..., $\beta_{n-2-j}$) couple to the line from $1$ to $n$.
}
\end{center}
\end{figure}
The Kleiss-Kuijf relations read
\bq
\label{Kleiss_Kuijf}
 A_n^{(0)}( 1, \alpha_1, ..., \alpha_j , n, \beta_1, ..., \beta_{n-2-j} )
 & = & 
 \left( -1 \right)^{n-2-j}
 \sum\limits_{\sigma \in \vec{\alpha} \; \shuffle \; \vec{\beta}^T}
 A_n^{(0)}( 1, \sigma_1, ..., \sigma_{n-2}, n ).
\eq
Here, $\vec{\alpha} \; \shuffle \; \vec{\beta}^T$ denotes the set of all shuffles of $\vec{\alpha}$ with $\vec{\beta}^T$, i.e.
the set of all permutations of the elements of $\vec{\alpha}$ and $\vec{\beta}^T$, which preserve the relative order of the
elements of $\vec{\alpha}$ and of the elements of $\vec{\beta}^T$.
The Kleiss-Kuijf relations reduce the number of independent primitive amplitudes to $(n-2)!$.

The Kleiss-Kuijf relations follow from the anti-symmetry of the colour-stripped trivalent vertices.
Let us first consider a theory with only trivalent vertices, which are anti-symmetric under the exchange of any two legs.
Consider as an example 
the situation shown in fig.~(\ref{figure_Kleiss_Kuijf}), corresponding to a specific contribution to the left-hand side 
of eq.~(\ref{Kleiss_Kuijf}).
Only currents consisting of particles from $\alpha_1$, ..., $\alpha_j$
or $\beta_1$, ..., $\beta_{n-2-j}$ couple to the line from $1$ to $n$.
No mixed currents, involving particles from $\alpha_1$, ..., $\alpha_j$ and $\beta_1$, ..., $\beta_{n-2-j}$, couple to the line from $1$ to $n$.
Next consider the right-hand side of eq.~(\ref{Kleiss_Kuijf}). Now all particles from 
$\{ \alpha_1, ..., \alpha_j  \} \cup \{ \beta_1, ..., \beta_{n-2-j} \}$ are above the line from $1$ to $n$.
Flipping the particles from the set $\{ \beta_1, ..., \beta_{n-2-j} \}$ will give the sign $(-1)^{n-2-j}$.
The shuffle product will cancel all contributions from mixed currents coupling to the line from $1$ to $n$.
To see this, consider a mixed current contributing to the right-hand side of eq.~(\ref{Kleiss_Kuijf}). 
Such a current will necessarily contain two sub-currents, one made out entirely of particles from
$\{ \alpha_1, ..., \alpha_j  \}$, the other made out entirely of particles from $\{ \beta_1, ..., \beta_{n-2-j} \}$ and coupled together through 
a trivalent vertex.
The shuffle product ensures that both cyclic orderings at this vertex contribute.
Since the trivalent vertex is anti-symmetric under the exchange of two legs, these contributions cancel.
Note that these arguments apply to any theory with anti-symmetric trivalent vertices only.

It remains to show that Yang-Mills theory can be cast into a form involving only anti-symmetric trivalent vertices.
The cyclic ordered three-gluon vertex is clearly anti-symmetric under the exchange of any two legs:
\bq
\begin{picture}(100,35)(0,55)
\Vertex(50,50){2}
\Gluon(50,50)(50,80){3}{4}
\Gluon(50,50)(76,35){3}{4}
\Gluon(50,50)(24,35){3}{4}
\Text(55,80)[lc]{$1$}
\Text(78,35)[lc]{$2$}
\Text(22,35)[rc]{$3$}
\end{picture}
 & = &
 -
\begin{picture}(100,35)(0,55)
\Vertex(50,50){2}
\Gluon(50,50)(50,80){3}{4}
\Gluon(50,50)(76,35){3}{4}
\Gluon(50,50)(24,35){3}{4}
\Text(55,80)[lc]{$1$}
\Text(78,35)[lc]{$3$}
\Text(22,35)[rc]{$2$}
\end{picture}
 \\
 \nonumber 
\\ \nonumber 
\eq
We have to eliminate the four-gluon vertex.
This can be done by introducing an auxiliary tensor field \cite{Draggiotis:1998gr,Duhr:2006iq}.
From eq.~(\ref{Lagrangian_QCD_expanded}) we read off that 
the four-gluon vertex part of the Lagrangian is
\bq
{\mathcal L}_{gggg} & = & 
 \frac{1}{2g^2} \; \mathrm{Tr} \; 
 \left[ A_\mu, A_\nu \right] \left[ A^\mu, A^\nu \right].
\eq
Let us introduce an auxiliary tensor field
\bq
 B_{[\mu\nu]} & = & \frac{g}{i} T^a B^a_{[\mu\nu]},
\eq
anti-symmetric in the indices $\mu$ and $\nu$
with the Lagrange density
\bq
\label{auxiliary_Lagrangian}
 {\mathcal L}_{\mathrm{aux}} & = & 
 \frac{2}{g^2} \; \mathrm{Tr} \; 
 \left(
  \frac{1}{2} B_{[\mu\nu]} B^{[\mu\nu]}
 - \frac{i}{\sqrt{2}} B_{[\mu\nu]} \left[ A^\mu, A^\nu \right]
 \right).
\eq
The auxiliary field $B_{[\mu\nu]}$ occurs at the most quadratically 
and can be integrated out
\bq
 \int {\mathcal D} B_{[\mu\nu]} \exp \left( i \int d^4x \; {\mathcal L}_{\mathrm{aux}} \right)
 & = &
 {\mathcal N}
 \exp \left( i \int d^4x \; {\mathcal L}_{\mathrm{gggg}} \right),
\eq
where the (irrelevant) prefactor ${\mathcal N}$ is given by
\bq
 {\mathcal N} & = &
 \int {\mathcal D} B_{[\mu\nu]} \exp \left( \frac{i}{g^2} \int d^4x \; 
  \mathrm{Tr} \;
 B_{[\mu\nu]} B^{[\mu\nu]}
 \right).
\eq
\bs
{\it {\bf Exercise \theexercise}: 
Let us investigate in more detail the procedure of ``integrating out'' a field $\phi$,
which appears at the most quadratically in the Lagrangian.
Consider the path integral
\bq
\int {\mathcal D}\phi \; \exp \; i \int d^4x \; \mathrm{Tr}\left( \frac{1}{2} \phi P \phi + \phi K \right).
\eq
Assume that $P$ is a pseudo-differential operator of even degree and independent of the other fields.
$K$ on the other hand may depend on some other fields.
Show that under the substitution
\bq
 \phi & \rightarrow & \phi+P^{-1}K,
\eq
where $P^{-1}$ denotes the inverse pseudo-differential operator, one obtains
\bq
\int {\mathcal D} \phi \; \exp \; i \int d^4x \; 
 \mathrm{Tr}\left( \frac{1}{2} \phi P \phi - \frac{1}{2} K P^{-1} K 
 \right).
\eq
The irrelevant factor
\bq
\int {\mathcal D}\phi \; \exp \; i \int d^4x \; \mathrm{Tr}\left( \frac{1}{2} \phi P \phi \right)
\eq
may be neglected, leaving
\bq
 \exp \; i \int d^4x \; 
 \mathrm{Tr}\left( - \frac{1}{2} K P^{-1} K \right)
\eq
as the result of integrating out the field $\phi$.
\stepcounter{exercise}
}
\es
\\
\\
From the Lagrangian in eq.~(\ref{auxiliary_Lagrangian}) we may derive the Feynman rules
for this tensor field.
The colour-stripped ``propagator'' for the tensor field is given by
\bq
 \begin{picture}(100,20)(0,5)
 \Line(20,8)(70,8)
 \Line(20,12)(70,12)
 \Text(15,12)[r]{\footnotesize $[\mu \nu]$}
 \Text(75,12)[l]{\footnotesize $[\rho \sigma]$}
\end{picture} 
& = &
 - \frac{i}{2} \left( g_{\mu\rho} g_{\nu\sigma} - g_{\mu\sigma} g_{\nu\rho} \right).
\eq
The cyclic-ordered vertex of the tensor field with the gluon fields is given by
\bq
\begin{picture}(100,35)(0,40)
\Vertex(50,50){2}
\Line(48,50)(48,20)
\Line(52,50)(52,20)
\Gluon(50,50)(76,65){3}{4}
\Gluon(50,50)(24,65){3}{4}
\Text(22,65)[rc]{$1, \mu$}
\Text(78,65)[lc]{$2, \nu$}
\Text(50,17)[t]{$3, [\rho \sigma]$}
\end{picture}
 & = &
 \frac{i}{\sqrt{2}}
 \left( g^{\mu\rho} g^{\nu\sigma} - g^{\mu\sigma} g^{\nu\rho} \right).
 \\
 \nonumber \\
 \nonumber 
\eq
This vertex is anti-symmetric;
\bq
\begin{picture}(100,35)(0,40)
\Vertex(50,50){2}
\Line(48,50)(48,20)
\Line(52,50)(52,20)
\Gluon(50,50)(76,65){3}{4}
\Gluon(50,50)(24,65){3}{4}
\Text(22,65)[rc]{$1$}
\Text(78,65)[lc]{$2$}
\Text(50,17)[t]{$3$}
\end{picture}
 & = &
 -
\begin{picture}(100,35)(0,40)
\Vertex(50,50){2}
\Line(48,50)(48,20)
\Line(52,50)(52,20)
\Gluon(50,50)(76,65){3}{4}
\Gluon(50,50)(24,65){3}{4}
\Text(22,65)[rc]{$2$}
\Text(78,65)[lc]{$1$}
\Text(50,17)[t]{$3$}
\end{picture}
 \\
 \nonumber \\ 
 \nonumber
\eq
\bs
{\it {\bf Exercise \theexercise}: 
Show that at the level of cyclic-ordered Feynman rules we have
\begin{center}
\includegraphics[scale=0.8]{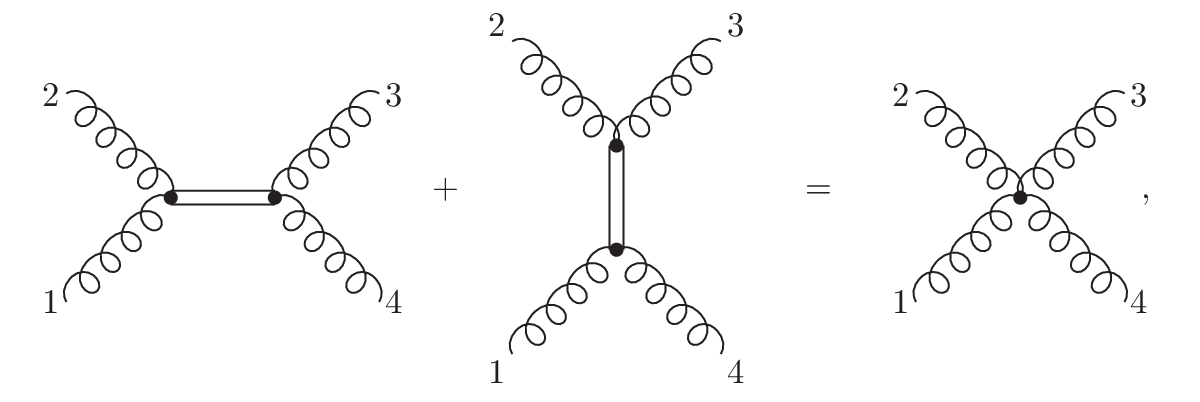}
\end{center}
where arbitrary sub-graphs may be attached to the legs $1$-$4$ (but always the same sub-graph
for a specific leg). 
This verifies at the level of diagrams that the four-gluon vertex can be cast into
trivalent vertices with the help of the auxiliary tensor field. 
\stepcounter{exercise}
}
\es
\\
\\
Let us continue with our investigation on the number of independent primitive amplitudes.
Apart from cyclic invariance and the Kleiss-Kuijf relations there
are in addition the {\bf Bern-Carrasco-Johansson relations} (BCJ relations) \cite{Bern:2008qj}.
The fundamental BCJ relations read
\bq
\label{fundamental_BCJ_relation}
 \sum\limits_{i=2}^{n-1} 
  \left( \sum\limits_{j=i+1}^n 2 p_2 p_j \right)
  A_n^{(0)}(1,3,...,i,2,i+1,...,n-1,n)
 & = & 0.
\eq
Cyclic invariance allows us to fix one external particle at a specified position, say position $1$.
The Kleiss-Kuijf relations allow us to fix a second external particle 
at another specified position, say position $n$.
The BCJ relations allow us to fix a third external particle
at a third specified position, say position $2$.
The BCJ relations reduce the number of independent primitive amplitudes to $(n-3)!$.
The full set of relations among primitive tree amplitudes in pure Yang-Mills theory is given by cyclic invariance,
Kleiss-Kuijf relations, and the fundamental BCJ relations.
Therefore a basis of independent primitive amplitudes consists of $(n-3)!$ elements.

The BCJ relations have been first conjectured by 
Bern, Carrasco and Johansson in \cite{Bern:2008qj}.
They have been proven first with methods from string theory \cite{BjerrumBohr:2009rd,Stieberger:2009hq}.
It should be mentioned that the BCJ relations do not rely on string theory and a proof within quantum field theory
exists \cite{Feng:2010my}. We will present this proof later in this report in section~(\ref{sect:BCFW_recursion}),
once we are familiar with on-shell recurrence relations.

\subsection{Colour-kinematics duality}
\label{sect_colour_kinematics_duality}

A Lie group $G$ is a group which is also an analytic manifold such that the mapping $(a,b) \rightarrow
ab^{-1}$ of the product manifold $G \times G$ into $G$ is analytic.
\\
A Lie algebra $\mathfrak{g}$ over a field $F$ is a vectorspace $A$ together with a mapping
$x \otimes y \rightarrow [x,y]$  such that for $x,y,z \in A$:
\bq
\label{def_Lie_algebra}
\left[x,x\right] & = & 0,
  \nonumber \\
   \left[ \left[ x, y \right], z \right] 
 + \left[ \left[ y, z \right], x \right] 
 + \left[ \left[ z, x \right], y \right] & = & 0.
\eq
\bs
{\it {\bf Exercise \theexercise}: 
Show that $[x, x]=0$ implies the anti-symmetry of the Lie bracket $[x,y]=-[y,x]$.
Show further that also the converse is true, provided $\mathrm{char}\; F \neq 2$.
Explain, why the argument does not work for $\mathrm{char}\; F = 2$.
\stepcounter{exercise}
}
\es
\\
\\
The second line in eq.~(\ref{def_Lie_algebra}) is called the Jacobi identity.
In physics we are usually concerned with Lie groups which are matrix groups.
The corresponding Lie algebra is then an algebra of matrices and the Lie bracket
corresponds to the commutator of matrices:
\bq
 \left[ T^a, T^b \right] & = &
 T^a \cdot T^b - T^b \cdot T^a.
\eq
It is an easy exercise to verify that the relations in eq.~(\ref{def_Lie_algebra}) are satisfied.
The Jacobi identity translates to
\bq
   \left[ \left[ T^a, T^b \right], T^c \right] 
 + \left[ \left[ T^b, T^c \right], T^a \right] 
 + \left[ \left[ T^c, T^a \right], T^b \right] & = & 0,
\eq
or equivalently
\bq
 \left( i f^{abe} \right) \left( i f^{ecd} \right) 
 +
 \left( i f^{bce} \right) \left( i f^{ead} \right) 
 +
 \left( i f^{cae} \right) \left( i f^{ebd} \right) 
 & = & 0.
\eq
A product in an algebra takes two elements of the algebra as an input and gives one
element of the algebra as an output.
This may be represented graphically by a trivalent vertex with two input lines on the top and
one output line at the bottom.
The anti-symmetry of the Lie bracket is expressed through
\bq
\label{antisymmetry_relation}
\begin{picture}(60,30)(0,15)
\Vertex(30,20){2}
\Line(30,5)(30,20)
\Line(30,20)(45,35)
\Line(30,20)(15,35)
\Text(15,40)[b]{$1$}
\Text(45,40)[b]{$2$}
\Text(30,0)[t]{$3$}
\end{picture}
 & = &
 -
\begin{picture}(60,30)(0,15)
\Vertex(30,20){2}
\Line(30,5)(30,20)
\Line(30,20)(45,35)
\Line(30,20)(15,35)
\Text(15,40)[b]{$2$}
\Text(45,40)[b]{$1$}
\Text(30,0)[t]{$3$}
\end{picture}
 \\ \nonumber 
 \\ \nonumber
\eq
and the Jacobi identity is visualised through
\bq
\label{visualise_Jacobi}
\begin{picture}(100,50)(0,25)
\Vertex(50,30){2}
\Vertex(35,45){2}
\Line(50,5)(50,30)
\Line(50,30)(80,60)
\Line(50,30)(20,60)
\Line(35,45)(50,60)
\Text(20,65)[b]{\footnotesize $1$}
\Text(50,65)[b]{\footnotesize $2$}
\Text(80,65)[b]{\footnotesize $3$}
\Text(50,0)[t]{\footnotesize $4$}
\end{picture}
 +
\begin{picture}(100,50)(0,25)
\Vertex(50,30){2}
\Vertex(35,45){2}
\Line(50,5)(50,30)
\Line(50,30)(80,60)
\Line(50,30)(20,60)
\Line(35,45)(50,60)
\Text(20,65)[b]{\footnotesize $2$}
\Text(50,65)[b]{\footnotesize $3$}
\Text(80,65)[b]{\footnotesize $1$}
\Text(50,0)[t]{\footnotesize $4$}
\end{picture}
 +
\begin{picture}(100,50)(0,25)
\Vertex(50,30){2}
\Vertex(35,45){2}
\Line(50,5)(50,30)
\Line(50,30)(80,60)
\Line(50,30)(20,60)
\Line(35,45)(50,60)
\Text(20,65)[b]{\footnotesize $3$}
\Text(50,65)[b]{\footnotesize $1$}
\Text(80,65)[b]{\footnotesize $2$}
\Text(50,0)[t]{\footnotesize $4$}
\end{picture}
 & = & 0.
 \\
 \nonumber \\
 \nonumber
\eq
In particular these two relations hold true, if we assume that 
each vertex of these graphs corresponds to a factor $(i f^{abc})$.
Let us now focus on the Jacobi identity.
We may attach subgraphs to the external legs $1$, $2$, $3$ and $4$.
The Jacobi identity remains true, as long as we always attach the same subgraph to leg $1$,
another subgraph to all occurrences of leg $2$, etc..

Let us consider a set of $n$ external momenta $p_1$, $p_2$, ..., $p_n$ and an internal edge $e$
of some tree graph.
Assume that the particles $\{\alpha_1,\alpha_2,...,\alpha_j\}$ are attached on the one side of the internal
edge $e$, while the remaining $(n-j)$ particles $\{\beta_1,...,\beta_{n-j}\}$ are attached
to the other side of the internal edge $e$.
We set
\bq
 s_e
 & = &
 \left( p_{\alpha_1} + ... + p_{\alpha_j} \right)^2
 \;\; = \;\;
 \left( p_{\beta_1} + ... + p_{\beta_{n-j}} \right)^2.
\eq
We are now in a position to discuss colour-kinematics duality \cite{Bern:2010ue}:
{\bf Colour-kinematics duality} states 
that Yang-Mills amplitudes can be brought into a form
\bq
\label{BCJ_form}
 {\mathcal A}_n^{(0)}
 & = &
 i g^{n-2}
 \sum\limits_{\mathrm{trivalent} \; \mathrm{graphs} \; G}
 \frac{C\left(G\right)N\left(G\right)}{D\left(G\right)},
 \;\;\;\;\;\;\;\;\;\;\;\;
 D\left(G\right) = \prod\limits_{\mathrm{edges} \; e} s_e,
\eq
where the summation is over all trivalent graphs $G$ (i.e. the sum does not involve graphs with
four-valent vertices),
$D(G)$ is the product of Lorentz invariants $s_e$ corresponding to the internal edges of the graph $G$
(i.e. up to factors of $i$, the quantity $1/D(G)$ is the product of the scalar propagators of the graph $G$)
and the quantity $C(G)$ is the colour factor of the graph
(i.e. a factor $if^{abc}$ for every vertex of the graph $G$).
The quantity $N(G)$ is called the kinematic numerator of the graph $G$.
We have seen that for graphs $G_1$, $G_2$ and $G_3$ as in eq.~(\ref{visualise_Jacobi})
the colour factors $C(G_1)$, $C(G_2)$ and $C(G_3)$ satisfy the Jacobi relation.
The non-trivial point of the representation of the Yang-Mills amplitude in the form of eq.~(\ref{BCJ_form})
is, that kinematic numerators $N(G)$ can be found, such that
the kinematical numerators $N(G)$ satisfy anti-symmetry and Jacobi-like relations, whenever the corresponding
colour factors $C(G)$ do:
\bq
\label{BCJ_condition}
 C\left(G_1\right) + C\left(G_2\right) = 0
 \hspace*{5mm}
 & \Rightarrow &
 \hspace*{5mm}
 N\left(G_1\right) + N\left(G_2\right) = 0,
 \nonumber \\
 C\left(G_1\right) + C\left(G_2\right) + C\left(G_3\right) = 0
 \hspace*{5mm}
 & \Rightarrow &
 \hspace*{5mm}
 N\left(G_1\right) + N\left(G_2\right) + N\left(G_3\right) = 0.
 \hspace*{6mm}
\eq
The anti-symmetry of the kinematic numerators follows trivially from the anti-symmetry of the trivalent vertices.
The non-trivial part are the Jacobi-like relations for the kinematic numerators.
Note that colour-kinematics duality does not state
that for any trivalent graph expansion as in eq.~(\ref{BCJ_form})
the kinematic numerators satisfy Jacobi-like relations.
Colour-kinematics duality states, that there exists kinematic numerators
such that eq.~(\ref{BCJ_form}) and eq.~(\ref{BCJ_condition}) are fulfilled.
In general, the kinematic numerators are not unique,
changing the kinematic numerators while leaving eq.~(\ref{BCJ_form}) and eq.~(\ref{BCJ_condition})
intact is referred to as a generalised gauge transformation.
The kinematic numerators may be found by solving a linear system of equations.
The basic idea is as follows:
If the vertices are anti-symmetric, we may 
rewrite eq.~(\ref{visualise_Jacobi}) as
\bq
\label{STU_relation}
\begin{picture}(110,30)(0,30)
\Line(10,10)(90,10)
\Vertex(50,30){2}
\Vertex(50,10){2}
\Line(50,10)(50,30)
\Line(50,30)(70,50)
\Line(50,30)(30,50)
\Text(5,10)[r]{$1$}
\Text(30,55)[b]{$2$}
\Text(70,55)[b]{$3$}
\Text(95,10)[l]{$4$}
\end{picture}
 & = & 
\begin{picture}(120,30)(-10,30)
\Line(10,10)(90,10)
\Vertex(35,10){2}
\Vertex(65,10){2}
\Line(35,10)(35,40)
\Line(65,10)(65,40)
\Text(5,10)[r]{$1$}
\Text(35,45)[b]{$2$}
\Text(65,45)[b]{$3$}
\Text(95,10)[l]{$4$}
\end{picture}
-
\begin{picture}(110,30)(-10,30)
\Line(10,10)(90,10)
\Vertex(35,10){2}
\Vertex(65,10){2}
\Line(35,10)(35,40)
\Line(65,10)(65,40)
\Text(5,10)[r]{$1$}
\Text(35,45)[b]{$3$}
\Text(65,45)[b]{$2$}
\Text(95,10)[l]{$4$}
\end{picture}.
 \\
 \nonumber \\
 \nonumber
\eq
Eq.~(\ref{STU_relation}) is called a STU-relation.
If eq.~(\ref{STU_relation}) holds we may reduce any tree graph with $n$ external legs
and containing only trivalent vertices to a multi-peripheral form with respect to $1$ and $n$.
We say that a graph is multi-peripheral with respect to $1$ and $n$,
if all other external legs connect directly to the line from $1$ to $n$,
i.e. there are no non-trivial sub-trees attached to this line.
A graph in multi-peripheral form can be drawn as
\bq
\begin{picture}(210,60)(0,0)
\Line(10,10)(200,10)
\Vertex(35,10){2}
\Vertex(65,10){2}
\Vertex(95,10){2}
\Vertex(145,10){2}
\Vertex(175,10){2}
\Line(35,10)(35,40)
\Line(65,10)(65,40)
\Line(95,10)(95,40)
\Line(145,10)(145,40)
\Line(175,10)(175,40)
\Text(5,10)[r]{$1$}
\Text(35,45)[b]{$\sigma_2$}
\Text(65,45)[b]{$\sigma_3$}
\Text(120,25)[b]{$...$}
\Text(175,45)[b]{$\sigma_{n-1}$}
\Text(205,10)[l]{$n$}
\end{picture}
\eq
Repeated use of eq.~(\ref{STU_relation}) reduces any tree graph with non-trivial sub-trees attached 
to the line from $1$ to $n$ to a multi-peripheral form.
We may therefore express any kinematic numerator as a linear combination of the $(n-2)!$ kinematic numerators
corresponding to the multi-peripheral graphs with leg $1$ and $n$ fixed.
Let us then consider cyclic-ordered amplitudes with legs $1$, $2$ and $n$ fixed.
On the one hand we may express these amplitudes in terms of the kinematic numerators corresponding to the
multi-peripheral graphs.
On the other hand, we may simply calculate these amplitudes by any method discussed in this report 
(for example Feynman diagrams will do).
This gives us a set of $(n-3)!$ linear equations for $(n-2)!$ unknown kinematic numerators, which may be solved
by linear algebra, determining $(n-3)!$ kinematic numerators and
leaving $(n-2)!-(n-3)!$ ``free'' kinematic numerators.
The kinematic numerators are not unique, we may give the ``free'' kinematic numerators any value.
In particular we may set them to zero.
The non-uniqueness of the kinematic numerators is referred to as {\bf generalised gauge-invariance}.
The existence of kinematic numerators satisfying anti-symmetry and Jacobi-like relations implies
the BCJ relations.
This can be seen by expressing an amplitude with legs $1$ and $n$ fixed, but $2$ arbitrary in terms
of kinematic numerators, which in turn may be expressed in terms of amplitudes with legs $1$, $2$ and $n$ fixed
and a priori the ``free'' kinematic numerators.
One then observes that the dependence on the ``free'' kinematic numerators drops out in this relation.

\subsection{The spinor helicity method}

Let us now return to the spin- and colour-summed matrix element squared,
\bq
 \sum\limits_{\mathrm{spins,colour}} 
 \left| {\mathcal A}_n^{(0)} \right|^2,
\eq
entering in leading order
the formula~(\ref{observable_master_hadron_hadron}) for the calculation of an observable.
Assume that ${\mathcal A}_n^{(0)}$ is given as the sum of $N_{\mathrm{terms}}$.
Given the number of diagrams in table~(\ref{table1})
and the fact that the Feynman rules for the vertices in eq.~(\ref{Feynman_rules_vertices})
imply that each diagram will contribute several terms, we anticipate that $N_{\mathrm{terms}}$
can be a rather large number.
Squaring the amplitude and summing over the spins will result in ${\mathcal O}(N_{\mathrm{terms}}^2)$ terms.
To be precise, we obtain by using the spin sum of eq.~(\ref{spin_sum_gluon})
with a light-like reference vector $n^\mu$ in the squared expression
$3^n N_{\mathrm{terms}}^2$ terms.
The factor $3^n$ comes from the three terms in the expression of the spin sum for a light-like reference vector
$n^\mu$ with $n^2=0$:
\bq
\sum\limits_{\lambda \in \{+,-\} } \left( \eps_\mu^\lambda(p) \right)^\ast \; \eps_\nu^\lambda(p)
 & = & 
 - g_{\mu\nu} + \frac{p_\mu n_\nu}{p \cdot n} + \frac{n_\mu p_\nu}{p \cdot n}.
\eq 
This ${\mathcal O}(N_{\mathrm{terms}}^2)$-behaviour is highly prohibitive.
Can be do better? Yes, we can.
Up to now we treated the polarisation vectors for the external gluons as abstract objects,
satisfying eqs.~(\ref{tranverse_pol_vectors})-(\ref{orthogonality_pol_vectors}) and eq.~(\ref{spin_sum_gluon}).
We may however use explicit expressions for the polarisation vectors.
Given numerical values for the external momenta $(p_1, ..., p_n)$ 
and -- for a specific helicity configuration $(\lambda_1,...,\lambda_n)$ --
numerical values for the polarisation vectors $(\eps_1, ... \eps_n)$, we may evaluate the $N_{\mathrm{terms}}$
terms of the amplitude and obtain a complex number, ignoring colour for the moment.
Taking the norm of a complex number is a ${\mathcal O}(1)$-operation.
We repeat this for every helicity configuration and sum up the individual contributions.
There are $2^n$ helicity configurations. In total we have to evaluate
$2^n N_{\mathrm{terms}}$ expressions and we arrive at an
${\mathcal O}(N_{\mathrm{terms}})$-behaviour.
In practice one even avoids the prefactor $2^n$ by Monte Carlo sampling over the helicity configurations
or a Monte Carlo integration over helicity angles \cite{Draggiotis:1998gr,Gotz:2012zz,Czakon:2009ss,Dittmaier:2008md}.

The spinor-helicity method \cite{Berends:1981rb,DeCausmaecker:1982bg,Gunion:1985vc,Kleiss:1986qc,Xu:1987xb,Gastmans_book}
gives us explicit expression for the polarisation vectors.
As the name suggests, it involves spinors.
This may seem at first sight a little bit strange, as we are dealing with particles which are bosons with spin $1$.
However, as we will soon see, the use of spinors is an elegant way to implement the transversality conditions
of the polarisation vectors.
Let us therefore discuss spinors in more detail.

\subsubsection{The Dirac equation}

The Lagrange density for a Dirac field depends on four-component spinors $\psi_\alpha(x)$ $(\alpha=1,2,3,4)$ and
$\bar{\psi}_\alpha(x) = \left( \psi^\dagger(x) \gamma^0 \right)_\alpha$:
\bq
{\mathcal L}(\psi, \bar{\psi}, \partial_\mu \psi ) & = & i \bar{\psi}(x) \gamma^\mu \partial_\mu \psi(x) - m \bar{\psi}(x) \psi(x)
\eq
Here, the $(4 \times 4)$-Dirac matrices satisfy the anti-commutation rules
\bq
 \{ \gamma^\mu, \gamma^\nu \} = 2 g^{\mu\nu} {\bf 1},
 \;\;\;
 \{ \gamma^\mu, \gamma_5 \} = 0,
 \;\;\;
 \gamma_5 = i \gamma^0 \gamma^1 \gamma^2 \gamma^3
          = \frac{i}{24} \eps_{\mu\nu\rho\sigma} \gamma^\mu \gamma^\nu \gamma^\rho \gamma^\sigma.
\eq
The Dirac equations read
\bq
 \left( i \gamma^\mu \partial_\mu - m \right) \psi(x) \;\; = \;\; 0,
 & &
 \bar{\psi}(x) \left( i \gamma^\mu \stackrel{\leftarrow}{\partial}_\mu + m \right ) \;\; = \;\; 0.
\eq
For computations it is useful to have an explicit representation of the Dirac matrices.
There are several widely used representations. A particular useful one is the {\bf Weyl representation} of the Dirac matrices:
\bq
 \gamma^{\mu} = \left(\begin{array}{cc}
 0 & \sigma^{\mu} \\
 \bar{\sigma}^{\mu} & 0 \\
\end{array} \right),
& &
\gamma_{5} = i \gamma^0 \gamma^1 \gamma^2 \gamma^3 
           = \left(\begin{array}{cc}
 1 & 0 \\
 0 & -1 \\
\end{array} \right)
\eq
Here, the 4-dimensional $\sigma^{\mu}$-matrices are defined by
\bq
\sigma_{A \dot{B}}^{\mu} = \left( 1 , - \vec{\sigma} \right), & &
\bar{\sigma}^{\mu \dot{A} B} = \left( 1 ,  \vec{\sigma} \right) .
\eq
and $\vec{\sigma}=(\sigma_x,\sigma_y,\sigma_z)$ are the standard
Pauli matrices:
\bq
\sigma_x = \left(\begin{array}{cc}
 0 & 1\\
 1 & 0 \\
\end{array} \right),
&
\sigma_y = \left(\begin{array}{cc}
 0 & -i\\
 i & 0 \\
\end{array} \right),
&
\sigma_z = \left(\begin{array}{cc}
 1 & 0\\
 0 & -1 \\
\end{array} \right).
\eq
The $\sigma^{\mu}$-matrices satisfy the Fierz identities
\bq
 \sigma^{\mu}_{A\dot{A}} \bar{\sigma}_{\mu}^{\dot{B}B} 
 = 
 2 \delta_{A}^{\; B} \delta_{\dot{A}}^{\; \dot{B}},
 \;\;\;\;\;\;\;\;\;
 \sigma^{\mu}_{A\dot{A}} \sigma_{\mu B \dot{B}} 
 = 
 2 \varepsilon_{AB} \varepsilon_{\dot{A}\dot{B}},
 \;\;\;\;\;\;\;\;\;
 \bar{\sigma}^{\mu\dot{A}A} \bar{\sigma}_{\mu}^{\dot{B}B} 
 = 
 2 \varepsilon^{\dot{A}\dot{B}} \varepsilon^{AB}.
\eq
Let us now look for plane wave solutions of the Dirac equation.
We make the ansatz
\bq
 \psi(x) & = & 
 \left\{ \begin{array}{llll}
  u(p) e^{-i p x}, & p^0>0, & p^2=m^2, & \mbox{incoming fermion}, \\
  v(p) e^{+i p x}, & p^0>0, & p^2=m^2, & \mbox{outgoing anti-fermion}. \\
  \end{array}
 \right.
\eq
$u(p)$ describes incoming particles, $v(p)$ describes outgoing anti-particles.
Similar,
\bq
 \bar{\psi}(x) & = & 
 \left\{ \begin{array}{llll}
  \bar{u}(p) e^{+i p x}, & p^0>0, & p^2=m^2, & \mbox{outgoing fermion}, \\
  \bar{v}(p) e^{-i p x}, & p^0>0, & p^2=m^2, & \mbox{incoming anti-fermion}, \\
  \end{array}
 \right.
\eq
where
\bq
 \bar{u}(p) = u^\dagger(p) \gamma^0,
 & &
 \bar{v}(p) = v^\dagger(p) \gamma^0.
\eq
$\bar{u}(p)$ describes outgoing particles, $\bar{v}(p)$ describes incoming anti-particles.
Then
\bq 
\label{Dirac_equations}
\left( p\!\!\!/ - m \right) u(p) = 0, & & \left( p\!\!\!/ + m \right) v(p) = 0, \nonumber \\
\bar{u}(p) \left( p\!\!\!/ - m \right) = 0, & & \bar{v}(p) \left( p\!\!\!/ + m \right) = 0,  
\eq
There are two solutions for $u(p)$ (and the other spinors $\bar{u}(p)$, $v(p)$, $\bar{v}(p)$).
We will label the various solutions with $\lambda$.
The degeneracy is related to the additional spin degree of freedom.
We require that the two solutions satisfy 
the orthogonality relations
\bq
\label{spinor_orthogonality}
\bar{u}(p,\bar{\lambda}) u(p,\lambda) & = & 2 m \delta_{\bar{\lambda}\lambda}, \nonumber \\
\bar{v}(p,\bar{\lambda}) v(p,\lambda) & = & -2 m \delta_{\bar{\lambda}\lambda}, \nonumber \\
\bar{u}(p,\bar{\lambda}) v(p,\lambda) & = & \bar{v}(\bar{\lambda}) u(\lambda) \;\; = \;\; 0,
\eq
and the completeness relations
\bq
\label{spinor_completeness_relations}
\sum\limits_{\lambda} u(p,\lambda) \bar{u}(p,\lambda) \;\; = \;\; p\!\!\!/ + m, 
 & &
\sum\limits_{\lambda} v(p,\lambda) \bar{v}(p,\lambda) \;\; = \;\; p\!\!\!/ - m.
\eq

\subsubsection{Massless spinors in the Weyl representation}
\label{sect_massless_spinors}

Let us now try to find explicit solutions for the spinors $u(p)$, $v(p)$, $\bar{u}(p)$ and $\bar{v}(p)$.
The simplest case is the one of a massless fermion:
\bq
 m & = & 0.
\eq
In this case the Dirac equation for the $u$- and the $v$-spinors are identical and it is sufficient to consider
\bq
 p\!\!\!/ u(p) = 0, & & 
\bar{u}(p) p\!\!\!/ = 0.
\eq
In the Weyl representation $p\!\!\!/$ is given by
\bq 
 p\!\!\!/ & = &
 \left(\begin{array}{cc}
 0 & p_\mu \sigma^{\mu} \\
 p_\mu \bar{\sigma}^{\mu} & 0 \\
\end{array} \right),
\eq
therefore the $4 \times 4$-matrix equation for $u(p)$ (or $\bar{u}(p)$) decouples into two
$2 \times 2$-matrix equations.
We introduce the following notation:
Four-component Dirac spinors are constructed out of two {\bf Weyl spinors} as follows:
\bq
\label{Weyl_ket}
u(p) & = & \left(\begin{array}{c} \left| p + \right\rangle \\ \left| p - \right\rangle \\ \end{array} \right) 
       =   \left(\begin{array}{c} \left| p \right\rangle \\ \left| p \right] \\ \end{array} \right)
       =   \left(\begin{array}{c} p_A \\ p^{\dot{B}} \\ \end{array} \right)
       =   \left(\begin{array}{c} u_+(p) \\ u_-(p) \\ \end{array} \right).
\eq
Bra-spinors are given by
\bq
\label{Weyl_bra}
\overline{u}(p) & = & \left( \; \left\langle p - \right|, \; \left\langle p + \right| \; \right)
                  =   \left( \; \left\langle p \right|, \; \left[ p \right| \; \right)
                  =   \left( \; p^A, \; p_{\dot{B}} \; \right)
                  =   \left( \; \bar{u}_-(p), \; \bar{u}_+(p) \; \right).
\eq
In the literature there exists various notations for Weyl spinors.
Eq.~(\ref{Weyl_ket}) and eq.~(\ref{Weyl_bra}) show four of them and the way how to translate from one
notation to another notation.
By a slight abuse of notation we will in the following not distinguish between 
a two-component Weyl spinor and a Dirac spinor, where either the upper two components or the lower two components
are zero.
If we define the chiral projection operators
\bq
P_+ \;\; = \;\; \frac{1}{2} \left( 1 + \gamma_5 \right) 
 = \left( \begin{array}{cc} 1 & 0 \\ 0 & 0 \\ \end{array} \right),
 & &
P_- \;\; = \;\; \frac{1}{2} \left( 1 - \gamma_5 \right) 
 = \left( \begin{array}{cc} 0 & 0 \\ 0 & 1 \\ \end{array} \right),
\eq
then (with the slight abuse of notation mentioned above)
\bq
u_\pm(p) = P_\pm u(p),
 & &
\bar{u}_\pm(p) = \bar{u}(p) P_\mp.
\eq
The two solutions of the Dirac equation
\bq
 p\!\!\!/ u(p,\lambda) & = & 0 
\eq
are then
\bq
 u(p,+) = u_+(p),
 & & 
 u(p,-) = u_-(p).
\eq
We now have to solve
\bq
p_\mu \bar{\sigma}^\mu \left| p+ \right\rangle = 0,
 & &
p_\mu \sigma^\mu \left| p- \right\rangle = 0,
 \nonumber \\
\left\langle p+ \right| p_\mu \bar{\sigma}^\mu  = 0,
 & &
\left\langle p- \right| p_\mu \sigma^\mu  = 0. 
\eq
It it convenient to express the four-vector $p^\mu=(p^0,p^1,p^2,p^3)$ in terms of light-cone
coordinates:
\bq
\label{def_contravariant_lightcone_coordinates}
 p^+=
 \frac{1}{\sqrt{2}}
 \left(p^0+p^3\right), 
 \;\;\;
 p^-=
 \frac{1}{\sqrt{2}}
 \left(p^0-p^3\right), 
 \;\;\; 
 p^\bot=
 \frac{1}{\sqrt{2}}
 \left(p^1+ip^2\right), 
 \;\;\;
 p^{\bot\ast}=
 \frac{1}{\sqrt{2}}
 \left(p^1-ip^2\right).
 \nonumber
\eq
Note that $p^{\bot\ast}$ does not involve a complex conjugation of $p^1$ or $p^2$.
For null-vectors one has
\bq
p^{\bot\ast} p^{\bot} & = & p^+ p^-.
\eq
Then the equation for the ket-spinors becomes
\bq
\left( \begin{array}{cc}
        p^- & -p^{\bot\ast} \\
        -p^\bot & p^+ \\ 
       \end{array}
\right)
 \left| p+ \right\rangle = 0,
 & &
\left( \begin{array}{cc}
        p^+ & p^{\bot\ast} \\
        p^\bot & p^- 
       \end{array}
\right)
 \left| p- \right\rangle = 0,
\eq
and similar equations can be written down for the bra-spinors.
This is a problem of linear algebra.
Solutions for ket-spinors are
\bq
 \left| p+ \right\rangle = p_A =
  c_1
  \left( \begin{array}{c} p^{\bot\ast} \\ p^- \end{array} \right),
 & &
 \left| p- \right\rangle = p^{\dot{A}} =
 c_2
 \left( \begin{array}{c} p^- \\ -p^\bot \end{array} \right),
\eq
with some yet unspecified multiplicative constants $c_1$ and $c_2$.
Solutions for bra-spinors are 
\bq
 \left\langle p+ \right| = p_{\dot{A}} =
 c_3
 \left( p^\bot, p^- \right),
 & &
 \left\langle p- \right| = p^A =
 c_4
 \left( p^-, -p^{\bot\ast} \right),
\eq
with some further constants $c_3$ and $c_4$.
Let us now introduce the 2-dimensional antisymmetric tensor:
\bq
\varepsilon_{AB} = \left(\begin{array}{cc}
 0 & 1\\
 -1 & 0 \\
\end{array} \right),
& &
\varepsilon_{BA} = - \varepsilon_{AB} 
\eq
Furthermore we set
\bq
\varepsilon^{AB} = \varepsilon^{\dot{A}\dot{B}} = \varepsilon_{AB} = \varepsilon_{\dot{A}\dot{B}}.
\eq
Note that these definitions imply
\bq
 \eps^{A C} \eps_{B C} = \delta^{A}_{\;B},
 & &
 \eps^{\dot{A} \dot{C}} \eps_{\dot{B} \dot{C}} = \delta^{\dot{A}}_{\;\dot{B}}.
\eq
We would like to have the following relations for raising and lowering a spinor index $A$ or $\dot{B}$:
\bq
\label{raising_and_lowering_spinor_indices}
 p^A = \varepsilon^{AB} p_B,
  & & 
 p^{\dot{A}} = \varepsilon^{\dot{A}\dot{B}} p_{\dot{B}},
  \nonumber\\
 p_{\dot{B}} = p^{\dot{A}} \varepsilon_{\dot{A}\dot{B}}, 
  & & 
 p_B = p^A \varepsilon_{AB}.
\eq
Note that raising an index is done by left-multiplication, whereas 
lowering is performed by right-multiplication.
Postulating these relations implies
\bq
 c_1 = c_4, & & c_2=c_3.
\eq
In addition we normalise the spinors according to
\bq
\label{normalisation_massless_spinors}
\langle p \pm | \gamma^{\mu} | p \pm \rangle = 2 p^{\mu}.
\eq
This implies
\bq
 c_1 c_3 = \frac{\sqrt{2}}{p^-},
 & &
 c_2 c_4 = \frac{\sqrt{2}}{p^-}.
\eq
Eq.~(\ref{raising_and_lowering_spinor_indices}) and eq.~(\ref{normalisation_massless_spinors})
determine the spinors only up to a scaling
\bq
 p_A \rightarrow \lambda p_A,
 & &
 p_{\dot{A}} \rightarrow \frac{1}{\lambda} p_{\dot{A}}.
\eq
This scaling freedom is referred to as {\bf little group scaling}.
Keeping the scaling freedom, we define the spinors as
\bq
\label{def_spinors}
\left| p+ \right\rangle = p_A = 
  \frac{\lambda_p 2^{\frac{1}{4}}}{\sqrt{p^-}} 
  \left( \begin{array}{c} p^{\bot\ast} \\ p^- \end{array} \right),
 & &
\left| p- \right\rangle = p^{\dot{A}} = 
 \frac{2^{\frac{1}{4}}}{\lambda_p\sqrt{p^-}} 
 \left( \begin{array}{c} p^- \\ -p^\bot \end{array} \right),
 \nonumber \\
\left\langle p+ \right| = p_{\dot{A}} = 
 \frac{2^{\frac{1}{4}}}{\lambda_p\sqrt{p^-}} 
 \left( p^\bot, p^- \right),
 & &
\left\langle p- \right| = p^A =
 \frac{\lambda_p 2^{\frac{1}{4}}}{\sqrt{p^-}} 
 \left( p^-, -p^{\bot\ast} \right).
\eq
Popular choices for $\lambda_p$ are
\bq
 \lambda_p = 1 & : & \mbox{symmetric},
 \nonumber \\
 \lambda_p = 2^{\frac{1}{4}} \sqrt{p^-} & : & \mbox{$p_A$ linear in $p^\mu$},
 \nonumber \\
 \lambda_p = \frac{1}{2^{\frac{1}{4}} \sqrt{p^-}} & : & \mbox{$p_{\dot{A}}$ linear in $p^\mu$}.
\eq
Note that all formul{\ae} in this sub-section~(\ref{sect_massless_spinors}) 
work not only for real momenta $p^\mu$
but also for complex momenta $p^\mu$.
This will be useful later on, where we encounter situations with complex momenta.
However there is one exception:
The relations
$p_A^\dagger = p_{\dot{A}}$ and $p^A{}^\dagger = p^{\dot{A}}$ (or equivalently $\bar{u}(p) = u(p)^\dagger \gamma^0$)
encountered in previous sub-sections are valid only for real momenta $p^\mu=(p^0,p^1,p^2,p^3)$.
If on the other hand the components $(p^0,p^1,p^2,p^3)$ are complex, these relations will in general not hold.
In the latter case $p_A$ and $p_{\dot{A}}$ are considered to be independent quantities.
The reason, why the relations $p_A^\dagger = p_{\dot{A}}$ and $p^A{}^\dagger = p^{\dot{A}}$
do not hold in the complex case lies in the definition of $p^{\bot\ast}$:
We defined $p^{\bot\ast}$ as $p^{\bot\ast}=(p^1-ip^2)/\sqrt{2}$, and not as $(p^1-i(p^2)^\ast)/\sqrt{2}$.
With the former definition $p^{\bot\ast}$ is a holomorphic function of 
$p^1$ and $p^2$.
There are applications where holomorphicity is more important than nice properties under
hermitian conjugation.
\\
\\
\bs
{\it {\bf Exercise \theexercise}: 
The helicity operator $h$ for particles is defined by
\bq
 h & = & \frac{\vec{p} \cdot \vec{S}}{\left|\vec{p}\right|},
\eq
where the components of $\vec{S}=(S_1,S_2,S_3)$ are given by
\bq
 S_i & = & \frac{1}{2} \eps_{ijk} S^{j k}
\eq
with
\bq
 S^{\mu \nu} & = & \frac{i}{4} \left[ \gamma^\mu, \gamma^\nu \right].
\eq
One finds in the Weyl representation
\bq
 h & = &
 \frac{1}{2 \left|\vec{p}\right|}
 \left( \begin{array}{cc}
 \vec{p} \cdot \vec{\sigma} & 0 \\
 0 & \vec{p} \cdot \vec{\sigma} \\
 \end{array} \right).
\eq
(The helicity operator for anti-particles is defined by $h=-\vec{p}\cdot\vec{S}/|\vec{p}|$.)
A particle with eigenvalue $h=+1/2$ is called right-handed, a particle with $h=-1/2$ is called left-handed.
Assume that $p^\mu$ is a real four-vector with positive energy ($p^0>0$).
Show that the spinors $| p+ \rangle = p_A$ and $\langle p+ | = p_{\dot{A}}$ have helicity $h=+1/2$, 
while the spinors $| p- \rangle = p^{\dot{A}}$ and $\langle p- | = p^A$ have helicity $h=-1/2$.
\stepcounter{exercise}
}
\es

\subsubsection{Spinor products}

Let us now make the symmetric choice $\lambda_p=1$.
Spinor products are defined by
\bq
\label{spinor_products}
 \langle p q \rangle 
 & = & 
 \langle p - | q + \rangle 
 = 
 p^A q_A
 =
 \frac{\sqrt{2}}{\sqrt{p^-} \sqrt{q^-}} \left( p^- q^{\bot\ast} - q^- p^{\bot\ast} \right),
 \nonumber \\
 \left[ q p \right] 
 & = & 
 \langle q + | p - \rangle 
 = 
 q_{\dot{A}} p^{\dot{A}}
 =
 \frac{\sqrt{2}}{\sqrt{p^-} \sqrt{q^-}} \left( p^- q^\bot - q^- p^\bot \right),
\eq 
where the last expression in each line used the choice $\lambda_p=\lambda_q=1$.
We have
\bq
 \langle p q \rangle \left[ q p \right]
 & = & 
 2 p q.
\eq
If $p^\mu$ and $q^\mu$ are real we have 
\bq
\left[ q p \right] 
 & = & \left\langle p q \right\rangle^\ast \; \mathrm{sign}(p^0) \; \mathrm{sign}(q^0).
\eq
The spinor products are anti-symmetric
\bq
 \langle q p \rangle = - \langle p q \rangle,
 & &
 [ p q ] = - [ q p ].
\eq
From the Schouten identity for the 2-dimensional antisymmetric tensor 
\bq
 \eps_{A B} \eps_{C D}
 +
 \eps_{B C} \eps_{A D}
 +
 \eps_{C A} \eps_{B D}
 & = &
 0.
\eq
one derives
\bq
 \langle p_1 p_2 \rangle \langle p_3 p_4 \rangle
 +
 \langle p_2 p_3 \rangle \langle p_1 p_4 \rangle
 +
 \langle p_3 p_1 \rangle \langle p_2 p_4 \rangle
 & = & 0,
 \nonumber \\
 \left[ p_1 p_2 \right] \left[ p_3 p_4 \right]
 +
 \left[ p_2 p_3 \right] \left[ p_1 p_4 \right]
 +
 \left[ p_3 p_1 \right] \left[ p_2 p_4 \right]
 & = & 0.
\eq
The Fierz identity reads
\bq
 \langle p_1+ | \gamma_{\mu} | p_2+ \rangle \langle p_3- | \gamma^{\mu} | p_4- \rangle 
 & = &
 2 [ p_1 p_4 ] \langle p_3 p_2 \rangle.
\eq
Note that with our slight abuse of notation we identify a two-component Weyl spinor
with a Dirac spinor, where the other two components are zero.
Therefore
\bq
 \langle p_1+ | \gamma_{\mu} | p_2+ \rangle
 \;\; = \;\;
 \langle p_1+ | \bar{\sigma}_{\mu} | p_2+ \rangle,
 & &
 \langle p_3- | \gamma^{\mu} | p_4- \rangle
 \;\, = \;\;
 \langle p_3- | \sigma^{\mu} | p_4- \rangle.
\eq
We further have the reflection identities
\bq
 \langle p\pm|\gamma^{\mu_1} ... \gamma^{\mu_{2n+1}} | q\pm \rangle 
 & = &
 \langle q\mp|\gamma^{\mu_{2n+1}} ... \gamma^{\mu_1} | p\mp \rangle,
 \nonumber \\
 \langle p\pm|\gamma^{\mu_1} ... \gamma^{\mu_{2n}} | q\mp \rangle 
 & = &
 - \langle q\pm|\gamma^{\mu_{2n}} ... \gamma^{\mu_1} | p\mp \rangle.
\eq
\bs
{\it {\bf Exercise \theexercise}: The field strength in spinor notation:
For a rank-$2$ tensor $F_{\mu\nu}$ we define the spinor representation by
\bq
 F_{A B \dot{A} \dot{B}} & = & F_{\mu\nu} \sigma^\mu_{A \dot{A}} \sigma^\nu_{B \dot{B}}.
\eq
On the other hand we may decompose $F_{\mu\nu}$ into a self dual part and an anti-self dual part,
\bq
 F_{\mu\nu} & = & F_{\mu\nu}^{\mathrm{self \; dual}} + F_{\mu\nu}^{\mathrm{anti-self \; dual}},
\eq
with
\bq
 F_{\mu\nu}^{\mathrm{self \; dual}}
 \;\; = \;\;
 \frac{1}{2} \left( F_{\mu\nu} + i \tilde{F}_{\mu\nu} \right),
 & &
 F_{\mu\nu}^{\mathrm{anti-self \; dual}}
 \;\; = \;\;
 \frac{1}{2} \left( F_{\mu\nu} - i \tilde{F}_{\mu\nu} \right).
\eq
The dual field strength $\tilde{F}_{\mu\nu}$ was defined in eq.~(\ref{def_dual_field_strength}).
Show that the spinor representations of the self dual part / anti-self dual part may be written as
\bq
 F_{A B \dot{A} \dot{B}}^{\mathrm{self \; dual}}
 \;\; = \;\;
 \eps_{A B} \bar{\phi}_{\dot{A}\dot{B}},
 & &
 F_{A B \dot{A} \dot{B}}^{\mathrm{anti-self \; dual}}
 \;\; = \;\;
 \phi_{AB} \eps_{\dot{A} \dot{B}},
\eq
where $\phi_{AB}$ and $\bar{\phi}_{\dot{A}\dot{B}}$ satisfy
\bq
 \phi_{AB} \;\; = \;\; \phi_{BA},
 & &
 \bar{\phi}_{\dot{A}\dot{B}} \;\; = \;\; \bar{\phi}_{\dot{B}\dot{A}}.
\eq
\stepcounter{exercise}
}
\es

\subsubsection{Polarisation vectors}

It was a major break-through,
when it was realised that also gluon polarisation vectors can be expressed in terms of two-component 
Weyl spinors \cite{Berends:1981rb,DeCausmaecker:1982bg,Gunion:1985vc,Kleiss:1986qc,Xu:1987xb,Gastmans_book}.
The polarisation vectors of external gluons can be chosen as
\bq
\label{def_polarisation_vectors}
\eps_{\mu}^{+}(p,q) = - \frac{\langle p+|\gamma_{\mu}|q+\rangle}{\sqrt{2} \langle p- | q + \rangle},
 & &
\eps_{\mu}^{-}(p,q) = \frac{\langle p-|\gamma_{\mu}|q-\rangle}{\sqrt{2} \langle p + | q - \rangle},
\eq
where $p$ is the momentum of the gluon and $q$ is an arbitrary light-like reference momentum.
The dependence on the reference spinors $| q+ \rangle$ and $| q- \rangle$ which enters through the gluon polarisation vectors
will drop out in gauge invariant quantities.
We have the relations
\bq
 \eps_{\mu}^{\pm}(p,q) p^\mu = 0,
  \;\;\;\;
 \eps_{\mu}^{\pm}(p,q) q^\mu = 0,
  \;\;\;\;
 \eps_{\mu}^{+}(p_1,q) \eps^{+ \mu}(p_2,q) = 0,
  \;\;\;\;
 \eps_{\mu}^{-}(p_1,q) \eps^{- \mu}(p_2,q) = 0.
\eq
The polarisation sum is
\bq
\sum\limits_{\lambda \in \{+,-\}} \varepsilon_\mu^{\lambda}(p,q) \varepsilon_\nu^{-\lambda}(p,q) 
 & = & 
 - g_{\mu \nu} + 2 \frac{p_\mu q_\nu + q_\mu p_\nu }{2 p q}.
\eq
If $p^\mu$ and $q^\mu$ are real we have
\bq
 \left( \eps_\mu^\lambda(p,q) \right)^\ast & = & \eps_\mu^{-\lambda}(p,q).
\eq
Changing the reference momentum will give a term proportional to the
momentum of the gluon:
\bq
 \eps^+_\mu(p,q_1) -\eps^+_\mu(p,q_2)
 & = & 
 \sqrt{2} \frac{\langle q_1 q_2 \rangle}{\langle q_1 p \rangle \langle p q_2 \rangle} p_\mu.
\eq
Eq.~(\ref{def_polarisation_vectors}) provides the desired explicit expression for the polarisation vectors.
For each external particle $i$ we may choose a light-like reference momentum $q_i$. 
When calculating gauge-invariant quantities like a primitive amplitude the reference momenta
$q_i$ and $q_j$ of two different external legs need not be equal.
When calculating several gauge-invariant quantities we may change the reference momentum $q_i$
of the external leg $i$ from one calculation to the other.
However we may not change the reference momentum $q_i$
of the external leg $i$ within one calculation of a gauge-invariant quantity.
Let $\lambda = (\lambda_1,...,\lambda_n)$ be a helicity configuration. We denote the primitive
helicity amplitude with cyclic order $(1,...,n)$ by
\bq
 A_n^{(0)}\left( 1^{\lambda_1}, 2^{\lambda_2}, ..., n^{\lambda_n} \right).
\eq
For the spin- and colour summed matrix element squared we have then
\bq
\label{matrix_element_squared_to_primitive_amplitudes}
\lefteqn{
 \sum\limits_{\mathrm{spins,colour}} 
 \left| {\mathcal A}_n \right|^2
 = 
 g^{2n-4} 
 \sum\limits_{\sigma \in S_{n}/Z_{n}} 
 \sum\limits_{\pi \in S_{n}/Z_{n}} 
 \;\;\;
 \sum\limits_{\lambda_1 \in \{+,-\}} ... \sum\limits_{\lambda_n \in \{+,-\}}
 } & & \\
 & &
 4 \; \mathrm{Tr} \left( T^{a_{\sigma_1}} ... T^{a_{\sigma_n}} \right)
 \; \mathrm{Tr} \left( T^{a_{\pi_n}} ... T^{a_{\pi_1}} \right)
 A_{n}^{(0)}\left( \sigma_1^{\lambda_{\sigma_1}}, ..., \sigma_n^{\lambda_{\sigma_n}} \right)
 \;\;
 A_{n}^{(0)}\left( \pi_1^{\lambda_{\pi_1}}, ..., \pi_n^{\lambda_{\pi_n}} \right)^\ast.
 \nonumber
\eq
Let us shortly discuss the singular limit of cyclic-ordered helicity amplitudes.
In the limit where one gluon $j$ becomes soft, the primitive amplitudes behave as
\bq
 \lim\limits_{p_j \rightarrow 0}
 A_n(p_1,...,p_j^+,...,p_n) 
 & = & 
  \sqrt{2} \frac{\langle p_{j-1} p_{j+1} \rangle}{\langle p_{j-1} p_j \rangle \langle p_j p_{j+1} \rangle} 
  A_{n-1}(p_1,...,p_{j-1},p_{j+1},...,p_n), 
 \nonumber \\
 \lim\limits_{p_j \rightarrow 0}
 A_n(p_1,...,p_j^-,...,p_n) 
 & = &
  \sqrt{2} \frac{[ p_{j+1} p_{j-1} ]}{[ p_{j+1} p_j ] [ p_j p_{j-1} ]} 
  A_{n-1}(p_1,...,p_{j-1},p_{j+1},...,p_n).
\eq
For a singular collinear limit the two collinear particles have to be {\bf adjacent} in the cyclic order.
Let us assume that the particles $i$ and $i+1$ become collinear as in eq.~(\ref{parametrisation_collinear}).
We then have
\bq
 \lim\limits_{p_i || p_{i+1}}
 A_n^{(0)}\left(...,p_i^{\lambda_i},p_{i+1}^{\lambda_{i+1}},...\right) 
 & = &
 \sum\limits_{\lambda_{\tilde{i}}} 
 \; 
 \mbox{Split}(p_{\tilde{i}},p_i,p_j,\lambda_{\tilde{i}},\lambda_i,\lambda_j) 
 \; 
 A_{n-1}^{(0)}\left(...,p_{\tilde{i}}^{\lambda_{\tilde{i}}},...\right).
\eq
For a singular contribution from the factorisation on propagator poles we have to divide
the external particles into two {\bf consecutive} subsets. Without loss of generality we may take
these subsets as $I_m=\{1,2,...,m\}$ and $J_{n-m}=\{m+1,m+2,...,n\}$.
We set $p=p_1+p_2+...,p_m$. In the limit, where $p^2$ vanishes we have
\bq
\label{factorisation_propagator_pole}
 \lim\limits_{p^2 \rightarrow 0}
 A_{n}^{(0)}\left(p_1^{\lambda_1},...,p_n^{\lambda_n}\right)
 =  
 \sum\limits_{\lambda}
 A_{m+1}^{(0)}\left(p_1^{\lambda_1},...,p_m^{\lambda_m},-p^{-\lambda}\right)
 \frac{i}{p^2}
 A_{n-m+1}^{(0)}\left(p^{\lambda},p_{m+1}^{\lambda_{m+1}},...,p_n^{\lambda_n}\right).
\eq
\bs
{\it {\bf Exercise \theexercise}: 
Calculate the primitive helicity amplitude $A_4^{(0)}(1^-,2^+,3^+,4^-)$.
\stepcounter{exercise}
}
\es

\subsubsection{Massive spinors}
\label{sect_massive_spinors}

As in the massless case, a massive spinor satisfying the Dirac equation has
a two-fold degeneracy. We will label the two different eigenvectors by ``+'' and
``-''.
Let $p$ be a massive four-vector with $p^2=m^2$, and let $q$ be an arbitrary
light-like four-vector.
With the help of $q$ we can construct a light-like vector $p^\flat$ associated to $p$:
\bq
 p^\flat & = & p - \frac{p^2}{2 p \cdot q} q.
\eq
We define \cite{Kleiss:1986qc,Schwinn:2005pi,Rodrigo:2005eu}
\bq
\label{def_u_spinor}
u(p,+) = \frac{1}{\langle p^\flat + | q - \rangle} \left( p\!\!\!/ + m \right) | q - \rangle,
& & 
v(p,-) = \frac{1}{\langle p^\flat + | q - \rangle} \left( p\!\!\!/ - m \right) | q - \rangle, \nonumber \\
u(p,-) = \frac{1}{\langle p^\flat - | q + \rangle} \left( p\!\!\!/ + m \right) | q + \rangle,
& & 
v(p,+) = \frac{1}{\langle p^\flat - | q + \rangle} \left( p\!\!\!/ - m \right) | q + \rangle.
\eq
For the conjugate spinors we have
\bq
\label{def_ubar_spinor}
\bar{u}(p,+) = \frac{1}{\langle q - | p^\flat + \rangle} \langle q - | \left( p\!\!\!/ + m \right), 
& & 
\bar{v}(p,-) = \frac{1}{\langle q - | p^\flat + \rangle}\langle q - | \left( p\!\!\!/ - m \right), \nonumber \\
\bar{u}(p,-) = \frac{1}{\langle q + | p^\flat - \rangle} \langle q + | \left( p\!\!\!/ + m \right),
& & 
\bar{v}(p,+) = \frac{1}{\langle q + | p^\flat - \rangle} \langle q + | \left( p\!\!\!/ - m \right). 
\eq
These spinors satisfy the Dirac equations of eq.~(\ref{Dirac_equations}),
the orthogonality relations of eq.~(\ref{spinor_orthogonality})
and the completeness relations of eq.~(\ref{spinor_completeness_relations}).
We further have
\bq
 \bar{u}(p,\bar{\lambda}) \gamma^\mu u(p,\lambda) = 2 p^\mu \delta_{\bar{\lambda} \lambda},
 & &
 \bar{v}(p,\bar{\lambda}) \gamma^\mu v(p,\lambda) = 2 p^\mu \delta_{\bar{\lambda} \lambda}.
\eq
In the massless limit the definition reduces to
\bq
 u(p,+) = v(p,-) = | p+ \rangle,
 &&
 \bar{u}(p,+) = \bar{v}(p,-) = \langle p+ |, 
 \nonumber \\
 u(p,-) = v(p,+) = | p- \rangle,
 & &
 \bar{u}(p,-) = \bar{v}(p,+) = \langle p- |,
\eq
and the spinors are independent of the reference spinors $|q+\rangle$ and $\langle q+|$.
\\
\\
\bs
{\it {\bf Exercise \theexercise}: 
Show that the spinors defined in eq.~(\ref{def_u_spinor}) and in eq.~(\ref{def_ubar_spinor})
satisfy the Dirac equations of eq.~(\ref{Dirac_equations}),
the orthogonality relations of eq.~(\ref{spinor_orthogonality})
and the completeness relations of eq.~(\ref{spinor_completeness_relations}).
\stepcounter{exercise}
}
\es

\subsubsection{The Majorana representation}

The Weyl spinors involve the light-cone coordinates 
$p^\bot=(p^1+ip^2)/\sqrt{2}$ and 
$p^{\bot\ast}=(p^1-ip^2)/\sqrt{2}$.
For real momenta $p^\mu=(p^0,p^1,p^2,p^3)$ the light-cone coordinates are in general complex numbers.
When we compute helicity amplitudes numerically this forces us to carry out all operations (like addition, multiplication)
with complex numbers.
It would be faster, if we could devise a method, which allows us to carry out a significant fraction
of the operations with real numbers only.
This can be achieved in the Majorana representation.
The Majorana representation is obtained from the Weyl representation by a unitary transformation
\bq
 \gamma^\mu_{\mathrm{Majorana}}
 & = & 
 U \gamma^\mu_{\mathrm{Weyl}} U^\dagger,
\eq
where
\bq
 U
 & = &
 \frac{1}{2}
 \left( \begin{array}{cc}
 1 + \sigma_2 & -i\left(1-\sigma_2\right) \\
 i\left(1-\sigma_2\right) & 1 + \sigma_2 \\
 \end{array} \right).
\eq
The unitary matrix $U$ has the additional property of being hermitian as well. Thus we have
\bq
 U^\dagger
 \;\; = \;\;
 U^{-1}
 \;\; = \;\;
 U.
\eq
Let us now write down the Dirac matrices in the Majorana representation explicitly.
We have
\bq
 \gamma^0_{\mathrm{Majorana}}
 \;\; = \;\;
 i
 \left( \begin{array}{rrrr}
 0 & 0 & 0 & -1 \\
 0 & 0 & 1 & 0 \\
 0 & -1 & 0 & 0 \\
 1 & 0 & 0 & 0 \\
 \end{array} \right),
 & &
 \gamma^1_{\mathrm{Majorana}}
 \;\; = \;\;
 i
 \left( \begin{array}{rrrr}
 0 & -1 & 0 & 0 \\
 -1 & 0 & 0 & 0 \\
 0 & 0 & 0 & 1 \\
 0 & 0 & 1 & 0 \\
 \end{array} \right),
 \nonumber \\
 \gamma^2_{\mathrm{Majorana}}
 \;\; = \;\;
 i
 \left( \begin{array}{rrrr}
 0 & 0 & 0 & 1 \\
 0 & 0 & -1 & 0 \\
 0 & -1 & 0 & 0 \\
 1 & 0 & 0 & 0 \\
 \end{array} \right),
 & &
 \gamma^3_{\mathrm{Majorana}}
 \;\; = \;\;
 i
 \left( \begin{array}{rrrr}
 -1 & 0 & 0 & 0 \\
 0 & 1 & 0 & 0 \\
 0 & 0 & -1 & 0 \\
 0 & 0 & 0 & 1 \\
 \end{array} \right).
\eq
The matrix $\gamma_5$ is given in the Majorana representation by
\bq
 \gamma_5^{\mathrm{Majorana}}
 & = &
 i
 \left( \begin{array}{rrrr}
 0 & -1 & 0 & 0 \\
 1 & 0 & 0 & 0 \\
 0 & 0 & 0 & 1 \\
 0 & 0 & -1 & 0 \\
 \end{array} \right).
\eq
Note that all matrices are purely imaginary.
Let us now consider massless spinors in the Majorana representation. In general we have
\bq
 \bar{u}_{\mathrm{Majorana}} = \bar{u}_{\mathrm{Weyl}} U^\dagger,
 & &
 u_{\mathrm{Majorana}} = U u_{\mathrm{Weyl}}.
\eq
As a basis for the two independent helicity states we could choose $u_+$ and $u_-$. However, it will be more
convenient to choose a basis given by
\bq
 \left( \bar{u}_a, \bar{u}_b \right) 
 \;\; = \;\; 
 \left( \bar{u}_+, \bar{u}_- \right) \cdot S^\dagger,
 & &
 \left( \begin{array}{c}
 u_a \\ u_b \\
 \end{array} \right)
 \;\; = \;\;
 S \cdot 
 \left( \begin{array}{c}
 u_+ \\ u_- \\
 \end{array} \right),
\eq
where
\bq
 S 
 & = & 
 \frac{1}{\sqrt{2}}
 \left( \begin{array}{rr}
 1 & 1 \\
 -i & i \\
 \end{array} \right).
\eq
The matrix $S$ is unitary:
\bq
 S^{-1}
 & = &
 \frac{1}{\sqrt{2}}
 \left( \begin{array}{rr}
 1 & i \\
 1 & -i \\
 \end{array} \right)
 \;\; = \;\; S^\dagger.
\eq
We then obtain (for $\lambda_p=1$)
\bq
 u_a^{\mathrm{Majorana}}
 \;\; = \;\; 
 \frac{2^{\frac{1}{4}} i }{2 \sqrt{p^-}}
 \left( \begin{array}{c}
 -p^0 - p^2 + p^3 \\
 p^1 \\
 p^1 \\
 p^0 - p^2 - p^3 \\
 \end{array} \right),
 & &
 u_b^{\mathrm{Majorana}}
 \;\; = \;\; 
 \frac{2^{\frac{1}{4}} i }{2 \sqrt{p^-}}
 \left( \begin{array}{c}
 -p^1 \\
 -p^0 - p^2 + p^3 \\
 p^0 - p^2 - p^3 \\
 -p^1 \\
 \end{array} \right),
 \nonumber
\eq
For the $\bar{u}$-spinors we find
\bq
 \bar{u}_a^{\mathrm{Majorana}}
 & = &
 \frac{2^{\frac{1}{4}} }{2 \sqrt{p^-}}
 \left( p^0 - p^2 - p^3, -p^1, p^1, p^0 + p^2 - p^3 \right),
 \nonumber \\
 \bar{u}_b^{\mathrm{Majorana}}
 & = &
 \frac{2^{\frac{1}{4}} }{2 \sqrt{p^-}}
 \left( -p^1, -p^0 + p^2 + p^3, -p^0 - p^2 + p^3, p^1 \right).
\eq
We recall that in the Majorana representation the Dirac matrices are purely imaginary.
In addition we may choose a spinor basis, where the $u$-spinors are purely imaginary and the $\bar{u}$-spinors are real.
The imaginary unit occurs only as a prefactor. We may separate this prefactor from the rest, perform numerically the
calculation with real numbers and multiply by the appropriate power of imaginary units $i$ in the end.
In this way the number of operations with complex numbers is reduced significantly.

\subsection{Twistors}

Twistors are closely related to spinors.
Although this is a more formal topic and not related to efficiency issues,
it is best discussed together with spinors.
We first delve directly into twistors in momentum space and the way they are used in connection with scattering amplitudes.
We then discuss twistors in position space and the twistor transform.

\subsubsection{Momentum twistors}

Let $p=(p_1,...,p_n)$ be a momentum configuration for the amplitude $A_n^{(0)}$. 
In physical applications all momenta are real, but nothing stops us to consider
the amplitude $A_n^{(0)}$ as a function of complex momenta $p_j^\mu$.
A momentum configuration is a set of $n$ real or complex four-vectors, 
subject to the constraints of momentum conservation
\bq
 p_1 + p_2 + ... + p_n & = & 0
\eq
and the on-shell conditions 
\bq
 p_j^2 & = & 0,
 \;\;\;\;\;\;
 j \in \{1,...,n\}.
\eq
Thus there are only $(3n-4)$ independent variables.
It is sometimes convenient to have a parametrisation of the momentum configurations of an $n$-point amplitude
free of constraints.
Momentum twistors allow us to do that.
Let us first focus on the on-shell conditions.
The $\sigma^\mu$- and $\bar{\sigma}^\mu$-matrices from the Weyl representation
can be used to convert any four-vector to a bispinor representation:
\bq
 p^{\dot{A}B} = p_{\mu} \bar{\sigma}^{\mu \dot{A} B} & \mbox{or} &
 p_{A\dot{B}} = p_{\mu} \sigma^{\mu}_{A\dot{B}}.
\eq
The inverse equations read
\bq
 p^\mu = \frac{1}{2} p^{\dot{A}B} \sigma^{\mu}_{B\dot{A}},
 & &
 p^\mu = \frac{1}{2} p_{A\dot{B}} \bar{\sigma}^{\mu \dot{B} A}.
\eq
If $p$ is light-like, the bispinor factorises into a dyad 
\bq
 p^{\dot{A}B}=p^{\dot{A}}p^{B},
 & &
 p_{A\dot{B}}=p_{A}p_{\dot{B}}
\eq
and we have
\bq
\label{fourvector_from_spinor}
 p^\mu & = &
 \frac{1}{2} p^{A} \sigma^{\mu}_{A\dot{B}} p^{\dot{B}} 
 =
 \frac{1}{2} p_{\dot{A}} \bar{\sigma}^{\mu \dot{A} B} p_{B}.
\eq 
Thus giving for each external particle a pair of spinors $p_A=|p+\rangle$ and $p_{\dot{A}}=\langle p+ |$
allow us to reconstruct the four-vector $p^\mu$ with the help of eq.~(\ref{fourvector_from_spinor}).
Furthermore, this four-vector will be automatically on-shell (i.e. light-like).
For an arbitrary pair of spinors $|p+\rangle$ and $\langle p+ |$ the resulting four-vector will be in general
complex.
We see that the use of spinor variables $(p_a,p_{\dot{A}})=(|p+ \rangle, \langle p+ |)$ for each external leg
trivialises the on-shell constraints.

We still have to think about the constraint of momentum conservation.
Here, momentum twistors enter the game \cite{Hodges:2005bf,Hodges:2005aj,Hodges:2009hk,Mason:2009sa}.
We denote a {\bf momentum twistor} by
\bq
 Z_\alpha & = & \left( p_A, \mu_{\dot{A}} \right) 
 = \left( |p+\rangle, \langle \mu+ | \right).
\eq
The index $\alpha$ takes the values $\alpha \in \{ 1, 2, \dot{1}, \dot{2} \}$.
The scaling behaviour of a momentum twistor is
\bq
 \left( p_A, \mu_{\dot{A}} \right) & \rightarrow & \left( \lambda p_A, \lambda \mu_{\dot{A}} \right),
\eq
and therefore $Z_\alpha \in {\mathbb C}{\mathbb P}^3$.
As with momentum vectors, we will not always write the index $\alpha$ explicitly.
Let us now consider $n$ momentum twistors $Z_1$, $Z_2$, ..., $Z_n$.
The ordered set $(Z_1, Z_2, ..., Z_n)$ defines a configuration of $n$ momentum vectors with associated spinors as
follows:
We first define the spinors by
\bq
\label{bra_ket_spinors_from_twistor}
 \left| p_i + \right\rangle & = & \left| p_i + \right\rangle,
 \\
 \left\langle p_i + \right|
 & = &
 - \frac{\left\langle p_i p_{i+1} \right\rangle}{\left\langle p_{i-1} p_{i} \right\rangle\left\langle p_i p_{i+1} \right\rangle}
   \left\langle \mu_{i-1} + \right|
 - \frac{\left\langle p_{i+1} p_{i-1} \right\rangle}{\left\langle p_{i-1} p_{i} \right\rangle\left\langle p_i p_{i+1} \right\rangle}
   \left\langle \mu_{i} + \right|
 - \frac{\left\langle p_{i-1} p_{i} \right\rangle}{\left\langle p_{i-1} p_{i} \right\rangle\left\langle p_i p_{i+1} \right\rangle}
   \left\langle \mu_{i+1} + \right|,
 \nonumber
\eq
where indices are understood modulo $n$.
The momentum vector $p_i^\mu$ is then given by
\bq
\label{fourvector_from_bra_ket_spinors}
 p_i^\mu & = & \frac{1}{2} \left\langle p_i+ \left| \bar{\sigma}^\mu \right| p_i+ \right\rangle.
\eq
Each momentum vector $p_i^\mu$ is massless:
\bq
 p_i^2 & = & 0.
\eq
The configuration of $n$ momentum vectors satisfies momentum conservation:
\bq
 \sum\limits_{i=1}^n p_i^\mu & = & 0.
\eq 
We see that the use of momentum twistors trivialises the on-shell constraints and the constraint from momentum
conservation.
Therefore we may start from a momentum twistor configuration $(Z_1,...,Z_n)$, where all momentum twistor variables can be chosen
freely without any constraints.
We then obtain the associated momentum configuration with the help of eq.~(\ref{bra_ket_spinors_from_twistor})
and eq.~(\ref{fourvector_from_bra_ket_spinors}).
This momentum configuration will automatically satisfy momentum conservation and the on-shell conditions.

Let us look into this construction in more detail:
Starting from an ordered set $(p_1,p_2,...,p_n)$ of light-like four-vectors satisfying 
momentum conservation $p_1+p_2+...+p_n=0$ and an arbitrary four-vector $Q$ we define
\begin{figure}
\begin{center}
\includegraphics[scale=1.0]{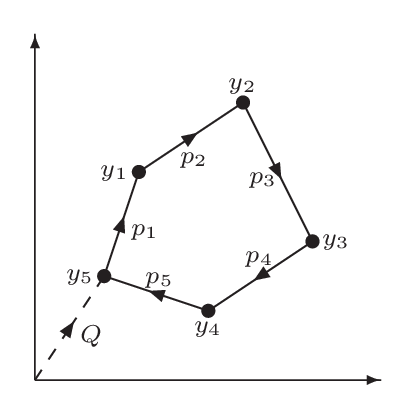}
\caption{\label{fig_y_space} 
Relation between 
the ordered set of $n$ four-vectors $(y_1,...,y_n)$
and 
the ordered set of $n$ four-vectors $(p_1,...,p_n)$ constrained by $p_1+...+p_n=0$ plus an additional four-vector $Q$.
Given $(y_1,...,y_n)$ we obtain $Q$ as $y_n$ and the $p_i$'s as $y_i-y_{i_1}$.
Given the $(p_1,...,p_n)$ alone we cannot construct the $(y_1,...,y_n)$, we need in addition the four-vector $Q$.
}
\end{center}
\end{figure}
four-vectors $y_j$ by
\bq
 y_j & = & Q + \sum\limits_{i=1}^j p_i.
\eq
In the reverse direction we have
\bq
 p_i = y_i - y_{i-1},
 & &
 Q = y_n.
\eq
The geometric situation is shown in fig.~(\ref{fig_y_space}).
We then define
\bq
 \left\langle \mu_i+ \right|
 & = &
 \left\langle p_i- \right| y\!\!\!\!/_i
 =
 \left\langle p_i- \right| y\!\!\!\!/_{i-1}.
\eq
In the reverse direction we have given 
$|p_i+\rangle$ and $\langle \mu_i+ |$
\bq
 \left| p_{i+1}+ \right\rangle \left\langle \mu_i+ \right| - \left| p_{i}+ \right\rangle \left\langle \mu_{i+1}+ \right|
 = 
 \left(  \left| p_{i+1}+ \right\rangle \left\langle p_i- \right| - \left| p_{i}+ \right\rangle \left\langle p_{i+1}- \right| \right)
 y\!\!\!\!/_i
 =
 \left\langle p_i p_{i+1} \right\rangle y\!\!\!\!/_i,
 \nonumber \\
\eq
and therefore
\bq
 y\!\!\!\!/_i
 & = &
 \frac{1}{\left\langle p_i p_{i+1} \right\rangle}
 \left(  \left| p_{i+1}+ \right\rangle \left\langle \mu_i+ \right| - \left| p_{i}+ \right\rangle \left\langle \mu_{i+1}+ \right| \right).
\eq
From
\bq
 p\!\!\!/_i & = & y\!\!\!\!/_i- y\!\!\!\!/_{i-1},
\eq
and writing the resulting expression as a dyad one finds the expression given in eq.~(\ref{bra_ket_spinors_from_twistor})
for $\langle p_i + |$ in terms of
$|p_i+\rangle$ and $\langle \mu_i+ |$.
\\
\\
For momentum twistors one defines a {\bf four-bracket} through
\bq
 \left\langle Z_1, Z_2, Z_3, Z_4 \right\rangle
 & = &
 \eps^{\alpha\beta\gamma\delta} Z_{1,\alpha} Z_{2,\beta} Z_{3,\gamma} Z_{4,\delta}
 \nonumber \\
 & = &
 - \left\langle p_1 p_2 \right\rangle \left[ \mu_3, \mu_4 \right]
 - \left\langle p_1 p_3 \right\rangle \left[ \mu_4, \mu_2 \right]
 - \left\langle p_1 p_4 \right\rangle \left[ \mu_2, \mu_3 \right]
 \nonumber \\
 & &
 - \left\langle p_3 p_4 \right\rangle \left[ \mu_1, \mu_2 \right]
 - \left\langle p_4 p_2 \right\rangle \left[ \mu_1, \mu_3 \right]
 - \left\langle p_2 p_3 \right\rangle \left[ \mu_1, \mu_4 \right],
\eq
with $\eps^{12\dot{1}\dot{2}}=1$.
The four-bracket equals the determinant of the $4 \times 4$-matrix, where
the rows (or equivalently the columns) are given by the momentum twistors $Z_j$:
\bq
 \left\langle Z_1, Z_2, Z_3, Z_4 \right\rangle
 & = &
 \det\left( Z_1, Z_2, Z_3, Z_4 \right).
\eq
Note that the spinor products $\langle p q \rangle$ and $[q p]$ can also be viewed
as determinants of $2 \times 2$-matrices:
\bq
 \left\langle p q \right\rangle 
 & = &
 - \eps^{AB} p_{A} q_{B}
 \;\; = \;\;
 - \det\left(p_A,q_B\right),
 \nonumber \\
 \left[ q p \right]
 & = &
 \eps^{\dot{A}\dot{B}} q_{\dot{A}} p_{\dot{B}}
 \;\; = \;\;
 \det\left( q_{\dot{A}}, p_{\dot{B}} \right).
\eq
We defined the momentum twistors as
\bq
 Z_\alpha & = & \left( p_A, \mu_{\dot{A}} \right) 
 = \left( |p+\rangle, \langle \mu+ | \right),
\eq
which scales as 
$( p_A, \mu_{\dot{A}} ) \rightarrow ( \lambda p_A, \lambda \mu_{\dot{A}} )$.
Of course there is a second possibility, transforming $p_A$ and keeping $p_{\dot{A}}$:
\bq
 W_\alpha & = & \left( \mu_A, p_{\dot{A}} \right) 
 = \left( |\mu+\rangle, \langle p+ | \right),
\eq
To recover the spinors we now have
\bq
\label{bra_ket_spinors_from_W_twistor}
 \left| p_i + \right\rangle
 & = &
 - \frac{\left[ p_{i+1} p_i \right]}{ \left[p_{i+1} p_i \right] \left[ p_{i} p_{i-1} \right]}
   \left| \mu_{i-1} + \right\rangle
 - \frac{\left[ p_{i-1} p_{i+1} \right]}{\left[ p_{i+1} p_i \right] \left[ p_{i} p_{i-1} \right]}
   \left| \mu_{i} + \right\rangle
 - \frac{\left[ p_{i} p_{i-1} \right]}{\left[ p_{i+1} p_i \right] \left[ p_{i} p_{i-1} \right]}
   \left| \mu_{i+1} + \right\rangle,
 \nonumber \\
 \left\langle p_i + \right| & = & \left\langle p_i + \right|.
\eq
The $W_\alpha$-momentum twistor scales as
\bq
 \left( \mu_A, p_{\dot{A}} \right) & \rightarrow & \left( \lambda^{-1} \mu_A, \lambda^{-1} p_{\dot{A}} \right).
\eq
For the four-bracket we now have
\bq
 \left\langle W_1, W_2, W_3, W_4 \right\rangle
 & = &
 \eps^{\alpha\beta\gamma\delta} W_{1,\alpha} W_{2,\beta} W_{3,\gamma} W_{4,\delta}
 \nonumber \\
 & = &
 - \left\langle \mu_1 \mu_2 \right\rangle \left[ p_3, p_4 \right]
 - \left\langle \mu_1 \mu_3 \right\rangle \left[ p_4, p_2 \right]
 - \left\langle \mu_1 \mu_4 \right\rangle \left[ p_2, p_3 \right]
 \nonumber \\
 & &
 - \left\langle \mu_3 \mu_4 \right\rangle \left[ p_1, p_2 \right]
 - \left\langle \mu_4 \mu_2 \right\rangle \left[ p_1, p_3 \right]
 - \left\langle \mu_2 \mu_3 \right\rangle \left[ p_1, p_4 \right].
\eq
\bs
{\it {\bf Exercise \theexercise}: 
Show that the momenta defined by eq.~(\ref{bra_ket_spinors_from_W_twistor}) satisfy momentum conservation.
\stepcounter{exercise}
}
\es

\subsubsection{The twistor transform}

Let us do a small excursion and discuss the original ideas of the twistor programme 
initiated by R. Penrose.
This paragraph is more mathematical. Readers not familiar with Grassmannians or flag varieties may skip
this paragraph in a first reading.
We will present a definition of Grassmannians and flag varieties in section~(\ref{sect:grassmannians}).
You may look up the definitions there.
The rest of this report will not depend on this paragraph.

In the previous paragraph we considered momentum twistors.
Originally, twistors were introduced in position space with the aim of 
relating residues of functions in projective twistor space to solutions of free field equations.
This relation is provided by the twistor transform.

Let us now review this construction \cite{Eastwood:1981jy,Penrose:1972ia}.
The twistor transform starts with a double fibration of a space $F$ over two spaces $P$ and $M$.
\bq
\begin{picture}(60,60)(0,0)
\LongArrow(55,50)(55,10)
\LongArrow(50,50)(5,10)
\Text(55,55)[cb]{$F$}
\Text(5,5)[rt]{$P$}
\Text(55,5)[ct]{$M$}
\Text(58,30)[l]{$\pi_M$}
\Text(25,34)[r]{$\pi_P$}
\end{picture} 
\eq
$F$ is called the {\bf correspondence space}.
We may think of the space $P$ as {\bf projective twistor space} 
and of the space $M$ as {\bf Minkowski space}.
To be precise, $M$ corresponds to a compactification of complex Minkowski space.
The two projections are denoted by $\pi_P$ and $\pi_M$.
In order to describe the correspondence space $F$, let us first introduce yet another space.
We set $T = {\mathbb C}^4$ and call $T$ {\bf twistor space}.
Sub-spaces of $T$ are denoted by $T_j$.
We define the correspondence space to be
\bq
 F 
 & = &
 \left\{ \left( T_1, T_2 \right) | T_1 \subset T_2 \subset T,
                                   \mathrm{dim}_{\mathbb C}\left(T_1\right) = 1,
                                   \mathrm{dim}_{\mathbb C}\left(T_2\right) = 2
 \right\}.
\eq
The space $F$ is a flag variety. Elements of $F$ are called flags and are given by pairs $(T_1,T_2)$,
where $T_2$ is a two-dimensional sub-space of $T$ and $T_1$ is a one-dimensional sub-space of $T_2$
(and of course of $T$).

We define the projections $\pi_P$ and $\pi_M$ by
\bq
 \pi_P\left( T_1, T_2 \right) \;\; = \;\; T_1,
 & &
 \pi_M\left( T_1, T_2 \right) \;\; = \;\; T_2.
\eq
Thus $P$ consists of all lines through the origin in $T$ and $M$ consists of all planes through the origin
in $T$.
In other words, we have
\bq
 P \;\; \cong \;\; \mathrm{Gr}_{1,4}\left({\mathbb C}\right) \;\; = \;\; {\mathbb C}{\mathbb P}^3,
 & &
 M \;\; \cong \;\; \mathrm{Gr}_{2,4}\left({\mathbb C}\right).
\eq
On $T$ we choose coordinates
\bq
 Z_\alpha
 & = &
 \left( \lambda_A, \pi_{\dot{B}} \right).
\eq
We apologise for the slightly confusing notation: 
$\pi_{\dot{1}}$ and $\pi_{\dot{2}}$ (both with a dot) denote two of the four coordinates on $T$,
while $\pi_P$ and $\pi_M$ denote projections.
We mentioned that $M$ can be considered as Minkowski space.
In order to see this, consider 
an affine chart of $\mathrm{Gr}_{2,4}({\mathbb C})$ with coordinates
\bq
\label{chart_G_2_4}
 \left( \begin{array}{rr}
 x_{1\dot{1}} & x_{1\dot{2}} \\
 x_{2\dot{1}} & x_{2\dot{2}} \\
 0 & -1 \\
 1 & 0 \\
  \end{array} \right).
\eq
The two columns of eq.~(\ref{chart_G_2_4}) correspond to two linearly independent vectors,
which span a plane $T_2$ in $T$. 
We have the usual isomorphisms
\bq
 x_{A\dot{B}} = x_\mu \sigma^{\mu}_{A\dot{B}},
 & &
 x^\mu = \frac{1}{2} x_{A\dot{B}} \bar{\sigma}^{\mu \dot{B}A},
\eq
giving us the relation between a point 
with coordinates $(x_{1\dot{1}}, x_{1\dot{2}}, x_{2\dot{1}}, x_{2\dot{2}})$ of $\mathrm{Gr}_{2,4}({\mathbb C})$
and a point $x^\mu$ in Minkowski space.
Let us  now consider 
a point $T_2 \in \mathrm{Gr}_{2,4}({\mathbb C})$ with coordinates as in eq.~(\ref{chart_G_2_4}).
Note that a point of $\mathrm{Gr}_{2,4}({\mathbb C})$ is a two-dimensional plane in $T$.
A point $( \lambda_A, \pi_{\dot{B}} ) \in T$ belongs to the plane $T_2$ if and only if
\bq
\label{incidence_relation}
 \lambda_A 
 \; = \;
 x_{A\dot{B}} \eps^{\dot{B}\dot{C}} \pi_{\dot{C}}
 & \mbox{or} &
 \lambda_A 
 \; = \; 
 x_{A\dot{B}} \pi^{\dot{B}}.
\eq
Eq.~(\ref{incidence_relation}) is called the {\bf incidence relation}.

Let us now consider the one-dimensional sub-spaces $T_1$ of the plane $T_2$.
Given a two-dimensional plane $T_2 \subset T$
the set of all one-dimensional lines $T_1 \subset T_2$ through the origin 
is isomorphic to ${\mathbb C} {\mathbb P}^1$.
We may use $[\pi_{\dot{B}}]$ as homogeneous coordinates for ${\mathbb C} {\mathbb P}^1$.
Putting everything together we have for the correspondence space $F$
\bq
 F & = & 
 \mathrm{Gr}_{1,2}\left({\mathbb C}\right) \times \mathrm{Gr}_{2,4}\left({\mathbb C}\right)
 \;\; = \;\;
 {\mathbb C} {\mathbb P}^1 \times \mathrm{Gr}_{2,4}\left({\mathbb C}\right).
\eq
We have $\dim_{\mathbb C} F = 5$.
For a point $w \in F$ we use the local coordinates
\bq
 w & = & 
 \left( \left[ \pi_{\dot{A}} \right], x_{B\dot{C}} \right).
\eq
The projections are then
\bq
 \pi_M & : & F \rightarrow \mathrm{Gr}_{2,4}({\mathbb C}), 
\nonumber \\
 & &
 \pi_M\left(w\right) = x_{B\dot{C}},
\eq
and
\bq
 \pi_P & : & F \rightarrow {\mathbb C} {\mathbb P}^3
 \nonumber \\
 & &
 \pi_P\left(w\right) =
 \left[  x_{A\dot{B}} \pi^{\dot{B}}, \pi_{\dot{C}} \right].
\eq
For $x_{A\dot{B}} \in M$ the fibre is
\bq
 \pi_M^{-1}\left(x_{A\dot{B}}\right)
 & \cong & 
 {\mathbb C} {\mathbb P}^1.
\eq
For $[\lambda_A : \pi_{\dot{B}} ] \in {\mathbb C} {\mathbb P}^3$ the fibre is
\bq
 \pi_P^{-1}\left( [\lambda_A : \pi_{\dot{B}} ] \right)
 & \cong &
 {\mathbb C} {\mathbb P}^2.
\eq
This can be seen as follows: 
A point $[\lambda_A : \pi_{\dot{B}} ] \in {\mathbb C} {\mathbb P}^3$ defines a 
one-dimensional sub-space $T_1$ of $T$. 
The fibre consists of all two-dimensional sub-spaces $T_2$ of $T$, which contain $T_1$.
These can be specified by choosing an independent vector not in $T_1$.
As only the direction matters, we see that the set of all possible choices corresponds
to ${\mathbb C} {\mathbb P}^2$.
\\
\\
Now let $f\left(\lambda_A,\pi_{\dot{B}}\right)$ be a function on $T$,
homogeneous of degree $(-2)$ 
and
holomorphic on $U_{\dot{1}} \cap U_{\dot{2}}$, where $U_{\dot{B}}$ is the chart with $\pi_{\dot{B}}\neq 0$.
We consider the contour integral
\bq
 \phi\left(x\right)
 & = &
 \frac{1}{2\pi i}
 \oint\limits_{{\mathcal C}} d\pi^{\dot{E}} \pi_{\dot{E}} \; f\left( x_{A\dot{B}} \eps^{\dot{B}\dot{C}} \pi_{\dot{C}}, \pi_{\dot{D}} \right),
\eq
where the contour ${\mathcal C}$ is a closed curve in $U_{\dot{1}} \cap U_{\dot{2}}$.
Since $f$ is required to be homogeneous of degree $(-2)$, the integral 
is actual an integral on projective twistor space $P$ and is
called the {\bf twistor transform} (or the {\bf Penrose transform}) of $f$.
The field $\phi$ is a scalar field (obviously with helicity $0$) and satisfies
\bq
\label{field_eq_Penrose_transform}
 \Box \phi\left(x\right) & = & 0.
\eq
The field $\phi$ is therefore a free field. Eq.~(\ref{field_eq_Penrose_transform}) is easily verified: With
\bq
 \Box & = &
 \partial_\mu \partial^\mu
 =
 \frac{1}{2} \eps_{A C} \eps_{\dot{B} \dot{D}} \frac{\partial}{\partial x_{A\dot{B}}} \frac{\partial}{\partial x_{C\dot{D}}}
\eq
we have
\bq
 \Box \phi\left(x\right) & = & 
 \frac{1}{2} \eps_{A C} \eps_{\dot{B} \dot{D}} 
 \frac{1}{2\pi i}
 \oint\limits_{{\mathcal C}} d\pi^{\dot{E}} \pi_{\dot{E}} \; \frac{\partial^2 f}{\partial \lambda_A \partial \lambda_C}
 \eps^{\dot{B}\dot{F}} \pi_{\dot{F}}
 \eps^{\dot{D}\dot{G}} \pi_{\dot{G}}
 \nonumber \\
 & = &
 \frac{1}{2}  
 \frac{1}{2\pi i}
 \oint\limits_{{\mathcal C}} d\pi^{\dot{E}} \pi_{\dot{E}} \; 
 \underbrace{ \left( \eps_{A C} \frac{\partial^2 f}{\partial \lambda_A \partial \lambda_C} \right) }_{0}
 \;
 \underbrace{ \vphantom{ \left( \eps_{A C} \frac{\partial^2 f}{\partial \lambda_A \partial \lambda_C} \right) }   
              \left[ \pi \pi \right] 
            }_{0}
 = 0.
\eq
Note that in this expressions we obtain a product of two zeros: One factor of zero comes from
the spinor product $[\pi\pi]$, the other from the second derivative, which is symmetric and vanishes
when contracted into the anti-symmetric tensor $\eps_{A C}$.

We may generalise this construction to free fields with non-zero helicities as follows:
Let us first consider
\bq
 \phi_{\dot{A}_1 ... \dot{A}_n} \left(x\right)
 & = &
 \frac{1}{2\pi i}
 \oint\limits_{{\mathcal C}} d\pi^{\dot{E}} \pi_{\dot{E}} 
 \; 
 \pi_{\dot{A}_1} ... \pi_{\dot{A}_n}
 \;
 f\left( x_{A\dot{B}} \eps^{\dot{B}\dot{C}} \pi_{\dot{C}}, \pi_{\dot{D}} \right),
\eq
where $f$ is homogeneous of degree $(-2-n)$.
This defines a field of helicity $+n/2$.
The field $\phi_{\dot{A}_1 ... \dot{A}_n}$ is symmetric in all indices $\{\dot{A}_1, ..., \dot{A}_n \}$.
The field satisfies
\bq
 \frac{\partial}{\partial x_{A \dot{A}_1}}
 \phi_{\dot{A}_1 ... \dot{A}_n} \left(x\right)
 & = & 0,
\eq
due to $[ \pi \pi ] = 0$.
Let us further consider
\bq
 \phi^{A_1 ... A_n} \left(x\right)
 & = &
 \frac{1}{2\pi i}
 \oint\limits_{{\mathcal C}} d\pi^{\dot{E}} \pi_{\dot{E}} 
 \; 
 \frac{\partial}{\partial \lambda_{A_1}} ... \frac{\partial}{\partial \lambda_{A_n}}
 \;
 f\left( x_{A\dot{B}} \eps^{\dot{B}\dot{C}} \pi_{\dot{C}}, \pi_{\dot{D}} \right),
\eq
where $f$ is now homogeneous of degree $(-2+n)$.
This defines a field of helicity $-n/2$.
Again, the field $\phi^{A_1 ... A_n}$ is symmetric in all indices $\{A_1, ..., A_n \}$.
We now have
\bq
 \eps_{A A_1}
 \frac{\partial}{\partial x_{A \dot{B}}}
 \phi^{A_1 ... A_n} \left(x\right)
 & = &
 0,
\eq
due to
\bq
 \eps_{A A_1} \frac{\partial^2 f}{\partial \lambda_{A} \partial \lambda_{A_1}} & = & 0.
\eq

\subsection{Off-shell recurrence relations}
\label{sect_off_shell_recursion}

Let us now return to the master formula for an observable, given in eq.~(\ref{observable_master_hadron_hadron}).
For physical particles the momenta are real and a phase space point 
(i.e. a momentum configuration of $n$ four-vectors, where two four-vectors correspond to the given incoming momenta,
all four-vectors are real and satisfy the on-shell conditions together with momentum conservation)
is generated with dedicated algorithms \cite{Byckling:1971vca,Kleiss:1986gy,vanHameren:2002tc,vanHameren:2010gg}.
The phase space integration is usually performed by Monte Carlo integration.
In eq.~(\ref{matrix_element_squared_to_primitive_amplitudes}) we expressed the spin- and colour-summed matrix element squared in terms
of primitive amplitudes.
Thus we need for a given momentum configuration $p=(p_1,...,p_n)$, a given helicity configuration 
$\lambda=(\lambda_1,...,\lambda_n)$ and a given cyclic order $\sigma=(\sigma_1,...,\sigma_n)$
the primitive amplitude
\bq
 A_{n}^{(0)}\left( \sigma_1^{\lambda_{\sigma_1}}, ..., \sigma_n^{\lambda_{\sigma_n}} \right).
\eq
This will evaluate to a complex number.
In this paragraph we will review an efficient algorithm to compute this complex number,
avoiding Feynman diagrams.
Without loss of generality we will take the cyclic order to be $(1,...,n)$.

Off-shell recurrence relations \cite{Berends:1987me}
build primitive amplitudes from smaller building blocks, usually
called cyclic-ordered off-shell currents.
{\bf Off-shell currents} are objects with $j$ on-shell legs and one additional leg off-shell, with $j$ ranging from $1$ 
to $(n-1)$.
Momentum conservation is satisfied. It should be noted that
off-shell currents are not gauge-invariant objects.
Recurrence relations relate off-shell currents with $j$ legs 
to off-shell currents with fewer legs.
The recursion starts with $j=1$:
\bq
J_\mu(p_1^{\lambda_1}) & = & \eps_\mu^{\lambda_1}(p_1,q_1).
\eq
$\eps_\mu^{\lambda_1}$ is the polarisation vector of the gluon with momentum $p_1$ and helicity $\lambda_1$.
$q_1$ an arbitrary light-like reference momentum.
The recursive relation states that a gluon couples to other gluons only via the three- or four-gluon vertices.
Off-shell currents with $j\ge 2$ are computed as
\bq
 J_\mu(p_1^{\lambda_1},...,p_j^{\lambda_j}) 
 & = & 
 \frac{-i}{p^2_{1,j}} 
 J^{\mathrm{trunc}}_\mu(p_1^{\lambda_1},...,p_j^{\lambda_j}) ,
\eq
where
\bq
 p_{i,j} & = & p_i + p_{i+1} + ... + p_j 
\eq
and $J^{\mathrm{trunc}}_\mu=g_{\mu\nu} J^{\nu,\mathrm{trunc}}$ is given by
\bq
\label{Berends_Giele_recursion}
 J^{\mu,\mathrm{trunc}}(p_1^{\lambda_1},...,p_j^{\lambda_j}) 
 & = &
 \sum\limits_{k=1}^{j-1} 
 V_3^{\mu\nu\rho}(-p_{1,j},p_{1,k},p_{k+1,j})
 J_\nu(p_1^{\lambda_1},...,p_k^{\lambda_k}) J_\rho(p_{k+1}^{\lambda_{k+1}},...,p_j^{\lambda_j}) 
 \\
 & & 
 + 
 \sum\limits_{k=1}^{j-2} \sum\limits_{l=k+1}^{j-1} 
 V_4^{\mu\nu\rho\sigma} 
 J_\nu(p_1^{\lambda_1},...,p_k^{\lambda_k}) J_\rho(p_{k+1}^{\lambda_{k+1}},...,p_l^{\lambda_l}) J_\sigma(p_{l+1}^{\lambda_{l+1}},...,p_j^{\lambda_j}),
 \nonumber
\eq
\begin{figure}
\begin{center}
\includegraphics[scale=1.0]{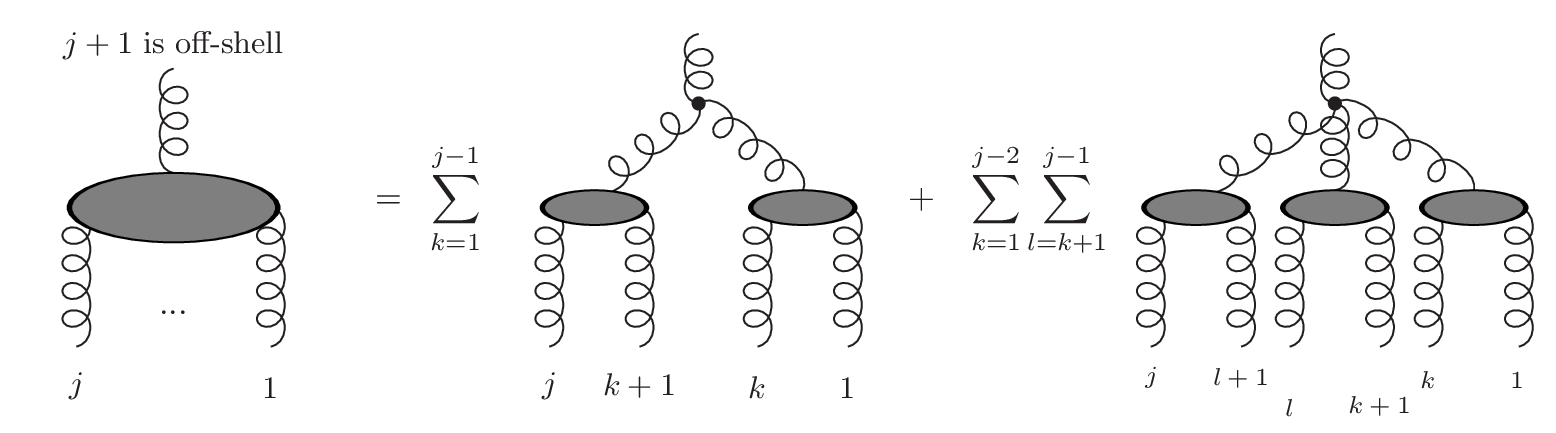}
\end{center}
\caption{Off-shell recurrence relation: In an off-shell current particle $n+1$ is kept off-shell.
This allows to express an off-shell current with $n$ on-shell legs in terms of currents with fewer legs.}
\label{figure_off_shell_current}
\end{figure}
where $V_3$ and $V_4$ are the cyclic-ordered three-gluon and four-gluon vertices
\bq
 V_3^{\mu\nu\rho}(p_1,p_2,p_3) 
 & = &
 i \left[
          g^{\mu\nu} \left( p_1^\rho - p_2^\rho \right)
        + g^{\nu\rho} \left( p_2^\mu - p_3^\mu \right)
        + g^{\rho\mu} \left( p_3^\nu - p_1^\nu \right)
   \right],
 \nonumber \\
 V_4^{\mu\nu\rho\sigma} & = & i \left( 2 g^{\mu\rho} g^{\nu\sigma} - g^{\mu\nu} g^{\rho\sigma} -g^{\mu\sigma} g^{\nu\rho} \right).
\eq
The recurrence relation is shown pictorially in fig.~\ref{figure_off_shell_current}.
The gluon current $J_\mu$ is {\bf conserved}:
\bq
\left( \sum\limits_{k=1}^j p_k^\mu \right) J_\mu\left(p_1,...,p_j\right) & = & 0.
\eq
The gluon current further satisfies the {\bf photon decoupling relation}
\bq
J_\mu(p_1,p_2,p_3,...,p_j) + J_\mu(p_2,p_1,p_3,...,p_j) + ... + J_\mu(p_2,p_3,...,p_j,p_1) & = & 0
\eq
and the {\bf reflection identity}
\bq
J_\mu(p_1,p_2,p_3,...,p_j) & = & (-1)^{j+1} J_\mu(p_j,...,p_3,p_2,p_1).
\eq
With the help of the off-shell currents
we obtain the following algorithm for the computation of a primitive amplitude:
\begin{algorithm}
Calculation of a primitive helicity amplitude $A_{n}^{(0)}( 1^{\lambda_1}, ..., n^{\lambda_n} )$
from off-shell currents.
\begin{enumerate}
\item For $j=1,2,...,(n-1)$ compute all off-shell currents for the ordered sequence of on-shell legs $(1,...,n-1)$,
starting from the one-currents $J_\mu(p_1^{\lambda_1})$, $J_\mu(p_2^{\lambda_2})$, ..., $J_\mu(p_{n-1}^{\lambda_{n-1}})$,
then the two-currents $J_\mu(p_1^{\lambda_1},p_2^{\lambda_2})$, ..., $J_\mu(p_{n-2}^{\lambda_{n-2}},p_{n-1}^{\lambda_{n-1}})$,
up to the $(n-1)$-current $J_\mu(p_1^{\lambda_1},...,p_{n-1}^{\lambda_{n-1}})$.
At each step re-use the results for the already computed lower-point currents.
\item The primitive amplitude is given by
\bq
 A_{n}^{(0)}\left( 1^{\lambda_1}, ..., n^{\lambda_n} \right)
 & = &
\eps_\mu^{\lambda_n}(p_n,q_n) J^{\mu,\mathrm{trunc}}\left(p_1^{\lambda_1},...,p_{n-1}^{\lambda_{n-1}}\right).
\eq
\end{enumerate}
\end{algorithm}
It can be shown that the scaling behaviour of this algorithm with the number of external particles $n$ is
$n^4$. This polynomial behaviour is much better than the factorial growth of the algorithm based on Feynman diagrams.
The re-use of the results for the already computed lower-point currents is essential in achieving this
polynomial behaviour.
We may even improve this scaling behaviour further.
Computing the contributions from the four-gluon vertex in the recurrence relations involves
the most operations and dominates the scaling behaviour.
We can reduce the scaling behaviour from $n^4$ down to $n^3$ by eliminating the four-gluon vertex
with the help of the auxiliary tensor field introduced in section~(\ref{sect:colour}).
Let us denote by
\bq
 V_4^{\mu\nu[\rho\sigma]} & = & 
  \frac{i}{\sqrt{2}}
 \left( g^{\mu\rho} g^{\nu\sigma} - g^{\mu\sigma} g^{\nu\rho} \right)
\eq
the cyclic-ordered gluon-gluon-tensor vertex.
For $j\ge 2$ we introduce a tensor current $J_{[\mu\nu]}$ through
\bq
 J_{[\mu\nu]}(p_1^{\lambda_1},...,p_j^{\lambda_j}) 
 & = &
 - i g_{\mu\rho} g_{\nu\sigma}
 \sum\limits_{k=1}^{j-1} 
 V_4^{\alpha\beta[\rho\sigma]}
 J_\alpha(p_1^{\lambda_1},...,p_k^{\lambda_k}) J_\beta(p_{k+1}^{\lambda_{k+1}},...,p_j^{\lambda_j}).
\eq
This equation is shown pictorially in fig.~(\ref{figure_off_shell_tensor_current}).
\begin{figure}
\begin{center}
\includegraphics[scale=1.0]{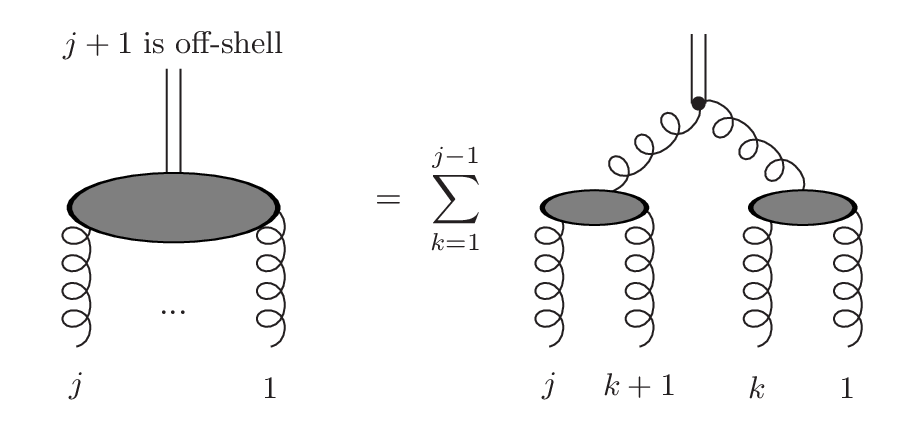}
\end{center}
\caption{The recurrence relation for the tensor current.}
\label{figure_off_shell_tensor_current}
\end{figure}
With the help of the auxiliary tensor current we may compute the gluon current as
\bq
\label{Berends_Giele_recursion_improved}
 J^{\mu,\mathrm{trunc}}(p_1^{\lambda_1},...,p_j^{\lambda_j}) 
 & = &
 \sum\limits_{k=1}^{j-1} 
 V_3^{\mu\nu\rho}(-p_{1,j},p_{1,k},p_{k+1,j})
 J_\nu(p_1^{\lambda_1},...,p_k^{\lambda_k}) J_\rho(p_{k+1}^{\lambda_{k+1}},...,p_j^{\lambda_j}) 
 \nonumber \\
 & &
 +
 \sum\limits_{k=1}^{j-2} 
 V_4^{\mu\nu[\rho\sigma]}
 J_\nu(p_1^{\lambda_1},...,p_k^{\lambda_k}) J_{[\rho\sigma]}(p_{k+1}^{\lambda_{k+1}},...,p_j^{\lambda_j}) 
 \nonumber \\
 & &
 +
 \sum\limits_{k=2}^{j-1} 
 V_4^{\nu\mu[\rho\sigma]}
 J_{[\rho\sigma]}(p_1^{\lambda_1},...,p_k^{\lambda_k}) J_\nu(p_{k+1}^{\lambda_{k+1}},...,p_j^{\lambda_j}).
\eq
Eq.~(\ref{Berends_Giele_recursion_improved}) 
\begin{figure}
\begin{center}
\includegraphics[scale=0.8]{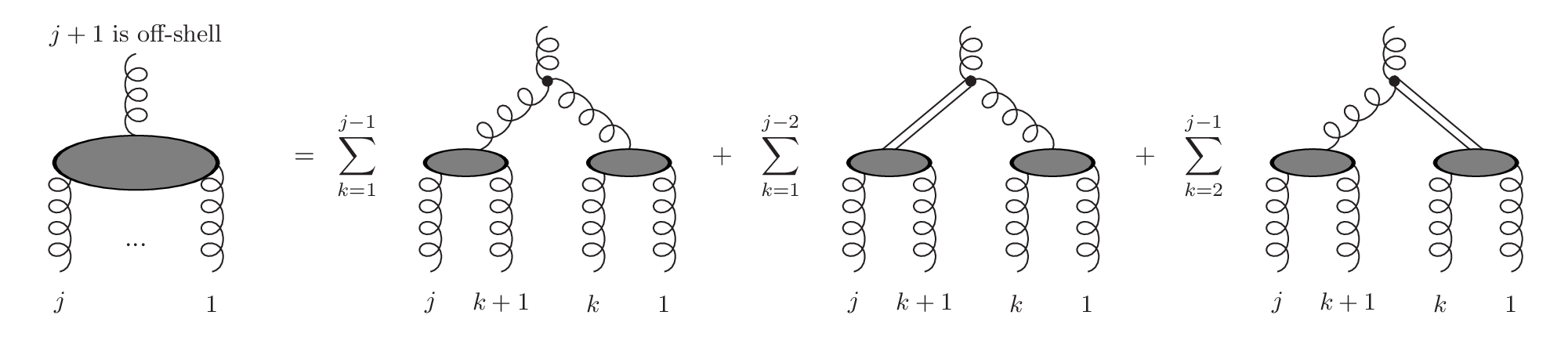}
\end{center}
\caption{The improved recurrence relation for the gluon current.}
\label{figure_improved_off_shell_current}
\end{figure}
is shown pictorially in fig.~(\ref{figure_improved_off_shell_current}).
Now all currents involve only trivalent vertices, giving us a $n^3$-scaling behaviour.
\\
\\
\bs
{\it {\bf Exercise \theexercise}: 
Recurrence relations are also useful for the following problem:
Count the number of Feynman diagrams contributing to the cyclic-ordered 
amplitude $A_n^{(0)}$.
\\
\\
Hint: Find a recurrence relation for the number of diagrams.
Consider a particular diagram: Follow the leg $n$ inwards into the diagram until you hit the first
vertex (either a three-gluon vertex or a four-gluon vertex).
Attached to this vertex are sub-graphs with fewer legs.
\stepcounter{exercise}
}
\es

% ----------------------------------------------------------------------------------
\newpage
\section{MHV amplitudes and MHV expansion}
\label{sect:mhv_amplitudes}

This section is centred around maximally helicity violating amplitudes.
These amplitudes have for any number of external particles an astonishing simple formula.

\subsection{The Parke-Taylor formul{\ae}}
\label{sect:parke_taylor}

Tree-level primitive amplitudes for specific helicity combinations are remarkably simple
or vanish altogether.
Parke and Taylor conjectured \cite{Parke:1986gb} that for $n\ge 4$
\bq
\label{mhv_amplitudes}
 A_{n}^{(0)}\left(1^+,2^+,...,n^+\right) 
 & = & 
 0,
 \nonumber \\
 A_{n}^{(0)}\left(1^+,2^+,...,j^-,...,n^+\right) 
 & = & 
 0,
 \nonumber \\
 A_{n}^{(0)}\left(1^+,2^+,...,j^-,...,k^-,...,n^+\right) 
 & = & 
 i \left( \sqrt{2} \right)^{n-2} 
 \frac{\langle j k \rangle^4}{\langle 1 2 \rangle ... \langle n 1 \rangle}.
\eq
The primitive amplitudes where all gluons have positive helicities vanish
and so do the primitive amplitudes where all gluons except one have positive helicities.
The first non-vanishing result is obtained for the amplitudes
with $n-2$ gluons of positive helicity and $2$ gluons of negative helicity.
It is given by a remarkable simple formula. Note that this formula holds for all $n$.
An amplitude with $n-2$ gluons of positive helicity and $2$ gluons of negative helicity
is called a {\bf maximally helicity violating amplitude} ({\bf MHV amplitude}).
As it depends only on the spinors $p_A = | p+ \rangle$ (and not on $p_{\dot{A}}$)  
one says that this amplitude is {\bf holomorphic}.
The name ``maximally helicity violating amplitude'' stems from the following fact: 
We use the convention that all particles are treated as outgoing. Crossing a particle to the initial state will
reverse the helicity.
Thus the the amplitude $A_n^{(0)}(1^+,2^+,...,n^+)$ corresponds -- when $1$ and $2$ are crossed to the initial state --
to the process $(-p_1)^- + (-p_2)^- \rightarrow p_3^+ + ... + p_n^+$.
In this process all final state particles have the opposite helicity of the two initial state particles.
This process would violate helicity conservation maximally. Actually the amplitude for this process is zero
and so is the amplitude for the process $(-p_1)^- + (-p_2)^- \rightarrow p_3^- + p_4^+ + ... + p_n^+$.
The first non-zero amplitude is obtained for the process
$(-p_1)^- + (-p_2)^- \rightarrow p_3^- + p_4^- + p_5^+ + ... + p_n^+$.
In this process we have -- starting with two negative helicity particles in the initial state -- the maximal number
of positive helicity particles in the final state such that the amplitude is non-zero.
The corresponding amplitude is therefore called a maximally helicity violating amplitude.

Obviously, we find similar formul{\ae} if we exchange all positive and negative helicities:
\bq
\label{anti_mhv_amplitudes}
 A_{n}^{(0)}\left(1^-,2^-,...,n^-\right) 
 & = & 
 0,
 \nonumber \\
 A_{n}^{(0)}\left(1^-,2^-,...,j^+,...,n^-\right) 
 & = & 
 0,
 \nonumber \\
 A_{n}^{(0)}\left(1^-,2^-,...,j^+,...,k^+,...,n^-\right) 
 & = & 
 i \left( \sqrt{2} \right)^{n-2} 
 \frac{[ k j ]^4}{[1 n ] [n (n-1)] ... [ 2 1 ]}.
\eq
Primitive amplitudes with $n-2$ gluons of negative helicity and $2$ gluons of positive helicity
are called {\bf anti-MHV amplitudes}.
These amplitudes depend only on the spinors $p_{\dot{A}} = \langle p+ |$ (and not on $p_A$) 
and are called {\bf anti-holomorphic}.

We may extend these formul{\ae} to $n=3$, by defining the amplitude with
one positive helicity and two negative helicities according to the MHV formula in eq.~(\ref{mhv_amplitudes})
and the amplitude with one negative helicity and two positive helicities 
according to the anti-MHV formula in eq.~(\ref{anti_mhv_amplitudes}).
Thus we have
\bq
\label{three_point_MHV_and_MHVbar}
 A_3^{(0)}\left(1^-,2^-,3^+\right)
 \;\; = \;\;
 i \sqrt{2}
 \frac{\langle 1 2 \rangle^3}{\langle 2 3  \rangle \langle 3 1  \rangle},
 & &
 A_3^{(0)}\left(1^+,2^+,3^-\right)
 \;\; = \;\;
 i \sqrt{2}
 \frac{[ 2 1 ]^3}{[1 3 ] [3 2]}.
\eq
The three-point amplitudes, where all helicities are equal vanishes.
\bq
 A_3^{(0)}\left(1^+,2^+,3^+\right)
 \;\; = \;\;
 A_3^{(0)}\left(1^-,2^-,3^-\right)
 \;\; = \;\;
 0.
\eq
Eq.~(\ref{mhv_amplitudes}) and eq.~(\ref{anti_mhv_amplitudes}) are called the Parke-Taylor formul{\ae}.
Before we proceed to a proof of these formul{\ae}, let us first look at the simple case $n=4$.
Amplitudes with four positive helicities or four negative  helicities vanish, as do amplitudes with
three helicities of one type and one helicity of the other type.
The only non-vanishing four-gluon helicitiy amplitudes are the ones with two positive helicities
and two negative helicities.
These are at the same time MHV amplitudes and anti-MHV amplitudes, 
or phrased differently at the same time holomorphic and anti-holomorphic.
As an example let us consider
the amplitude $A_4^{(0)}(1^-,2^-,3^+,4^+)$.
The MHV representation (or holomorphic representation) is given by
\bq
A_4^{(0)}\left(1^-,2^-,3^+,4^+\right) 
 & = & 2 i 
 \frac{\langle 1 2 \rangle^4}{\langle 1 2 \rangle \langle 2 3 \rangle \langle 3 4 \rangle \langle 4 1 \rangle}.
\eq
The anti-MHV represenation (or anti-holomorphic representation) is given by
\bq
A_4^{(0)}\left(1^-,2^-,3^+,4^+\right) 
 & = & 2 i 
 \frac{[ 3 4 ]^4}{[ 1 2 ] [ 2 3 ] [ 3 4 ] [ 4 1 ]}.
\eq
Of course, these two expressions have to be equal. Let's see how this works out:
From momentum conservation it follows that $2 p_1 p_2 = 2 p_3 p_4$ and therefore
\bq
 \frac{\langle 1 2 \rangle}{\langle 3 4 \rangle} & = & \frac{[ 3 4 ]}{[ 1 2 ]}.
\eq
Momentum conservation implies also that $p_2 = - p_1 - p_3 - p_4$ and therefore
\bq
 \frac{\langle 1 2 \rangle}{\langle 4 1 \rangle} & = & - \frac{[ 3 4 ]}{[ 2 3 ]}.
\eq
A similar relation follows from $p_1 = - p_2 - p_3 - p_4$:
\bq
 \frac{\langle 1 2 \rangle}{\langle 2 3 \rangle} & = & - \frac{[ 3 4 ]}{[ 4 1 ]}.
\eq
Therefore, the two representations are equivalent.

Now let us turn to the proof of the Parke-Taylor formul{\ae}:
The proof of the Parke-Taylor formul{\ae} has been given by
Berends and Giele \cite{Berends:1987me}
and uses the off-shell currents discussed in section~(\ref{sect_off_shell_recursion}).
In some cases the recurrence relations can be solved in closed form. 
If all the gluons have the same helicity one obtains
\bq
\label{off_shell_current_all_plus}
 J_\mu\left(p_1^+,p_2^+,...,p_n^+\right) 
 & = & 
 \left( \sqrt{2} \right)^{n-2}
 \frac{\langle q- | \gamma_\mu p\!\!\!/_{1,n}|q+ \rangle}
      {\langle q1 \rangle \langle 1 2 \rangle ... \langle (n-1) n \rangle \langle n q \rangle},
\eq
if a common reference momentum $q$ is chosen for all gluons. 
The backslash notation stands for contraction with $\gamma_\mu$, e.g.
\bq
 p\!\!\!/_{1,n} & = & \gamma_\nu \left( p_1^\nu + ... + p_n^\nu \right).
\eq
\bs
{\it {\bf Exercise \theexercise}: 
Prove eq.~(\ref{off_shell_current_all_plus}) by induction.
\stepcounter{exercise}
}
\es
\\
\\
If one gluon has opposite helicity, let's say gluon $1$, one finds
\bq
\label{off_shell_current_one_minus}
 J_\mu\left(p_1^-,p_2^+,...,p_n^+\right) 
 & = & 
 \left( \sqrt{2} \right)^{n-2}
 \frac{\langle 1 -| \gamma_\mu p\!\!\!/_{2,n}|1+ \rangle}
      {\langle 1 2 \rangle ... \langle n 1 \rangle } 
 \sum\limits_{m=3}^n \frac{\langle 1 - | p\!\!\!/_m p\!\!\!/_{1,m} | 1+ \rangle }{p^2_{1,m-1} p^2_{1,m}},
\eq
where the reference momentum choice is $q_1 = p_2$, $q_2 = ... = q_n = p_1$.
\\
\\
From the off-shell currents in eq.~(\ref{off_shell_current_all_plus}) and in eq.~(\ref{off_shell_current_one_minus})
we obtain the amplitudes $A_{n+1}^{(0)}$ by multiplying with $i p_{1,n}^2$ and contracting with $\eps^{\lambda_{n+1}}_\mu(p_{n+1})$.
From eq.~(\ref{off_shell_current_all_plus}) we see immediately that the amplitudes with all plus helicities 
and the amplitudes with one minus helicity and the rest plus helicities must vanish, as the off-shell current
$J_\mu(p_1^+,p_2^+,...,p_n^+)$ does not have a pole at $p_{1,n}^2=0$.
From eq.~(\ref{off_shell_current_one_minus})
we obtain the MHV amplitude where the two negative helicity particles are adjacent
by multiplying with $i p_{1,n}^2$ and contracting with $\eps^-_\mu(p_{n+1})$.
Proving the correctness of the Parke-Taylor formula for two non-adjacent negative helicities will
be an exercise in section~(\ref{sect:application_on_shell_recursion}), after we introduced on-shell recursion relations.

\subsection{The CSW construction}
\label{sect:CSW_construction}

The Parke-Taylor formul{\ae} give us compact expressions for amplitudes with zero, one or two 
gluons of negative helicity and all remaining gluons having positive helicity.
An obvious question is what happens if we go to amplitudes with a higher number of negative
helicity gluons.
Amplitudes with three gluons of negative helicitiy are called next-to-maximally helicity violating
amplitudes (or NMHV amplitudes for short), 
amplitudes with four gluons of negative helicitiy are called next-to-next-to-maximally helicity violating
amplitudes (or N${}^2$MHV amplitudes for short).
In general, a N${}^{k-2}$MHV amplitude contains $k$ gluons with negative helicity.

After the discovery of the Parke-Taylor formul{\ae} in 1986 it took almost twenty years (until 2004)
to find the pattern for the general case.
This generalisation triggered a very rapid development of the field following the years after 2004.
We will now review the Cachazo-Svrcek-Witten (CSW) construction \cite{Cachazo:2004kj}.
The basic idea of the CSW construction is to obtain 
tree amplitudes in Yang-Mills theory from tree graphs in which the vertices are tree-level MHV
scattering amplitudes, continued off-shell in a particular fashion.

Let us first discuss the off-shell continuation.
Let $q$ be a light-like four-vector, which will be kept fixed throughout the discussion.
Using $q$, any massive vector $p$ can be written as 
a sum of two light-like four-vectors $p^\flat$ and $q$ \cite{Kosower:2004yz}:
\bq
\label{offshellcont}
 p & = & p^\flat + \frac{p^2}{2pq} q.
\eq
Obviously, if $p^2=0$, we have $p = p^\flat$. Note further that $2pq = 2p^\flat q$.
This construction appeared already in section~(\ref{sect_massive_spinors}).
Using eq.~(\ref{offshellcont}) we may associate a massless four-vector $p^\flat$
to any four-vector $p$.
Using the projection onto $p^\flat$ we define the off-shell
continuation of Weyl spinors as
\bq
\label{offshellcontspinor}
 | p \pm \rangle & \rightarrow & | p^\flat \pm \rangle,
 \nonumber \\
 \langle p \pm | & \rightarrow & \langle p^\flat \pm |.
\eq
The basic building blocks of the CSW construction are the {\bf off-shell continued MHV amplitudes}, which serve as new vertices:
\bq
\label{mhv_vertex}
V_n(1^+,...,j^-,...,k^-,...,n^+) 
 & = & i \left( \sqrt{2} \right)^{n-2} 
 \frac{\langle j^\flat k^\flat \rangle^4}{\langle 1^\flat 2^\flat \rangle ... \langle n^\flat 1^\flat \rangle}.
\eq
Each MHV vertex has exactly two lines carrying negative helicity and at least one line carrying positive helicity.
Each internal line has a positive helicity label on one side and a negative helicity label on the other side.
The propagator for each internal line is the propagator of a scalar particle:
\bq
 \frac{i}{p^2}
\eq
The number of MHV vertices is related to the number of negative helicity gluons.
To see this, consider a tree diagram with $v$ vertices.
This diagram 
\begin{description}
\item{-} has $(v-1)$ propagators,
\item{-} has in total $2v$ negative helicity labels (two per vertex),
\item{-} has $(v+1)$ external negative helicity label (since $(v-1)$ labels are used by the internal propagators).
\end{description}
Therefore an amplitude with $k$ negative helicity gluons has $(k-1)$ MHV vertices.

Let us now consider an example. 
\begin{figure}
\begin{center}
\includegraphics[scale=1.0]{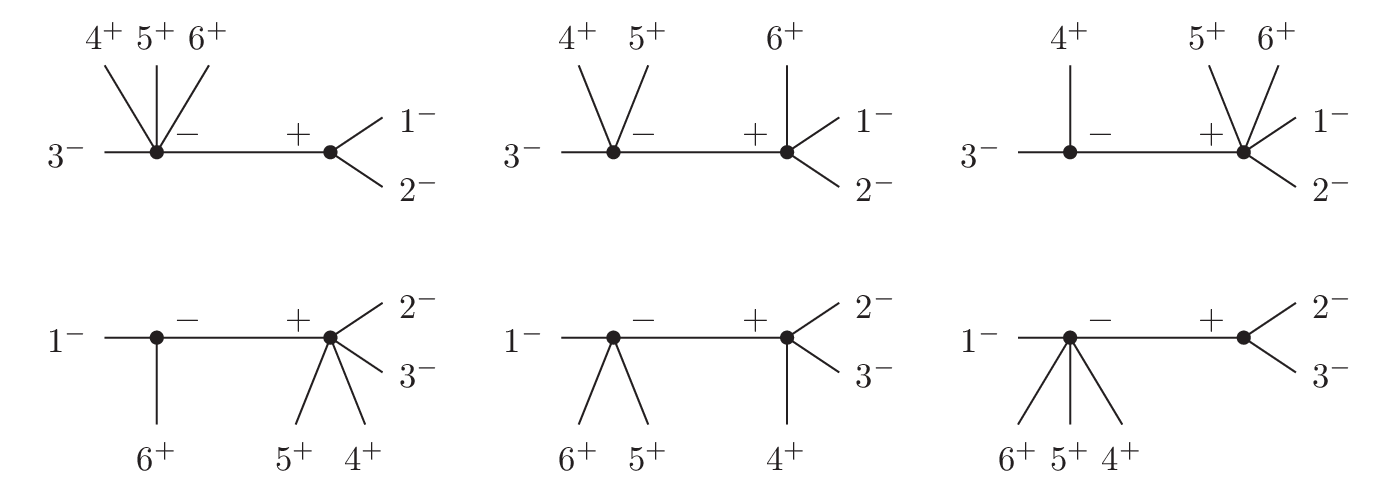}
\end{center}
\caption{MHV diagrams contributing to the tree-level six-gluon amplitude $A_6(1^-,2^-,3^-,4^+,5^+,6^+)$.}
\label{fig_mhv_example}
\end{figure}
The NMHV amplitude $A_6(1^-,2^-,3^-,4^+,5^+,6^+)$ has three gluons of positive helicity
and three gluons of negative helicity and is one of the first non-trivial amplitudes, which are non-zero and which are not
MHV amplitudes.
It will be given by diagrams with two MHV vertices.
Fig. \ref{fig_mhv_example} shows the six MHV diagrams contributing to this amplitude.
We obtain the set of MHV diagrams by first considering the stripped diagrams with only external negative-helicity 
gluons attached to it, and then adding the positive-helicity gluons in all possible ways consistent with the cyclic
ordering.
For the amplitude $A_6(1^-,2^-,3^-,4^+,5^+,6^+)$ the three possible stripped diagrams are shown in fig.~(\ref{fig_stripped_diagrams}).
Note that the second stripped diagram will be dressed with all
positive helicty gluons inserted between leg $3$ and leg $1$. Therefore one MHV vertex with two negative
helicity gluons and zero positive helicity gluons remains. 
\begin{figure}
\begin{center}
\includegraphics[scale=1.0]{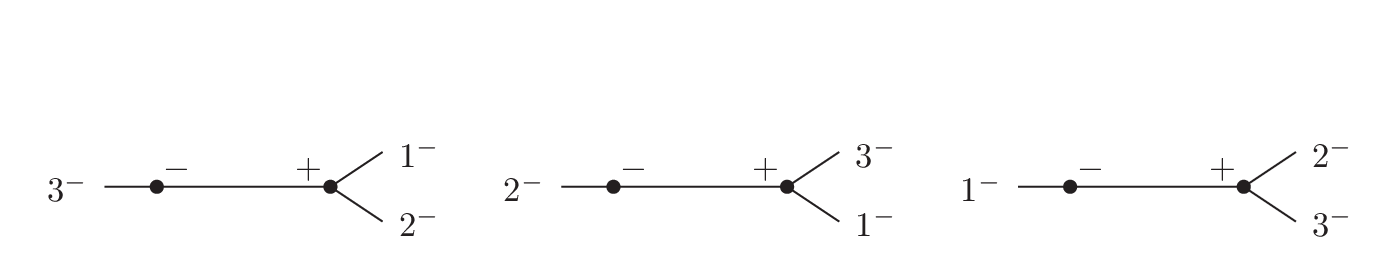}
\end{center}
\caption{Stripped diagrams for the amplitude $A_6(1^-,2^-,3^-,4^+,5^+,6^+)$.
In stripped diagrams only external negative-helicity
gluons are drawn.}
\label{fig_stripped_diagrams}
\end{figure}
Therefore this diagram does not give a contribution and we obtain the six MHV diagrams shown in fig.~(\ref{fig_mhv_example}).

Let us consider the first MHV diagram of fig.~(\ref{fig_mhv_example}).
This diagram yields
\bq
\lefteqn{
\begin{picture}(120,60)(0,40)
 \Vertex(25,50){2}
 \Vertex(75,50){2}
 \Line(25,50)(75,50)
 \Line(10,50)(25,50)
 \Line(75,50)(90,60)
 \Line(75,50)(90,40)
 \Text(95,60)[l]{\scriptsize $1^-$}
 \Text(95,40)[l]{\scriptsize $2^-$}
 \Text(5,50)[r]{\scriptsize $3^-$}
 \Text(30,52)[lb]{\scriptsize $-$}
 \Text(70,52)[rb]{\scriptsize $+$}
 \Line(25,50)(10,75)
 \Line(25,50)(25,75)
 \Line(25,50)(40,75)
 \Text(10,80)[b]{\scriptsize $4^+$}
 \Text(25,80)[b]{\scriptsize $5^+$}
 \Text(40,80)[b]{\scriptsize $6^+$}
\end{picture}
 = }
 \nonumber \\
 & & \nonumber \\
 & &
 \left[
 i \sqrt{2} 
 \frac{\langle 1 2 \rangle^4}{\langle 1 2 \rangle \langle 2 \left(-p_{12}\right)^\flat \rangle
 \langle \left(-p_{12}\right)^\flat 1 \rangle}
 \right]
 \;\;\;
 \frac{i}{p_{12}^2}
 \;\;\;
 \left[
 i \left( \sqrt{2} \right)^{3} 
 \frac{\langle 3 p_{12}^\flat \rangle^4}{\langle 3 4 \rangle 
 \langle 4 5\rangle \langle 5 6 \rangle 
 \langle 6 p_{12}^\flat \rangle \langle p_{12}^\flat 3 \rangle }
 \right],
 \nonumber
\eq
and similar expressions are obtained for the five other diagrams.
In this expression, $p_{12}=p_1+p_2$ is the momentum flowing through the internal line and $p_{12}^\flat$ is the
projection onto a light-like four-vector as in eq.~(\ref{offshellcont}).

Bena, Bern and Kosower~\cite{Bena:2004ry} derived a recursive formulation, which allows to obtain vertices
with more gluons of negative helicity from simpler building blocks:
\bq
\label{MHV_recursion}
\lefteqn{
 V_n(p_1^{\lambda_1},...,p_n^{\lambda_n}) 
 = 
 \frac{1}{(k-2)}
 \sum\limits_{j=1}^n
 \sum\limits_{l=j+1}^{j-3}
  \frac{i}{p_{j,l}^2} 
  V_{(l-j+2) \;\mbox{\scriptsize mod}\; n}(p_{j}^{\lambda_{j}},\ldots,p_{l}^{\lambda_{l}},(-p_{j,l})^{-}) 
}
 & & \nonumber \\
 & & 
 \hspace*{45mm}
 \times
 V_{(j-l) \;\mbox{\scriptsize mod}\; n}(p_{l+1}^{\lambda_{l+1}},\ldots,p_{j-1}^{\lambda_{j-1}},(-p_{(l+1),(j-1)})^+),
\hspace*{5mm}
\eq
where $k$ is the number of negative helicity gluons. 
The recursion stops if $k$ is less or equal than two. 
For $k=0$ or $k=1$ the quantity $V_n(p_1^{\lambda_1},\ldots,p_n^{\lambda_n})$ vanishes. 
For $k=2$ it is given by eq. (\ref{mhv_vertex}).
The primitive amplitude $A_n^{(0)}$ coincides with $V_n$,
if all gluons are on-shell:
\bq
 A_n^{(0)}(p_1^{\lambda_1},...,p_n^{\lambda_n}) 
 & = &
 V_n(p_1^{\lambda_1},...,p_n^{\lambda_n}).
\eq
This gives us the following algorithm:
\begin{algorithm}
Calculation of the primitive helicity amplitude $A_{n}^{(0)}( p_1^{\lambda_1}, ..., p_n^{\lambda_n} )$
with $k$ gluons of negative helicity
from MHV vertices:
\begin{enumerate}
\item Calculate recursively the N${}^{k-2}$MHV vertex $V_n(p_1^{\lambda_1},...,p_n^{\lambda_n})$, 
using eq.~(\ref{MHV_recursion}).
\item The N${}^{k-2}$MHV amplitude $A_{n}^{(0)}( p_1^{\lambda_1}, ..., p_n^{\lambda_n} )$ is then given by
\bq
 A_n^{(0)}(p_1^{\lambda_1},...,p_n^{\lambda_n}) 
 & = &
 V_n(p_1^{\lambda_1},...,p_n^{\lambda_n}).
\eq
\end{enumerate}
\end{algorithm}
\noindent 
Note that this algorithm over-counts each contribution $(k-2)$ times.
This over-counting is compensated by the explicit factor $1/(k-2)$ in front.

\subsection{The Lagrangian of the MHV expansion}

Let us now consider a proof of the MHV expansion.
There are various possibilities to prove the MHV expansion.
The first approach makes use of a canonical transformation in the 
field variables~\cite{Gorsky:2005sf,Mansfield:2005yd,Ettle:2006bw,Ettle:2007qc,Ettle:2008ey,Buchta:2010qr}.
A second approach starts from an action in 
twistor space~\cite{Mason:2005kn,Mason:2005zm,Boels:2007qn,Boels:2006ir,Boels:2007gv,Boels:2008fc}.
The action in twistor space has an extended gauge symmetry.
The conventional Lagrangian and the MHV Lagrangian are then obtained from the action in twistor space for different gauge
choices.
A third approach proves the MHV expansion with the help of on-shell recursion relations \cite{Risager:2005vk},
which will be discussed in section~(\ref{sect:BCFW_recursion}).

Here, we will discuss the approach based on a canonical transformation.
The exposition follows \cite{Buchta:2010qr}.
The derivation proceeds through four steps.
We will be working in light-cone coordinates.
We defined contra-variant light-cone coordinates in eq.~(\ref{def_contravariant_lightcone_coordinates}).
For the co-variant light-cone coordinates we use the convention
\bq
 x_+=\frac{1}{\sqrt{2}}\left(x_0+x_3\right), 
 \;\;\;
 x_-=\frac{1}{\sqrt{2}}\left(x_0-x_3\right), 
 \;\;\;
 x_{\bot}=\frac{1}{\sqrt{2}}\left(x_1+ix_2\right), 
 \;\;\;
 x_{\bot\ast}=\frac{1}{\sqrt{2}}\left(x_1-ix_2\right).
\eq
With these conventions we have
\bq
 p^\mu x_\mu & = & p^+ x_+ + p^- x_- + p^\bot x_{\bot\ast} + p^{\bot\ast} x_\bot.
\eq
For the vector $\vec{x}=(x_-,x_{\bot},x_{\bot\ast})$ we set in this section
\bq
 \vec{p} \cdot \vec{x} & = & p^- x_- + p^\bot x_{\bot\ast} + p^{\bot\ast} x_\bot.
\eq
We define the spinors as in eq.~(\ref{def_spinors}) with $\lambda_p=1$.
This definition applies to all four-vectors $p^\mu$. 
If the four-vector $p^\mu$ is light-like, the spinors are the eigenstates of the Dirac equation
with eigenvalue zero.
If the four-vector $p^\mu$ is not light-like, eq.~(\ref{def_spinors}) defines the off-shell continuation of the
spinors.

{\bf Step 1: Light-cone gauge}.
Our starting point is the Lagrangian of Yang-Mills theory, given in
eq.~(\ref{Lagrangian_YM}). We can re-write this Lagrangian as
\bq
{\mathcal L}_{\mathrm{YM}} & = & 
 \frac{1}{2g^2} \; \mathrm{Tr} \; F_{\mu\nu} F^{\mu\nu}
 \;\; = \;\;
 {\mathcal L}_2 + {\mathcal L}_3 + {\mathcal L}_4,
\eq
such that ${\mathcal L}_2$ contains all terms bilinear in the gauge fields.
Terms with three or four gauge fields are collected in ${\mathcal L}_3$ and ${\mathcal L}_4$, respectively.
We choose the light-cone gauge
\bq
A_- & = & 0.
\eq
In this gauge we have
\bq
\label{lagrangian_light_cone}
 {\mathcal L}_2 \! & = & \! 
 \frac{1}{g^2} 
 \mathrm{Tr} \left[ 
                  A_+ \partial_-^2 A_+
                - 2 A_+ \partial_- \partial_{\bot} A_{\bot\ast}
                - 2 A_+ \partial_- \partial_{\bot\ast} A_{\bot}
                + A_{\bot} \partial_{\bot\ast}^2 A_{\bot}
                + A_{\bot\ast} \partial_{\bot}^2 A_{\bot\ast}
 \right. \nonumber \\
 \! & & \left. \!
                + 2 A_{\bot\ast} \left( 2 \partial_- \partial_+ - \partial_{\bot} \partial_{\bot\ast} \right) A_{\bot}
           \right],
 \nonumber \\
 {\mathcal L}_3 \! & = & \! 
 \frac{2}{g^2} 
 \mathrm{Tr} \left[ 
                 \left( \partial_{\bot} A_{\bot\ast} \right) \left[ A_{\bot\ast}, A_{\bot} \right]
               + \left( \partial_{\bot\ast} A_{\bot} \right) \left[ A_{\bot}, A_{\bot\ast} \right]
               - \left( \partial_- A_{\bot} \right) \left[ A_+, A_{\bot\ast} \right]
               - \left( \partial_- A_{\bot\ast} \right) \left[ A_+, A_{\bot} \right]
            \right],
 \nonumber \\
 {\mathcal L}_4 \! & = & \! 
 -\frac{1}{g^2} 
 \mathrm{Tr} \left[ A_{\bot}, A_{\bot\ast} \right] \left[ A_{\bot}, A_{\bot\ast} \right].
\eq

{\bf Step 2: Integrating out $A_+$}.
We observe that the field $A_+$ 
occurs only quadratically or linearly in eq.~(\ref{lagrangian_light_cone}). 
We can therefore integrate this field out, 
similar to way we treated the field $B_{[\mu\nu]}$ in section~(\ref{sect:colour}).
After integrating out $A_+$ we can write the Lagrange density as
\bq
\label{Lagrangian_transverse}
 {\mathcal L}_{\mathrm{YM}} & = & 
 {\mathcal L}_{+-} + {\mathcal L}_{++-} + {\mathcal L}_{+--} + {\mathcal L}_{++--},
\eq
with
\bq
\label{Lagrangian_transverse2}
 {\mathcal L}_{+-} & = & 
  \frac{4}{g^2} \mathrm{Tr}
  A_{\bot\ast} \left( \partial_- \partial_+ - \partial_{\bot} \partial_{\bot\ast} \right) A_{\bot},
 \nonumber \\
 {\mathcal L}_{++-} & = & 
  \frac{4}{g^2} \mathrm{Tr} \;
     \left( \partial_{\bot\ast} A_{\bot} \right) \partial_-^{-1} \left[ A_{\bot}, \partial_- A_{\bot\ast} \right],
 \nonumber \\
 {\mathcal L}_{+--} & = & 
  \frac{4}{g^2} \mathrm{Tr} \;
     \left( \partial_{\bot} A_{\bot\ast} \right) \partial_-^{-1} \left[ A_{\bot\ast}, \partial_- A_{\bot} \right],
 \nonumber \\
 {\mathcal L}_{++--} & = & 
  -\frac{4}{g^2} \mathrm{Tr} \;
          \left[ A_{\bot\ast}, \partial_- A_{\bot} \right] \partial_-^{-2} \left[ A_{\bot}, \partial_- A_{\bot\ast} \right].
\eq
The Lagrange density contains now only the transverse degrees of freedom for the field $A$.

{\bf Step 3: Canonical transformation}.
In the third step we eliminate the non-MHV vertices contained in ${\mathcal L}_{++-}$ by 
a canonical change of the field variables.

To motivate the canonical transformation we treat the variable $x_+$ as a time variable
and collect the remaining three variables in a vector $\vec{x}=(x_-,x_{\bot},x_{\bot\ast})$. 
In order to simplify the notation we will suppress the dependence of the fields on $x_+$ and write
$A(\vec{x})$ instead of $A(x_+,\vec{x})$.
We will denote the new field after the canonical transformation with a tilde, e.g.
\bq
 A \rightarrow \tilde{A}.
\eq
Now let us look again at eq.~(\ref{Lagrangian_transverse}) and eq.~(\ref{Lagrangian_transverse2}).
The ``momentum'' conjugate to $A_\bot^a$ is
\bq
 \frac{\delta {\mathcal L}_{\mathrm{YM}}}{\delta \partial_+ A_{\bot}^a}
 =
 2 \partial_- A_{\bot\ast}^a.
\eq
We look for a canonical transformation, where the generating function of  the transformation depends 
on the new ``coordinates'' $\tilde{A}_{\bot}$
and the old ``momenta'' $\partial_- A_{\bot\ast}$:
\bq
 G\left[\tilde{A}_{\bot}, \partial_- A_{\bot\ast} \right]
 & = &  
 \int d^3y 
 \;
  A^a_{\bot}\left[\tilde{A}_{\bot}(\vec{y})\right] 
 \;
 \partial_- A_{\bot\ast}^a(\vec{y})
\eq
The new ``momenta'' are then given by
\bq
\label{new_momenta}
 \partial_- \tilde{A}_{\bot\ast}^a(\vec{x})
 & = & \int d^3y \; \frac{\delta A_{\bot}^b(\vec{y})}{\delta \tilde{A}_{\bot}^a(\vec{x})} 
                 \; \partial_- A_{\bot\ast}^b(\vec{y}).
\eq
The transformation should eliminate the unwanted ${\mathcal L}_{++-}$ term, therefore we require
\bq
\label{elimination_++-}
 {\mathcal L}_{+-}\left[\tilde{A}_{\bot},\tilde{A}_{\bot\ast}\right]
 & = &
 {\mathcal L}_{+-}\left[A_{\bot},A_{\bot\ast}\right]
 +
 {\mathcal L}_{++-}\left[A_{\bot},A_{\bot\ast}\right].
\eq
The fact that the transformation is canonical implies
\bq
\label{tranformation_kinetic_terms}
 \int d^3x \; 
 2 \left( \partial_- A_{\bot\ast}^a \right) \left( \partial_+ A_{\bot}^a \right) 
 & = &
 \int d^3x \; 
 2 \left( \partial_- \tilde{A}_{\bot\ast}^a \right) \left( \partial_+ \tilde{A}_{\bot}^a \right).
\eq
We then plug the expressions in eq.~(\ref{new_momenta}) into eq.~(\ref{elimination_++-}) and use
eq.~(\ref{tranformation_kinetic_terms}).
It is convenient to introduce the following two differential operators
\bq
 \omega = \frac{\partial_{\bot} \partial_{\bot\ast}}{\partial_-},
 & &
 \zeta = \frac{\partial_{\bot\ast}}{\partial_-}.
\eq
From the coefficient of $\partial_- A_{\bot\ast}$
we find the integro-differential equation
\bq
\label{eq_canonical_trafo}
 \omega A_{\bot}^a(\vec{x}) - g f^{abc} \left( \zeta A_{\bot}^b(\vec{x}) \right) A_{\bot}^c(\vec{x}) 
 & = &
 \int d^3y \frac{\delta A_{\bot}^a(\vec{x})}{\delta \tilde{A}_{\bot}^b(\vec{y})} 
 \omega_y \tilde{A}_{\bot}^b(\vec{y}).
\eq
The solution to the integro-differential equation~(\ref{eq_canonical_trafo}) is given by
\bq
\label{solution_canonical_trafo}
 A_{\bot}^a\left(\vec{x}\right)
 & = &
 \sum\limits_{n=1}^\infty 
 2 \; \mathrm{Tr} \left( T^a T^{a_1} ... T^{a_n} \right)
 \int \frac{d^3p_1}{(2\pi)^3} ... \frac{d^3p_n}{(2\pi)^3}
 e^{-i (\vec{p}_1 + ... + \vec{p}_n ) \cdot \vec{x}} 
 \\
 & &
 \Upsilon\left(\vec{p}_1,...,\vec{p}_n\right)
 \tilde{A}_{\bot}^{a_1}\left(\vec{p}_1\right) ... \tilde{A}_{\bot}^{a_n}\left(\vec{p}_n\right),
 \nonumber \\
 A_{\bot\ast}^a\left(\vec{x}\right)
 & = & 
 \sum\limits_{n=1}^\infty 
 \sum\limits_{r=1}^n
 2 \; \mathrm{Tr} \left( T^a T^{a_1} ... T^{a_n} \right)
 \int \frac{d^3p_1}{(2\pi)^3} ... \frac{d^3p_n}{(2\pi)^3}
 e^{-i (\vec{p}_1 + ... + \vec{p}_n ) \cdot \vec{x}} 
 \nonumber \\
 & &
 \Xi_r\left(\vec{p}_1,...,\vec{p}_n\right)
 \tilde{A}_{\bot}^{a_1}\left(\vec{p}_1\right) 
 ... 
 \tilde{A}_{\bot}^{a_{r-1}}\left(\vec{p}_{r-1}\right) 
 \tilde{A}_{\bot\ast}^{a_r}\left(\vec{p}_r\right) 
 \tilde{A}_{\bot}^{a_{r+1}}\left(\vec{p}_{r+1}\right) 
 ... 
 \tilde{A}_{\bot}^{a_n}\left(\vec{p}_n\right).
 \nonumber
\eq
The coefficient functions are given by
\bq
\label{coefficient_functions}
 \Upsilon\left(\vec{p}_1,...,\vec{p}_n\right) 
 & = &
 \frac{\left(  \sqrt{2} g \right)^{n-1}}{\langle p_1 p_2 \rangle ... \langle p_{n-1} p_n \rangle} \frac{p_1^-+...+p_n^-}{\sqrt{p_1^- p_n^-}},
 \nonumber \\
 \Xi_r\left(\vec{p}_1,...,\vec{p}_n\right) 
 & = & \left( \frac{p_r^-}{p_1^-+...+p_n^-} \right)^2 \Upsilon\left(\vec{p}_1,...,\vec{p}_n\right).
\eq
\bs
{\it {\bf Exercise \theexercise}: 
Derive eq.~(\ref{solution_canonical_trafo}) together with eq.~(\ref{coefficient_functions}) from eq.~(\ref{eq_canonical_trafo}).
\stepcounter{exercise}
}
\es
\\
\\
We remark that the field $A_{\bot}^a\left(\vec{x}\right)$ is expressed in terms of the fields
$\tilde{A}_{\bot}^{a}\left(\vec{p}\right)$ alone, while the field
$A_{\bot\ast}^a\left(\vec{x}\right)$ involves the fields 
$\tilde{A}_{\bot\ast}^{a}\left(\vec{p}\right)$ and
$\tilde{A}_{\bot}^{a}\left(\vec{p}\right)$.
The new fields agree with the old fields to leading order in $g$:
\bq
\label{simple_equivalence}
 A_{\bot}^a\left(\vec{x}\right)
 = 
 \tilde{A}_{\bot}^{a}\left(\vec{x}\right) + {\mathcal O}\left(g\right),
 & &
 A_{\bot\ast}^a\left(\vec{x}\right)
 =  
 \tilde{A}_{\bot\ast}^{a}\left(\vec{x}\right) + {\mathcal O}\left(g\right).
\eq

{\bf Step 5: Assembling the pieces}.
We are now in a position to put all the pieces together.
Inserting the solutions~(\ref{solution_canonical_trafo}) 
of the canonical transformation into the Lagrange density~(\ref{Lagrangian_transverse})
one finds that the Lagrange density can be written in the following form:
\bq
\label{Lagrangian_MHV_assembled}
 {\mathcal L}_{\mathrm{YM}} & = &
 {\mathcal L}_{\mathrm{kin}} 
 + \sum\limits_{n=3}^\infty {\mathcal L}^{(n)}.
\eq
The first term ${\mathcal L}_{\mathrm{kin}}$ is rather simple and contains the kinetic term:
\bq
\label{Lagrangian_kinetic}
 {\mathcal L}_{\mathrm{kin}}
 & = &
 - \tilde{A}^a_{\bot\ast}(x) \Box \tilde{A}^a_{\bot}(x).
\eq
We further obtain an ascending tower of interaction vertices.
Each interaction vertex is most conveniently expressed with the help of the Fourier transforms.
One finds for ${\mathcal L}^{(n)}$
\bq
 {\mathcal L}^{(n)}
 & = & 
 \frac{1}{2}
   \sum\limits_{j=2}^n
   \;
   \int \frac{d^4p_1}{(2\pi)^4} ... \frac{d^4p_n}{(2\pi)^4}
   e^{-i\left(p_1+...+p_n\right)\cdot x}
   \;
   \alpha_{j}\left(p_1,...,p_n\right)
 \nonumber \\
 & &
  2 \mathrm{Tr}\left( 
   \tilde{A}_{\bot\ast}(p_1) \tilde{A}_{\bot}(p_2) ... \tilde{A}_{\bot}(p_{j-1}) 
   \tilde{A}_{\bot\ast}(p_j) 
   \tilde{A}_{\bot}(p_{j+1}) ... \tilde{A}_{\bot}(p_{n}) \right).
\eq
The vertex function $\alpha_{j}\left(p_1,...,p_n\right)$ is given 
\bq
 \alpha_{j}\left(p_1,...,p_n\right) 
 & = &
 - \frac{1}{g^2}
 \left(  i \sqrt{2} \right)^{n-2}
 \frac{\langle p_1 p_j \rangle^4}{\langle p_1 p_2 \rangle \langle p_2 p_3 \rangle ... \langle p_{n-1} p_n \rangle \langle p_n p_1 \rangle}
\eq
and corresponds exactly to the MHV formula.
Note that the vertex function depends only on the light-cone coordinates $p^{\bot\ast}$ and $p^-$, but not on $p^{\bot}$ and
$p^+$.
Each vertex contains two fields $\tilde{A}_{\bot\ast}$ with indices $1$ and $j$ and an arbitrary number of fields $\tilde{A}_{\bot}$.
Since the trace is cyclic, we have
\bq
 \mathrm{Tr}\left( T^{a_1} ... T^{a_{j-1}} T^{a_j} ... T^{a_n} \right) & = &
  \mathrm{Tr}\left( T^{a_j} ... T^{a_n} T^{a_1} ... T^{a_{j-1}} \right).
\eq
The factor $1/2$ takes into account that we are summing twice over identical traces.

Finally let us comment on a subtle point: We are interested in the scattering amplitudes
involving the original fields $A^a_\bot$ and $A^a_{\bot\ast}$.
Let us denotes these amplitudes by $A_n^{(0)}(p_1,...,p_n)$ and the amplitudes
involving the new fields $\tilde{A}^a_\bot$ and $\tilde{A}^a_{\bot\ast}$
by $\tilde{A}_n^{(0)}(p_1,...,p_n)$.
In order to obtain the amplitude $A_n^{(0)}(p_1,...,p_n)$ from the theory with the new
fields $\tilde{A}^a_\bot$ and $\tilde{A}^a_{\bot\ast}$ we have to go one step
back in quantum field theory 
and recall that $A_n^{(0)}$ is obtained from un-amputated Green's function through
the LSZ reduction procedure. 
At tree-level the LSZ reduction procedure corresponds to a multiplication with the inverse propagator
$(-i p_j^2)$ for each external leg $j$.
In the un-amputated Green's functions we replace the original fields 
$A^a_\bot$ and $A^a_{\bot\ast}$ by the new fields $\tilde{A}^a_\bot$ and $\tilde{A}^a_{\bot\ast}$,
using eq.~(\ref{solution_canonical_trafo}).
We have seen in eq.~(\ref{simple_equivalence}) that the new fields and the old fields agree
to leading order in $g$.
The additional higher-order terms introduced through eq.~(\ref{solution_canonical_trafo})
are called MHV completion vertices.
For $n\ge 4$ and non-exceptional kinematics the terms with MHV completion vertices miss at least
one singular propagator and are therefore killed by the LSZ reduction procedure.
In this case we simply have $A_n^{(0)}=\tilde{A}_n^{(0)}$.

This argumentation fails if the coefficient functions in eq.~(\ref{coefficient_functions})
become singular. This may happen for the three-point amplitudes.
In this case terms with MHV completion vertices may give a finite non-zero result
after the LSZ reduction procedure.
In particular this is the way how the three-point anti-MHV amplitude $A_3^{(0)}(1^+,2^+,3^-)$ is obtained
from the Lagrangian in eq.~(\ref{Lagrangian_MHV_assembled}).

This mechanism is sometimes called evasion of the $S$-matrix equivalence theorem \cite{Ettle:2007qc}.

% ----------------------------------------------------------------------------------
\newpage
\section{On-shell recursion}
\label{sect:BCFW_recursion}

In section~(\ref{sect_off_shell_recursion}) we discussed off-shell recurrence relations, which
allow us to compute an amplitude recursively from off-shell currents.
The off-shell currents are gauge-dependent objects.
There is nothing wrong with the fact that at intermediate stages we deal with
gauge-dependent objects, even in a Feynman diagram based calculation the evaluation
of an individual Feynman diagrams gives in general a gauge-dependent expression.
The amplitude itself is gauge-independent and all gauge-dependence cancels in the expression
for the amplitude.

Nevertheless we may ask, if there is a way to deal at all stages with on-shell gauge-invariant
objects only.
This is indeed possible and we will now discuss on-shell recursion relations.
On-shell recursion relations compute the primitive tree amplitude $A_n^{(0)}$ with $n$
external particles recursively from primitive tree amplitudes with fewer legs.
Obviously, the primitive tree amplitudes with fewer legs entering this calculation
satisfy momentum conservation and the on-shell conditions.
However we do not require that the momenta of the external particles are real.
We will allow complex external momenta.
This may seem strange at first sight, but a tree-level primitive amplitude is a rational
function of the external momenta and nothing stops us to evaluate this function for complex
external momenta.
Allowing complex momenta opens the door to the tools of complex analysis and Cauchy's theorem
in particular.

A special role is played by the three-point amplitudes.
The equal helicity ones vanish
$A_3^{(0)}(1^+,2^+,3^+) = A_3^{(0)}(1^-,2^-,3^-) = 0$, the ones with mixed helicities are given
by
\bq
\label{three_point_amplitudes}
 A_3^{(0)}\left(1^-,2^-,3^+\right)
 \;\; = \;\;
 i \sqrt{2}
 \frac{\langle 1 2 \rangle^3}{\langle 2 3  \rangle \langle 3 1  \rangle},
 & &
 A_3^{(0)}\left(1^+,2^+,3^-\right)
 \;\; = \;\;
 i \sqrt{2}
 \frac{[ 2 1 ]^3}{[1 3 ] [3 2]},
\eq
and cyclic permutation thereof.
Momentum conservation and the on-shell conditions read
\bq
 p_1+p_2+p_3 \;\; = \;\; 0,
 & &
 p_1^2
 \; = \;
 p_2^2
 \; = \;
 p_3^2
 \; = \;
 0.
\eq
Therefore 
\bq
 2 p_1 \cdot p_2 
 \;\; = \;\; 
 \left( p_1 + p_2 \right)^2
 \;\; = \;\; 
 p_3^2
 \;\; = \;\; 
 0.
\eq
On the other hand we have
\bq
 2 p_1 \cdot p_2 
 \;\; = \;\; 
 \left\langle p_1 p_2 \right\rangle \left[ p_2 p_1 \right],
\eq
and it follows that either $\langle p_1 p_2 \rangle = 0$ {\bf or} $[ p_1 p_2 ] = 0$.
The same conclusion is reached for the other spinor products.
For real momenta we have in addition
\bq
\label{real_restriction}
 \left| \left\langle p_1 p_2 \right\rangle \right|
  & = &
 \left| \left[ p_2 p_1 \right] \right|,
\eq
and therefore it follows that $\langle p_1 p_2 \rangle = 0$ {\bf and} $[ p_1 p_2 ] = 0$.
In the real case all three-point amplitudes vanish. 
A rough argument uses the fact, that in eq.~(\ref{three_point_amplitudes})
there are three spinor products in the numerator, but only two in the denominator.
A more careful analysis uses a parametrisation of the momenta like in eq.~(\ref{parametrisation_collinear}) and shows that the amplitudes vanish in the limit 
$k_\perp \rightarrow 0$.

However for complex momenta we do not have eq.~(\ref{real_restriction})
and the three-point amplitudes may be non-zero.

\subsection{BCFW recursion}

\begin{figure}
\begin{center}
\includegraphics[scale=1.0]{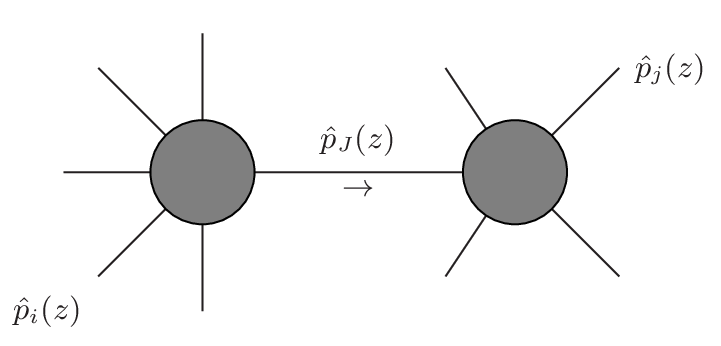}
\end{center}
\caption{Contribution to the residues from an on-shell propagator. For a $z$-dependent momentum flow 
through the indicated internal propagator, particle $i$ has to be on the one side and particle $j$ has to 
be on the other side.}
\label{fig_on_shell_residue}
\end{figure}

Let us now derive the on-shell recursion relation. This relation is also known as 
Britto-Cachazo-Feng-Witten (BCFW) recursion relation \cite{Britto:2004ap,Britto:2005fq}.
We consider a tree-level primitive amplitude $A_n^{(0)}(p_1^{\lambda_1}, ...,p_n^{\lambda_n})$.
The basic idea is to pick two momenta, say $p_i$ and $p_j$, and to {\bf deform} them as
\bq
\label{shift_1}
 \hat{p}_i^\mu \;\; = \;\; p_i^\mu - z n^\mu,
 & &
 \hat{p}_j^\mu \;\; = \;\; p_j^\mu + z n^\mu,
\eq
where $z$ is a complex variable and $n$ a light-like four-vector ($n^2=0$).
It is clear that this deformation respects momentum conservation.
If in addition we require
\bq
\label{on_shell_conditions_shift}
 2 p_i \cdot n \;\; = \;\; 0,
 & &
 2 p_j \cdot n \;\; = \;\; 0,
\eq
also the on-shell conditions $\hat{p}_i^2 = \hat{p}_j^2 = 0$ are satisfied.
The first question to be asked is the following:
Does such a light-like four-vector $n^\mu$ satisfying
eq.~(\ref{on_shell_conditions_shift}) exist?
It does, a possible choice (not the only one) is given by
\bq
\label{shift_2}
 n^\mu
 & = &
 \frac{1}{2}
 \left\langle i+ \left| \gamma^\mu \right| j+ \right\rangle.
\eq
The four-vector $n^\mu$ is light-like
\bq
 n^2
 & = &
 \frac{1}{4}
 \left\langle i+ \left| \gamma^\mu \right| j+ \right\rangle
 \left\langle i+ \left| \gamma_\mu \right| j+ \right\rangle
 \;\; = \;\;
 0,
\eq
due to the Fierz identity.
The four-vector $n^\mu$ satisfies in addition the conditions in eq.~(\ref{on_shell_conditions_shift}):
\bq
 2 p_i \cdot n 
 & = & 
 \left\langle i+ \left| p\!\!\!/_i \right| j+ \right\rangle
 \;\; = \;\;
 0,
 \nonumber \\
 2 p_j \cdot n 
 & = & 
 \left\langle i+ \left| p\!\!\!/_j \right| j+ \right\rangle
 \;\; = \;\;
 0.
\eq
Let us now consider the amplitude with the two deformed momenta. We may view this amplitude
as a function of $z$:
\bq
 A\left(z\right)
 & = &
 A_n^{(0)}\left(p_1^{\lambda_1}, ..., \hat{p}_i(z)^{\lambda_i}, ..., \hat{p}_j(z)^{\lambda_j}, ...., p_n^{\lambda_n}\right).
\eq
$A(z)$ is a rational function of $z$, since tree amplitudes are rational functions of the momenta
variables.
Let us assume that $A(z)$ falls off at least like $1/z$ for $|z|\rightarrow \infty$.
We will discuss the detailed conditions for this to happen in the next paragraph.
With this assumption we have
\bq
 \frac{1}{2\pi i} \oint dz \frac{A\left(z\right)}{z}
 & = &
 0,
\eq
where the contour is a large circle at $|z| = \infty$ oriented counter-clockwise.
On the other hand we may evaluate this integral with the help of {\bf Cauchy's residue theorem}.
There is one residue at $z=0$ due to the explicit factor of $1/z$ in the integrand.
This residue gives 
\bq
 A\left(0\right)
 & = &
 A_n^{(0)}\left(p_1^{\lambda_1}, ..., p_i^{\lambda_i}, ..., p_j^{\lambda_j}, ...., p_n^{\lambda_n}\right),
\eq
which is the undeformed amplitude we want to calculate.
There cannot be any residue coming from the explicit representation of the polarisation vectors.
This would depend on the reference momenta $q_i$, however the amplitude is gauge-invariant and therefore
independent of the choice of the reference momenta $q_i$.
Therefore all other residues come from internal propagators of the amplitude.
We have to consider only the propagators, which are $z$-dependent.
Let us consider a situation as shown in fig.~(\ref{fig_on_shell_residue}).
It is clear that in order to have a $z$-dependent propagator
particle $i$ has to be on the one side and particle $j$ on the other side.
Let us denote the set of external legs on the one side by $I$ and 
the set of external legs on the other side by $J$.
Let us further assume that $|I|=m$ and $|J|=n-m$.
We set
\bq
 p_J & = & \sum\limits_{j \in J} p_j.
\eq
The momentum flowing through the internal propagator is
\bq
 \hat{p}_J & = & p_J + z n.
\eq
The internal propagator goes on-shell for
\bq
 \hat{p}_J^2 \; = \; 0
 & \Rightarrow &
 z_J \; = \;
 - \frac{p_J^2}{2 p_J \cdot n}.
\eq
In this limit the amplitude factorises as in eq.~(\ref{factorisation_propagator_pole})
and the residue is given by
\bq
 - 
 \sum\limits_{\lambda}
 A_{m+1}^{(0)}\left(..., \hat{p}_i(z_J)^{\lambda_i}, ..., \hat{p}_J(z_J)^{\lambda}\right)
 \frac{i}{p_J^2}
 A_{n-m+1}^{(0)}\left(..., \hat{p}_j(z_J)^{\lambda_j}, ...., -\hat{p}_J(z_J)^{-\lambda}\right).
\eq
Summing over all residues we obtain the {\bf on-shell recursion relation}:
\bq
\label{on_shell_recursion}
\lefteqn{
 A_n^{(0)}\left(p_1^{\lambda_1}, ..., p_i^{\lambda_i}, ..., p_j^{\lambda_j}, ...., p_n^{\lambda_n}\right)
 = } & & 
 \\
 & &
 \sum\limits_{\mathrm{partitions}\; (I,J)}
 \sum\limits_{\lambda}
 A_{m+1}^{(0)}\left(..., \hat{p}_i(z_J)^{\lambda_i}, ..., \hat{p}_J(z_J)^{\lambda}\right)
 \frac{i}{p_J^2}
 A_{n-m+1}^{(0)}\left(..., \hat{p}_j(z_J)^{\lambda_j}, ...., -\hat{p}_J(z_J)^{-\lambda}\right),
 \nonumber
\eq
where the sum is over all partitions $(I,J)$ such that particle $i \in I$ and particle $j \in J$.
The momentum $\hat{p}_J(z_J)$ is again on-shell ($\hat{p}_J(z_J))^2=0$), the momentum $p_J$ appearing in the
denominator is in general not on-shell ($p_J^2 \neq 0$).
Eq.~(\ref{on_shell_recursion}) allows us to compute the $n$-particle amplitude $A_n^{(0)}$ recursively through
on-shell amplitudes with fewer external legs.
In order to apply this recursion relation we have to ensure that the amplitude $A(z)$ vanishes
as $|z|\rightarrow \infty$.

\subsection{Momenta shifts and the behaviour at infinity}

Let us now consider the momenta shifts and the behaviour at $|z| = \infty$ in more detail.
Usually we view the amplitude $A_n^{(0)}$ as a function of the four-momenta 
$p_j^\mu$. However, we may replace each four-vector by a pair of two-component
Weyl spinors.
In detail this is done as follows:
Each four-vector $p_\mu$ has a bi-spinor representation, given by
\bq
\label{bispinor_representation}
 p_{A\dot{B}} = p_\mu \sigma^\mu_{A\dot{B}},
 & &
 p_\mu = \frac{1}{2} p_{A\dot{B}} \bar{\sigma}_\mu^{\dot{B}A}.
\eq
For light-like vectors this bi-spinor representation factorises into a dyad of Weyl spinors:
\bq
\label{dyad}
 p_\mu p^\mu = 0
 & \Leftrightarrow &
 p_{A\dot{B}} = p_{A} p_{\dot{B}}.
\eq
The equations~(\ref{bispinor_representation}) and~(\ref{dyad})
allow us to convert any light-like four-vector into a dyad of Weyl spinors and vice versa.
Therefore the partial amplitude $A_n^{(0)}$, 
being originally a function of the momenta $p_j$ and helicities
$\lambda_j$,
can equally be viewed as a function of the Weyl spinors $p_A^j$, $p_{\dot{B}}^j$ and the helicities
$\lambda_j$:
\bq
 A_n^{(0)}\left(p_1^{\lambda_1},...,p_n^{\lambda_n}\right) 
 & = & 
 A_n^{(0)}\left( p_A^1, p_{\dot{B}}^1, \lambda_1, ..., p_A^n, p_{\dot{B}}^n, \lambda_n \right).
\eq
Note that for an arbitrary pair of Weyl spinors, the corresponding four-vector will in general be complex-valued.
The shift defined in eq.~(\ref{shift_1}) combined with eq.~(\ref{shift_2}) reads
\bq
\label{shift_12}
 \hat{p}_i^\mu \;\; = \;\; p_i^\mu - \frac{1}{2} z \left\langle i+ \left| \gamma^\mu \right| j+ \right\rangle,
 & &
 \hat{p}_j^\mu \;\; = \;\; p_j^\mu + \frac{1}{2} z \left\langle i+ \left| \gamma^\mu \right| j+ \right\rangle,
\eq
and corresponds to the following shift in the Weyl spinors:
\bq
\label{spinor_shift}
 \hat{p}_A^i = p_A^i  - z p_A^j,
 & &
 \hat{p}_{\dot{B}}^i = p_{\dot{B}}^i,
 \nonumber \\
 \hat{p}_A^j = p_A^j,
 & &
 \hat{p}_{\dot{B}}^j = p_{\dot{B}}^j + z p_{\dot{B}}^i.
\eq
Eq.~(\ref{spinor_shift}) deforms the spinors $p_A^i$ and $p_{\dot{B}}^j$.
The spinors $p_{\dot{B}}^i$ and $p_A^j$ remain unchanged.
Let us now study the effect of the shift on the polarisation vectors
of the gluons.
We have
\bq
\eps_{\mu}^{+}(p,q) = -\frac{\langle p+|\gamma_{\mu}|q+\rangle}{\sqrt{2} \langle p- | q + \rangle}
 =
 - \frac{p_{\dot{A}} \bar{\sigma}_\mu^{\dot{A}B} q_B}{\sqrt{2} p_C \eps^{CD} q_D},
 & &
\eps_{\mu}^{-}(p,q) = \frac{\langle p-|\gamma_{\mu}|q-\rangle}{\sqrt{2} \langle p + | q - \rangle}
 = 
 \frac{q_{\dot{A}} \bar{\sigma}_\mu^{\dot{A}B} p_B}{\sqrt{2} p_{\dot{C}} \eps^{\dot{C}\dot{D}} q_{\dot{D}}}.
 \nonumber
\eq
We may now read-off the large-$z$ behaviour of the polarisation vectors:
\bq
\begin{array}{lccl}
 \eps_\mu^+\left(\hat{p}_i,q_i\right) \; : \; z^{-1},
 &&&
 \eps_\mu^+\left(\hat{p}_j,q_j\right) \; : \; z, 
 \\
 \eps_\mu^-\left(\hat{p}_i,q_i\right) \; : \; z,
 &&&
 \eps_\mu^-\left(\hat{p}_j,q_j\right) \; : \; z^{-1}.
 \end{array}
\eq
Let us now consider the amplitude
\bq
 A_n^{(0)}\left( ..., \hat{p}_i^+, ..., \hat{p}_j^-, ... \right)
\eq
For this helicity configuration $(\lambda_i,\lambda_j)=(+,-)$ 
and the shift as in eq.~(\ref{shift_12}) each individual Feynman diagram 
vanishes for $z \rightarrow \infty$. 
In order to see this consider the flow of the $z$-dependence in a particular diagram.
The most dangerous contribution comes from a path, where all vertices are three-gluon-vertices. 
For a path made of $n$ propagators we have $n+1$ vertices and the product of propagators
and vertices behaves therefore like $z$ for large $z$. 
This statement remains true for a path containing only one vertex and no propagators.
The polarisation vectors for the helicity combination $(\lambda_i,\lambda_j)=(+,-)$ 
contribute a factor $1/z^2$, therefore the complete diagram
behaves like $1/z$ and vanishes therefore for $z \rightarrow \infty$.
\\
\\
\bs
{\it {\bf Exercise \theexercise}: 
Consider the MHV amplitude $A_n^{(0)}(...,\hat{p}_i^+,...,\hat{p}_j^-,...,p_k^-,...)$, where the
particles $j$ and $k$ have negative helicity, while all other particles have positive helicity.
Consider the shift of eq.~(\ref{shift_12}), which deforms the momenta of the particles $i$ and $j$.
What is the large-$z$ behaviour of the amplitude?
\stepcounter{exercise}
}
\es
\\
\\
What about the other helicity configurations? There is an alternative approach
to study the large-$z$ behaviour.
Physically, the large-$z$ limit corresponds to a hard particle moving
through a soft background \cite{ArkaniHamed:2008yf}.
In this limit we may use eikonal approximations to study the large-$z$ behaviour and one finds
that under the shift defined in eq.~(\ref{shift_12}) the amplitude $A_n^{(0)}(...,\hat{p}_i^{\lambda_i},...,\hat{p}_j^{\lambda_j},...)$ behaves at least as follows
for the various helicity configurations
\bq
\begin{array}{lccl}
 \left( \lambda_i, \lambda_j \right) = (+,+) \; : \; z^{-1},
 &&&
 \left( \lambda_i, \lambda_j \right) = (+,-) \; : \; z^{-1},
 \\
 \left( \lambda_i, \lambda_j \right) = (-,+) \; : \; z^3,
 &&&
 \left( \lambda_i, \lambda_j \right) = (-,-) \; : \; z^{-1}.
 \\
\end{array}
\eq
We see that we may use the momentum shift of eq.~(\ref{shift_12}) for all helicitiy configurations
except $(\lambda_i,\lambda_j)=(-,+)$.

What about the helicity configuration $(\lambda_i,\lambda_j)=(-,+)$?
The answer is simple: Instead of the momentum shift of eq.~(\ref{shift_12}) consider
the shift
\bq
\label{shift_12bar}
 \hat{p}_i^\mu \;\; = \;\; p_i^\mu - \frac{1}{2} z \left\langle j+ \left| \gamma^\mu \right| i+ \right\rangle,
 & &
 \hat{p}_j^\mu \;\; = \;\; p_j^\mu + \frac{1}{2} z \left\langle j+ \left| \gamma^\mu \right| i+ \right\rangle.
\eq
This shift corresponds to the following shift in the Weyl spinors:
\bq
 \hat{p}_A^i = p_A^i,
 & &
 \hat{p}_{\dot{B}}^i = p_{\dot{B}}^i - z p_{\dot{B}}^j,
 \nonumber \\
 \hat{p}_A^j = p_A^j + z p_A^i,
 & &
 \hat{p}_{\dot{B}}^j = p_{\dot{B}}^j.
\eq
We now shift the spinors $p_{\dot{B}}^i$ and $p_A^j$ and leave the spinors 
$p_A^i$ and $p_{\dot{B}}^j$ untouched.
The large-$z$ behaviour of the polarisation vectors is now given by
\bq
\begin{array}{lccl}
 \eps_\mu^+\left(\hat{p}_i,q_i\right) \; : \; z,
 &&&
 \eps_\mu^+\left(\hat{p}_j,q_j\right) \; : \; z^{-1}, 
 \\
 \eps_\mu^-\left(\hat{p}_i,q_i\right) \; : \; z^{-1},
 &&&
 \eps_\mu^-\left(\hat{p}_j,q_j\right) \; : \; z,
 \end{array}
\eq
and one finds that the amplitude behaves for large $z$ at least as
\bq
\begin{array}{lccl}
 \left( \lambda_i, \lambda_j \right) = (+,+) \; : \; z^{-1},
 &&&
 \left( \lambda_i, \lambda_j \right) = (+,-) \; : \; z^3,
 \\
 \left( \lambda_i, \lambda_j \right) = (-,+) \; : \; z^{-1},
 &&&
 \left( \lambda_i, \lambda_j \right) = (-,-) \; : \; z^{-1}.
 \\
\end{array}
\eq
Therefore we may use the shift of eq.~(\ref{shift_12bar}) for all helicity configurations
except $(\lambda_i,\lambda_j)=(+,-)$
and we see that for each helicity configuration there is at least one possible momenta shift,
such that the amplitude vanishes for $|z| \rightarrow \infty$.

Before we close this paragraph let us mention that there are more possibilities of deforming the
momenta, while keeping the on-shell conditions and momentum conservation.
Up to now we only considered shifts involving two particles.
As an alternative, we may deform the momenta (or spinors) of three particles.
Consider for example the three-particle shift \cite{Risager:2005vk}
\bq
\label{three_particle_shift_schouten}
 \hat{p}_A^i & = & p_A^i - z \left[ j k \right] \; \eta_A,
 \nonumber \\
 \hat{p}_A^j & = & p_A^j - z \left[ k i \right] \; \eta_A,
 \nonumber \\
 \hat{p}_A^k & = & p_A^k - z \left[ i j \right] \; \eta_A,
\eq
where $\eta_A$ is an arbitrary spinor.
The deformed spinors correspond to the following on-shell momenta
\bq
 \hat{p}_i^\mu & = & p_i^\mu - \frac{1}{2} z \left[ j k \right] \left\langle i+ \left| \gamma^\mu \right| \eta+ \right\rangle,
 \nonumber \\
 \hat{p}_j^\mu & = & p_j^\mu - \frac{1}{2} z \left[ k i \right] \left\langle j+ \left| \gamma^\mu \right| \eta+ \right\rangle,
 \nonumber \\
 \hat{p}_k^\mu & = & p_k^\mu - \frac{1}{2} z \left[ i j \right] \left\langle k+ \left| \gamma^\mu \right| \eta+ \right\rangle.
\eq
Momentum conservation is satisfied due to the Schouten identity.
Such a shift can be advantageous in situations, where particles $i$, $j$ and $k$ have positive
helicities. The polarisation vectors contribute then a factor $1/z^3$ to the large-$z$ behaviour.
We will see an application of this shift later on.

Shifts, which deform the momenta of more than two or three particles are also possible.
Ref.~\cite{Cheung:2015cba} contains an extensive discussion of generalised shifts.
As a rule of thumb, deforming more momenta may help to improve the large-$z$ behaviour.

\subsection{The on-shell recursion algorithm}

Let us now look at an example.
\begin{figure}
\begin{center}
\includegraphics[scale=1.0]{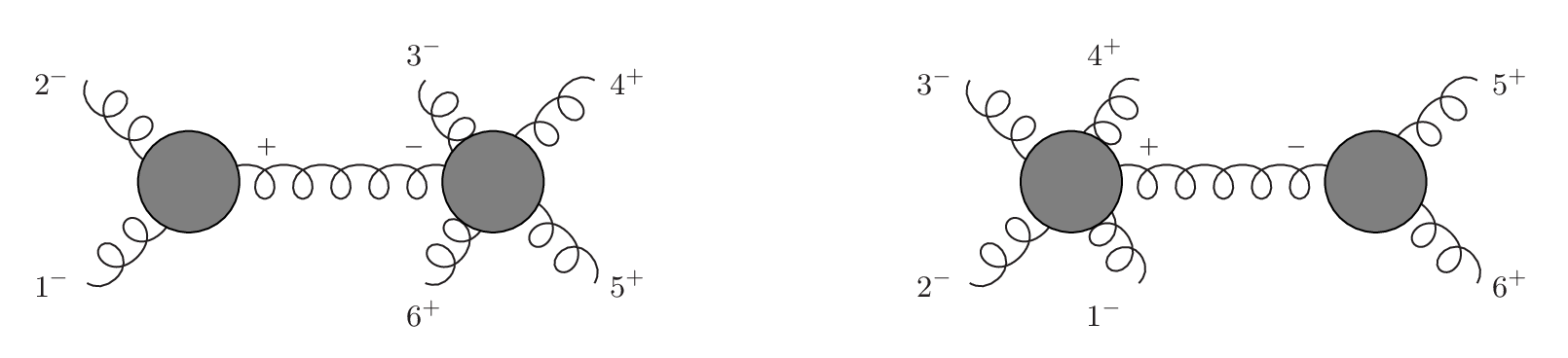}
\end{center}
\caption{Diagrams contributing to the tree-level six-gluon amplitude $A_6(1^-,2^-,3^-,4^+,5^+,6^+)$ in the on-shell 
recursive approach. The vertices are on-shell amplitudes.}
\label{fig_example_on_shell_recursion}
\end{figure}
We consider again the amplitude $A_6(1^-,2^-,3^-,4^+,5^+,6^+)$.
The non-vanishing diagrams from the on-shell recursion approach 
are shown in fig.~\ref{fig_example_on_shell_recursion}.
Note that we may stop the recursion as soon as a sub-amplitude is a MHV amplitude and use the 
Parke-Taylor formula for this sub-amplitude.
One obtains as a result for the amplitude
\bq
\label{result_on_shell_example}
\lefteqn{
A_6(1^-,2^-,3^-,4^+,5^+,6^+) = } & & 
 \\
 & &
 4 i \left[
           \frac{\langle 6+ | 1+2 | 3+ \rangle^3}{[ 61 ] [ 12 ] \langle 34 \rangle \langle 45 \rangle s_{126} \langle 2+ | 1+6 | 5+ \rangle}
         + \frac{\langle 4+ | 5+6 | 1+ \rangle^3}{[ 23 ] [ 34 ] \langle 56 \rangle \langle 61 \rangle s_{156} \langle 2+ | 1+6 | 5+ \rangle}
     \right],
 \nonumber
\eq
where we used the notation
\bq
 s_{ijk} \;\; = \;\; \left(p_i+p_j+p_k\right)^2,
 & &
 \langle i+ | j+k | l+ \rangle
 \;\; = \;\;
 \langle p_i+ | p\!\!\!/_j + p\!\!\!/_k | p_j+ \rangle.
\eq
Note that there are only two diagrams, which need to be calculated.
We recall from table~\ref{table2} that a brute force approach
would require the calculation of $220$ Feynman diagrams.
Restricting ourselves to a primitive amplitude with a fixed cyclic order 
reduces this number to $38$ diagrams.
In the approach based on MHV vertices there are only six diagrams.
The on-shell recursion method brings the number of diagrams further down to two diagrams.
Table~\ref{table3} shows a comparison of the number of diagrams
contributing to the cyclic-ordered 
\begin{table}
\begin{center}
\begin{tabular}{l|r}
method & diagrams \\
\hline
brute force approach &  220 \\
colour-ordered amplitudes & 38 \\
MHV vertices & 6 \\
on-shell recursion & 2 \\
\end{tabular}
\end{center}
\caption{The number of diagrams contributing to the cyclic-ordered 
six-gluon amplitude $A_6(1^-,2^-,3^-,4^+,5^+,6^+)$ using various methods.}
\label{table3}
\end{table}
six-gluon amplitude in the various approaches.

Let us formulate an algorithm based on the on-shell recursion approach. We consider the primitive $A_n^{(0)}$ with
$n_+$ gluons of positive helicity and $n_-=n-n_+$ gluons of negative helicity.
We discuss the case of mostly plus helicities $n_- \le n_+$, the case mostly minus case ($n_->n_+$)
can be obtained be exchanging dotted with un-dotted indices and $+ \leftrightarrow -$.
We have the following algorithm:
\begin{algorithm}
Calculation of the primitive helicity amplitude $A_{n}^{(0)}( p_1^{\lambda_1}, ..., p_n^{\lambda_n} )$
from on-shell recursion relations.
\begin{enumerate}
\item If $n_- \in \{0,1,2\}$, the amplitude $A_{n}^{(0)}$ is given by the Parke-Taylor formul{\ae}.
\item Pick a pair of adjacent particles, such that the helicities of this pair in the cyclic
order are not $(+,-)$. Label these adjacent particles $n$ and $1$.
It is always possible to find such a pair, 
since on-shell amplitudes, where all gluons have the same helicity, vanish.
We may therefore assume that $(\lambda_n,\lambda_1) \neq (+,-)$.
\item Consider the shift
\bq
 \hat{p}_A^1 = p_A^1  - z p_A^n,
 & &
 \hat{p}_{\dot{B}}^n = p_{\dot{B}}^n + z p_{\dot{B}}^1.
\eq
\item The amplitude is given by the recurrence relation
\bq
A_n^{(0)}\left( p_1^{\lambda_1}, ..., p_n^{\lambda_n} \right)
 & = &
 \sum\limits_{j=2}^{n-2} 
 \sum\limits_{\lambda=\pm}
  A_{j+1}^{(0)}\left( \hat{p}_1(z_j)^{\lambda_1}, p_2^{\lambda_2}, ..., p_j^{\lambda_j}, -\hat{p}_{1,j}(z_j)^{-\lambda} \right)
 \nonumber \\
 & &
 \times
  \frac{i}{p_{1,j}^2} 
  A_{n-j+1}^{(0)}\left( \hat{p}_{1,j}(z_j)^{\lambda}, p_{j+1}^{\lambda_{j+1}}, ..., p_{n-1}^{\lambda_{n-1}}, \hat{p}_n(z_j)^{\lambda_n} \right),
\eq
where
\bq
 \hat{p}^\mu_{1,j}(z) & = & \sum\limits_{k=1}^j p^\mu_k - \frac{z}{2} \langle 1+ | \gamma^\mu | n+ \rangle,
\eq
and
\bq 
 z_j & = & \frac{p_{1,j}^2}{\left\langle 1+ \left| p\!\!\!/_{1,j} \right| n+ \right\rangle}.
\eq
The four-vector $\hat{p}_{1,j}^\mu(z_j)$ is on-shell, the corresponding Weyl spinors are given by
\bq
 \left| \hat{p}_{1,j} + \right\rangle
 \;\; = \;\;
 \frac{ p\!\!\!/_{1,j} \left| 1- \right\rangle}{\sqrt{\left\langle 1+ \left| p\!\!\!/_{1,j} \right| n+ \right\rangle}},
 & &
 \left\langle \hat{p}_{1,j} + \right|
 \;\; = \;\;
 \frac{ \left\langle n- \right| p\!\!\!/_{1,j}}{\sqrt{\left\langle 1+ \left| p\!\!\!/_{1,j} \right| n+ \right\rangle}}.
\eq
\end{enumerate}
\end{algorithm}
\noindent
The on-shell recursion algorithm is a very powerful tool to obtain 
compact analytical results for helicity amplitudes and -- as we will see --
to facilitate proofs.
However, when used numerically, the on-shell recursion algorithm does not beat
the off-shell recursion algorithm discussed in section~(\ref{sect_off_shell_recursion}),
both in speed and numerical stability \cite{Dinsdale:2006sq,Duhr:2006iq,Badger:2012uz}.
This can be understood as follows:
The on-shell recursion algorithm introduces at each step new momenta (or pairs of Weyl spinors).
The off-shell algorithm operates on the same set of momenta from the very beginning.
Furthermore, we see from eq.~(\ref{result_on_shell_example}) 
that the on-shell algorithm introduces spurious singularities like $1/\langle 2+ | 1+6 | 5+ \rangle$,
which affect the numerical stability.

\subsection{Three-point amplitudes and on-shell constructible theories}

The on-shell algorithm of the previous paragraph uses the Parke-Taylor formula 
as soon as an MHV amplitude is encountered.
This is practical and efficient, since the MHV amplitudes are given by the compact Parke-Taylor formula.
On the other hand it is in principle possible to use the BCFW recursion relation also for MHV amplitudes. 
This possibility implies that all primitive tree-level Yang-Mills amplitudes can be brought down to 
three-point amplitudes.
In other words, the three-point amplitudes together with the on-shell recursion relations
determine the full set of primitive tree-level Yang-Mills amplitudes.
The same holds true for any theory, whose tree-level amplitudes are constructible
through on-shell recursion relations.
For massless particles there aren't too many theories of this type, since Lorentz invariance and locality
put tight constraints on the possible three-point amplitudes \cite{Benincasa:2007xk}.
Let us denote the three-point amplitudes in a general massless on-shell constructible theory by 
${\mathcal M}_3^{(0)}(1^{h_1},2^{h_2},3^{h_3})$, where $h_j$ denotes the helicity of particle $j$.
We recall that for the three-point kinematics momentum conservation and the on-shell conditions forces either
\bq
 \langle p_1 p_2 \rangle
 \;\; = \;\;
 \langle p_2 p_3 \rangle
 \;\; = \;\;
 \langle p_3 p_1 \rangle
 \;\; = \;\;
 0,
\eq
or
\bq
 [ p_1 p_2 ]
 \;\; = \;\;
 [ p_2 p_3 ]
 \;\; = \;\;
 [ p_3 p_1 ]
 \;\; = \;\;
 0.
\eq
A mixed case, where two spinor products of one type and one spinor product of the other type are zero, implies
that the third spinor product of the type where already two are zero, is zero as well.
It follows that the three-point amplitude is a sum of two functions, one which depends only
on the spinor products $\langle p_1 p_2 \rangle$, $\langle p_2 p_3 \rangle$ and $\langle p_3 p_1 \rangle$,
while the other function depends only on the spinor products
$[ p_1 p_2]$, $[ p_2 p_3 ]$ and $[ p_3 p_1 ]$.
Under little group scaling $p_A^j \rightarrow \lambda p_A^j$, $p_{\dot{A}}^j \rightarrow \lambda^{-1} p_{\dot{A}}^j$
the amplitude scales as $\lambda^{-2 h_j}$ for a particle of helicity $h_j$.
This implies that the holomorphic function must be proportional to
\bq
\label{general_holomorphic_three_point}
 \left\langle p_1 p_2 \right\rangle^{h_3-h_1-h_2}
 \left\langle p_2 p_3 \right\rangle^{h_1-h_2-h_3}
 \left\langle p_3 p_1 \right\rangle^{h_2-h_3-h_1},
\eq
while the anti-holomorphic one must be proportional to
\bq
 \left[ p_1 p_2 \right]^{-h_3+h_1+h_2}
 \left[ p_2 p_3 \right]^{-h_1+h_2+h_3}
 \left[ p_3 p_1 \right]^{-h_2+h_3+h_1}.
\eq
If we now restrict ourselves to theories where all particles have helicity $\pm h$, with $h$ being a positive integer,
one finds
\bq
 {\mathcal M}_3^{(0)}\left(1^{-h}_a, 2^{-h}_b, 3^{+h}_c \right)
 = 
 i \kappa_{abc} \left( \frac{\langle 1 2 \rangle^3}{\langle 2 3  \rangle \langle 3 1  \rangle} \right)^h,
 & &
 {\mathcal M}_3^{(0)}\left(1^{+h}_a, 2^{+h}_b, 3^{-h}_c \right)
 = 
 i \kappa_{abc} \left(  \frac{[ 2 1 ]^3}{[1 3 ] [3 2]} \right)^h,
 \nonumber \\
 {\mathcal M}_3^{(0)}\left(1^{-h}_a, 2^{-h}_b, 3^{-h}_c \right)
 = 
 i \kappa'_{abc} \left( \langle 1 2 \rangle \langle 2 3  \rangle \langle 3 1 \rangle \right)^h,
 & &
 {\mathcal M}_3^{(0)}\left(1^{+h}_a, 2^{+h}_b, 3^{+h}_c \right)
 = 
 i \kappa'_{abc} \left( [3 2] [ 2 1 ] [1 3 ] \right)^h.
 \nonumber
\eq
The indices $a$, $b$, $c$ denote other quantum numbers of the particles (like colour).
The couplings $\kappa$ and $\kappa'$ may depend on these.
Since the amplitude ${\mathcal M}_3^{(0)}$ must be symmetric under the exchange of two particles, it follows
that $\kappa_{abc}$ and $\kappa'_{abc}$ are anti-symmetric if $h$ is odd and symmetric if $h$ is even.
\\
\\
\bs
{\it {\bf Exercise \theexercise}: 
Show that a term of the form
\bq
 \left\langle p_1 p_2 \right\rangle^{\nu_3}
 \left\langle p_2 p_3 \right\rangle^{\nu_1}
 \left\langle p_3 p_1 \right\rangle^{\nu_2}
\eq
and contributing to ${\mathcal M}_3^{(0)}(1^{h_1},2^{h_2},3^{h_3})$ must have the exponents as in eq.~(\ref{general_holomorphic_three_point}).
\stepcounter{exercise}
}
\es

\subsection{Application: Proof of the fundamental BCJ relations}
\label{sect:application_on_shell_recursion}

Let us come back to the fundamental BCJ relation, stated in eq.~(\ref{fundamental_BCJ_relation}):
\bq
\label{fundamental_BCJ_relation_2}
 \sum\limits_{i=2}^{n-1} 
  \left( \sum\limits_{j=i+1}^n 2 p_2 p_j \right)
  A_n^{(0)}(1,3,...,i,2,i+1,...,n-1,n)
 & = & 0.
\eq
We promised a proof of this relation based on on-shell recursion relations \cite{Feng:2010my,Jia:2010nz,Chen:2011jxa}. The exposition follows \cite{delaCruz:2015dpa}.
We prove the fundamental BCJ relations by induction.
For $n=3$ the fundamental BCJ-relation reduces to
\bq
  2 p_2 p_3 \; A_3^{(0)}\left(1,2,3\right)
 & = & 0.
\eq
For generic external momenta $A_3^{(0)}(1,2,3)$ is finite and
\bq
 2 p_2 p_3 
 \;\; = \;\;
 \left(p_2+p_3\right)^2
 \;\; = \;\;
 p_1^2 
 \;\; = \;\;
 0.
\eq
For the induction step let us consider a three-particle shift, 
where we deform the momenta of the particles $1$, $2$ and $n$:
\bq
 \hat{p}_1(z), 
 \;\;\;\;\;\;
 \hat{p}_2(z), 
 \;\;\;\;\;\;
 \hat{p}_n(z).
\eq
We require
\bq
 \hat{p}_1(0) = p_1, 
 \;\;\;\;\;\;
 \hat{p}_2(0) = p_2, 
 \;\;\;\;\;\;
 \hat{p}_n(0) = p_n.
\eq
For $j \neq 1,2,n$ we simply set $\hat{p}_j(z)=p_j$.
Let us consider the quantity
\bq
\label{I_n}
 I_n\left(z\right)
 & = &
 \sum\limits_{i=2}^{n-1}
  \left( \sum\limits_{j=i+1}^n 2 \hat{p}_2 \hat{p}_j \right)
  A_n^{(0)}\left(\hat{1},3,...,i,\hat{2},i+1,...,n-1,\hat{n}\right).
\eq
For $z=0$ the expression $I_n(z)$ reduces to the left-hand-side of eq.~(\ref{fundamental_BCJ_relation_2}).
$I_n(z)$ is clearly a rational function of $z$.
We have to show that
\bq
\label{eq_to_prove}
 I_n\left(0\right) & = & 0,
\eq
or equivalently
\bq
\label{eq_to_prove_2}
 \frac{1}{2\pi i}
 \oint\limits_{z=0} \frac{dz}{z} I_n\left(z\right) & = & 0,
\eq
where the contour is a small counter-clockwise circle around $z=0$.
We may assume that $I_j(0)=0$ for $j<n$.
Let us assume that $I_n(z)$ falls off for large $z$ at least with $1/z$, we will show in a minute
that we can always deform the momenta in such a way to achieve this.
Deforming the contour to a large circle at infinity
and the residues at the finite poles $z_\alpha \neq 0$ we obtain
\bq
 I_n\left(0\right) 
 & = & 
 - \sum\limits_{\alpha} \mathrm{res}\left(\frac{I_n\left(z\right)}{z}\right)_{z_\alpha}.
\eq
Since we assumed that $I_n(z)$ falls off for large $z$ at least with $1/z$,
there is no  contribution from the large circle at infinity.
It will be convenient to introduce the following notation for the various factorisation channels:
\bq
\label{notation_factorisation_channels}
\lefteqn{
 A_n^{(0)}\left(\hat{1},2,...,k,\hat{p} | -\hat{p}, k+1,...,n-1,\hat{n}\right)
 =
} & & \nonumber \\
 & &
 \sum\limits_\lambda
 A_{k+1}^{(0)}\left(\hat{1},2,...,k,\hat{p}\right)
 \frac{i}{p^2}
 A_{n-k+1}^{(0)}\left(-\hat{p},k+1,...,n-1,\hat{n}\right),
\eq
together with the convention that the hatted quantities are evaluated at $z=z_\alpha$.
The sum is over the helicity of the intermediate particle.
Let us look at the $z$-momentum flow for a three-particle BCFW-shift.
For each diagram we may divide the $z$-dependent propagators into three segments.
Each segment starts at the common vertex,
where the $z$-dependent momentum flow meets and
goes outwards towards the particles $1$, $2$ and $n$.
We may use these segments to divide the finite residues into three groups and we write
\bq
 I_n\left(0\right)
 & = &
 R_1 + R_2 + R_n,
\eq
with
\bq
 R_1
 & = &
 \sum\limits_{i=2}^{n-1}
  \left( \sum\limits_{j=i+1}^n 2 \hat{p}_2 \hat{p}_j \right)
        \sum\limits_{k=3}^i A_n^{(0)}\left(\hat{1},3,...,k,\hat{p}|-\hat{p},k+1,...,i,\hat{2},i+1,...,n-1,\hat{n}\right),
 \nonumber \\
 R_2
 & = &
 \sum\limits_{i=2}^{n-1}
  \left( \sum\limits_{j=i+1}^n 2 \hat{p}_2 \hat{p}_j \right)
 \nonumber \\
 & &
 \times
        \sum\limits_{k=2}^i 
        \sum\limits_{\substack{l=i \\ (k,l) \neq (i,i)}}^{n-1}
        A_n^{(0)}\left(k+1,...,i,\hat{2},i+1,...,l,\hat{p}|-\hat{p},l+1,...,n-1,\hat{n},\hat{1},3,...,k\right),
 \nonumber \\
 R_n
 & = &
 \sum\limits_{i=2}^{n-1}
  \left( \sum\limits_{j=i+1}^n 2 \hat{p}_2 \hat{p}_j \right)
  \sum\limits_{k=i}^{n-2} A_n^{(0)}\left(\hat{1},3,...,i,\hat{2},i+1,...,k,\hat{p}|-\hat{p},k+1,...,n-1,\hat{n}\right).
 \nonumber \\
\eq
Let us first look at $R_1$.
We may exchange the summation over $i$ and $k$ as
\bq
 \sum\limits_{i=2}^{n-1} \sum\limits_{k=3}^i f\left(i,k\right)
 & = &
 \sum\limits_{k=3}^{n-1} \sum\limits_{i=k}^{n-1} f\left(i,k\right).
\eq
One obtains
\bq
 R_1 & = &
 \sum\limits_{k=3}^{n-1} \sum\limits_{i=k}^{n-1}
  \left( \sum\limits_{j=i+1}^n 2 \hat{p}_2 \hat{p}_j \right)
   A_n^{(0)}\left(\hat{1},3,...,k,\hat{p}|-\hat{p},k+1,...,i,\hat{2},i+1,...,n-1,\hat{n}\right).
 \;\;\;
\eq
We recognise the fundamental BCJ relation for $(n-k+2)<n$ external particles.
We may therefore use the induction hypothesis and we conclude
\bq
 R_1 & = & 0.
\eq
The argument for $R_n$ is very similar. By exchanging the summation over $i$ and $k$ 
and by using momentum conservation in the sum over $j$ one shows $R_n=0$. 

For the contribution from $R_2$ we have to work a little bit more.
Exchanging the summation indices for $R_2$ one obtains
\bq
 R_2
 & = &
 \sum\limits_{k=2}^{n-2}
 \sum\limits_{l=k+1}^{n-1}
 \\
 & & 
 \times
 \sum\limits_{i=k}^{l}
  \left( \sum\limits_{j=i+1}^n 2 \hat{p}_2 \hat{p}_j \right)
        A_n^{(0)}\left(k+1,...,i,\hat{2},i+1,...,l,\hat{p}|-\hat{p},l+1,...,n-1,\hat{n},\hat{1},3,...,k\right).
 \nonumber
\eq
We may split the sum over $j$ as
\bq
 \sum\limits_{j=i+1}^n 2 \hat{p}_2 \hat{p}_j
 & = &
 \underbrace{\sum\limits_{j=i+1}^l 2 \hat{p}_2 \hat{p}_j}_{A}
 +
 \underbrace{\sum\limits_{j=l+1}^n 2 \hat{p}_2 \hat{p}_j}_{B}
\eq
The terms of type $A$ vanish again by the induction hypothesis
\bq
 \sum\limits_{i=k}^{l-1}
  \left( \sum\limits_{j=i+1}^l 2 \hat{p}_2 \hat{p}_j \right)
        A_{l-k+2}^{(0)}\left(\hat{p},k+1,...,i,\hat{2},i+1,...,l\right)
 & = & 0.
\eq
Note that the sum over $i$ extends only to $(l-1)$, the case $i=l$ contributes only to the terms of type $B$.

For the terms of type $B$ the sum over $j$ is independent of $i$ and may be taken outside the sum over $i$.
The sum over $i$ vanishes then due to the $U(1)$-decoupling relation.
\bq
  \left( \sum\limits_{j=l+1}^n 2 \hat{p}_2 \hat{p}_j \right)
 \sum\limits_{i=k}^{l}
        A_{l-k+2}^{(0)}\left(\hat{p},k+1,...,i,\hat{2},i+1,...,l\right)
 & = & 0.
\eq
The $U(1)$-decoupling relation is a special case of the Kleiss-Kuijf relations given in eq.~(\ref{Kleiss_Kuijf}), corresponding to the case where $\vec{\beta}$ consists only of a single element.
We therefore conclude that $R_2 = 0$.
Putting the partial results for $R_1$, $R_2$ and $R_n$ together we find that
\bq
 I_n\left(0\right)
 & = &
 0.
\eq
It remains to show that we can always find a momenta shift such 
that $I_n(z)$ falls off for large $z$ at least with $1/z$.
Our strategy is as follows: We will choose a deformation such that
the external polarisation vectors contribute a factor $z^{-3}$.
The contribution from internal propagators and vertices is at worst $z^1$.
Together with another factor $z^1$ from the prefactor $(2 \hat{p}_2 \hat{p}_j)$
we then achieve a total $1/z$-behaviour for $I_n(z)$.

The explicit expressions for the momenta shifts will depend on the helicities of the particles $1$, $2$
and $n$.
For $(\lambda_1,\lambda_2,\lambda_n)=(+,+,+)$ we may choose the shift defined in eq.~(\ref{three_particle_shift_schouten}).
For $(\lambda_1,\lambda_2,\lambda_n)=(+,+,-)$ we may choose 
\bq
\label{shift_++-}
 \hat{p}_A^1 = p_A^1 - z y_1 p_A^n,
 & &
 \hat{p}_{\dot{B}}^n = p_{\dot{B}}^n + z y_1 p_{\dot{B}}^1 + z y_2 p_{\dot{B}}^2,
 \nonumber \\
 \hat{p}_A^2 = p_A^2 - z y_2 p_A^n,
 & &
\eq
where $y_1$ and $y_2$ are two non-zero constants.
The cases $(\lambda_1,\lambda_2,\lambda_n)=(+,-,+)$
and $(\lambda_1,\lambda_2,\lambda_n)=(-,+,+)$ may be obtained from eq.~(\ref{shift_++-})
by permutation.
Finally, the shifts for the helicity configurations 
\bq
 (-,-,-), \;\; (-,-,+), \;\; (-,+,-), \;\; (+,-,-)
\eq
can be obtained from the helicity configurations
\bq
 (+,+,+), \;\; (+,+,-), \;\; (+,-,+), \;\; (-,+,+)
\eq
by exchanging holomorphic and anti-holomorphic spinors.
This completes the proof of the fundamental BCJ relations.

On-shell recursion relations may also be used to prove the 
MHV expansion discussed in section~(\ref{sect:CSW_construction}), the interested reader is referred to
\cite{Risager:2005vk}.

As another application of the on-shell recursion relations consider the following exercise:
\\
\\
\bs
{\it {\bf Exercise \theexercise}: 
In section~(\ref{sect:parke_taylor}) we used off-shell recurrence relations to prove the 
Parke-Taylor formula
for two adjacent negative helicities.
Consider now the general case of two non-adjacent negative helicities.
Prove the Parke-Taylor formula for two non-adjacent negative helicities from the formula of the adjacent case and
by using on-shell recursion relations.
\stepcounter{exercise}
}
\es

% ----------------------------------------------------------------------------------
\newpage
\section{Grassmannian geometry}
\label{sect:grassmannian}

In this section we explore the relation between scattering amplitudes and Grassmannian manifolds.
We first give a short introduction to projective spaces, Grassmannian manifolds and flag varieties
and present then the link representation of a primitive tree-level amplitude as a residue on a Grassmannian manifold.
In addition we introduce the notation related to the ${\mathcal N}=4$ supersymmetric extension of Yang-Mills theory
and discuss in this context the scattering amplitude as a volume of the amplituhedron.

\subsection{Projective spaces, Grassmannians and flag varieties}
\label{sect:grassmannians}

The {\bf complex projective space} ${\mathbb C} {\mathbb P}^n$ is the set of lines 
through the origin in ${\mathbb C}^{n+1}$.
Let us consider points $( x_0, x_1, ..., x_n ) \in {\mathbb C}^{n+1} \backslash \{0\}$. 
We call two points of ${\mathbb C}^{n+1} \backslash \{0\}$ equivalent,
if there is a $\lambda \neq 0$ such that
\bq
 \left( x_0, x_1, ..., x_n \right) & = & \left( \lambda y_0, \lambda y_1, ..., \lambda y_n \right).
\eq
We have
\bq
 {\mathbb C} {\mathbb P}^n 
 & = & 
 \left( {\mathbb C}^{n+1} \backslash \{0\} \right) / {\mathbb C}^\ast,
\eq
where ${\mathbb C}^\ast = {\mathbb C}\backslash \{0\}$.
Points in ${\mathbb C} {\mathbb P}^n$ will be denoted in {\bf homogeneous coordinates}
by
\bq
\label{homogeneous_coordinates_projective_space}
 \left[ z_0 : z_1 : ... : z_n \right].
\eq
{\bf Affine charts} are defined by
\bq
 U_j & = &
 \left\{ \left( z_0,...,z_{j-1},1,z_{j+1},...,z_n \right) \right\},
 \;\;\;\;\;\;
 0 \le j \le n.
\eq
The {\bf cell decomposition} of ${\mathbb C} {\mathbb P}^n$ is
\bq
\label{cell_decomposition_projective_space}
 {\mathbb C} {\mathbb P}^n
 & = &
 {\mathbb C}^n \sqcup {\mathbb C}^{n-1} \sqcup ... \sqcup {\mathbb C}^1 \sqcup {\mathbb C}^0,
\eq
where $\sqcup$ denotes the disjoint union.
This can be seen as follows: Consider ${\mathbb C} {\mathbb P}^n$ with homogeneous coordinates $[z_0:z_1:...:z_n]$.
Divide the space into the regions $z_n \neq 0$ and $z_n = 0$.
The first region is homeomorphic to ${\mathbb C}^n$, the second region is homeomorphic to ${\mathbb C} {\mathbb P}^{n-1}$.
We therefore have the recursion
\bq
 {\mathbb C} {\mathbb P}^n
 & = &
 {\mathbb C}^n \sqcup {\mathbb C} {\mathbb P}^{n-1},
\eq
from which the result in eq.~(\ref{cell_decomposition_projective_space}) follows immediately.

Projective spaces have a generalisation:
In order to obtain the projective space, we considered lines through the origin in ${\mathbb C}^{n+1}$.
Instead of lines we may consider higher dimensional sub-spaces.
This brings us to the definition of a Grassmannian manifold.
The {\bf Grassmannian manifold} $\mathrm{Gr}_{k,n}({\mathbb C})$ is the set of $k$-dimensional planes in ${\mathbb C}^n$.
Let $\mathrm{M}_{k,n}({\mathbb C})$ be the set of $k \times n$ matrices of rank $k$ (with $k \le n$). 
An element of $\mathrm{M}_{k,n}({\mathbb C})$ may be written as $k$ linearly independent row vectors
\bq
 \left( \begin{array}{ccc}
        a_{11} & ... & a_{1n} \\
        ... &  & ... \\
        a_{k1} & ... & a_{kn} \\
 \end{array} \right).
\eq
Two matrices $A_1, A_2 \in \mathrm{M}_{k,n}({\mathbb C})$ are called equivalent, 
if there is a $\Lambda \in \mathrm{GL}(k,{\mathbb C})$ such that
\bq
 A_1 & = & \Lambda \; A_2.
\eq
The Grassmannian is then defined as
\bq
 \mathrm{Gr}_{k,n}({\mathbb C}) 
 & = & 
 \mathrm{M}_{k,n}({\mathbb C}) / \mathrm{GL}(k,{\mathbb C}).
\eq
The projective space ${\mathbb C} {\mathbb P}^n$ is a special case of a Grassmannian manifold:
\bq
 {\mathbb C} {\mathbb P}^n
 & = & 
 \mathrm{Gr}_{1,n+1}({\mathbb C}).
\eq
Let us discuss {\bf affine charts} for Grassmannian manifolds.
Let $I=(i_1,i_2,...,i_k)$ be a set of $k$ column indices (with $1 \le i_1 < i_2 < ... < i_k \le n$), such that
the corresponding $k \times k$-minor $A_I$ of $A \in \mathrm{M}_{k,n}({\mathbb C})$ has $\det A_I \neq 0$.
We may then use $\Lambda \in \mathrm{GL}(k,{\mathbb C})$ to convert $A_I$ to the $k \times k$ unit matrix.
This defines an affine chart of $\mathrm{Gr}_{k,n}({\mathbb C})$.
For example, in the case $I=(n-k+1,n-k+2,...,n)$ 
the elements of $\mathrm{Gr}_{k,n}({\mathbb C})$ in this chart are parametrised as
\bq
 \left( \begin{array}{ccccccc}
        a_{11} & ... & a_{1 (n-k)} & 1 & 0 & ... & 0 \\
        a_{21} & ... & a_{2 (n-k)} & 0 & 1 & ... & 0 \\
        ... &  & ... & ... & ... & & ... \\
        a_{k1} & ... & a_{k (n-k)} & 0 & 0 & ... & 1 \\
 \end{array} \right).
\eq
This shows that
\bq
 \mathrm{dim}_{\mathbb C} \; \mathrm{Gr}_{k,n}({\mathbb C})
 & = &
 k \left( n - k \right).
\eq
A Grassmannian manifold may be embedded as a sub-manifold in a higher dimensional
complex projective space.
Set
\bq
 d & = & \left(\begin{array}{c}n\\k\\ \end{array}\right).
\eq
This is the number of different sets of $k$ indices $I=(i_1,i_2,...,i_k)$.
The {\bf Pl\"ucker embedding} is the map \cite{Postnikov:2013}
\bq
 \mathrm{Gr}_{k,n}({\mathbb C}) & \rightarrow & {\mathbb C} {\mathbb P}^{d-1},
 \nonumber \\
 A & \rightarrow & \left[ \det A_{I_1} : \det A_{I_2} : ... : \det A_{I_d} \right].
\eq
The numbers 
$\det A_{I_1}$, $\det A_{I_2}$, etc. are referred to as {\bf Pl\"ucker coordinates}.

Up to now we considered complex Grassmannian manifolds.
There is an analogue definition for real Grassmannian manifolds:
\bq
 \mathrm{Gr}_{k,n}({\mathbb R}) 
 & = & 
 \mathrm{M}_{k,n}({\mathbb R}) / \mathrm{GL}(k,{\mathbb R}).
\eq
Over the real numbers we may define the {\bf positive Grassmannian} $\mathrm{Gr}^+_{k,n}({\mathbb R})$
by the condition that all Pl\"ucker coordinates are positive.
Note that we required the set of column indices $(i_1,i_2,...,i_k)$ to be ordered:
$1 \le i_1 < i_2 < ... < i_k \le n$.
This reduces the definition of positivity of points of the Grassmannian $\mathrm{Gr}_{k,n}({\mathbb R})$
to the definition of
positivity of points in the projective space ${\mathbb R} {\mathbb P}^{d-1}$.

Let us return to the complex case.
We learned about projective spaces and Grassmannians, 
but we haven't reached the end of sophistication yet.
Consider a plane through the origin in ${\mathbb C}^4$. The plane is isomorphic to ${\mathbb C}^2$
and within this plane we may again consider lines through the origin.
This brings us to the definition of a flag.
\begin{figure}
\begin{center}
\includegraphics[scale=0.8]{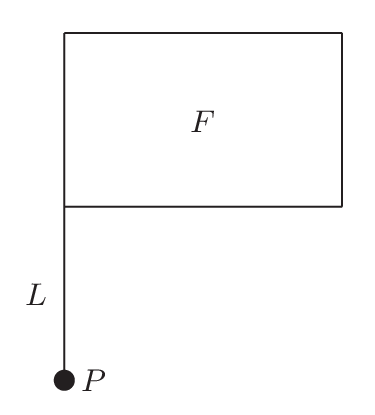}
\caption{\label{fig_flag_variety} The name flag variety derives from 
the example with a point, a line and a plane.
Note that in the figure only a line segment and a finite area inside the plane
are shown.
The point $P$ and the line $L$ lie in the plane.
}
\end{center}
\end{figure}
A {\bf flag} is an increasing sequence of sub-spaces
\bq
 \{0\} \; \subset \; V_1 \; \subset \; V_2 \; \subset \; ... \; \subset \; V_k \; = \; {\mathbb C}^n.
\eq
The motivation for the name ``flag'' is shown in fig.~(\ref{fig_flag_variety}).
We set
\bq
 d_j & = & \dim V_j.
\eq
The signature of the flag is the sequence
\bq
 \left( d_1, d_2, ..., d_k \right).
\eq
Note that by definition we always have $d_k=n$.
The {\bf flag variety} $\mathrm{F}_{d_1,d_2,...,d_k}\left({\mathbb C}\right)$ is the space
of all flags of signature $(d_1,d_2,...,d_k)$ in the vector space ${\mathbb C}^n$.
Note that we have $d_k=n$.
For $k=2$ this definition reduces to the Grassmannian manifold:
\bq
 \mathrm{F}_{d,n}\left({\mathbb C}\right)
 & = & 
 \mathrm{Gr}_{d,n}\left({\mathbb C}\right).
\eq
A flag is called a complete flag if $k=n$ and consequently $(d_1,d_2,...,d_n)=(1,2,...,n)$, 
otherwise it is called a partial flag.
A partial flag can be obtained from a complete flag by deleting 
some of the sub-spaces $V_j \neq V_n$.
Let us first consider the complete flag variety.
Let $(e_1,e_2,...,e_n)$ be an ordered basis of ${\mathbb C}^n$.
The standard flag associated to this basis is given by the vector spaces
\bq
 V_j & = & \left\langle e_1, ..., e_j \right\rangle.
\eq
The group $\mathrm{GL}(n,{\mathbb C})$ acts transitively on $\mathrm{F}_{1,2,...,n}({\mathbb C})$.
The stabiliser group of the standard flag is the group of all non-singular upper triangular matrices.
Multiples of the identity matrix act trivially on all flags and we can restrict ourselves to
$\mathrm{SL}(n,{\mathbb C})$ and the group of upper triangular matrices of determinant $1$.
We denote the latter by $B_n$. (The letter $B$ stands for Borel sub-group.)
The complete flag variety is given as the homogeneous space
\bq
 \mathrm{F}_{1,2,...,n}\left({\mathbb C}\right)
 & = &
 \mathrm{SL}\left(n,{\mathbb C}\right) / B_n.
\eq
We have
\bq
 \dim_{\mathbb C} \mathrm{F}_{1,2,...,n}\left({\mathbb C}\right)
 & = & 
 \frac{1}{2} \left(n-1\right) n.
\eq
Let us now discuss partial flag varieties.
As an example of a partial flag variety we consider the Grassmannian $\mathrm{Gr}_{d,n}({\mathbb C})$.
The stabiliser group of $V_d=\langle e_1, ..., e_d \rangle$ 
consists of $A \in \mathrm{SL}(n,{\mathbb C})$ of the form
\bq
 A & = &
 \left( \begin{array}{llllll}
 a_{1,1} & ... & a_{1,d} & a_{1,d+1} & ... & a_{1,n} \\
 ...     &     & ...     & ...       &     & ... \\
 a_{d,1} & ... & a_{d,d} & a_{d,d+1} & ... & a_{d,n} \\
 0       & ... & 0       & a_{d+1,d+1} & ... & a_{d+1,n} \\
 ...     &     & ...     & ...       &     & ... \\
 0       & ... & 0       & a_{n,d+1} & ... & a_{n,n} \\
 \end{array} \right).
\eq
Matrices of this form are called block upper triangular matrices.
For a flag variety with signature $(d_1,d_2,...,d_k)$ the stabiliser group of the standard flag consists 
of block upper triangular matrices of unit determinant, where the sizes of the blocks on the diagonal are $n_i=d_i-d_{i-1}$
(with the convention that $d_0=0$).
These block upper triangular matrices define a sub-group $P$ of $\mathrm{SL}(n,{\mathbb C})$.
(The letter $P$ stands for parabolic sub-group.)
The partial flag variety is then given as the homogeneous space
\bq
 \mathrm{F}_{d_1,d_2,...,d_k}\left({\mathbb C}\right)
 & = &
 \mathrm{SL}\left(n,{\mathbb C}\right) / P.
\eq
The dimension of $\mathrm{F}_{d_1,d_2,...,d_k}\left({\mathbb C}\right)$ is 
(again with the convention $d_0=0$) \cite{Brion:2004}
\bq
 \mathrm{dim}_{\mathbb C} \mathrm{F}_{d_1,d_2,...,d_k}\left({\mathbb C}\right)
 & = &
 \sum\limits_{1 \le i < j \le k} \left( d_i - d_{i-1} \right) \left( d_j - d_{j-1} \right).
\eq

\subsection{The link representation}
\label{sect:link_representation}

We may express momentum conservation in various ways:
\bq
\label{momentum_conservation_link}
 \sum\limits_{j=1}^n p_j^\mu \;\; = \;\; 0
 & \Leftrightarrow &
 \sum\limits_{j=1}^n p_A^j p_{\dot{A}}^j \;\; = \;\; 0.
\eq
Let us now consider the $n$-dimensional vectors 
\bq
 P_1=(p_1^1,p_1^2,...,p_1^n), & & P_{\dot{1}} = (p_{\dot{1}}^1,p_{\dot{1}}^2,...,p_{\dot{1}}^n),
 \nonumber \\
 P_2=(p_2^1,p_2^2,...,p_2^n), & & P_{\dot{2}} = (p_{\dot{2}}^1,p_{\dot{2}}^2,...,p_{\dot{2}}^n).
\eq
The last equation in eq.~(\ref{momentum_conservation_link}) 
can be interpreted as an orthogonality condition:
The plane spanned by the vectors $(P_1,P_2)$ is orthogonal to the plane
spanned by the vectors $(P_{\dot{1}},P_{\dot{2}})$.
Now consider a $k$-dimensional plane $C$ in ${\mathbb C}^n$, together with its 
$(n-k)$-dimensional orthogonal complement $C^\bot$.
Assume now that $P_{\dot{1}}$ and $P_{\dot{2}}$ are orthogonal to $C$ and
that $P_1$ and $P_2$ are orthogonal to $C^\bot$.
In formul\ae:
\bq
\label{Grassmann_condition}
 C^\bot \cdot P_A \;\; = \;\; 0,
 & &
 C \cdot P_{\dot{A}} \;\; = \;\; 0.
\eq
Of course, these conditions imply $P_A \in C$ and $P_{\dot{A}} \in C^\bot$.
In general, there are many planes $C \in \mathrm{Gr}_{k,n}({\mathbb C})$ subject
to the orthogonality condition in eq.~(\ref{Grassmann_condition}).
The idea is now to integrate over all possibilities with a suitable measure.
This idea leads us to the link representation for the primitive tree amplitude $A_n^{(0)}$
and ties the amplitude to Grassmannian geometry.

Let us see how this is done:
Consider the primitive tree-amplitude $A_n^{(0)}(1^{\lambda_1},...,n^{\lambda_n})$
with the cyclic order $(1,2,...,n)$.
Assume that $k$ gluons have negative helicities, while the remaining $(n-k)$ gluons
have positive helicity.
Let us denote the indices of the $k$ negative helicity gluons by
$J=(j_1,...,j_k)$ with $j_1<j_2<...<j_k$.
The ordered set $J$ defines a chart of the Grassmannian $\mathrm{Gr}_{k,n}({\mathbb C})$: We gauge-fix the 
$\mathrm{GL}(k)$-redundancy by requiring that for $C \in \mathrm{Gr}_{k,n}({\mathbb C})$ we have
\bq
\label{gauge_fix_Grassmannian}
 C_{i j_l} & = & \delta_{i l},
 \;\;\;\;\;\; l=1,...,k.
\eq
Thus, the $j_l$-th column of $C$ is given by the $l$-th unit vector.
We denote the remaining $(n-k)$ indices not in $J$ by $K=(k_1,...,k_{n-k})$.
The {\bf link representation} of the (non-supersymmetric) helicity amplitude multiplied by the momentum-conserving delta-function
is given by \cite{ArkaniHamed:2009si,ArkaniHamed:2009dn,ArkaniHamed:2009sx,ArkaniHamed:2009dg,ArkaniHamed:2012nw}
\bq
\label{link_representation}
\lefteqn{
 A_n^{(0)}
 \delta^4\left( \sum\limits_{j=1}^n p_A^j p_{\dot{A}}^j \right)
 = 
 \frac{i \left( \sqrt{2} \right)^{n-2}}{\left(2\pi i\right)^{(k-2)(n-k-2)}}
} 
 & & \\
 & &
 \int\limits_{\mathcal C}
 \frac{d^{k \times n}C 
       \; 
       \prod\limits_{l=1}^k \delta^k\left( C_{ij_l} - \delta_{il} \right)
       \prod\limits_{l=1}^k \delta^2\left( p_{\dot{A}}^{j_l} + C_{j_l k_i} p_{\dot{A}}^{k_i} \right)
       \prod\limits_{m=1}^{n-k} \delta^2\left( p_{A}^{k_m} - C_{j_i k_m} p_{A}^{j_i} \right)
      }
      {\left(1,2,..,k\right)\left(2,3,...,k+1\right) ... \left(n,1,...,(k-1)\right)},
 \nonumber
\eq
where $(i_1,...,i_k)$ denotes the $k \times k$-minor of the matrix $C$ 
made out of the columns $i_1$, ..., $i_k$:
\bq
 \left(i_1,...,i_k\right) 
 & = &
 \det C_{(i_1,...,i_k)}
 \;\; = \;\;
 \eps^{h_1 ... h_k}
 C_{h_1 i_1} ... C_{h_k i_k}.
\eq
The first product of delta-functions fixes the gauge according to eq.~(\ref{gauge_fix_Grassmannian}).
The other delta-functions ensure
\bq
 p_{A}^{k} & = & C_{j_i k} p_{A}^{j_i}, \;\;\;\;\;\;\;\; k \in K,
 \nonumber \\
 p_{\dot{A}}^{j} & = & - C_{j k_i} p_{\dot{A}}^{k_i} \;\;\;\;\;\; j \in J.
\eq
These equations imply momentum conservation:
\bq
 \sum\limits_{j=1}^n p_A^j p_{\dot{A}}^j 
 & = &
 \sum\limits_{k \in K} p_A^k p_{\dot{A}}^k 
 +
 \sum\limits_{j \in J} p_A^j p_{\dot{A}}^j 
 \;\; = \;\;
 \sum\limits_{j \in J} \sum\limits_{k \in K} C_{j k} p_{A}^{j} p_{\dot{A}}^k 
 -
 \sum\limits_{j \in J} \sum\limits_{k \in K} C_{j k} p_A^j p_{\dot{A}}^{k}
 \;\; = \;\;
 0.
\eq
There are $k \cdot n$ integrations in eq.~(\ref{link_representation}), the delta-functions
remove $k^2 + 2 k + 2 (n-k) - 4$ integrations, since four delta-functions should remain for the 
four-dimensional delta-function expressing momentum conservation on the left-hand-side of eq.~(\ref{link_representation}).
This leaves us with
\bq
 k n - k^2 -2 k - 2 \left( n - k \right) + 4
 & = &
 \left(k-2\right) \left(n-k-2\right)
\eq
integrations, which are understood as residues.
Multi-variable residues are defined as follows:
Consider a map 
\bq
 p & : & {\mathbb C}^n \rightarrow {\mathbb C}^n,
 \nonumber \\
 & & z \; = \; \left( z_1, ..., z_n \right) \; \rightarrow \; \left( p_1(z), ..., p_n(z) \right).
\eq
Let us further assume that the system of equations $p(z)=0$ has as solutions a finite number
of isolated points $z^{(j)}=(z_1^{(j)}, ..., z_n^{(j)})$, where $j$ labels the individual solutions.
Let us further consider a function $g : {\mathbb C}^n \rightarrow {\mathbb C}$, regular at the solutions
$z^{(j)}$.
We define the {\bf local residue} of $g$ with respect to $p_1$, ..., $p_n$ at $z^{(j)}$ by
\bq
\label{local_residue_1}
 \mathrm{Res}_{\{p_1,...,p_n\}}\left( g, z^{(j)} \right)
 & = &
\frac{1}{\left(2\pi i\right)^{n}}
 \oint\limits_{\Gamma_\delta}
 \frac{g\left(z\right) \; dz_1 \wedge ... \wedge dz_n}{p_1\left(z\right) ... p_n\left(z\right)}.
\eq
The integration in eq.~(\ref{local_residue_1}) is around a small $n$-torus
\bq
 \Gamma_\delta & = &
 \left\{
   \left( z_1, ..., z_n \right) \in {\mathbb C}^n | \left| p_i\left(z\right)\right| = \delta
 \right\},
\eq
encircling $z^{(j)}$ with orientation
\bq
 d \arg p_1 \wedge d \arg p_2 \wedge ... \wedge d \arg p_n \ge 0.
\eq
The {\bf global residue} of $g$ with respect to $p_1$, ..., $p_n$ is defined as
\bq
\label{global_residue_1}
 \mathrm{Res}_{\{p_1,...,p_n\}}\left( g \right)
 & = &
 \sum\limits_{\mathrm{solutions} \; j}
 \mathrm{Res}_{\{p_1,...,p_n\}}\left( g, z^{(j)} \right).
\eq
In order to evaluate a local residue it is advantageous to perform a change of variables
\bq
 z_i' & = & p_i\left(z\right),
 \;\;\;\;\;\; i=1,...,n.
\eq
Let us denote the Jacobian of this transformation by
\bq
 J\left(z\right) & = &
 \frac{1}{\det\left( \frac{\partial\left(p_1,...,p_n\right)}{\partial\left(z_1,...,z_n\right)} \right)}.
\eq
The local residue at $z^{(j)}$ is then given by
\bq
 \mathrm{Res}_{\{p_1,...,p_n\}}\left( g, z^{(j)} \right)
 & = &
\frac{1}{\left(2\pi i\right)^{n}}
 \oint\limits_{\Gamma_\delta}
 \frac{g\left(z\right) \; dz_1 \wedge ... \wedge dz_n}{p_1\left(z\right) ... p_n\left(z\right)}
 \;\; = \;\; 
 J\left(z^{(j)}\right) g\left(z^{(j)}\right),
\eq
and the global residue as the sum over all local residues.

Let us now return to eq.~(\ref{link_representation}).
There are $(k-2)(n-k-2)$ integrations and in order to specify the integration contour
we have to give a map
$F : {\mathbb C}^{(k-2)(n-k-2)} \rightarrow {\mathbb C}^{(k-2)(n-k-2)}$,
or equivalently to give $(k-2)(n-k-2)$ maps
$f_l^j : {\mathbb C}^{(k-2)(n-k-2)} \rightarrow {\mathbb C}$
for $j=1,..,(k-2)$ and $l=3,...,n$.
The appropriate maps have been derived in \cite{Bourjaily:2010kw} and 
are given as a product of three minors
\bq
 f_l^j & : & {\mathbb C}^{(k-2)(n-k-2)} \rightarrow {\mathbb C},
 \;\;\;\;\;\;\;\;\;\;\;\;\;\;\;\;\;\;\;\;\;\;\;\;\;\;\;\;\;
 j=1,...,(k-2),
 \;\;\;
 l=3,...,n,
 \nonumber \\
 & & f_l^j \;\; = \;\; 
                      \left(\sigma_l^j,l-2,l-1,l \right)
                      \left(\sigma_l^j,l,j,j+1 \right)
                      \left(\sigma_l^j,j+1,j+2,l-2 \right),
\eq
where $\sigma_l^j$ denotes the $(k-3)$ columns $(1,...,j-1) \cup (j+l-k,...,l-3)$.
It can be shown that we may re-write the integrand of eq.~(\ref{link_representation}) in a form, where the
product
\bq
 \prod\limits_{l=k+3}^n \left( \prod_{j=1}^{k-2} f_l^j \right)
\eq
appears in the denominator and a regular function in the numerator.
The contour integration is then defined as a global residue of the zero locus $f_l^j=0$.
The link representation may be proved with the help of the on-shell recursion relations.

Let us look at an example. We consider the $n$-point MHV amplitude $A_n^{(0)}(1^-,2^-,3^+,...,n^+)$.
In this case we have $(k-2)(n-k-2)=0$ and there are no residues to be taken.
We have $J=(1,2)$ and $K=(2,3,...,n)$.
Thus we have
\bq
\lefteqn{
 A_n^{(0)}\left(1^-,2^-,3^+,...,n^+\right)
 \delta^4\left( \sum\limits_{j=1}^n p_A^j p_{\dot{A}}^j \right)
 = 
 i \left( \sqrt{2} \right)^{n-2}
} 
 & & \\
 & &
 \int\limits_{\mathcal C}
 \frac{d^{2 \times n}C 
       \; 
       \prod\limits_{i,l=1}^2 \delta\left( C_{i l} - \delta_{il} \right)
       \prod\limits_{l=1}^2 \delta^2\left( p_{\dot{A}}^{l} + C_{l k} p_{\dot{A}}^{k} \right)
       \prod\limits_{m=3}^{n} \delta^2\left( p_{A}^{m} - p_{A}^{j} C_{j m} \right)
      }
      {\left(1,2\right)\left(2,3\right) ... \left(n,1\right)},
 \nonumber
\eq
with $j \in J$ and $k \in K$.
For the case at hand it is the simplest to perform a $\mathrm{GL}(2,{\mathbb C})$-variable transformation
\bq
 C' & = & g \cdot C,
\eq
with 
\bq
 g \;\; = \;\;
 \left( \begin{array}{rr}
  p_1^1 & p_1^2 \\
  p_2^1 & p_2^2 \\
 \end{array} \right),
 \;\;\;
 g^{-1} \;\; = \;\;
 - \frac{1}{\langle 1 2 \rangle}
 \left( \begin{array}{rr}
  p_2^2 & -p_1^2 \\
  -p_2^1 & p_1^1 \\
 \end{array} \right),
 \;\;\;
 \det g \;\; = \;\; - \left\langle 1 2 \right\rangle.
\eq
We then have
\bq
\lefteqn{
 A_n^{(0)}\left(1^-,2^-,3^+,...,n^+\right)
 \delta^4\left( \sum\limits_{j=1}^n p_A^j p_{\dot{A}}^j \right)
 = 
 i \left( \sqrt{2} \right)^{n-2}
 \left\langle 1 2 \right\rangle^4
} 
 & & \\
 & &
 \int\limits_{\mathcal C}
 \frac{d^{2 \times n}C'
       \; 
       \prod\limits_{A,l=1}^2 \delta\left( C_{A l}' - g_{Al} \right)
       \prod\limits_{B=1}^2 \delta^2\left( g_{Bj} p_{\dot{A}}^{j} + C_{B k}' p_{\dot{A}}^{k} \right)
       \prod\limits_{m=3}^{n} \delta^2\left( p_{A}^{m} - C_{A m}' \right)
      }
      {\left(1,2\right)\left(2,3\right) ... \left(n,1\right)},
 \nonumber
\eq
where the minors are now with respect to the matrix $C'$. The matrix $C'$ is completely fixed by
the delta-functions:
\bq
 C' & = &
 \left( \begin{array}{rrrrr}
  p_1^1 & p_1^2 & p_1^3 & ... & p_1^n \\
  p_2^1 & p_2^2 & p_2^3 & ... & p_2^n \\
 \end{array} \right),
\eq
The minors are then
\bq
 \left(i,j\right) & = & - \left\langle i j \right\rangle.
\eq
We further have
\bq
 \prod\limits_{B=1}^2 \delta^2\left( g_{Bj} p_{\dot{A}}^{j} + C_{B k}' p_{\dot{A}}^{k} \right)
 & = &
 \delta^4\left( \sum\limits_{j=1}^n p_A^j p_{\dot{A}}^{j} \right),
\eq
giving the momentum-conserving delta-function.
Thus
\bq
 A_n^{(0)}\left(1^-,2^-,3^+,...,n^+\right)
 \delta^4\left( \sum\limits_{j=1}^n p_A^j p_{\dot{A}}^j \right)
 & = &
 i \left( \sqrt{2} \right)^{n-2}
 \frac{\left\langle 1 2 \right\rangle^4}{\left\langle 1 2 \right\rangle ... \left\langle n 1 \right\rangle}
 \delta^4\left( \sum\limits_{j=1}^n p_A^j p_{\dot{A}}^j \right).
\eq

\subsection{Supersymmetric notation}

Up to now we considered individual helicity amplitudes.
Is there a simple way to write a formal expression, covering all possible helicity 
configurations in a single formula?
Yes, there is a possibility.
Let's look at one specific external gluon. 
This gluon can have either helicity $+1$ or $-1$.
The amplitude is linear in the polarisation vector for this
particle.
Thus we may consider to substitute
\bq
\label{simple_general_polarisation}
 \eps^+_\mu\left(p\right) + \eta \eps^-_\mu\left(p\right)
\eq
for the polarisation vector.
One recovers the result for $h=+1$ by setting $\eta=0$,
and the one for $h=-1$ by first differentiating with respect to $\eta$ and setting afterwards $\eta=0$.
(Since eq.~(\ref{simple_general_polarisation}) is linear in $\eta$, the last operation has no effect.)
This will be the basic idea.
We are going to use a fancy version of this basic idea: Instead of one number $\eta$ we are going to use
a product of four numbers $\eta_1 \eta_2 \eta_3 \eta_4$.
We will write these numbers as $\eta_I$ with $I=1,...,4$.
Instead of being ordinary numbers, we will take the $\eta_I$'s to be anti-commuting Grassmann numbers.
Finally, instead of just considering a field with the two helicity states $h=\pm 1$, we will
be  considering a field with one state of helicity $+1$, four states of helicity $+1/2$, six states
of helicity $0$, four states of helicity $-1/2$ and one state of helicity $-1$.
Thus we write for the field \cite{Nair:1988bq,ArkaniHamed:2008gz}
\bq
\lefteqn{
 \Phi
 = } & & \\
 & &
 \!\!
| +1 \rangle
 \; + \; \eta_I | + 1/2 \rangle_I
 \; + \; \frac{1}{2} \eta_I \eta_J | 0 \rangle_{IJ}
 \; + \; \frac{1}{6} \eps_{IJKL} \eta_I \eta_J \eta_K | -1/2 \rangle_{L}
 \; + \; \frac{1}{24} \eps_{IJKL} \eta_I \eta_J \eta_K \eta_L | -1 \rangle.
 \nonumber 
\eq
Of course, this is just the description of the ${\mathcal N}=4$ supersymmetric Yang-Mills theory \cite{Wess:1992cp}.
The index $I$ refers to the four supersymmetric generators.
We are not going into detail on supersymmetric generalisations of Yang-Mills theory,
which we will use in this report merely as a book-keeping device.
The $|+1\rangle$-state is projected out by setting all $\eta_I=0$,
the $|-1\rangle$-state is projected out by applying first the differential operator
\bq
 \frac{\partial}{\partial \eta_1} \frac{\partial}{\partial \eta_2} \frac{\partial}{\partial \eta_3} \frac{\partial}{\partial \eta_4}
 & = &
 \frac{\partial}{\partial \eta_4} \frac{\partial}{\partial \eta_3} \frac{\partial}{\partial \eta_2} \frac{\partial}{\partial \eta_1}
\eq
and then setting all $\eta_I$'s to zero.
Note that derivatives with respect to Grassmann numbers anti-commute.
The primitive tree amplitudes with all external particles having helicities $\pm 1$ agree between
non-supersymmetric Yang-Mills theory and ${\mathcal N}=4$ supersymmetric Yang-Mills theory.
This statement is no longer true if we go to loop amplitudes.
The reason is that in tree amplitudes with all external states restricted to be spin-one states, the additional
states of spin $1/2$ or $0$ present in ${\mathcal N}=4$ theory
but not in the non-supersymmetric version
cannot propagate internally. In loop amplitudes they may propagate in closed loops.

We associate to each external particle 
four Grassmann numbers $\eta_I^j$ with $I=1,...,4$ and $j=1,...,n$.
We may specify the momentum of each particle by two spinors $p_A^j$ and $p_{\dot{A}}^j$.
We call for a given $j$ the quantity
\bq
 p_A^{j} \eta_I^j
\eq
the {\bf fermionic part of the super-momentum}, the bosonic part is as usual
\bq
 p_\mu^j & = & \frac{1}{2} p_{\dot{A}}^j \bar{\sigma}_\mu^{\dot{A}A} p_A^j.
\eq
Momentum conservation for the fermionic part of the super-momentum reads
\bq
 \sum\limits_{j=1}^n p_A^{j} \eta_I^j
 & = & 0.
\eq
Note that these are eight equations, since $A\in\{1,2\}$ and $I\in\{1,2,3,4\}$.

Let us define a fermionic delta-function. We first consider a single Grassmann variable $\eta$.
Recall that the Taylor expansion of a function $f(\eta)$ stops after the linear term:
$f(\eta)=f_0 + f_1 \eta$.
Integration over Grassmann variables is very simple and defined by the two rules
\bq
 \int d\eta \;\; = \;\; 0,
 & &
 \int d\eta \cdot \eta \;\; = \;\; 1.
\eq
Thus we have
\bq
 \int d\eta \; \delta\left(\eta\right) \; f\left(\eta\right)
 \;\; = \;\;
 f_0
 \;\; = \;\;
 \int d\eta \; \eta \; f\left(\eta\right)
\eq
and it follows that $\delta(\eta)=\eta$.
Conservation of super-momentum is enforced by four bosonic and eight ($=2{\mathcal N}$) fermionic
delta-functions:
\bq
 \delta^4\left( \sum\limits_{j=1}^n p_A^{j} p_{\dot{A}}^j \right)
 \delta^8\left( \sum\limits_{j=1}^n p_A^{j} \eta_I^j \right).
\eq
It is convenient to set
\bq 
\label{fermionic_momentum}
 Q_{A I} 
 & = &
 \sum\limits_{j=1}^n p_A^{j} \eta_I^j.
\eq
Within a product of fermionic delta-functions the order matters. We define the eight-fold
fermionic delta-function as
\bq
 \delta^8\left(Q_{A I}\right)
 & = &
 \delta\left(Q_{1 1}\right)
 \delta\left(Q_{1 2}\right)
 \delta\left(Q_{1 3}\right)
 \delta\left(Q_{1 4}\right)
 \delta\left(Q_{2 1}\right)
 \delta\left(Q_{2 2}\right)
 \delta\left(Q_{2 3}\right)
 \delta\left(Q_{2 4}\right)
 \nonumber \\
 & = &
 \delta\left(Q_{1 1}\right)
 \delta\left(Q_{2 1}\right)
 \delta\left(Q_{1 2}\right)
 \delta\left(Q_{2 2}\right)
 \delta\left(Q_{1 3}\right)
 \delta\left(Q_{2 3}\right)
 \delta\left(Q_{1 4}\right)
 \delta\left(Q_{2 4}\right).
\eq
With the help of the supersymmetric notation we may write the three-point MHV amplitudes as
\bq
\label{three_point_SYM_MHV}
 A_3^{(0)}\left(1,2,3\right)
 \;\; = \;\;
 i \sqrt{2}
 \frac{\delta^8\left( p_A^1 \eta_I^1 + p_A^2 \eta_I^2 + p_A^3 \eta_I^3 \right)}{\langle 1 2 \rangle \langle 2 3  \rangle \langle 3 1  \rangle}.
\eq
The three-point anti-MHV amplitudes are
\bq
\label{three_point_SYM_MHVbar}
 A_3^{(0)}\left(1,2,3\right)
 \;\; = \;\;
 i \sqrt{2}
 \frac{\delta^4\left( [ 3 2 ] \eta_I^1 + [ 1 3 ] \eta_I^2 + [ 2 1 ] \eta_I^3 \right) }{[3 2] [2 1] [1 3]}.
\eq
\bs
{\it {\bf Exercise \theexercise}: 
Re-derive the expression for the MHV amplitude $A_3^{(0)}\left(1^-,2^-,3^+\right)$
and the anti-MHV amplitude $A_3^{(0)}\left(1^+,2^+,3^-\right)$ given in eq.~(\ref{three_point_MHV_and_MHVbar})
from eq.~(\ref{three_point_SYM_MHV}) and eq.~(\ref{three_point_SYM_MHVbar}), respectively.
\stepcounter{exercise}
}
\es
\\
\\
The link representation of the
${\mathcal N}=4$ supersymmetric primitive tree-level N${}^{k-2}$MHV amplitude
is
\bq
\label{link_SYM_representation}
\lefteqn{
 A_n^{(0)}
 \delta^4\left( \sum\limits_{j=1}^n p_A^j p_{\dot{A}}^j \right)
 \delta^8\left(Q_{A I} \right)
 = 
 \frac{i \left( \sqrt{2} \right)^{n-2}}{\left(2\pi i\right)^{(k-2)(n-k-2)}}
} 
 & & \\
 & &
 \int\limits_{\mathcal C}
 \frac{d^{k \times n}C 
       \; 
       \prod\limits_{l=1}^k \delta^k\left( C_{ij_l} - \delta_{il} \right)
       \prod\limits_{l=1}^k \delta^2\left( p_{\dot{A}}^{j_l} + C_{j_l k_i} p_{\dot{A}}^{k_i} \right)
       \prod\limits_{m=1}^{n-k} \delta^2\left( p_{A}^{k_m} - C_{j_i k_m} p_{A}^{j_i} \right)
      }
      {\left(1,2,..,k\right)\left(2,3,...,k+1\right) ... \left(n,1,...,(k-1)\right)}
 \nonumber \\
 & & 
 \times 
        \prod\limits_{o=1}^k \delta^4\left( \eta_I^{j_o} + C_{j_o k_i} \eta_I^{k_i} \right).
 \nonumber
\eq
Eq.~(\ref{link_SYM_representation}) differs from eq.~(\ref{link_representation})
only by an additional product of $4k$ fermionic delta-functions.
Note that eq.~(\ref{link_SYM_representation}) gives all helicity configurations in the
N${}^{k-2}$MHV sector. This implies that the gauge fixing is arbitrary, we may choose
any set of $k$ columns $J=(j_1,...,j_k)$.
In particular we may compute a specific non-supersymmetric helicity amplitude with
any gauge fixing.
However, projecting out the specific helicity amplitude from the super-amplitude
is trivial if we choose $J=(j_1,...,j_k)$ to be the set of negative helicity states.
For other gauge-fixing conditions the projection operation is more involved.
Of course it will give identical results.

\subsection{The amplituhedron}

We would like to close this chapter by giving an introduction
to recent ideas, which express the scattering
amplitude as the volume of a geometric object, called the
amplituhedron \cite{Arkani-Hamed:2013jha,Arkani-Hamed:2013kca,Arkani-Hamed:2014dca,Bai:2014cna,Franco:2014csa,Bern:2015ple}.

The development of this method originated from a study of spurious singularities appearing in the on-shell recursion
relations.
In section~(\ref{sect:BCFW_recursion}) we gave in eq.~(\ref{result_on_shell_example}) an example
for a helicity amplitude calculated with the help of the on-shell recursion relations:
\bq
\label{six_point_example}
\lefteqn{
A_6(1^-,2^-,3^-,4^+,5^+,6^+) = } & & 
 \\
 & &
 4 i \left[
           \frac{\langle 6+ | 1+2 | 3+ \rangle^3}{[ 61 ] [ 12 ] \langle 34 \rangle \langle 45 \rangle s_{126} \langle 2+ | 1+6 | 5+ \rangle}
         + \frac{\langle 4+ | 5+6 | 1+ \rangle^3}{[ 23 ] [ 34 ] \langle 56 \rangle \langle 61 \rangle s_{156} \langle 2+ | 1+6 | 5+ \rangle}
     \right].
 \nonumber
\eq
Let us look at the denominators: The spinor products like $[ 1 2 ]$ or $\langle 3 4 \rangle$ correspond
to collinear singularities, the Lorentz invariants $s_{126}$ and $s_{156}$ to three-particle poles.
These are all physical.
However, the singularity at $\langle 2+ | 1+6 | 5+ \rangle=0$ is spurious and cancels between the two terms.
Had we used in the recursion relation a different momentum shift, we would have obtained
a representation with different spurious singularities.
\begin{figure}
\begin{center}
\includegraphics[scale=0.8]{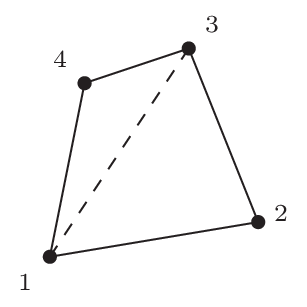}
\hspace*{8mm}
\includegraphics[scale=0.8]{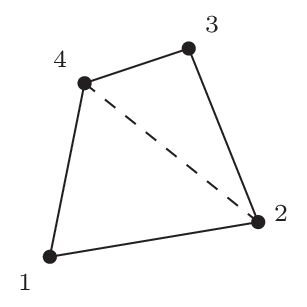}
\caption{\label{fig_triangulation} Two triangulations of a quadrangle.}
\end{center}
\end{figure}
It turns out, that the NMHV amplitude in eq.~(\ref{six_point_example})
corresponds to a volume of a polytope \cite{Hodges:2009hk}, and the two terms in eq.~(\ref{six_point_example}) 
to the volumes
of two smaller building blocks in a decomposition of the polytope.
The situation is similar to the one shown in fig.~(\ref{fig_triangulation}):
The area of the quadrangle can be computed from a triangulation as the sum over the areas of the triangles.
Different triangulations yields the same area, but the expressions may look different.

Let us now formalise this idea: 
In this paragraph we consider ${\mathcal N}=4$ supersymmetric primitive tree amplitudes
of Grassmann weight $4k$.
These will give us the non-supersymmetric Yang-Mills amplitudes with $k$ negative helicity gauge bosons. 
It is convenient to factor out the Parke-Taylor amplitude and to consider the remainder:
\bq
 A_{n,k}^{(0)}\left(1,...,n\right)
 \delta^4\left( \sum\limits_{j=1}^n p_A^j p_{\dot{A}}^j \right)
 \delta^8\left(Q_{A I} \right)
 & = &
 \frac{i \left( \sqrt{2} \right)^{n-2} \delta^4\left( \sum\limits_{j=1}^n p_A^j p_{\dot{A}}^j \right) \delta^8\left(Q_{A I} \right)}{\langle 1 2 \rangle ... \langle n 1 \rangle}
 B_{n,k}^{(0)}\left(1,...,n\right).
 \nonumber \\
\eq
The remainder $B_{n,k}^{(0)}$ has Grassmann weight $4(k-2)$.
We will also assume in this paragraph that the signature of the space-time metric is $(2,2)$.
This has the advantage that spinors and twistors can be chosen as real variables.
Working with real numbers allows us to talk about positive and negative numbers.
In particular we have the notion of the positive Grassmannian $\mathrm{Gr}_{k,n}^+({\mathbb R})$, defined
by the condition that all Pl\"ucker coordinates (i.e. all ordered $k \times k$-minors) are positive.
We are actually interested in the closure of this object, allowing some coordinates to be zero.
This defines the non-negative Grassmannian $\overline{\mathrm{Gr}}_{k,n}^+({\mathbb R})$, 
where all Pl\"ucker coordinates are non-negative.
In a similar way we may introduce positive $k \times n$-matrices $m \in \mathrm{M}_{k,n}^+({\mathbb R})$,
defined by the condition that all ordered minors are positive,
the difference with the Grasmannian $\mathrm{Gr}_{k,n}^+({\mathbb R})$ being that for 
$\mathrm{M}_{k,n}^+({\mathbb R})$ the $\mathrm{GL}(k,{\mathbb R})$-action has not been modded out.
At the end of day, when an analytic formula has been obtained, we may relax the restriction to real numbers
and continue to complex numbers.

The {\bf tree amplituhedron} is a region in the Grassmannian 
$\mathrm{Gr}_{k-2,D+k-2}({\mathbb R})=\mathrm{Gr}_{k-2,k+2}({\mathbb R})$, where $D=4$ denotes the dimension of space-time,
obtained as the image of the non-negative Grassmannian $\overline{\mathrm{Gr}}_{k-2,n}^+({\mathbb R})$
under the map
\bq
 \overline{\mathrm{Gr}}_{k-2,n}^+({\mathbb R}) \stackrel{Z}{\longrightarrow} \mathrm{Gr}_{k-2,k+2}({\mathbb R}).
\eq
This map is given for $Y \in \mathrm{Gr}_{k-2,k+2}({\mathbb R})$,
$C \in \overline{\mathrm{Gr}}_{k-2,n}^+({\mathbb R})$ and
$Z \in \mathrm{M}_{k+2,n}^+({\mathbb R})$ by
\bq
 Y & = & C  \cdot Z^T.
\eq
The $(k+2)\times n$-matrix $Z$ depends on the external data for the scattering process.
This matrix is made out of $n$ column vectors $Z^j$, one for each particle $j=1,...,n$.
The vectors $Z^j$ are $(k+2)$-dimensional, the first four entries are the momentum twistor variables for particle $j$,
the remaining $(k-2)$ entries are denoted by $c_i^j$, 
\bq
 Z^j
 & = &
 \left( \begin{array}{c}
 p_1^j \\
 p_2^j \\
 \mu_{\dot{1}}^j \\
 \mu_{\dot{2}}^j \\
 c_1^j \\
 ... \\
 c_{k-2}^j \\
 \end{array} \right).
\eq
The $c_i^j$'s should be thought of as infinitesimal variables, in the sense that some power of them can be neglected.
This can be achieved by writing them as
\bq
\label{def_additional_components}
 c_i^j
 & = & 
 \sum\limits_{I=1}^4
 \phi_{i,I} \eta_I^j,
\eq
where $\phi_{i,I}$ and $\eta_I^i$ are Grassmann numbers. We then have $(c_i^j)^5=0$.
Note that the $c_i^j$'s have even Grassmann degree and commute like ordinary numbers.
In the end we will integrate over the auxiliary Grassmann numbers $\phi_{i,I}$.
The dimension of the Grassmannian $\mathrm{Gr}_{k-2,k+2}({\mathbb R})$ is
\bq
 \dim_{\mathbb R} \mathrm{Gr}_{k-2,k+2}({\mathbb R})
 & = & 
 4 \left(k-2\right).
\eq
Up to now we defined the region in $\mathrm{Gr}_{k-2,k+2}({\mathbb R})$, which makes up the amplituhedron.
We still need a volume form, which we can integrate over this region to give the volume.
Let us start with a brief discussion of {\bf top forms} on a Grassmannian manifold $\mathrm{Gr}_{k,n}({\mathbb R})$.
These are differential forms of degree $k(n-k)$.
We start with the projective space ${\mathbb R} {\mathbb P}^{n-1} = \mathrm{Gr}_{1,n}({\mathbb R})$
with homogeneous coordinates $x=[x_1:...:x_n]$.
Any top form on ${\mathbb R} {\mathbb P}^{n-1}$ can be written as
\bq
\label{def_Omega_projective_space}
 \Omega & = & \frac{\omega}{f\left(x\right)},
\eq
where the differential $(n-1)$-form $\omega$ is given by
\bq
\label{def_omega_k_eq_1}
 \omega & = & 
 \left(n-1\right)!
 \sum\limits_{j=1}^n (-1)^{j-1}
  \; x_j \; dx_1 \wedge ... \wedge \widehat{dx_j} \wedge ... \wedge dx_n.
\eq
The hat indicates that the corresponding term is omitted.
The factor $(n-1)!$ appears in order to have a consistent notation with later generalisations.
The function $f(x)$ must be homogeneous of degree $n$:
\bq
 f\left(\lambda x \right) & = & \lambda^n f\left(x\right).
\eq
This ensures that $\Omega$ is invariant under the $\mathrm{GL}(1,{\mathbb R})$-action.

Let us now move on to the Grassmannian manifold $\mathrm{Gr}_{k,n}({\mathbb R})$
and consider $C \in \mathrm{Gr}_{k,n}({\mathbb R})$.
Any top form on $\mathrm{Gr}_{k,n}({\mathbb R})$ can be written as
\bq
 \Omega & = & \frac{\omega}{f\left(C\right)},
\eq
where $f(C)$ is a function of the $k \times k$-minors of $C$ and homogeneous under scaling as
\bq
 f\left( \lambda C \right) 
 & = &
 \lambda^{k\cdot n} f\left( C \right).
\eq
The $k(n-k)$-form $\omega$ is given by
\bq
\label{def_omega_general_k}
 \omega & = &
 \left\langle C^1, ..., C^k, \left( dC^1 \right)^{(n-k)} \right\rangle
 \wedge ... \wedge
 \left\langle C^1, ..., C^k, \left( dC^k \right)^{(n-k)} \right\rangle,
\eq
and
\bq
  \left\langle C^1, ..., C^k, \left( dC^i \right)^{(n-k)} \right\rangle
 & = &
 \eps^{a_1 a_2 ... a_n}
 C_{1 a_1} ... C_{k a_k} dC_{i,a_{k+1}} \wedge ... \wedge dC_{i,a_n}.
\eq
For example,
\bq
\label{example_top_form}
 \Omega & = &
 \frac{\omega}{\left(1,2,..,k\right)\left(2,3,...,k+1\right) ... \left(n,1,...,(k-1)\right)}
\eq
would be a top form on $\mathrm{Gr}_{k,n}({\mathbb R})$.
We encountered this form in the link representation discussed in section~(\ref{sect:link_representation}).
Now consider the top-cell $\mathrm{Gr}_{k,n}^+({\mathbb R})$.
It can be shown that this cell has $n$ co-dimension-one boundaries, corresponding to the cases
where one of the $n$ consecutive minors vanishes.
Thus we may associate to the top-cell the differential form $\Omega$, given in
eq.~(\ref{example_top_form}). This differential form is characterised by the fact that it has
logarithmic singularities on the boundary of the cell.
\\
\\
\bs
{\it {\bf Exercise \theexercise}: 
Show that for $k=1$ the form $\omega$ given in eq.~(\ref{def_omega_general_k}) reduces to the expression
given in eq.~(\ref{def_omega_k_eq_1}).
\stepcounter{exercise}
}
\es
\\
\\
Now let us return to the problem of finding a suitable volume form for the amplituhedron.
Consider a cell decomposition of the amplituhedron.
In each cell $\Gamma$ associated with positive coordinates $\alpha_1^\Gamma$, ..., $\alpha_{4(k-2)}^\Gamma$ 
we may associate a differential form with {\bf logarithmic singularities on the boundary on the cell}:
\bq
 \Omega^{\Gamma}_{n,k}
 & = & 
 \prod\limits_{i=1}^{4(k-2)}
 \frac{d\alpha_i^\Gamma}{\alpha_i^\Gamma}.
\eq
Consider now a collection $T$ of cells covering the amplituhedron.
The volume form for the amplituhedron is then
\bq
 \Omega_{n,k}
 & = & 
 \sum\limits_{\Gamma \in T}
 \Omega^{\Gamma}_{n,k}.
\eq
By a $\mathrm{GL}(k+2,{\mathbb R})$ transformation on the input data $Z$ we may send $Y$ to 
\bq
 Y_0 & = &
 \left( {\bf 0}_{(k-2) \times 4}, {\bf 1}_{(k-2) \times (k-2)} \right).
\eq
We now have all ingredients to give an expression for $B_{n,k}^{(0)}$ from the amplituhedron.
$B_{n,k}^{(0)}$ is obtained by localising $Y$ to $Y_0$ 
and by integrating over the Grassmann variables $\phi_{i,I}$ \cite{Arkani-Hamed:2013jha}:
\bq
\label{B_from amplituhedron}
 B_{n,k}^{(0)}
 & = &
 \int d^{{\mathcal N}}\phi_1
 ...
 \int d^{{\mathcal N}}\phi_{k-2}
 \int \Omega_{n,k}
 \delta^{4(k-2)}\left(Y;Y_0\right).
\eq
The delta-function is defined by
\bq
 \delta^{4(k-2)}\left(Y;Y_0\right)
 & = &
 \int d^{(k-2)(k-2)}g_{ij} \det\left(g\right)^4 \delta^{(k-2)(k+2)}\left( Y - g Y_0 \right).
\eq
Note that there are in eq.~(\ref{B_from amplituhedron})
as many (bosonic) delta-functions as there are (bosonic) integrations.
In addition one integrates over ${\mathcal N}(k-2)=4(k-2)$ auxiliary Grassmann variables $\phi_{i,I}$,
leaving a homogeneous function of degree $4(k-2)$ in the Grassmann variables $\eta_I^j$.
Therefore all integrations are trivial.
The non-trivial information is contained in the differential form $\Omega_{n,k}$.
Let us illustrate this with an example.
Since the dimension of the amplituhedron is $4(k-2)$, we have for $k=2$ a zero-dimensional object,
for $k=3$ a four-dimensional object.
The former is trivial, the latter already too complicated to draw.
We will therefore consider the case, where
\bq
 Y \in \mathrm{Gr}_{1,3}({\mathbb R}),
 \;\;\;\;\;\;
 C \in \overline{\mathrm{Gr}}_{1,4}^+({\mathbb R})
 \;\;\;\;\;\; \mbox{and}
 \;\;\;\;\;\;
 Z \in \mathrm{M}_{3,4}^+({\mathbb R}).
\eq
This case corresponds to $D=2$, $n=4$ and $k=3$.
As external data we take as an example
\bq
 Z & = &
 \left( \begin{array}{cccc}
 2 & 8 & 6 & 3 \\
 2 & 3 & 8 & 7 \\
 1 & 1 & 1 & 1 \\
 \end{array} \right).
\eq
It is easily checked that all ordered minors, obtained by deleting one column, are positive.
Thus we have $Z \in \mathrm{M}_{3,4}^+({\mathbb R})$.
The Grassmannian $\mathrm{Gr}_{1,4}({\mathbb R})$ is identical to ${\mathbb R} {\mathbb P}^3$ and points
in the Grassmannian $\mathrm{Gr}_{1,4}({\mathbb R})$ can be denoted by homogeneous coordinates
$[c_1:c_2:c_3:c_4]$.
The non-negative Grassmannian $\overline{\mathrm{Gr}}_{1,4}^+({\mathbb R})$ is the set of points,
whose homogeneous coordinates satisfy 
\bq
 c_j \ge 0 & \mbox{for} & j=1,...,4.
\eq
This is a three-dimensional simplex, whose corners are given by the four points
\bq
\label{corner_points}
 \left[1:0:0:0\right],
 \;\;\;
 \left[0:1:0:0\right],
 \;\;\;
 \left[0:0:1:0\right],
 \;\;\;
 \left[0:0:0:1\right].
\eq
For $Y \in \mathrm{Gr}_{1,3}({\mathbb R})$ we will use the homogeneous coordinates $Y=[y_1:y_2:y_3]$.
The amplituhedron is then the set of points with the coordinates
\bq
\label{def_example_amplituhedron}
 y_1 & = & 2 c_1 + 8 c_2 + 6 c_3 + 3 c_4,
 \nonumber \\
 y_2 & = & 2 c_1 + 3 c_2 + 8 c_3 + 7 c_4,
 \;\;\;\;\;\;\;\;\; c_j \ge 0,
 \nonumber \\
 y_3 & = & c_1 + c_2 + c_3 + c_4.
\eq
Let us look at the images of the four corner points given in eq.~(\ref{corner_points}).
These are mapped to the four points
\bq
\label{images_corner_points}
 Y_1 \;\; = \;\; \left[2:2:1\right],
 \;\;\;
 Y_2 \;\; = \;\; \left[8:3:1\right],
 \;\;\;
 Y_3 \;\; = \;\; \left[6:8:1\right],
 \;\;\;
 Y_4 \;\; = \;\; \left[3:7:1\right].
\eq
We may now draw the amplituhedron in the chart $y_3=1$.
The result is shown in fig.~(\ref{fig_triangulation}).
Thus for this example the amplituhedron is a quadrangle.
How do we obtain the differential form $\Omega_{4,3}$?
Let us first assume that the amplituhedorn would be the 
non-negative Grassmannian $\overline{\mathrm{Gr}}_{1,3}^+({\mathbb R})$,
parametrised with homogeneous coordinates
\bq
\label{def_standard_traingle}
 \left[ \tilde{y}_1 : \tilde{y}_2 : \tilde{y}_3 \right],
 \;\;\;\;\;\; \tilde{y}_j \ge 0.
\eq
Geometrically, this is a triangle.
We have to find a differential two-form with logarithmic singularities on the boundary.
For the non-negative Grassmannian $\overline{\mathrm{Gr}}_{1,3}^+({\mathbb R})$
we know the answer from eq.~(\ref{def_Omega_projective_space}):
\bq
\label{def_Omega_tilde}
 \tilde{\Omega} & = & \frac{\tilde{y}_1 d\tilde{y}_2 \wedge d\tilde{y}_3 + \tilde{y}_2 d\tilde{y}_3 \wedge d\tilde{y}_1 + \tilde{y}_3 d\tilde{y}_1 \wedge d\tilde{y}_2  }{\tilde{y}_1 \tilde{y}_2 \tilde{y}_3}.
\eq
In the chart $\tilde{y}_3=1$ this form reduces to
\bq
 \tilde{\Omega} & = & \frac{d\tilde{y}_1 \wedge d\tilde{y}_2  }{\tilde{y}_1 \tilde{y}_2}.
\eq
In order to find the form $\Omega_{4,3}$ for amplituhedron defined in eq.~(\ref{def_example_amplituhedron})
we first triangulate the quadrangle.
There are two possible triangulations, both are shown in fig.~(\ref{fig_triangulation}).
Let us take for concreteness the triangulation
\bq
 \Gamma_1 \;\; = \;\; \mathrm{triangle}\left( Y_1,Y_2,Y_3 \right),
 & &
 \Gamma_2 \;\; = \;\; \mathrm{triangle}\left( Y_1,Y_3,Y_4 \right).
\eq
We may map each of these two triangles to the triangle of the non-negative Grassmannian $\overline{\mathrm{Gr}}_{1,3}^+({\mathbb R})$ defined in eq.~(\ref{def_standard_traingle}).
Thus we have maps
\bq
 \varphi_i & : & \Gamma_i \rightarrow \overline{\mathrm{Gr}}_{1,3}^+({\mathbb R}),
 \;\;\;\;\;\;
 i=1,2.
\eq
On $\overline{\mathrm{Gr}}_{1,3}^+({\mathbb R})$ we have a form with logarithmic singularities on the boundaries,
given in eq.~(\ref{def_Omega_tilde}).
Pulling this form back with $\varphi_1$ or $\varphi_2$ gives us a form on $\Gamma_1$ or $\Gamma_2$ with logarithmic
singularities on the boundaries:
\bq
 \Omega_{4,3}^{\Gamma_1} \;\; = \;\; \varphi_1^\ast \tilde{\Omega},
 & &
 \Omega_{4,3}^{\Gamma_2} \;\; = \;\; \varphi_2^\ast \tilde{\Omega},
\eq
We therefore have
\bq
 \Omega_{4,3}
 & = & 
 \Omega_{4,3}^{\Gamma_1} + \Omega_{4,3}^{\Gamma_2}
 \;\; = \;\; 
 \varphi_1^\ast \tilde{\Omega} + \varphi_2^\ast \tilde{\Omega}.
\eq
The virtue of the representation of $B_{n,k}^{(0)}$ in eq.~(\ref{B_from amplituhedron})
in terms of the amplituhedron lies in the fact
that although it might take a while to digest this representation, the basic 
ingredients of this representation are of simple geometric origin.
In this paragraph we discussed the tree amplituhedron.
It is important to mention that the geometric concepts underlying
the amplituhedron 
generalise to loop amplitudes \cite{Arkani-Hamed:2013jha}.

\subsection{Infinity twistors}

In this paragraph we would like to motivate the occurrence of $Y_0$
and of the Grassmann numbers $\phi_{i,I}$
in the formula for the amplituhedron in eq.~(\ref{B_from amplituhedron}). 
This will involve a discussion of infinity twistors.

We have given the cell decomposition of ${\mathbb C} {\mathbb P}^n$ in eq.~(\ref{cell_decomposition_projective_space}), and a similar cell decomposition holds for ${\mathbb R} {\mathbb P}^n$.
Let us consider ${\mathbb R} {\mathbb P}^1$ and let us denote a point $p \in {\mathbb R} {\mathbb P}^1$
by homogeneous coordinates $[x_1:x_2]$.
For ${\mathbb R} {\mathbb P}^1$ we have the cell decomposition
\bq
 {\mathbb R} {\mathbb P}^1
 & = &
 {\mathbb R}^1 \sqcup {\mathbb R}^0.
\eq
More concretely we may write
\bq
 {\mathbb R} {\mathbb P}^1
 & = &
 \left\{
 \; \left[x_1:1\right] \; | \; x_1 \in {\mathbb R} \;
 \right\}
 \;\; \sqcup \;\;
 \left\{ \; \left[1:0\right] \; \right\}.
\eq
The point $[1:0]$ is called the {\bf point at infinity}.
We may think of this point as being obtained from $[x_1:1]$ in the limit $x_1\rightarrow \infty$, where we may neglect the second entry $1$ against the large number $x_1$.

Let us now consider momentum twistors
\bq
 Z_{\alpha j} & = & \left( p_A^j, \mu_{\dot{A}}^j \right),
\eq
where the index $\alpha$ takes the values $\alpha \in \{ 1, 2, \dot{1}, \dot{2} \}$
and $j \in \{1,...,n\}$.
In addition we define two {\bf infinity twistors}, corresponding to the two cases
where either $\mu_{\dot{2}}$ or $\mu_{\dot{1}}$ is large compared to all other components.
We may take these two infinity twistors as $I_{\alpha 3} = ( 0,0,0,-1)$ and $I_{\alpha 4}=(0,0,1,0)$.
Note that since we are in projective space, only the direction matters, not the sign.
If we pack these two infinity twistors in a $4 \times 2$-matrix we have
\bq
 \left( I_{\alpha 3}, I_{\alpha 4} \right)
 & = &
 \left( \begin{array}{rr}
 0 & 0 \\
 0 & 0 \\
 0 & 1 \\
 -1 & 0 \\
 \end{array} \right)
 \;\; = \;\;
 \left( \begin{array}{l}
 {\bf 0}_{2 \times 2} \\
 \eps_{\dot{A} \dot{B}} \\
 \end{array} \right),
\eq
which motivates the specific choice of the two infinity twistors.
Let us now look at a four-bracket with two infinity twistors.
We have
\bq
 \left\langle Z_1, Z_2, I_3, I_4 \right\rangle
 & = &
 \eps^{\alpha\beta\gamma\delta} Z_{\alpha 1} Z_{\beta 2} I_{\gamma 3} I_{\delta 4}
 \;\; = \;\;
 - \left\langle p_1 p_2 \right\rangle.
\eq
Thus the four-bracket reduces to the spinor product.

We may proceed in complete analogy for the momentum twistors
\bq
 W_{\alpha j} & = & \left( \mu_A^j, p_{\dot{A}}^j \right).
\eq
We define two infinity twistors $J_{\alpha 1}=(0,-1,0,0)$ and $J_{\alpha 2}=(1,0,0,0)$.
If we pack them again into a $4 \times 2$-matrix we now have
\bq
 \left( J_{\alpha 1}, J_{\alpha 2} \right)
 & = &
 \left( \begin{array}{rr}
 0 & 1 \\
 -1 & 0 \\
 0 & 0 \\
 0 & 0 \\
 \end{array} \right)
 \;\; = \;\;
 \left( \begin{array}{l}
 \eps_{A B} \\
 {\bf 0}_{2 \times 2} \\
 \end{array} \right).
\eq
For a four-bracket with the two infinity twistors $J_1$ and $J_2$ we obtain
\bq
 \left\langle J_1, J_2, W_3, W_4 \right\rangle
 & = &
 \eps^{\alpha\beta\gamma\delta} J_{\alpha 1} J_{\beta 2} W_{\gamma 3} W_{\delta 4}
 \;\; = \;\;
 \left[ p_3, p_4 \right].
\eq
Again, we see that the four-bracket reduces to a spinor product.

Let us now go to higher dimensions: If we have five five-dimensional vectors $Z_\alpha^{\;j}$
with $\alpha=1,...,5$ and $j=1,...,5$,
we may construct a five-bracket by taking the determinant of the matrix formed by the five
vectors $Z^1$, ..., $Z^5$:
\bq
 \left\langle Z^1, Z^2, Z^3, Z^4, Z^5 \right\rangle
 & = &
 \eps^{\alpha\beta\gamma\delta\epsilon} \; Z_{\alpha}^{\;1} \; Z_{\beta}^{\;2} \; Z_{\gamma}^{\;3} \; Z_{\delta}^{\;4} \; Z_{\epsilon}^{\;5}.
\eq
More generally one defines a $n$-bracket for $n$ vectors $Z^1$, ..., $Z^n$ of dimension $n$ by
\bq
 \left\langle Z^1, Z^2, ..., Z^n\right\rangle
 & = &
 \eps^{\alpha_1 \alpha_2 ... \alpha_n} \; Z_{\alpha_1}^{\;1} \; Z_{\alpha_2}^{\;2} \; ... \; Z_{\alpha_n}^{\;n}.
\eq
Suppose now, that our initial input data lies in a four-dimensional subspace. Thus we may assume that
$Z_5^{\; j}=0$ for all $j=1,...,5$.
The five-bracket vanishes in this case.
There are two possible ways out to obtain a non-vanishing five-bracket.
The first option is to introduce an additional infinity vector
\bq
 I_\alpha^{\;0} & = &
 \left(\begin{array}{c}
 0 \\
 0 \\
 0 \\
 0 \\
 1 \\
 \end{array} \right).
\eq
A five-bracket with four of the five vectors $Z^j$
from above will in general now be non-zero.
It will not depend on the fifth component (i.e. the component with $\alpha=5$)
of any $Z^j$, since $I_\alpha^{\; 0}=0$ for
$\alpha=1,...,4$.
Of course, this five-bracket will depend only on four out of the five vectors $Z^j$
and will be equal to the four-bracket of these four vectors.

In order to obtain an expression, which depends on all five vectors, let us consider a second
option: For each vector, we introduce a fifth component as in eq.~(\ref{def_additional_components})
\bq
\label{fifth_component}
 Z_5^{\;j}
 & = &
 \sum\limits_{I=1}^4
 \phi_I \eta_I^j.
\eq
Let us now assume, that the fifth components of the $Z^j$'s are given by eq.~(\ref{fifth_component}).
Let us now consider the expression
\bq
\label{def_five_bracket}
\lefteqn{
 \left[ Z^1, Z^2, Z^3, Z^4, Z^5 \right]
 =
 } & & \\
 & &
 \frac{1}{4!}
 \int d^4\phi
 \frac{\left\langle 1, 2, 3, 4, 5 \right\rangle^4}
      {
       \left\langle 0, 1, 2, 3, 4 \right\rangle
       \left\langle 0, 2, 3, 4, 5 \right\rangle
       \left\langle 0, 3, 4, 5, 1 \right\rangle
       \left\langle 0, 4, 5, 1, 2 \right\rangle
       \left\langle 0, 5, 1, 2, 3 \right\rangle
      },
 \nonumber
\eq
with
\bq
 \left\langle 1, 2, 3, 4, 5 \right\rangle
 \;\; = \;\; 
 \left\langle Z^1, Z^2, Z^3, Z^4, Z^5 \right\rangle,
 & &
 \left\langle 0, i, j, k, l \right\rangle
 \;\; = \;\; 
 \left\langle I^0, Z^i, Z^j, Z^k, Z^l \right\rangle.
\eq
Note that in eq.~(\ref{def_five_bracket}) no Grassmann number occurs in the denominator.
Eq.~(\ref{def_five_bracket}) is equivalent to
\bq
\label{def_five_bracket_2}
 \left[ Z^1, Z^2, Z^3, Z^4, Z^5 \right]
 & = &
 \frac{\delta^4\left( \left\langle 1, 2, 3, 4 \right\rangle \eta_I^5 + \mbox{cyclic} \right)}
      {
       \left\langle 1, 2, 3, 4 \right\rangle
       \left\langle 2, 3, 4, 5 \right\rangle
       \left\langle 3, 4, 5, 1 \right\rangle
       \left\langle 4, 5, 1, 2 \right\rangle
       \left\langle 5, 1, 2, 3 \right\rangle
      }.
\eq
Expressions of this type are encountered when one evaluates NMHV amplitudes with the help
of the link representation in eq.~(\ref{link_SYM_representation})
or with the help of the amplituhedron in eq.~(\ref{B_from amplituhedron}).
\\
\\
\bs
{\it {\bf Exercise \theexercise}: 
Derive the expression in eq.~(\ref{def_five_bracket_2}) from eq.~(\ref{def_five_bracket}).
\stepcounter{exercise}
}
\es
\\
\\
Can we give a geometric meaning to the integrand in eq.~(\ref{def_five_bracket})?
We will now show that it corresponds to the volume of a polytope.
Let us start with the simplest polytope, a triangle.
Suppose that the corners of the triangles are given in ${\mathbb R}{\mathbb P}^2$ by
\bq
 X_\alpha^{\;1}
 \;\; = \;\;
 \left( \begin{array}{c}
   X_1^{\;1} \\
   X_2^{\;1} \\
   1 \\
  \end{array} \right),
 \;\;\;\;\;\;
 X_\alpha^{\;2}
 \;\; = \;\;
 \left( \begin{array}{c}
   X_1^{\;2} \\
   X_2^{\;2} \\
   1 \\
  \end{array} \right),
 \;\;\;\;\;\;
 X_\alpha^{\;2}
 \;\; = \;\;
 \left( \begin{array}{c}
   X_1^{\;3} \\
   X_2^{\;3} \\
   1 \\
  \end{array} \right).
\eq
The ``volume'' (i.e. area) of the triangle is then
\bq
 \mathrm{vol}\left(\mathrm{triangle}\right)
 & = &
 \frac{1}{2}
 \left| \begin{array}{ccc}
  X_1^{\;1} & X_1^{\;2} & X_1^{\;3} \\
  X_2^{\;1} & X_2^{\;2} & X_2^{\;3} \\
  1 & 1 & 1 \\
 \end{array} \right|
 \\
 & = &
 \frac{1}{2} \left[
 \left( X_1^{\;2} - X_1^{\;1} \right) \left( X_2^{\;3} - X_2^{\;1} \right)
 -
 \left( X_2^{\;2} - X_2^{\;1} \right) \left( X_1^{\;3} - X_1^{\;1} \right)
 \right].
 \nonumber
\eq
This is most easily seen by first considering the area of parallelogram
spanned by the two vectors $(X_\alpha^{\;2}-X_\alpha^{\;1})$ and $(X_\alpha^{\;3}-X_\alpha^{\;1})$
and then taking half of this area.
The area of the parallelogram is
\bq
 \left(\vec{X}^{2}-\vec{X}^{1}\right) \times \left(\vec{X}^{3}-\vec{X}^{1}\right)
 & = &
 \left( X_1^{\;2} - X_1^{\;1} \right) \left( X_2^{\;3} - X_2^{\;1} \right)
 -
 \left( X_2^{\;2} - X_2^{\;1} \right) \left( X_1^{\;3} - X_1^{\;1} \right).
\eq
Instead of specifying the triangle by its corners, we may equally well specify the boundary lines.
Each boundary line can be characterised by a vector orthogonal to it.
Let's call this vector $Z_\alpha^{\;j}$ for the boundary line which does not contain the corner $j$.
Points $X_\alpha$ on the boundary line $j$ satisfy the incidence relation
\bq
 \sum\limits_{\alpha=1}^3
 Z_\alpha^{\;j} X_\alpha & = & 0.
\eq
Given the $Z^j$'s, we may reconstruct the $X^j$'s as
\bq
 X_\alpha^{\;1}
 =
 \eps_{\alpha\beta\gamma} Z_\beta^{\;2} Z_\gamma^{\;3},
 \;\;\;\;\;\;
 X_\alpha^{\;2}
 =
 \eps_{\alpha\beta\gamma} Z_\beta^{\;3} Z_\gamma^{\;1},
 \;\;\;\;\;\;
 X_\alpha^{\;3}
 =
 \eps_{\alpha\beta\gamma} Z_\beta^{\;1} Z_\gamma^{\;2}.
\eq
In terms of the $Z^j$'s, the volume of the triangle is given by
\bq
 \mathrm{vol}\left(\mathrm{triangle}\right)
 & = &
 \frac{1}{2}
 \frac{\left\langle Z^1, Z^2, Z^3 \right\rangle^2}
      {\left\langle I^0, Z^1, Z^2 \right\rangle \left\langle I^0, Z^2, Z^3 \right\rangle \left\langle I^0, Z^3, Z^1 \right\rangle},
\eq
where $I^0$ is an infinity vector
\bq
 I^0 & = &
 \left(\begin{array}{c}
 0 \\
 0 \\
 1 \\
 \end{array} \right).
\eq
We may generalise this to an $n$-simplex in ${\mathbb R}{\mathbb P}^n$, 
specified by $(n+1)$ vectors $Z_\alpha^{\;j}$, with $\alpha=1,...,(n+1)$ and $j=1,...,(n+1)$.
These vectors define the $(n+1)$ faces (or boundaries) of the simplex.
A point $X_\alpha$ lies on the $j$-th boundary if
\bq
 \sum\limits_{\alpha=1}^{n+1}
 Z_\alpha^{\;j} X_\alpha & = & 0.
\eq
The {\bf volume of the simplex} is given by
\bq
\label{volume_simlex}
 \left[ Z^1, ..., Z^{n+1} \right]
 & = &
 \frac{1}{n!}
 \frac{\left\langle Z^1, ..., Z^{n+1} \right\rangle^n}
      {\left\langle I^0, Z^1, ..., Z^{n} \right\rangle \left\langle I^0, Z^2, ..., Z^{n+1} \right\rangle ... \left\langle I^0, Z^{n+1}, Z^1, ..., Z^{n-1} \right\rangle},
\eq
where $I^0$ is again an infinity vector or reference vector, which we may take
as $I_{\alpha}^0=0$ for $\alpha=1,...,n$ and $I_{n+1}^0=1$.
We recognise in this formula the integrand of eq.~(\ref{def_five_bracket}).

% ----------------------------------------------------------------------------------
\newpage
\section{The CHY representation}
\label{sect:chy_representation}

In this section we return to pure (non-supersymmetric) Yang-Mills theory
and introduce the CHY representation for the primitive tree-level amplitudes.
This representation expresses the amplitude as a global residue at the zeros of the scattering equations
and separates the information on the external cyclic order and on the polarisations in two different functions.
The formul{\ae} in this section are not specific to four space-time dimensions and we may take
Minkowski space to be $D$ dimensional.

\subsection{The scattering equations}
\label{sect_scattering_equations}

We denote by ${\mathbb C}M$ the complexified Minkowski space, i.e. a complex vector space of dimension $D$.
The Minkowski metric $g_{\mu\nu}=\mathrm{diag}(1,-1,-1,-1,...)$ is extended by linearity.
We further denote by $\Phi_n$ the (momentum) configuration space of $n$ external massless particles:
\bq
 \Phi_n & = &
 \left\{ \left(p_1,p_2,...,p_n\right) \in \left({\mathbb C} M\right)^n | p_1+p_2+...+p_n=0, p_1^2 = p_2^2 = ... = p_n^2 = 0 \right\}.
\eq
In other words, a $n$-tuple $p=(p_1, p_2, ..., p_n)$ of momentum vectors belongs to $\Phi_n$ if this $n$-tuple satisfies momentum conservation
and the mass-shell conditions $p_i^2=0$ for massless particles. 
It will be convenient to use the notation $p$  without any index to denote such an $n$-tuple.
In the same spirit we denote 
the $n$-tuple of external polarisations by $\eps=(\eps_1^{\lambda_1},...,\eps_n^{\lambda_n})$
with $\lambda_j \in \{+,-\}$.
A primitive tree amplitude $A_{n}^{(0)}$ has a fixed cyclic order of the external legs,
which can be specified by a permutation $\sigma=(\sigma_1,...,\sigma_n)$.
In this section it will be convenient to specify a primitive tree amplitude by 
the three $n$-tuples $\sigma$, $p$ and $\eps$:
\bq
 A_n^{(0)}\left( \sigma, p, \eps \right)
 & = &
 A_n^{(0)}\left( p_{\sigma_1}^{\lambda_{\sigma_1}}, ..., p_{\sigma_n}^{\lambda_{\sigma_n}} \right).
\eq
We denote by $\hat{\mathbb C} = {\mathbb C} \cup \{\infty\}$.
The space $\hat{\mathbb C}$ is equivalent to the complex projective space ${\mathbb C}{\mathbb P}^1$.
For amplitudes with $n$ external particles we consider the space $\hat{\mathbb C}^n$.
Points in $\hat{\mathbb C}^n$ will be denoted by $z=(z_1,z_2,...,z_n)$.
Again we use the convention that $z$ without any index denotes an $n$-tuple.

We set for $1\le i \le n$
\bq
 f_i\left(z,p\right) & = & 
 \sum\limits_{j=1, j \neq i}^n \frac{ 2 p_i \cdot p_j}{z_i - z_j}.
\eq
The {\bf scattering equations} read \cite{Cachazo:2013gna,Cachazo:2013hca,Cachazo:2013iea}
\bq
\label{scattering_equations}
 f_i\left(z,p\right) & = & 0,
 \;\;\;\;\;\;
 i \in \{1,...,n\}.
\eq
For a fixed $p \in \Phi_n$
a solution of the scattering equation is a point $z \in \hat{\mathbb C}^n$,
such that the scattering equations in eq.~(\ref{scattering_equations}) are satisfied.

The scattering equations are invariant under the {\bf projective special linear group}
$\mathrm{PSL}(2,{\mathbb C})=\mathrm{SL}(2,{\mathbb C})/{\mathbb Z}_2$.
Here, ${\mathbb Z}_2$ is given by $\{ {\bf 1}, -{\bf 1} \}$, with ${\bf 1}$ denoting
the unit matrix.
Let
\bq
 g = \left(\begin{array}{cc} a & b \\ c & d \\ \end{array} \right) & \in & \mathrm{PSL}(2,{\mathbb C}).
\eq
Each $g \in \mathrm{PSL}(2,{\mathbb C})$ defines an automorphism of $\hat{\mathbb C}$ as follows:
\bq
\label{moebius_trafo}
 g \cdot z_i & = &
 \frac{a z_i + b}{c z_i + d},
 \;\;\;\;\;\; z_i \in \hat{\mathbb C}.
\eq
Conversely, since every automorphism of $\hat{\mathbb C}$ is of the form above, it follows that
each automorphism of $\hat{\mathbb C}$ defines an element $g \in \mathrm{PSL}(2,{\mathbb C})$.
We therefore have a group isomorphism
\bq
\label{group_homomorphism}
 \mathrm{PSL}(2,{\mathbb C}) & \rightarrow & \mathrm{Aut}\left(\hat{\mathbb C}\right),
\eq
where we denoted by $\mathrm{Aut}(\hat{\mathbb C})$ the automorphism group of $\hat{\mathbb C}$.
The transformation in eq.~(\ref{moebius_trafo}) is called a {\bf M\"obius transformation}.
We further set
\bq
 g \cdot \left(z_1, z_2, ..., z_n \right)
 & = &
 \left(g \cdot z_1, g \cdot z_2, ..., g \cdot z_n \right).
\eq
If $(z_1,z_2, ..., z_n)$ is a solution of eq.~(\ref{scattering_equations}), then also
$(z_1',z_2', ..., z_n') = g \cdot (z_1,z_2, ..., z_n)$ is a solution.
\\
\\
\bs
{\it {\bf Exercise \theexercise}: 
Let $g \in \mathrm{PSL}(2,{\mathbb C})$ and let $(z_1,z_2, ..., z_n)$ be a solution
of the scattering equations.
Show that $(z_1',z_2', ..., z_n') = g \cdot (z_1,z_2, ..., z_n)$ is also a solution
of the scattering equations.
\stepcounter{exercise}
}
\es
\\
\\
We call two solutions which are related by a $\mathrm{PSL}(2,{\mathbb C})$-transformation {\bf equivalent solutions}.

The $n$ scattering equations in eq.~(\ref{scattering_equations}) are not independent, only $(n-3)$ of them are.
The M\"obius invariance implies the relations
\bq
\label{relations_scattering_equations}
 \sum\limits_{i=1}^n f_i\left(z,p\right) = 0,
 \;\;\;\;\;\;
 \sum\limits_{i=1}^n z_i f_i\left(z,p\right) = 0,
 \;\;\;\;\;\;
 \sum\limits_{i=1}^n z_i^2 f_i\left(z,p\right) = 0.
\eq
\bs
{\it {\bf Exercise \theexercise}: Prove the relations in eq.~(\ref{relations_scattering_equations}).
\stepcounter{exercise}
}
\es
\\
\\
There is an alternative formulation of the scattering equations in {\bf polynomial form} \cite{Dolan:2014ega}.
We set $I = \{ 1, 2, ..., n \}$ and for any subset $S \subseteq I$ we define
\bq
 p_S = \sum\limits_{i \in S} p_i,
 & &
 z_S = \prod\limits_{i \in S} z_i.
\eq
For the empty set $\emptyset$ we define $p_\emptyset = 0$ and $z_\emptyset = 1$.
For $0 \le m \le n$ we define the polynomials $h_m(z,p)$ in the variables $z=(z_1,...,z_n)$ by
\bq
\label{def_polynomials}
 h_m\left(z,p\right)
 & = &
 \sum\limits_{S\subset I, \left|S\right|=m} p_S^2 \; z_S.
\eq
The polynomial $h_0$ is trivially zero, due to $p_\emptyset=0$.
The polynomials $h_1$, $h_{n-1}$ and $h_n$ vanish trivially due to momentum conservation and the on-shell conditions:
\bq
 h_1\left(z,p\right)
 & = &
 \sum\limits_{j=1}^n p_j^2 z_j
 \;\; = \;\; 0,
 \nonumber \\
 h_{n-1}\left(z,p\right)
 & = &
 \sum\limits_{j=1}^n p_j^2 \; z_1 ... z_{j-1} z_{j+1} ... z_n
 \;\; = \;\; 0,
 \nonumber \\
 h_n\left(z,p\right)
 & = &
 \left( p_1 + ... + p_n \right)^2 z_1 ... z_n
 \;\; = \;\; 0.
\eq
The non-vanishing polynomials $h_m$ are homogeneous polynomials in the variables $z_1$, $z_2$, ..., $z_n$ of degree $m$.
They are linear in each variable $z_j$.
The scattering equations are equivalent to the set of equations
\bq
\label{polynomial_form}
 h_m\left(z,p\right)
 & = &
 0,
 \;\;\;\;\;\; 2 \le m \le n-2.
\eq
Let us denote by
\bq
 C^{n-3} & = & \hat{\mathbb C}^n / \mathrm{Aut}\left(\hat{\mathbb C}\right),
\eq
i.e. $\hat{\mathbb C}^n$ modulo the diagonal action of $\mathrm{Aut}\left(\hat{\mathbb C}\right)$.
$C^{n-3}$ is a space of complex dimension $(n-3)$.
The polynomials in eq.~(\ref{def_polynomials}) define an algebraic variety, called scattering variety \cite{He:2014wua}:
\bq
 V_n\left(p\right) & = &
 \left\{ z \in C^{n-3} | \; h_m\left(z,p\right)=0; \; 2 \le m \le n-2 \right\}.
\eq
Since there are $(n-3)$ non-vanishing polynomials $h_m(z,p)$, the algebraic variety $V_n$ is of dimension zero.
In other words, $V_n$ consists of individual points.
There is a theorem in algebraic geometry by B\'ezout \cite{Griffiths:book}
which tells us that $V_n$ consists of $(n-3)!$ points.
Therefore there are $(n-3)!$ different solutions of the scattering equations 
not related by a $\mathrm{PSL}(2,{\mathbb C})$-transformation.
We will denote a solution by 
\bq
 z^{(j)} & = & \left( z_1^{(j)}, ..., z_n^{(j)} \right)
\eq
and a sum over the $(n-3)!$ inequivalent solutions by
\bq
 \sum\limits_{\mathrm{solution}\;j}
\eq
Let us add the following remark: Consider a Riemann sphere (i.e. an algebraic curve of genus zero)
with $n$ distinct marked points.
The moduli space of genus $0$ curves with $n$ distinct marked points is denoted
by
\bq
 {\mathcal M}_{0,n}
 & = &
 \left\{ z \in \left( {\mathbb C} {\mathbb P}^1 \right)^n : z_i \neq z_j \right\}/\mathrm{PSL}\left(2,{\mathbb C}\right).
\eq
${\mathcal M}_{0,n}$ is an affine algebraic variety of dimension $(n-3)$.
We see that any solution of the scattering equations with $z_i\neq z_j$ for all $i \neq j$ corresponds
to a point $z \in {\mathcal M}_{0,n}$.
\\
\\
\bs
{\it {\bf Exercise \theexercise}: 
Consider the Koba-Nielsen function
\bq
 U\left(z,p\right)
 & = &
 \prod\limits_{i<j} \left( z_i - z_j \right)^{2 p_i \cdot p_j }.
\eq
Show
\bq
 U^{-1} \frac{\partial}{\partial z_i} U
 & = &
 f_i\left(z,p\right).
\eq
\stepcounter{exercise}
}
\es

\subsection{Polarisation factors and Parke-Taylor factors}

There are two essential ingredients for the CHY representation of $A_n^{(0)}$:
A polarisation factor $E(p,\eps,z)$ and a cyclic factor (or Parke-Taylor factor) $C(\sigma,z)$,
which we will both define in this paragraph.
We start by defining a $(2n)\times(2n)$ anti-symmetric matrix $\Psi$ through 
\bq
 \Psi 
 & = &
 \left( \begin{array}{cc}
 A & - C^T \\
 C & B \\
 \end{array} \right)
\eq
with
\bq
 A_{ab}
 = 
 \left\{ \begin{array}{cc}
 \frac{2 p_a \cdot p_b}{z_a-z_b} & a \neq b, \\
 0 & a = b, \\
 \end{array} \right.
 & &
 B_{ab}
 = 
 \left\{ \begin{array}{cc}
 \frac{2 \eps_a \cdot \eps_b}{z_a-z_b} & a \neq b, \\
 0 & a = b, \\
 \end{array} \right.
\eq
and
\bq
 C_{ab}
 & = &
 \left\{ \begin{array}{cc}
 \frac{2 \eps_a \cdot p_b}{z_a-z_b} & a \neq b, \\
 - \sum\limits_{j=1, j\neq a}^n \frac{2 \eps_a \cdot p_j}{z_a-z_j}  & a = b. \\
 \end{array} \right.
\eq
Let $1 \le i < j \le n$.
One denotes by $\Psi^{ij}_{ij}$ the $(2n-2)\times(2n-2)$-matrix 
where the rows and columns $i$ and $j$ of $\Psi$ have been deleted.
$\Psi^{ij}_{ij}$ has a non-vanishing Pfaffian and one sets
as {\bf polarisation factor}
\bq
 E\left(p,\eps,z\right)
 & = &
 \frac{\left(-1\right)^{i+j}}{2 \left(z_i-z_j\right)} \mathrm{Pf} \; \Psi^{ij}_{ij}.
\eq
$E(p,\eps,z)$ is independent of the choice of $i$ and $j$ on the solutions of the scattering equations.

Let us now consider a permutation $\sigma=(\sigma_1,...,\sigma_n)$.
The {\bf cyclic factor} (or {\bf Parke-Taylor factor}) is given by
\bq
\label{def_cyclic_factor}
 C\left(\sigma,z\right)
 & = & 
 \frac{1}{\left(z_{\sigma_1} - z_{\sigma_2} \right) \left( z_{\sigma_2} - z_{\sigma_3} \right) ... \left( z_{\sigma_r} - z_{\sigma_1} \right)}.
\eq
The cyclic factor depends on the cyclic order $\sigma$, but not on the polarisations $\eps$.
On the other hand, the polarisation factor depends on the polarisations $\eps$, but not on the cyclic order $\sigma$.
Under a $\mathrm{PSL}(2,{\mathbb C})$ transformation we have
\bq
\label{PSL2_trafo_E_C}
 E\left(p,\eps,g \cdot z\right)
 & = &
 \left( \prod\limits_{j=1}^n \left(c z_j + d\right)^2 \right)
 E\left(p,\eps,z\right),
 \nonumber \\
 C\left(\sigma, g \cdot z \right)
 & = &
 \left( \prod\limits_{j=1}^n \left(c z_j + d\right)^2 \right)
 C\left(\sigma, z \right).
\eq

\subsection{The two forms of the CHY representation}

There are two equivalent forms 
of the Cachazo-He-Yuan (CHY) representation \cite{Cachazo:2013gna,Cachazo:2013hca,Cachazo:2013iea}.
The first form expresses the primitive tree amplitude $A_n^{(0)}$
as a {\bf multidimensional complex contour integral} 
in the  auxiliary space $\hat{\mathbb C}^n$:
\bq
\label{CHY_int_representation_A_n}
 A_n^{(0)}\left(\sigma,p,\eps\right)
 & = &
 i \oint\limits_{\mathcal C} d\Omega_{\mathrm{CHY}} \; C\left(\sigma,z\right) \; E\left(z,p,\eps\right),
\eq
where the cyclic factor $C(\sigma,z)$ and the polarisation factor $E(z,p,\eps)$ have been
defined in the previous paragraph.
The measure $d\Omega_{\mathrm{CHY}}$ is defined by
\bq
 d\Omega_{\mathrm{CHY}}
 & = &
 \frac{1}{\left(2\pi i\right)^{n-3}}
 \frac{d^nz}{d\omega}
 \;
 \prod{}' \frac{1}{f_a\left(z,p\right)},
\eq
where
\bq
 \prod{}' \frac{1}{f_a\left(z,p\right)}
 & = & 
 \left(-1\right)^{i+j+k}
 \left( z_i - z_j \right) \left( z_j - z_k \right) \left( z_k - z_i \right)
 \prod\limits_{a \neq i,j,k} \frac{1}{f_a\left(z,p\right)},
\eq
and
\bq
 d\omega
 & = &
 \left(-1\right)^{p+q+r}
 \frac{dz_p dz_q dz_r}{\left( z_p - z_q \right) \left( z_q - z_r \right) \left( z_r - z_q \right)}.
\eq
The primed product of $1/f_a$ is independent of the choice of $i$, $j$, $k$
and takes into account that only $(n-3)$ of the scattering equations are independent.
The quantity $d\omega$ is independent of the choice of $p$, $q$, $r$ and corresponds to the invariant measure
on $\mathrm{PSL}(2,{\mathbb C})$.
Dividing by $d\omega$ ensures that from each class of equivalent solutions only one representative is taken.
The integration contour ${\mathcal C}$ encircles the inequivalent solutions of the scattering equations.
Let us investigate how the various pieces transform 
under a $\mathrm{PSL}(2,{\mathbb C})$-transformation $z' = g \cdot z$.
We have
\bq
 z_i' - z_j'
 \;\; = \;\;
 \frac{z_i-z_j}{\left(c z_i+d\right)\left(c z_j+d\right)}
 & \mbox{and} &
 d z_j' 
 \;\; = \;\;
 \frac{d z_j}{\left(c z_j + d \right)^2}.
\eq
The measure $d^nz$ transforms as
\bq
 d^nz' & = &
 \left( \prod\limits_{j=1}^n \frac{1}{\left(c z_j + d\right)^2} \right)
 d^n z.
\eq
The measure $d\omega$ is invariant:
\bq
 d\omega' & = & d\omega.
\eq
The primed product of $1/f_a$ transforms as
\bq
 \prod{}' \frac{1}{f_a\left( z',p\right)}
 & = &
 \left( \prod\limits_{j=1}^n \frac{1}{\left(c z_j + d\right)^2} \right)
 \left( \prod{}' \frac{1}{f_a\left(z,p\right)} \right).
\eq
With the transformation properties of $C(\sigma,z)$ and $E(z,p,\eps)$,
given in eq.~(\ref{PSL2_trafo_E_C}), we see that the integrand is 
$\mathrm{PSL}(2,{\mathbb C})$-invariant.
In eq.~(\ref{CHY_int_representation_A_n}) there are $(n-3)$ contour integrations encircling the poles
given by the zeros of the independent scattering equations.
We may equally well express the quantity of eq.~(\ref{CHY_int_representation_A_n})
as a {\bf sum over the inequivalent zeros of the scattering equations}.
Doing so will introduce a Jacobian factor $J(z,p)$ and we find
\bq
\label{CHY_sum_representation_A_n}
 A_n^{(0)}\left(\sigma,p,\eps\right)
 & = &
 i
 \sum\limits_{\mathrm{solutions} \; j} J\left(z^{(j)},p\right) \; C\left(\sigma, z^{(j)}\right) \; E\left(z^{(j)},p,\eps\right).
\eq
The Jacobian is obtained as follows:
We define a $n \times n$-matrix $\Phi$ with entries
\bq
 \Phi_{ab}
 & = &
 \frac{\partial f_a}{\partial z_b}
 \;\; = \;\;
 \left\{
 \begin{array}{cc}
 \frac{2 p_a \cdot p_b}{\left(z_a-z_b\right)^2} & a \neq b, \\
 - \sum\limits_{j=1, j\neq a}^n \frac{2 p_a \cdot p_j}{\left(z_a-z_j\right)^2} & a=b. \\
 \end{array}
 \right.
\eq
Let $\Phi^{ijk}_{rst}$ denote the $(n-3)\times(n-3)$-matrix, where the rows $\{i,j,k\}$ and the columns $\{r,s,t\}$ have
been deleted.
We set
\bq
 \det{}' \; \Phi
 & = &
 \left(-1\right)^{i+j+k+r+s+t}
 \frac{\left|\Phi^{ijk}_{rst}\right|}{\left(z_{ij}z_{jk}z_{ki}\right)\left(z_{rs}z_{st}z_{tr}\right)},
\eq
where we used the abbreviation $z_{ab}=z_a-z_b$.
With the above sign included, the quantity
$\det{}' \; \Phi$ is independent of the choice of $\{i,j,k\}$ and $\{r,s,t\}$
on the solutions of the scattering equations.
We then have
\bq
 J\left(z,p\right) & = &
 \frac{1}{\det{}' \; \Phi}.
\eq
Under $\mathrm{PSL}(2,{\mathbb C})$-transformations the Jacobian $J(z,p)$ transforms
as
\bq
 J\left(g \cdot z, p\right)
 & = &
 \left( \prod\limits_{j=1}^n \frac{1}{\left(c z_j + d\right)^4} \right)
 J\left(z,p\right)
\eq
and each summand in eq.~(\ref{CHY_sum_representation_A_n})
is $\mathrm{PSL}(2,{\mathbb C})$-invariant.
Let us briefly look at the mass dimensions of the various ingredients.
The dimension of $A_n^{(0)}$ is 
\bq
 \dim A_n^{(0)} 
 & = &
 4 - n.
\eq
The ingredients of eq.~(\ref{CHY_sum_representation_A_n})
have the mass dimensions
\bq
 \dim J\left(z,p\right) 
 \; = \;
 - 2 \left( n-3\right),
 \;\;\;
 \dim C\left(\sigma,z\right) 
 \; = \; 
 0,
 \;\;\;
 \dim E\left(z,p,\eps\right) 
 \; = \;
 n-2,
\eq
adding up to $4-n=\dim A_n^{(0)}$ as they should.

Eq.~(\ref{CHY_int_representation_A_n}) and eq.~(\ref{CHY_sum_representation_A_n})
are the two forms of the CHY representation of the primitive tree amplitude $A_n^{(0)}$.
Note that the cyclic order enters only through the Parke-Taylor factor $C(\sigma,z)$,
whereas the helicity configuration enters only through the polarisation factor $E(z,p,\eps)$.
The CHY representation separates the information on the cyclic order from the information on
the helicity configuration.

How do we know that eq.~(\ref{CHY_int_representation_A_n}) or eq.~(\ref{CHY_sum_representation_A_n})
actually compute the primitive tree amplitude $A_n^{(0)}$?
We can prove this by induction.
The induction start for $n=3$ will be shown below,
the induction step uses once again the powerful method of on-shell recursion relations
and is given in \cite{Dolan:2013isa}.

Let us now look at a few examples.
We first discuss the case $n=3$. 
In general we have $(n-3)!$ inequivalent solutions of the scattering equations.
For $n=3$ we thus have $0!=1$ inequivalent solution.
Due to M\"obius invariance we may fix in general three $z$-variables at some chosen values.
For $n=3$ we have in total only three variables $(z_1,z_2,z_3)$.
Therefore we may take without loss of generality as representative for the one inequivalent solution
\bq
 z_1^{(1)} = 0,
 \;\;\;
 \;\;\;
 z_2^{(1)} = 1,
 \;\;\;
 \;\;\;
 z_3^{(1)} = \infty.
\eq
Let us consider the cyclic order $\sigma=(1,2,3)$.
We find for the Jacobian, the Parke-Taylor factor and the polarisation factor
\bq
 J\left(z,p\right)
 & = &
 \left( z_1 - z_2 \right)^2 \left( z_2 - z_3 \right)^2 \left( z_3 - z_1 \right)^2,
 \nonumber \\
 C\left(\sigma,z\right) & = &
 \frac{1}{\left( z_1 - z_2 \right) \left( z_2 - z_3 \right) \left( z_3 - z_1 \right)},
 \nonumber \\
 E\left(z,p,\eps\right) & = &
 \frac{2 \left[ 
                \left( \eps_1 \cdot \eps_2 \right) \left( \eps_3 \cdot p_1 \right)
              + \left( \eps_2 \cdot \eps_3 \right) \left( \eps_1 \cdot p_2 \right)
              + \left( \eps_3 \cdot \eps_1 \right) \left( \eps_2 \cdot p_3 \right)
         \right]}{\left( z_1 - z_2 \right) \left( z_2 - z_3 \right) \left( z_3 - z_1 \right)}.
\eq
Putting everything together we obtain for $A_3^{(0)}$
\bq
 A_3^{(0)}\left(\sigma,p,\eps\right)
 & = &
 i J\left(z^{(1)},p\right) \; C\left(\sigma, z^{(1)}\right) \; E\left(z^{(1)},p,\eps\right)
 \\
 & = &
      2 i \left[ 
                \left( \eps_1 \cdot \eps_2 \right) \left( \eps_3 \cdot p_1 \right)
              + \left( \eps_2 \cdot \eps_3 \right) \left( \eps_1 \cdot p_2 \right)
              + \left( \eps_3 \cdot \eps_1 \right) \left( \eps_2 \cdot p_3 \right)
         \right]
 \nonumber \\
 & = &
 i
 \left[
   g_{\mu_1\mu_2} \left(p_1-p_2\right)_{\mu_3}
 + g_{\mu_2\mu_3} \left(p_2-p_3\right)_{\mu_1}
 + g_{\mu_3\mu_1} \left(p_3-p_1\right)_{\mu_2}
 \right]
 \eps_1^{\mu_1} \eps_2^{\mu_2} \eps_3^{\mu_3},
 \nonumber
\eq
which is the correct result.

Let us also look at the case $n=4$. We consider the cyclic order $\sigma=(1,2,3,4)$.
The scattering equations read
\begin{alignat}{5}
                      && \frac{s}{z_1-z_2} && +\frac{u}{z_1-z_3} && +\frac{t}{z_1-z_4} && = 0,
 \nonumber \\
 -\frac{s}{z_1-z_2}   &&                    && +\frac{t}{z_2-z_3} && +\frac{u}{z_2-z_4} && = 0,
 \nonumber \\
 -\frac{u}{z_1-z_3}   && -\frac{t}{z_2-z_3}  &&                   && +\frac{s}{z_3-z_4} && = 0,
 \nonumber \\
 -\frac{t}{z_1-z_4}   && -\frac{u}{z_2-z_4}  && -\frac{s}{z_3-z_4} &&                   && = 0.
\end{alignat}
We immediately obtain the following cross-ratios:
\bq
 \frac{\left(z_1-z_2\right)\left(z_3-z_4\right)}{\left(z_1-z_4\right)\left(z_3-z_2\right)} = - \frac{s}{t},
 \;\;\;\;\;\;
 \frac{\left(z_1-z_4\right)\left(z_2-z_3\right)}{\left(z_1-z_3\right)\left(z_2-z_4\right)} = - \frac{t}{u},
 \;\;\;\;\;\;
 \frac{\left(z_1-z_3\right)\left(z_4-z_2\right)}{\left(z_1-z_2\right)\left(z_4-z_3\right)} = - \frac{u}{s}.
\eq
There is one solution to the scattering equations, which can be taken without loss of generality as
\bq
 z_1^{(1)} = -\frac{s}{t},
 \;\;\;
 \;\;\;
 z_2^{(1)} = 0,
 \;\;\;
 \;\;\;
 z_3^{(1)} = 1,
 \;\;\;
 \;\;\;
 z_4^{(1)} = \infty.
\eq
For $n=4$ we have as usual the Mandelstam relation
\bq
\label{mandelstam_relation}
 s + t + u & = & 0.
\eq
In addition, we have from the polynomial form of the scattering equations the relation
\bq
\label{scattering_relation}
 \left( z_1 z_2 + z_3 z_4 \right) s
 +
 \left( z_2 z_3 + z_1 z_4 \right) t
 +
 \left( z_1 z_3 + z_2 z_4 \right) u
 & = & 0.
\eq
There are several equivalent forms for $J(z,p)$, due to the choice of rows and columns to be deleted from the
matrix $\Phi$.
All these forms can be related with the help
of eq.~(\ref{mandelstam_relation}) and eq.~(\ref{scattering_relation}).
We find
\bq
 J\left(z,p\right)
 & = &
 - \frac{1}{s}
   \left( z_1 - z_2 \right)^2 \left( z_1 - z_3 \right) \left( z_1 - z_4 \right) 
   \left( z_2 - z_3 \right) \left( z_2 - z_4 \right) \left( z_3 - z_4 \right)^2.
\eq
The cyclic factor is given by
\bq
 C\left(\sigma,z\right) & = &
 \frac{1}{\left( z_1 - z_2 \right) \left( z_2 - z_3 \right) \left( z_3 - z_4 \right) \left( z_4 - z_1 \right)}.
\eq
For the polarisation factor we find
\bq
\lefteqn{
 E\left(z,p,\eps\right) 
 = } & &
 \\
 & &
 \frac{1}{z_{12} z_{23} z_{34} z_{41}}
 \left[ 
        4 \left( \eps_1 \cdot \eps_2 \right) \left( \eps_3 \cdot p_2 \right) \left( \eps_4 \cdot p_1 \right)
        -4 \left( \eps_1 \cdot \eps_4 \right) \left( \eps_3 \cdot p_2 \right) \left( \eps_2 \cdot p_1 \right)
 \right. \nonumber \\
 & & \left.
        -4 \left( \eps_3 \cdot \eps_4 \right) \left( \eps_1 \cdot p_2 \right) \left( \eps_2 \cdot p_3 \right)
        +4 \left( \eps_2 \cdot \eps_3 \right) \left( \eps_1 \cdot p_2 \right) \left( \eps_4 \cdot p_3 \right)
        +4 \left( \eps_2 \cdot \eps_4 \right) \left( \eps_1 \cdot p_2 \right) \left( \eps_3 \cdot p_2 \right)
 \right. \nonumber \\
 & & \left.
        +t \left( \eps_1 \cdot \eps_2 \right) \left( \eps_3 \cdot \eps_4 \right)
        +s \left( \eps_2 \cdot \eps_3 \right) \left( \eps_1 \cdot \eps_4 \right)
 \right]
 \nonumber \\
 & &
 + \frac{1}{z_{23}z_{31}z_{14}z_{42}}
 \left[
        4 \left( \eps_2 \cdot \eps_3 \right) \left( \eps_1 \cdot p_3 \right) \left( \eps_4 \cdot p_2 \right)
        -4 \left( \eps_2 \cdot \eps_4 \right) \left( \eps_1 \cdot p_3 \right) \left( \eps_3 \cdot p_2 \right)
 \right. \nonumber \\
 & & \left.
        -4 \left( \eps_1 \cdot \eps_4 \right) \left( \eps_2 \cdot p_3 \right) \left( \eps_3 \cdot p_1 \right)
        +4 \left( \eps_3 \cdot \eps_1 \right) \left( \eps_2 \cdot p_3 \right) \left( \eps_4 \cdot p_1 \right)
        +4 \left( \eps_3 \cdot \eps_4 \right) \left( \eps_2 \cdot p_3 \right) \left( \eps_1 \cdot p_3 \right)
 \right. \nonumber \\
 & & \left.
        +u \left( \eps_2 \cdot \eps_3 \right) \left( \eps_1 \cdot \eps_4 \right)
        +t \left( \eps_3 \cdot \eps_1 \right) \left( \eps_2 \cdot \eps_4 \right)
 \right]
 \nonumber \\
 & &
 + \frac{1}{z_{31}z_{12}z_{24}z_{43}}
 \left[
        4 \left( \eps_3 \cdot \eps_1 \right) \left( \eps_2 \cdot p_1 \right) \left( \eps_4 \cdot p_3 \right)
        -4 \left( \eps_3 \cdot \eps_4 \right) \left( \eps_2 \cdot p_1 \right) \left( \eps_1 \cdot p_3 \right)
 \right. \nonumber \\
 & & \left.
        -4 \left( \eps_2 \cdot \eps_4 \right) \left( \eps_3 \cdot p_1 \right) \left( \eps_1 \cdot p_2 \right)
        +4 \left( \eps_1 \cdot \eps_2 \right) \left( \eps_3 \cdot p_1 \right) \left( \eps_4 \cdot p_2 \right)
        +4 \left( \eps_1 \cdot \eps_4 \right) \left( \eps_3 \cdot p_1 \right) \left( \eps_2 \cdot p_1 \right)
 \right. \nonumber \\
 & & \left.
        +s \left( \eps_3 \cdot \eps_1 \right) \left( \eps_2 \cdot \eps_4 \right)
        +u \left( \eps_1 \cdot \eps_2 \right) \left( \eps_3 \cdot \eps_4 \right)
 \right].
 \nonumber
\eq
Let us now specialise to the helicity configuration $1^-, 2^-, 3^+, 4^+$.
Putting everything together we obtain
\bq
 A_4^{(0)}\left(1^-, 2^-, 3^+, 4^+\right)
 \; = \;
 i J\left(z^{(1)},p\right) \; C\left(\sigma, z^{(1)}\right) \; E\left(z^{(1)},p,\eps\right)
 \; = \;
 2 i \frac{\langle 12 \rangle^4}{\langle 1 2 \rangle \langle 2 3 \rangle \langle 3 4 \rangle \langle 4 1 \rangle}. 
 \;\;
\eq

\subsection{Global residues}

The CHY representation has a mathematical interpretation 
as a global residue \cite{Sogaard:2015dba,Bosma:2016ttj}.
This allows us to compute the primitive tree amplitude $A_n^{(0)}$ from the CHY representation without
the need to know the solutions of the scattering equations.
Note that we already encountered global and local residues in the context 
of the link representation in section~(\ref{sect:link_representation}).

In eq.~(\ref{CHY_int_representation_A_n}) we presented 
the primitive tree amplitude $A_n^{(0)}$ as a multidimensional complex contour integral:
\bq
 A_n^{(0)}\left(\sigma,p,\eps\right)
 & = &
 i
 \frac{\left(-1\right)^{i+j+k}}{\left(2\pi i\right)^{n-3}}
 \oint\limits_{\mathcal C} 
 \frac{d^nz}{d\omega}
 \;
 \frac{z_{ij} z_{jk} z_{ki}}{\prod\limits_{a \neq i,j,k} f_a\left(z,p\right)}
 \; 
 C\left(\sigma,z\right) \; E\left(z,p,\eps\right).
\eq
Switching to the polynomial form of the scattering equations we have
\bq
 A_n^{(0)}\left(\sigma,p,\eps\right)
 & = &
 i
 \frac{\left(-1\right)^{n}}{\left(2\pi i\right)^{n-3}}
 \oint\limits_{\mathcal C} 
 \frac{d^nz}{d\omega}
 \;
 \frac{\prod\limits_{i<j} \left(z_i-z_j\right)}{\prod\limits_{m=2}^{n-2} h_m\left(z,p\right) }
 \; 
 C\left(\sigma,z\right) \; E\left(z,p,\eps\right),
\eq
where the polynomials $h_2(z,p)$, ..., $h_{n-2}(z,p)$ have been defined in eq.~(\ref{def_polynomials}).
Let us now use the $\mathrm{PSL}(2,{\mathbb C})$-invariance and gauge-fix
three variables 
\bq
 z_1=0, \;\;\; z_{n-1}=1, \;\;\; z_n=\infty. 
\eq
We then have
\bq
 A_n^{(0)}\left(\sigma,p,\eps\right)
 & = &
 i
 \frac{1}{\left(2\pi i\right)^{n-3}}
 \oint\limits_{\mathcal C} 
 \frac{R\left(z,\sigma,p,\eps\right) \; dz_2 \wedge ... \wedge dz_{n-2}}{h_2'\left(z,p\right) ... h_{n-2}'\left(z,p\right)},
\eq
with
\bq
\label{def_rational_function_global_residue}
 R\left(z,\sigma,p,\eps\right)
 & = & 
 \left.
 -
 z_n^{4}
 \left( \prod\limits_{i<j<n} z_{ij} \right)
 C\left(\sigma,z\right) \; E\left(z,p,\eps\right)
 \right|_{z_1=0,z_{n-1}=1,z_n=\infty}
\eq
and
\bq
 h_m'\left(z,p\right) 
 & = & 
 \left. \frac{d h_m\left(z,p\right) }{dz_n} \right|_{z_n=0}.
\eq
The polynomials $h_m(z,p)$ are linear in each variable $z_j$. Therefore $h_m'(z,p)$ gives
the coefficient of $z_n$ in the polynomial $h_m(z,p)$.
In the following we will simply write $R(z)=R(z,\sigma,p,\eps)$ and $h_m'(z)=h_m'(z,p)$.
The quantity $R(z)$ is a rational function of the variables $(z_2,..,z_{n-2})$.

Let $f(z)=f(z_2,...,z_{n-2})$ be a meromorphic function, regular at the solutions $z^{(j)}$ of the scattering equations.
We define the {\bf local residue} at $z^{(j)}$ with respect to the divisors $h_2'$, ..., $h_{n-2}'$ as
\bq
\label{local_residue}
 \mathrm{Res}_{\{h_2',...,h_{n-2}'\}}\left( f, z^{(j)} \right)
 & = &
\frac{1}{\left(2\pi i\right)^{n-3}}
 \oint\limits_{\Gamma_\delta}
 \frac{f\left(z\right) \; dz_2 \wedge ... \wedge dz_{n-2}}{h_2'\left(z\right) ... h_{n-2}'\left(z\right)}.
\eq
The integration in eq.~(\ref{local_residue}) is around a small $(n-3)$-torus
\bq
 \Gamma_\delta & = &
 \left\{
   \left( z_2, ..., z_{n-2} \right) \in {\mathbb C}^{n-3} | \left| h_m'\left(z\right)\right| = \delta
 \right\},
\eq
encircling $z^{(j)}$ with orientation
\bq
 d \arg h_2 \wedge d \arg h_3 \wedge ... \wedge d \arg h_{n-2} \ge 0.
\eq
The {\bf global residue} with respect to the divisors $h_2'$, ..., $h_{n-2}'$ is defined as
\bq
 \mathrm{Res}_{\{h_2',...,h_{n-2}'\}}\left( f \right)
 & = &
 \sum\limits_{\mathrm{solutions} \; j}
 \mathrm{Res}_{\{h_2',...,h_{n-2}'\}}\left( f, z^{(j)} \right).
\eq
If it is clear which divisors are meant, we will simply write 
$\mathrm{Res}( f ) = \mathrm{Res}_{\{h_2',...,h_{n-2}'\}}( f )$.
Thus
\bq
\label{amplitude_as_global_residue}
 A_n^{(0)}\left(\sigma,p,\eps\right)
 & = &
 i \; \mathrm{Res}\left( \; R\left(z,\sigma,p,\eps\right) \; \right).
\eq
Let us now consider the ring $\mathrm{R}={\mathbb C}[z_2,...,z_{n-2}]$
and the ideal $I=\langle h_2', ..., h_{n-2}' \rangle$.
The zero locus of $h_2'=...=h_{n-2}'=0$ is a zero-dimensional variety.
For the case at hand the zero locus consists of $(n-3)!$ points.
It follows that he quotient ring $\mathrm{R}/I$ is a finite-dimensional ${\mathbb C}$-vectorspace.
Let $\{e_i\}$ be a basis of this vectorspace and $P_1, P_2 \in \mathrm{R}/I$
two polynomials (i.e. vectors) in this vectorspace.
A theorem of algebraic geometry states that the global residue defines
a {\bf symmetric non-degenerate inner product}:
\bq
 \left\langle P_1, P_2 \right\rangle
 & = &
 \mathrm{Res}\left( \; P_1 \cdot P_2 \right).
\eq
Since the inner product is non-degenerate there exists a {\bf dual basis} $\{\Delta_i\}$ 
with the property
\bq
 \left\langle e_i, \Delta_j \right\rangle & = & \delta_{ij}.
\eq
To compute the global residue of a polynomial $P(z)$ we therefore obtain the following method:
We express $P$ in the basis $\{e_i\}$ and $1$ in the dual basis $\{\Delta_i\}$:
\bq
 P \; = \; \sum\limits_i \alpha_i e_i, 
 \;\;\;
 1 \; = \; \sum\limits_i \beta_i \Delta_i, 
 \;\;\;
 \;\;\;
 \alpha_i, \beta_i \in {\mathbb C}.
\eq
We then have
\bq
\label{global_residue_polynomial}
 \mathrm{Res}\left( \; P \; \right)
 \;\; = \;\;
 \mathrm{Res}\left( \; P \cdot 1 \; \right)
 \;\; = \;\;
 \sum\limits_i \sum\limits_j \alpha_i \beta_j \left\langle e_i, \Delta_j \right\rangle
 \;\; = \;\;
 \sum\limits_i \alpha_i \beta_i.
\eq
Given a basis $\{e_i\}$ and the associated dual basis $\{\Delta_i\}$, eq.~(\ref{global_residue_polynomial}) allows us to compute the global residue of a polynomial $P$ without knowing the solutions
of the scattering equations.
Eq.~(\ref{global_residue_polynomial}) simplifies, if the dual basis contains
a constant polynomial $\Delta_{i_0}=c$.
We then have
\bq
 \mathrm{Res}\left( \; P \; \right)
 & = &
 \frac{\alpha_{i_0}}{c}.
\eq
Eq.~(\ref{global_residue_polynomial}) is not yet directly applicable to our problem, 
since in eq.~(\ref{amplitude_as_global_residue})
we have a rational function $R(z)$, not a polynomial.
We write $R(z)=P(z)/Q(z)$. We may assume that $\{h_2',...,h_{n-2}',Q\}$ have no common zeros,
otherwise we would have in the contour integrals a double pole.
Hilbert's Nullstellensatz guarantees 
that there exist polynomials $p_2, ..., p_{n-2}, \tilde{Q} \in \mathrm{R}$, such that
\bq
 p_2 h_2' + ... +p_{n-2} h_{n-2}' + \tilde{Q} Q & = & 1.
\eq
We call $\tilde{Q}$ the polynomial inverse of $Q$ with 
respect to $\langle h_2', ..., h_{n-2}' \rangle$.
For the global residue we have
\bq
 \mathrm{Res}\left( \; R \; \right)
 \;\; = \;\;
 \mathrm{Res}\left( \; \frac{P}{Q} \; \right)
 \;\; = \;\;
 \mathrm{Res}\left( \; P \tilde{Q} \; \right).
\eq
Note $P \tilde{Q}$ is a polynomial. We have therefore reduced the case of a rational function
$R(z)$ to the polynomial case $P(z) \tilde{Q}(z)$.

The above calculations can be carried out with the help of a 
Gr\"obner basis for the ideal $I$ \cite{Sogaard:2015dba}.
However, the determination of a Gr\"obner basis is by itself computationally intensive
and actually not needed.
It is sufficient to use a coarser analogue, known as a {\bf Macaulay H-basis}.
A set of polynomials $\{b_1,...,b_k\} \in I$ is an H-basis for the ideal $I$, 
if for all $P \in I$ there exists polynomials $q_1,...,q_k \in \mathrm{R}$ 
with $\deg q_j \le \deg P - \deg b_j$ and
\bq
 P & = & \sum\limits_{j=1}^k q_j b_j.
\eq
The essential requirement is that the degrees of the polynomials $q_j$ are bounded.
There is an alternative definition of an H-basis:
For any polynomial $P \in \mathrm{R}$ one denotes by $\mathrm{in}(P)$ the homogeneous part of $P$
of degree $\deg P$.
The polynomial $\mathrm{in}(P)$ is called the {\bf initial form} of $P$.
To give an example
\bq
 \mathrm{in}\left( x_1^2 x_2^3 + 5 x_3^5 + x_2 x_3 + x_1 \right)
 & = &
 x_1^2 x_2^3 + 5 x_3^5.
\eq
A set of polynomials $\{b_1,...,b_k\} \in I$ is an H-basis for the ideal $I$, 
if
\bq
 \left\langle \mathrm{in}\left(I\right) \right\rangle
 & = &
 \left\langle
  \mathrm{in}\left(b_1\right), ..., \mathrm{in}\left(b_k\right)
 \right\rangle.
\eq
Bosma, S{\o}gaard and Zhang \cite{Bosma:2016ttj} have shown that the polynomials
$\{h_2',...,h_{n-2}'\}$ form an H-basis for the ideal $I$.
Furthermore, any polynomial $P \in \mathrm{R}$ may be reduced with respect to $\{h_2',...,h_{n-2}'\}$
as
\bq
 P & = & \sum\limits_{j=2}^{n-2} q_j h_j' + \tilde{P},
\eq
where the remainder $\tilde{P}$ is bounded in degree by
\bq
 \deg \tilde{P} & \le & d^\ast \;\; = \;\; \frac{1}{2} \left(n-4\right) \left(n-3\right).
\eq
It is obvious that the terms proportional to $h_j'$ do not contribute to the global residue.
It is less obvious that neither the terms in $\tilde{P}$ of degree strictly less than $d^\ast$
contribute to the global residue.
Thus
\bq
 \mathrm{Res}_{\{h_2',...,h_{n-2}'\}}\left( \; P \; \right)
 & = &
 \mathrm{Res}_{\{h_2',...,h_{n-2}'\}}\left( \; \tilde{P} \; \right)
 \;\; = \;\;
 \mathrm{Res}_{\{h_2',...,h_{n-2}'\}}\left( \; \mathrm{in}\left(\tilde{P}\right) \; \right)
 \nonumber \\
 & = &
 \mathrm{Res}_{\{\mathrm{in}(h_2'),...,\mathrm{in}(h_{n-2}')\}}\left( \; \mathrm{in}\left(\tilde{P}\right) \; \right).
\eq
The last equality follows from the proper map theorem for H-bases.

Let us now look at the vectorspace $V_{d^\ast}$
of homogeneous polynomials $P \in \mathrm{R}$ with degree $d^\ast$.
It can be shown that
\bq
 \dim_{\mathbb C} V_{d^\ast}
 -
 \dim_{\mathbb C}\left( V_{d^\ast} \cap \left\langle \mathrm{in}\left(I\right) \right\rangle \right)
 & = & 1.
\eq
Therefore we may choose a basis of $V_{d^\ast}$, where only a single element is not in $\langle \mathrm{in}(I)\rangle$.
Such an element is easily found:
\bq
 M & = & \prod\limits_{m=2}^{n-2} z_m^{m-2}.
\eq
Thus we may write
\bq
 \mathrm{in}\left(\tilde{P}\right)
 & = & 
 \alpha M + \sum\limits_{j=2}^{n-2} q_j' \; \mathrm{in}\left(h_j'\right),
\eq
with $\alpha \in {\mathbb C}$ and $q_j' \in \mathrm{R}$.
We then have
\bq
 \mathrm{Res}_{\{h_2',...,h_{n-2}'\}}\left( \; P \; \right)
 & = &
 \alpha \; \mathrm{Res}_{\{h_2',...,h_{n-2}'\}}\left( \; M \; \right).
\eq
The global residue of $M$ is easily computed:
\bq
 \mathrm{Res}\left( \; M \; \right)
 & = &
 \frac{\left(-1\right)^{\frac{(n-4)(n-3)}{2}}}{\prod\limits_{j=2}^{n-2} \left( p_{j,n-2} + p_n \right)^2}.
\eq
This gives us the following algorithm:
\begin{algorithm}
Calculation of a primitive helicity amplitude $A_{n}^{(0)}( \sigma, p, \eps )$
as a global residue.
\begin{enumerate}
\item Start with the rational function $R(z)$ given in eq.~(\ref{def_rational_function_global_residue}):
\bq
 R\left(z,\sigma,p,\eps\right)
 & = & 
 \left.
 -
 z_n^{4}
 \left( \prod\limits_{i<j<n} z_{ij} \right)
 C\left(\sigma,z\right) \; E\left(z,p,\eps\right)
 \right|_{z_1=0,z_{n-1}=1,z_n=\infty}
\eq
and write $R(z)=P(z)/Q(z)$, where $P(z)$ and $Q(z)$ are polynomials with $\mathrm{gcd}(P,Q)=1$.
\item Determine the polynomial inverse $\tilde{Q}(z)$ of $Q(z)$ with respect to 
$\{h_2',...,h_{n-2}'\}$ and set $P_1(z) = P(z) \tilde{Q}(z)$.
\item Reduce $P_1(z)$ with respect to $h_2'$, ..., $h_{n-2}'$:
\bq
 P_1(z) & = & p_2 h_2' + ... + p_{n-2} h_{n-2}' + \tilde{P}(z).
\eq
The remainder $\tilde{P}(z)$ is a polynomial of degree at most $(n-3)(n-4)/2$.
\item Express the initial form $\mathrm{in}(\tilde{P})$ as
\bq
 \mathrm{in}\left(\tilde{P}\right)
 & = & 
 \alpha M + \sum\limits_{j=2}^{n-2} q_j' \; \mathrm{in}\left(h_j'\right).
\eq
\item The amplitude is given by
\bq
 A_n^{(0)}
 & = &
 \frac{i \left(-1\right)^{\frac{(n-4)(n-3)}{2}} \alpha}{\prod\limits_{j=2}^{n-2} \left( p_{j,n-2} + p_n \right)^2}.
\eq
\end{enumerate}
\end{algorithm}
Let us look at an example.
In order to keep things simple we will replace in the CHY representation the polarisation factor
with another copy of the Parke-Taylor factor.
We consider
\bq
 m_n^{(0)}\left(\sigma,\tilde{\sigma},p\right)
 & = &
 i \oint\limits_{\mathcal C} d\Omega_{\mathrm{CHY}} \; C\left(\sigma,z\right) \; C\left(\tilde{\sigma},z\right).
\eq
In the next section we will discuss amplitudes of this type in more detail.
Let us now take $n=4$ and $\sigma=\tilde{\sigma}=(1,2,3,4)$.
We gauge-fix $z_1=0$, $z_3=1$ and $z_4=\infty$.
The rational function $R(z)$ is then given by
\bq
 R\left(z\right)
 & = &
 \frac{1}{z_2\left(1-z_2\right)}.
\eq
The polynomial $h_2'$ is given by
\bq
 h_2' & = & u z_2 + s.
\eq
The degree zero polynomial $\tilde{Q}=u^2/(s t)$ (the degree refers to the variable $z_2$) 
is a polynomial inverse to $z_2(1-z_2)$ with respect to $h_2'$, since
\bq
 \frac{u^2}{s t} z_2 \left(1-z_2\right) + \left( \frac{u}{s t} z_2 + \frac{1}{s} \right) h_2'
 & = & 1.
\eq
We further have $M=1$ and $\alpha=u^2/(s t)$.
Hence
\bq
 m_n^{(0)}\left(\sigma,\tilde{\sigma},p\right)
 & = &
 \frac{i \alpha}{\left(p_2+p_4\right)^2}
 \;\; = \;\;
 i \frac{u}{s t}
 \;\; = \;\;
 - i \left( \frac{1}{s} + \frac{1}{t} \right).
\eq

% ----------------------------------------------------------------------------------
\newpage
\section{Perturbative quantum gravity}
\label{sect:gravity}

Up to now we only considered Yang-Mills theory. Let us now turn to a second theory:
In this section we will discuss gravity.
At the classical level gravity is described by Einstein's theory of general relativity.
Einstein's equations (for the case without any matter)
are derived from the {\bf Einstein-Hilbert Lagrange density}
\bq
 {\mathcal L}_{\mathrm{EH}}
 & = &
 - \frac{2}{\kappa^2} 
 \sqrt{-g} \left( R + 2 \Lambda \right).
\eq
We continue to use the convention that
\bq
 c = 1,
 & &
 \hbar = 1.
\eq
The parameter $\kappa$ is a dimensionful constant of
dimension $\mathrm{mass}^{-1}$ given, 
by $\kappa =\sqrt{8 G_N}$ in natural units, 
where $G_N$ is Newton's constant.
In the Gau{\ss} unit system we have $\kappa=\sqrt{32\pi G_N}$.
The {\bf Planck mass} is defined by $m_{\mathrm{Pl}}=1/\sqrt{G_N}$.
We denote the full metric in general relativity by $g_{\mu\nu}$ and write
\bq
 g & = & \det\left( g_{\mu\nu} \right).
\eq
The quantity $R$ denotes the {\bf scalar curvature} and is given by
\bq
 R & = &
 g^{\mu\nu} \mathrm{Ric}_{\mu\nu}
 \;\; = \;\;
 g^{\mu\nu} R^{\lambda}_{\;\mu \lambda \nu},
\eq
where $\mathrm{Ric}_{\mu\nu}$ denotes the Ricci tensor and $R^{\kappa}_{\;\lambda \mu \nu}$
the Riemann curvature tensor.
The parameter $\Lambda$ denotes the {\bf cosmological constant}.

Let us denote the metric of flat Minkowski space by
\bq
 \eta_{\mu\nu} & = & 
 \left( \begin{array}{rrrr} 
 1 & 0 & 0 & 0 \nonumber \\
 0 & -1 & 0 & 0 \nonumber \\
 0 & 0 & -1 & 0 \nonumber \\
 0 & 0 & 0 & -1 \nonumber \\
 \end{array} \right).
\eq
The metric $\eta_{\mu\nu}$ of flat Minkowski space is a solution of Einstein's fields equations
without a cosmological constant,
\bq
 \mathrm{Ric}_{\mu\nu} - \frac{1}{2} g_{\mu\nu} R & = & 0.
\eq
Note that $\eta_{\mu\nu}$ is not a solution of Einstein's field equations for a non-zero value of the
cosmological constant.

We will now discuss {\bf perturbative quantum gravity} \cite{DeWitt:1967yk,DeWitt:1967ub,DeWitt:1967uc,Veltman:1975vx}.
We consider small fluctuations of the gravitational field around the flat Minkowski metric
and weight every fluctuation with $\exp( i S)$, where $S$ is the action.
We further assume that the coupling $\kappa/4$ is small. 
Within the energy range accessible at current collider experiments (i.e. up to $\mathrm{TeV}$ scales),
the gravitational coupling is incredible small.
This set-up leads us directly to a perturbative description of quantum gravity
and allows us to describe gravitational waves and their scattering.
Perturbative quantum gravity is an effective quantum field theory, 
the ``true'' quantum theory of gravity has to agree with perturbative quantum gravity at low energies.
They may differ at higher energies, where
perturbative quantum gravity is expected to break down.

We write
\bq
 g_{\mu\nu} & = &
 \eta_{\mu\nu} + \kappa h_{\mu\nu}
\eq
and treat the term $\kappa h_{\mu\nu}$ as a small perturbation.
The tensor $h_{\mu\nu}$ describes the field of a {\bf graviton}.
Expanding around the flat Minkowski metric $\eta_{\mu\nu}$ and requiring that the background metric
is a solution of the field equations, implies that we are considering the case of a vanishing cosmological constant.
In the sequel we assume therefore $\Lambda=0$.
The Einstein-Hilbert action without a cosmological constant is given by
\bq
\label{Einstein_Hilbert_action}
 S_{\mathrm{EH}}
 = 
 \int d^4x \; {\mathcal L}_{\mathrm{EH}},
 & & 
 {\mathcal L}_{\mathrm{EH}}
 =
 - \frac{2}{\kappa^2} 
 \sqrt{-g} R.
\eq

\subsection{Gauge invariance of gravity}

The Einstein-Hilbert action in eq.~(\ref{Einstein_Hilbert_action}) is invariant
under {\bf general coordinate transformations}
\bq
\label{general_coordinate_transformations}
 {x'}^\mu & = & f^\mu\left( x \right).
\eq
In fact, one of Einstein's original motivations was to find a theory invariant under these transformations.
We may view the general coordinate transformations in eq.~(\ref{general_coordinate_transformations})
as (generalised) gauge transformations.
We write an infinitesimal general coordinate transformation as
\bq
 {x'}^\mu & = &
 x^\mu - \eps \xi^\mu\left( x \right).
\eq
The minus sign has no particular importance and is just a convention.
The infinitesimal inverse transformation is given by
\bq
 x^\mu & = &
 {x'}^\mu + \eps \xi^\mu\left(x'\right)
 + {\mathcal O}\left(\eps^2\right).
\eq
Let us now work out the metric in the transformed system:
\bq
 g'_{\mu'\nu'}\left(x'\right)
 & = &
 \frac{\partial x^\mu}{\partial {x'}^{\mu'}}
 \frac{\partial x^\nu}{\partial {x'}^{\nu'}}
 g_{\mu\nu}\left( x\left( x' \right) \right)
 \nonumber \\
 & = &
 \left( \delta^\mu_{\mu'} + \eps \partial_{\mu'} \xi^\mu\left(x'\right) \right)
 \left( \delta^\nu_{\nu'} + \eps \partial_{\nu'} \xi^\nu\left(x'\right) \right)
 \left( g_{\mu\nu}\left(x'\right) + \eps \xi^\rho\left(x'\right) \partial_\rho g_{\mu\nu}\left(x'\right) \right)
 + {\mathcal O}\left(\eps^2\right)
 \nonumber \\
 & = &
 g_{\mu'\nu'}\left(x'\right)
 + \eps \left[
                \left( \partial_{\mu'} \xi^\mu\left(x'\right) \right) g_{\mu \nu'}\left(x'\right)
              + \left( \partial_{\nu'} \xi^\nu\left(x'\right) \right) g_{\mu' \nu}\left(x'\right)
              + \xi^\rho\left(x'\right) \partial_\rho g_{\mu'\nu'}\left(x'\right)
        \right]
 \nonumber \\
 & &
 + {\mathcal O}\left(\eps^2\right).
\eq
We may write this in a shortened form as
\bq
 g'_{\mu\nu}
 & = &
 g_{\mu\nu}
 + \eps \left[
                \left( \partial_{\mu} \xi^\rho \right) g_{\rho \nu}
              + \left( \partial_{\nu} \xi^\rho \right) g_{\mu \rho}
              + \xi^\rho \partial_\rho g_{\mu\nu}
        \right]
 + {\mathcal O}\left(\eps^2\right).
\eq
Let us now specialise to an expansion around the flat Minkowski metric. With
\bq
 g_{\mu\nu}\left(x\right)
 & = & 
 \eta_{\mu\nu}
 + \kappa h_{\mu\nu}\left(x\right),
\eq
we find for $h'_{\mu\nu}$:
\bq
 h'_{\mu\nu}
 & = & 
 h_{\mu\nu}
 + \frac{\eps}{\kappa} 
   \left[ \left( \partial_{\mu} \xi^\rho \right) \eta_{\rho \nu}
        + \left( \partial_{\nu} \xi^\rho \right) \eta_{\mu \rho} \right]
 \nonumber \\
 & &
 + \eps
   \left[ \left( \partial_{\mu} \xi^\rho \right) h_{\rho \nu}
        + \left( \partial_{\nu} \xi^\rho \right) h_{\mu \rho}
        + \xi^\rho \partial_\rho h_{\mu \nu} \right]
 + {\mathcal O}\left(\eps^2\right).
\eq
This expression can be simplified and we find 
\bq
 h'_{\mu\nu}
 & = & 
 h_{\mu\nu}
 + \frac{\eps}{\kappa} 
   \left[
         \nabla_{\mu} \xi_\nu + \nabla_{\nu} \xi_\mu
   \right]
 + {\mathcal O}\left(\eps^2\right),
\eq
where
$\xi_\mu = g_{\mu\nu} \xi^\nu = \eta_{\mu\nu} \xi^\nu + \kappa h_{\mu\nu} \xi^\nu$.
We may view the transformation from $h_{\mu\nu}$ to $h'_{\mu\nu}$ as
an {\bf infinitesimal gauge transformation}.

\subsection{Gravity amplitudes from Feynman diagrams}

We may use the textbook formalism of quantum field theory to compute
perturbatively scattering amplitudes for gravitons.
In the previous section we have seen that the action is invariant under general coordinate
transformations.
In the path integral formalism we would like to restrict the path integral to configurations
not equivalent by general coordinate transformations.
As in section~(\ref{sect_gauge_fixing}) this can be achieved with the technique of gauge fixing and
the Faddeev-Popov method.

Let us now consider for $h_{\mu\nu}$ the effective (not necessarily renormalisable) quantum field theory
described by the generating functional
\bq
\label{generating_functional}
 Z\left[J^{\mu\nu}\right]
 & = &
 \int {\mathcal D}h_{\mu\nu}
 \;
 \exp\left[ i \int d^4x \; {\mathcal L}_{\mathrm{EH}} + {\mathcal L}_{\mathrm{GF}} + {\mathcal L}_{\mathrm{FP}} + J^{\mu\nu} h_{\mu\nu} \right],
\eq
where ${\mathcal L}_{\mathrm{GF}}$ denotes the gauge-fixing term and ${\mathcal L}_{\mathrm{FP}}$
the corresponding Faddeev-Popov term.
We will give an expression for the gauge-fixing term later on. 
The Faddeev-Popov term will again only contribute to loop amplitudes.
We will treat the quantum field theory defined by eq.~(\ref{generating_functional})
perturbatively.
Our first goal is the expansion of the Lagrange density in powers of 
$h_{\mu\nu}$ (or equivalently in powers of $\kappa$).
Let us introduce the following abbreviations:
\bq
 \left( \eta h \eta \right)^{\mu\nu}
 & = &
 \eta^{\mu \mu_1} h_{\mu_1 \mu_2} \eta^{\mu_2 \nu},
 \nonumber \\
 \left( \eta h \eta h \eta \right)^{\mu\nu}
 & = &
 \eta^{\mu \mu_1} h_{\mu_1 \mu_2} \eta^{\mu_2 \mu_3} h_{\mu_3 \mu_4} \eta^{\mu_4 \nu},
 \nonumber \\
 \left( \eta h \eta h \eta h \eta \right)^{\mu\nu}
 & = &
 \eta^{\mu \mu_1} h_{\mu_1 \mu_2} \eta^{\mu_2 \mu_3} h_{\mu_3 \mu_4} \eta^{\mu_4 \mu_5} h_{\mu_5 \mu_6} \eta^{\mu_6 \nu}.
\eq
With the help of these abbreviations we may express the inverse metric tensor $g^{\mu\nu}$
through $h_{\mu\nu}$:
\bq
 g^{\mu\nu}
 & = &
 \eta^{\mu\nu}
 - \kappa \left( \eta h \eta \right)^{\mu\nu}
 + \kappa^2 \left( \eta h \eta h \eta \right)^{\mu\nu}
 - \kappa^3 \left( \eta h \eta h \eta h \eta \right)^{\mu\nu}
 + {\mathcal O}\left(\kappa^4\right).
\eq
The inverse metric tensor is an infinite power series in $\kappa$.
Let us now turn to the determinant $g=\det(g_{\mu\nu})$.
Also here we introduce a few abbreviations:
\bq
 \left( \eta h \right) & = & 
 \eta^{\mu_1 \mu_2} h_{\mu_2 \mu_1},
 \nonumber \\
 \left( \eta h \eta h \right) & = &
 \eta^{\mu_1 \mu_2} h_{\mu_2 \mu_3} \eta^{\mu_3 \mu_4} h_{\mu_4 \mu_1},
 \nonumber \\
 \left( \eta h \eta h \eta h \right) & = &
 \eta^{\mu_1 \mu_2} h_{\mu_2 \mu_3} \eta^{\mu_3 \mu_4} h_{\mu_4 \mu_5} \eta^{\mu_5 \mu_6} h_{\mu_6 \mu_1},
 \nonumber \\
 \left( \eta h \eta h \eta h \eta h \right) & = &
 \eta^{\mu_1 \mu_2} h_{\mu_2 \mu_3} \eta^{\mu_3 \mu_4} h_{\mu_4 \mu_5} \eta^{\mu_5 \mu_6} h_{\mu_6 \mu_7} \eta^{\mu_7 \mu_8} h_{\mu_8 \mu_1}.
\eq
We then find for the determinant:
\bq
\lefteqn{
 - \det\left(g_{\mu\nu}\right)
 = 
} & &  
 \\
 & &
 1
 + \kappa \left( \eta h \right)
 + \kappa^2 \left[ 
                   \frac{1}{2} \left( \eta h \right)^2
                 - \frac{1}{2} \left( \eta h \eta h \right)
            \right]
 + \kappa^3 \left[
                   \frac{1}{6} \left( \eta h \right)^3
                 - \frac{1}{2} \left( \eta h \eta h \right) \left( \eta h \right)
                 + \frac{1}{3} \left( \eta h \eta h \eta h \right)
            \right]
 \nonumber \\
 & &
 + \kappa^4 \left[
                   \frac{1}{24} \left( \eta h \right)^4
                 - \frac{1}{4} \left( \eta h \eta h \right) \left( \eta h \right)^2
                 + \frac{1}{8} \left( \eta h \eta h \right)^2
                 + \frac{1}{3} \left( \eta h \eta h \eta h \right) \left( \eta h \right)
                 - \frac{1}{4} \left( \eta h \eta h \eta h \eta h \right)
            \right].
 \nonumber
\eq
Note that this expression is a polynomial in $\kappa$ and terminates with the $\kappa^4$-term.
However, by taking the square root of this expression we again obtain an infinite power series in $\kappa$:
\bq
 \sqrt{-g}
 & = &
 1
 + \frac{\kappa}{2} \left( \eta h \right)
 + \frac{\kappa^2}{8} \left[ \left( \eta h \right)^2 - 2 \left( \eta h \eta h \right)\right]
 + \frac{\kappa^3}{48} \left[ \left( \eta h \right)^3 - 6 \left( \eta h \eta h \right) \left( \eta h \right) + 8 \left( \eta h \eta h \eta h \right) \right]
 \nonumber \\
 & &
 + {\mathcal O}\left(\kappa^4\right).
\eq
In order to find the expression for the scalar curvature $R$ let us first consider the Christoffel symbols
\bq
 \Gamma_{\kappa\mu\nu}
 & = &
 \frac{1}{2} \left( \partial_{\mu} g_{\nu \kappa} + \partial_{\nu} g_{\mu \kappa} - \partial_{\kappa} g_{\mu \nu} \right)
 \;\; = \;\;
 \frac{\kappa}{2} \left( \partial_{\mu} h_{\nu \kappa} + \partial_{\nu} h_{\mu \kappa} - \partial_{\kappa} h_{\mu \nu} \right).
\eq
Here we used $\partial_\alpha \eta_{\beta\gamma}=0$.
The Riemann curvature tensor is then given by
\bq
R_{\kappa\lambda \mu \nu} =  
 \frac{\kappa}{2} \left(
 \partial_\lambda \partial_\mu h_{\kappa \nu}
- \partial_\kappa \partial_\mu h_{\lambda \nu}
+ \partial_\kappa \partial_\nu h_{\lambda \mu}
- \partial_\lambda \partial_\nu h_{\kappa \mu}
 \right)
 + g^{\xi\eta} \left( \Gamma_{\xi\kappa\nu} \Gamma_{\eta\lambda\mu}
                    - \Gamma_{\xi\kappa\mu} \Gamma_{\eta\lambda\nu}
               \right).
\eq
The first term is linear in $h_{\mu\nu}$, while the second term is at least quadratic in $h_{\mu\nu}$.
For the scalar curvature we have then
\bq
 R & = &
 g^{\kappa \mu} g^{\lambda \nu}
 R_{\kappa\lambda \mu \nu}.
\eq
Since both $g^{\mu\nu}$ and $\sqrt{-g}$ are infinite power series in $\kappa$
we obtain for the Lagrange density an infinite power series in $\kappa$ as well.
We write
\bq
\label{Einstein_Hilbert_expansion}
 {\mathcal L}_{\mathrm{EH}} + {\mathcal L}_{\mathrm{GF}}
 & = &
 \sum\limits_{j=1}^\infty
 {\mathcal L}^{(j)},
\eq
where the term ${\mathcal L}^{(j)}$ contains the field $h_{\mu\nu}$ exactly $j$ times.
In this way we obtain a theory with an infinite tower of vertices, ordered by the number of the fields.
The term ${\mathcal L}^{(1)}$ is given by
\bq
 {\mathcal L}^{(1)}
 & = &
 -\frac{2}{\kappa} 
 \eta^{\kappa \mu} \eta^{\lambda \nu}
 \partial_\lambda 
 \left( \partial_\mu h_{\kappa \nu} - \partial_\nu h_{\kappa \mu} \right).
\eq
This term is a total derivative and vanishes in the action after partial integration:
\bq
 -\frac{2}{\kappa} 
 \eta^{\kappa \mu} \eta^{\lambda \nu}
 \int d^4x \;
 \partial_\lambda 
 \left( \partial_\mu h_{\kappa \nu} - \partial_\nu h_{\kappa \mu} \right)
 & = & 0.
\eq
We may therefore ignore this term and start the expansion of the Lagrange density in powers of $\kappa$
with the term quadratic in $h_{\mu\nu}$.
\\
\\
Let us add the following remark: If we would have expanded naively the Einstein-Hilbert action with
a cosmological constant $\Lambda \neq 0$ around the flat Minkowski metric $\eta_{\mu\nu}$, we
would have picked up an additional term 
\bq
 - \frac{2\Lambda}{\kappa} \eta^{\mu\nu} h_{\mu\nu}
\eq
contributing to ${\mathcal L}^{(1)}$, coming from the expansion of $\sqrt{-g}$.
This additional term is not a total derivative and does not vanish.
Terms of this type are called tadpoles and indicate that we expanded around the wrong background field.
\\
\\
Let us now return to the case $\Lambda=0$. We consider the term ${\mathcal L}^{(2)}$, bilinear in $h_{\mu\nu}$.
The gauge-fixing term ${\mathcal L}_{\mathrm{GF}}$ gives a contribution to ${\mathcal L}^{(2)}$.
A popular gauge choice for gravity is {\bf de Donder gauge}.
This gauge is defined by
\bq
 {\mathcal L}_{\mathrm{GF}}
 & = &
 \frac{1}{\kappa^2} C_\mu \eta^{\mu\nu} C_\nu,
\eq
where $C_\mu$ is given by
\bq
 C_\mu & = & \eta^{\alpha\beta} \Gamma_{\mu \alpha \beta}
 =
 \frac{\kappa}{2} \eta^{\alpha\beta} \left( \partial_{\alpha} h_{\beta \mu} + \partial_{\beta} h_{\alpha \mu} - \partial_{\mu} h_{\alpha \beta} \right)
 =
 \kappa \eta^{\alpha\beta} \left( \partial_{\alpha} h_{\beta \mu} - \frac{1}{2} \partial_{\mu} h_{\alpha \beta} \right).
\eq
In this gauge one finds
\bq
 {\mathcal L}^{(2)}
 & = &
 \frac{1}{2}
 h_{\mu_1\mu_2} 
 \left( 
 \frac{1}{2} \eta^{\mu_1\mu_2} \eta^{\nu_1\nu_2} 
 - \frac{1}{2} \eta^{\mu_1\nu_1} \eta^{\mu_2\nu_2} 
 - \frac{1}{2} \eta^{\mu_1\nu_2} \eta^{\mu_2\nu_1} 
 \right) \Box 
 h_{\nu_1\nu_2}.
\eq
Here, we symmetrised the expression in the bracket in $(\mu_1,\mu_2)$ and $(\nu_1,\nu_2)$.
We are free to do this, since $h_{\mu\nu}$ is symmetric under an exchange of $\mu$ and $\nu$.
Let us first consider the tensor structure (in $D$ space-time dimensions).
For
\bq
 M^{\mu_1\mu_2\nu_1\nu_2}
 & = &
 \frac{1}{2} \eta^{\mu_1\nu_1} \eta^{\mu_2\nu_2} 
 + \frac{1}{2} \eta^{\mu_1\nu_2} \eta^{\mu_2\nu_1}
 -  \frac{1}{2} \eta^{\mu_1\mu_2} \eta^{\nu_1\nu_2},
 \nonumber \\
 N_{\mu_1\mu_2\nu_1\nu_2}
 & = &
 \frac{1}{2} \left(
 \eta_{\mu_1\nu_1} \eta_{\mu_2\nu_2} + \eta_{\mu_1\nu_2} \eta_{\mu_2\nu_1} 
 -  \frac{2}{D-2} \eta_{\mu_1\mu_2} \eta_{\nu_1\nu_2}
 \right)
\eq
we have
\bq
 M^{\mu_1\mu_2\rho_1\rho_2}
 N_{\rho_1\rho_2\nu_1\nu_2}
 & = &
 \frac{1}{2} \left( \delta^{\mu_1}_{\nu_1} \delta^{\mu_2}_{\nu_2} + \delta^{\mu_1}_{\nu_2} \delta^{\mu_2}_{\nu_1} \right).
\eq 
The propagator of the graviton is therefore given by
\bq
 \frac{1}{2}
 \left( \eta_{\mu_1\nu_1} \eta_{\mu_2\nu_2} + \eta_{\mu_1\nu_2} \eta_{\mu_2\nu_1} - \frac{2}{D-2} \eta_{\mu_1\mu_2} \eta_{\nu_1\nu_2} \right)
 \frac{i}{p^2}.
\eq
Let us now turn to the three-graviton vertex.
The three-graviton vertex is determined by ${\mathcal L}^{(3)}$.
After a longer calculation and by using integration-by-parts one finds
\bq
\label{Lagrangian_three_graviton_vertex}
 {\mathcal L}^{(3)}
 & = &
 \kappa
 \left[
%  - 1/4*n(nu2,nu3)*n(i1,j1)*n(i2,j2)*n(i3,j3)
%  + 1/4*n(nu2,nu3)*n(i1,j1)*n(i2,j3)*n(i3,j2)
   - \frac{1}{4} \eta^{\mu_1\nu_1} \eta^{\mu_2\nu_2} \eta^{\mu_3\nu_3} \eta^{\rho_2\rho_3}
   + \frac{1}{4} \eta^{\mu_1\nu_1} \eta^{\mu_2\nu_3} \eta^{\mu_3\nu_2} \eta^{\rho_2\rho_3}
%  + n(nu2,nu3)*n(i1,j2)*n(i2,j1)*n(i3,j3)
%  - n(nu2,nu3)*n(i1,j2)*n(i2,j3)*n(i3,j1)
   + \eta^{\mu_1\nu_2} \eta^{\mu_2\nu_1} \eta^{\mu_3\nu_3} \eta^{\rho_2\rho_3}
 \right. \nonumber \\
 & & \left.
   - \eta^{\mu_1\nu_2} \eta^{\mu_2\nu_3} \eta^{\mu_3\nu_1} \eta^{\rho_2\rho_3}
%  + 1/2*n(nu3,j1)*n(i1,nu2)*n(i2,j2)*n(i3,j3)
%  - 1/2*n(nu3,j1)*n(i1,nu2)*n(i2,j3)*n(i3,j2)
   + \frac{1}{2} \eta^{\mu_1\rho_2} \eta^{\rho_3\nu_1} \eta^{\mu_2\nu_2} \eta^{\mu_3\nu_3}
   - \frac{1}{2} \eta^{\mu_1\rho_2} \eta^{\rho_3\nu_1} \eta^{\mu_2\nu_3} \eta^{\mu_3\nu_2}
 \right. \nonumber \\
 & & \left.
%  + 2*n(nu3,j2)*n(i1,nu2)*n(i2,j3)*n(i3,j1)
%  - n(nu3,j2)*n(i1,nu2)*n(i2,j1)*n(i3,j3)
   + 2 \eta^{\mu_1\rho_2} \eta^{\rho_3\nu_2} \eta^{\mu_2\nu_3} \eta^{\mu_3\nu_1}
   -   \eta^{\mu_1\rho_2} \eta^{\rho_3\nu_2} \eta^{\mu_2\nu_1} \eta^{\mu_3\nu_3}
%  - 1/2*n(nu3,j2)*n(i1,j1)*n(i2,j3)*n(i3,nu2)
%  + n(nu3,j2)*n(i1,j3)*n(i2,j1)*n(i3,nu2)
   - \frac{1}{2} \eta^{\mu_3\rho_2} \eta^{\rho_3\nu_2} \eta^{\mu_1\nu_1} \eta^{\mu_2\nu_3}
 \right. \nonumber \\
 & & \left.
   +             \eta^{\mu_3\rho_2} \eta^{\rho_3\nu_2} \eta^{\mu_1\nu_3} \eta^{\mu_2\nu_1}
%  - n(nu3,j3)*n(i1,nu2)*n(i2,j2)*n(i3,j1)
%  - n(nu3,j3)*n(i1,j2)*n(i2,j1)*n(i3,nu2)
%  + 1/2*n(nu3,j3)*n(i1,j1)*n(i2,j2)*n(i3,nu2)
   -             \eta^{\mu_1\rho_2} \eta^{\rho_3\nu_3} \eta^{\mu_2\nu_2} \eta^{\mu_3\nu_1}
   -             \eta^{\mu_3\rho_2} \eta^{\rho_3\nu_3} \eta^{\mu_1\nu_2} \eta^{\mu_2\nu_1}
 \right. \nonumber \\
 & & \left.
   + \frac{1}{2} \eta^{\mu_3\rho_2} \eta^{\rho_3\nu_3} \eta^{\mu_1\nu_1} \eta^{\mu_2\nu_2}
 \right]
 h_{\mu_1 \nu_1} \left( \partial _{\rho_2} h_{\mu_2 \nu_2} \right) \left( \partial_{\rho_3} h_{\mu_3 \nu_3} \right).
\eq
Let us write ${\mathcal L}^{(3)}$ as
\bq
 {\mathcal L}^{(3)}
 & = &
 O^{\mu_1\mu_2\mu_3 \nu_1\nu_2\nu_3}\left(\partial_1, \partial_2, \partial_3\right)
 h_{\mu_1 \nu_1} h_{\mu_2 \nu_2} h_{\mu_3 \nu_3},
\eq
where $O^{\mu_1\mu_2\mu_3 \nu_1\nu_2\nu_3}\left(\partial_1, \partial_2, \partial_3\right)$
is defined by comparison with the previous equation~(\ref{Lagrangian_three_graviton_vertex}).
The notation $\partial_j$ denotes a derivative acting on the field $h_{\mu_j \nu_j}$.
The Feynman rule for the three-graviton vertex is then
\bq
 V^{\mu_1\mu_2\mu_3 \nu_1\nu_2\nu_3}\left(p_1,p_2,p_3\right) & = &
 i \sum\limits_{\sigma \in S_3} 
 O^{\mu_{\sigma(1)}\mu_{\sigma(2)}\mu_{\sigma(3)} \nu_{\sigma(1)}\nu_{\sigma(2)}\nu_{\sigma(3)}}\left(i p_{\sigma(1)}, i p_{\sigma(2)}, i p_{\sigma(3)}\right).
\eq
The explicit expression for $V^{\mu_1\mu_2\mu_3 \nu_1\nu_2\nu_3}$ is rather long and not given here.
However, one interesting property should be mentioned:
The three-graviton vertex can be written as
\bq
 V^{\mu_1\mu_2\mu_3 \nu_1\nu_2\nu_3}\left(p_1,p_2,p_3\right) & = &
 i \; \frac{\kappa}{4} \;
 V^{\mu_1\mu_2\mu_3}\left(p_1,p_2,p_3\right)
 V^{\nu_1\nu_2\nu_3}\left(p_1,p_2,p_3\right)
 + ...,
\eq
where the dots denote terms, which vanish in the on-shell limit.
The expression $V^{\mu_1\mu_2\mu_3}\left(p_1,p_2,p_3\right)$ is the Feynman rule for the colour-stripped
cyclic-order three-gluon vertex (see eq.~(\ref{cyclic_ordered_Feynman_rules})), given by
\bq
 V^{\mu_1\mu_2\mu_3}\left(p_1,p_2,p_3\right)
 & = &
 i \left[ g^{\mu_1\mu_2} \left( p_1^{\mu_3} - p_2^{\mu_3} \right)
         +g^{\mu_2\mu_3} \left( p_2^{\mu_1} - p_3^{\mu_1} \right)
         +g^{\mu_3\mu_1} \left( p_3^{\mu_2} - p_1^{\mu_2} \right)
   \right].
\eq
We see that the three-graviton vertex in the on-shell limit is given (up to a prefactor involving the coupling)
as the square of the cyclic-ordered three gluon vertex.
This relates gravity with non-abelian gauge theories.
We will explore this connection in more detail in the next sub-section.

In principle it is possible to derive from the Lagrange density in eq.~(\ref{Einstein_Hilbert_expansion})
systematically the additional Feynman rules for vertices with four, five, ..., $n$ gravitons.
In addition we need a rule for the external graviton states. This rule is rather simple.
A graviton is a spin $2$ particle with two polarisation states, corresponding to the helicities $h=+2$ and $h=-2$.
We label these states by $++$ and $--$.
We may describe the polarisation tensor of an external graviton by a product of two polarisation vectors for gauge bosons:
\bq
\label{polarisation_graviton}
 \eps_{\mu\nu}^{++}\left(p\right)
 \;\; = \;\;
 \eps_\mu^+\left(p\right) \eps_\nu^+\left(p\right),
 & &
 \eps_{\mu\nu}^{--}\left(p\right)
 \;\; = \;\;
 \eps_\mu^-\left(p\right) \eps_\nu^-\left(p\right).
\eq
Let us denote the tree-level scattering amplitude for $n$ gravitons by
\bq
 {\mathcal M}_n^{(0)}\left(p_1^{\lambda_1 \tilde{\lambda}_1}, ..., p_n^{\lambda_n \tilde{\lambda}_n} \right),
\eq
with $(\lambda_j,\tilde{\lambda}_j)$ either $(+,+)$ or $(-,-)$.
It will be convenient to factor of the gravitational coupling and we define $M_n^{(0)}$ by
\bq
\label{coupling_convention}
 {\mathcal M}_n^{(0)}\left(p_1^{\lambda_1 \tilde{\lambda}_1}, ..., p_n^{\lambda_n \tilde{\lambda}_n} \right)
 & = &
 \left( \frac{\kappa}{4} \right)^{n-2}
 M_n^{(0)}\left(p_1^{\lambda_1 \tilde{\lambda}_1}, ..., p_n^{\lambda_n \tilde{\lambda}_n} \right).
\eq
For the calculation of the scattering amplitude $M_n^{(0)}$ with $n$ gravitons we will need all vertices with up to $n$
gravitons.
The scattering amplitude may then be computed through Feynman diagrams.
However, this approach is rather tedious.
In the next sub-section we will discuss more efficient methods for the computation of the $n$-graviton amplitude.

\subsection{Gravity amplitudes from the CHY representation}

In this paragraph we discuss the CHY representation of tree-level graviton amplitudes.
This will give us a first method to compute the $n$-graviton amplitude without recourse
to Feynman diagrams.
We recall that in the context of the CHY representation for primitive gauge amplitudes we used an alternative notation
for the arguments of the amplitude:
\bq
 A_{n}^{(0)}\left( \sigma, p, \eps \right)
 & = &
 A_{n}^{(0)}\left( p_{\sigma_1}^{\lambda_{\sigma_1}}, ..., p_{\sigma_n}^{\lambda_{\sigma_n}}\right),
\eq
where $\sigma=(\sigma_1,...,\sigma_n)$ denotes the external cyclic order, $p=(p_1,...,p_n)$ the $n$-tuple
of the external momenta and $\eps=(\eps_1,...,\eps_n)$ the $n$-tuple of the external polarisation vectors.
We have seen in eq.~(\ref{polarisation_graviton}) that the polarisation of an external graviton is described by a product of
two polarisation vectors
\bq
 \eps_{\mu_j\nu_j}^{\lambda_j \tilde{\lambda}_j}\left(p_j\right)
 & = &
 \eps_{\mu_j}^{\lambda_j}\left(p_j\right) \eps_{\nu_j}^{\tilde{\lambda}_j}\left(p_j\right),
\eq
therefore we may describe the polarisation configuration of $n$ external gravitons by two $n$-tuples
\bq
 \eps \;\; = \;\; \left( \eps_1^{\lambda_1}, ..., \eps_n^{\lambda_n} \right),
 & &
 \tilde{\eps} \;\; = \;\; \left( \eps_1^{\tilde{\lambda}_1}, ..., \eps_n^{\tilde{\lambda}_n} \right),
\eq
where for each graviton the $n$-tuple $\eps$ contains one polarisation vector and the $n$-tuple $\tilde{\eps}$ the other
polarisation vector. 
Of course, since either $(\lambda_j,\tilde{\lambda}_j)=(+,+)$ or $(\lambda_j,\tilde{\lambda}_j)=(-,-)$
we have $\eps=\tilde{\eps}$ for gravitons.
This motivates us to introduce also for the graviton amplitudes an alternative notation:
\bq
 M_n^{(0)}\left(p,\eps,\tilde{\eps}\right)
 & = &
 M_n^{(0)}\left(p_1^{\lambda_1 \tilde{\lambda}_1}, ..., p_n^{\lambda_n \tilde{\lambda}_n} \right).
\eq
Let us recall the CHY representation for gauge amplitudes. We may express the primitive tree-level gauge amplitude
either as a multidimensional contour integral around the zeros of the scattering equations or as a sum over
the solutions of the scattering equations:
\bq
\label{CHY_representation_YM}
 A_n^{(0)}\left(\sigma,p,\eps\right)
 & = &
 i \oint\limits_{\mathcal C} d\Omega_{\mathrm{CHY}} \; C\left(\sigma,z\right) \; E\left(z,p,\eps\right)
 \nonumber \\
 & = &
 i
 \sum\limits_{\mathrm{solutions} \; j} J\left(z^{(j)},p\right) \; C\left(\sigma, z^{(j)}\right) \; E\left(z^{(j)},p,\eps\right).
\eq
What happens if we replace the Parke-Taylor factor $C(\sigma,z)$ by another copy of $E(z,p,\eps)$?
It turns out that this gives the $n$-graviton amplitude $M_n^{(0)}$.
We thus arrive at the {\bf CHY representation of the $n$-graviton amplitude}
\bq
\label{CHY_representation_gravity}
 M_n^{(0)}\left(p,\eps,\tilde{\eps}\right)
 & = &
 i \oint\limits_{\mathcal C} d\Omega_{\mathrm{CHY}} \; E\left(z,p,\eps\right) \; E\left(z,p,\tilde{\eps}\right)
 \nonumber \\
 & = &
 i
 \sum\limits_{\mathrm{solutions} \; j} J\left(z^{(j)},p\right) \; E\left(z^{(j)},p,\eps\right) \; E\left(z^{(j)},p,\tilde{\eps}\right).
\eq
Note that for gravitons we have $\eps=\tilde{\eps}$, which derives from the fact 
that the external graviton states are
described by a product of equal-helicity polarisation vectors $\eps^\pm_\mu\eps^\pm_\nu$.
This brings up two questions:
What states do the opposite helicity combinations $\eps^\pm_\mu \eps^\mp_\nu$ describe
and what does eq.~(\ref{CHY_representation_gravity}) compute, 
if we evaluate this formula with some opposite helicity combinations?
The answer is the following: 
The opposite helicity combinations correspond to a linear combination of a {\bf dilaton} 
and an {\bf anti-symmetric tensor field},
\bq
 \eps_{\mu\nu}^{\mathrm{dilaton}}
 \;\; = \;\; \frac{1}{\sqrt{2}} \left( \eps^+_\mu \eps^-_\nu + \eps^-_\mu \eps^+_\nu \right),
 & &
 \eps_{\mu\nu}^{\mathrm{anti-symm}}
 \;\; = \;\; \frac{1}{\sqrt{2}} \left( \eps^+_\mu \eps^-_\nu - \eps^-_\mu \eps^+_\nu \right).
\eq
The anti-symmetric tensor field can be related through Hodge duality to an axion field.
If some opposite helicity combinations are plugged into eq.~(\ref{CHY_representation_gravity}),
this formula computes the corresponding tree amplitude in a theory
with a graviton, a dilaton and an anti-symmetric tensor field.
If we restrict ourselves to $\eps=\tilde{\eps}$ we obtain the pure graviton amplitude.
It can be shown for tree amplitudes, that in the case where all external particles are gravitons,
the dilaton and anti-symmetric tensor modes in the extended theory do not propagate internally.
Thus the tree graviton amplitudes coincide in Einstein gravity and in a theory consisting of
Einstein gravity plus a dilaton plus an anti-symmetric tensor field.
This statement is no longer true for loop amplitudes.

In going from eq.~(\ref{CHY_representation_YM}) to eq.~(\ref{CHY_representation_gravity})
we replaced the cyclic factor $C(\sigma,z)$ with another copy of the polarisation factor $E(z,p,\eps)$.
What happens, if we do the reverse, i.e. if we replace the polarisation factor $E(z,p,\eps)$ with
another copy of a Parke-Taylor factor $C,\tilde{\sigma},z)$?
Let us consider
\bq
\label{CHY_representation_biadjoint}
 m_n^{(0)}\left(\sigma,\tilde{\sigma},p\right)
 & = &
 i \oint\limits_{\mathcal C} d\Omega_{\mathrm{CHY}} \; C\left(\sigma,z\right) \; C\left(\tilde{\sigma},z\right)
 \nonumber \\
 & = &
 i
 \sum\limits_{\mathrm{solutions} \; j} J\left(z^{(j)},p\right) \; C\left(\sigma, z^{(j)}\right) \; C\left(\tilde{\sigma}, z^{(j)}\right).
\eq
The quantity $m_n^{(0)}(\sigma,\tilde{\sigma},p)$ depends on two external orders $\sigma$ and $\tilde{\sigma}$.
It turns out that $m_n^{(0)}(\sigma,\tilde{\sigma},p)$ computes 
the {\bf double-ordered amplitudes of a bi-adjoint scalar theory with trivalent vertices} \cite{Cachazo:2013iea,White:2016jzc}.
This theory consists of a scalar field $\phi^{ab}$ in adjoint representation of two Lie groups
$G$ and $\tilde{G}$.
We will denote indices referring to $G$ by $a$, indices referring to $\tilde{G}$ by $b$.
The theory is described by the Lagrange density
\bq
 {\mathcal L}_{\mathrm{bi-adjoint \; scalar}}
 & = &
 \frac{1}{2} \left( \partial_\mu \phi^{ab} \right) \left( \partial^\mu \phi^{ab} \right)
 - \frac{\lambda}{3!} f^{a_1a_2a_3} \tilde{f}^{b_1b_2b_3} \phi^{a_1b_1} \phi^{a_2b_2} \phi^{a_3b_3}.
\eq
The Feynman rule for the propagator is
\bq
 \frac{i}{p^2} \delta^{a_1 a_2} \delta^{b_1 b_2},
\eq
the Feynman rule for the trivalent vertex is
\bq
 i \lambda \left( i f^{a_1a_2a_3} \right) \left( i \tilde{f}^{b_1b_2b_3} \right).
\eq
Amplitudes in this theory have a double colour decomposition, similar to the (single)
colour decomposition of gauge amplitudes in eq.~(\ref{colour_decomposition}):
\bq
{\mathcal m}_{n}^{(0)}\left(p\right) 
 & = & 
 \lambda^{n-2} 
 \sum\limits_{\sigma \in S_{n}/Z_{n}} 
 \sum\limits_{\tilde{\sigma} \in S_{n}/Z_{n}} 
 2 \; \mathrm{Tr} \left( T^{a_{\sigma(1)}} ... T^{a_{\sigma(n)}} \right)
 \;\;
 2 \; \mathrm{Tr} \left( \tilde{T}^{b_{\tilde{\sigma}(1)}} ... \tilde{T}^{b_{\tilde{\sigma}(n)}} \right)
 \;\;
 m_{n}^{(0)}\left( \sigma, \tilde{\sigma}, p \right).
 \nonumber \\
\eq
The double-ordered amplitudes $m_{n}^{(0)}( \sigma, \tilde{\sigma}, p )$
are computed by eq.~(\ref{CHY_representation_biadjoint}).
Of course we may also compute the amplitudes $m_{n}^{(0)}( \sigma, \tilde{\sigma}, p )$
from Feynman diagrams, which for a scalar theory is not so complicated.
We denote by ${\mathcal T}_n(\sigma)$ the set of all ordered tree diagrams with trivalent vertices and 
external order $\sigma$.
The number of diagrams in this set is given by
\bq
 \left| {\mathcal T}_n(\sigma) \right|
 & = &
 \frac{(2n-4)!}{(n-2)!(n-1)!}.
\eq
Two diagrams with different external orders are considered to be equivalent,
if we can transform one diagram into the other by a sequence of flips.
Under a flip operation one exchanges at a vertex two branches.
We denote by ${\mathcal T}_n(\sigma) \cap {\mathcal T}_n(\tilde{\sigma})$ the set of diagrams compatible 
with the external orders $\sigma$ and $\tilde{\sigma}$
and by $n_{\mathrm{flip}}(\sigma,\tilde{\sigma})$ the number of flips needed to transform any diagram 
from ${\mathcal T}_n(\sigma) \cap {\mathcal T}_n(\tilde{\sigma})$ with the external order
$\sigma$ into a diagram with the external order $\tilde{\sigma}$.
The number $n_{\mathrm{flip}}(\sigma,\tilde{\sigma})$ will be the same for all diagrams from 
${\mathcal T}_n(\sigma) \cap {\mathcal T}_n(\tilde{\sigma})$.
For a diagram $G$ we denote by $E(G)$ the set of the internal edges and by $s_e$ the Lorentz invariant 
corresponding to the internal edge $e$.
We then have
\bq
\label{def_m_n}
 m_{n}^{(0)}\left( \sigma, \tilde{\sigma}, p \right)
 & = &
 i \left(-1\right)^{n-3+n_{\mathrm{flip}}(\sigma,\tilde{\sigma})}
 \sum\limits_{G \in {\mathcal T}_n(\sigma) \cap {\mathcal T}_n(\tilde{\sigma})} 
 \;\;\;
 \prod\limits_{e \in E(G)} \frac{1}{s_e}.
\eq
The prefactor $i (-1)^{n-3}$ collects all the factors of $i$ from the vertices and propagators: 
We have $(n-3)$ factors of $i$ from the propagators and $(n-2)$ factors of $i$ from the vertices.
Let us give an example: For $n=4$ and $\sigma=(1,2,3,4)$ and $\tilde{\sigma}=(1,3,2,4)$ we have
\bq
 m_{n}^{(0)}\left( \sigma, \sigma, p \right)
 \;\; = \;\;
 - i \left( \frac{1}{s} + \frac{1}{t} \right),
 & &
 m_{n}^{(0)}\left( \sigma, \tilde{\sigma}, p \right)
 \;\; = \;\;
 \frac{i}{t}.
\eq
We may equally well express the full amplitude ${\mathcal m}_{n}^{(0)}(p)$ as a sum over Feynman diagrams.
To this aim we denote by ${\mathcal U}_n$ the set of all unordered tree diagrams with trivalent vertices.
The number of diagrams in this set is given by
\bq
 \left| {\mathcal U}_n \right| 
 & = & 
(2n-5)!!.
\eq
We have
\bq
\label{full_bi_adjoint_amplitude}
{\mathcal m}_{n}^{(0)}\left(p\right) 
 & = &
 i \left(-1\right)^{n-3}
 \lambda^{n-2} 
 \sum\limits_{G \in {\mathcal U}_n} 
 \frac{C\left(G\right) \tilde{C}\left(G\right)}{D\left(G\right)},
\eq
where the factor in the denominator $D(G)$ and 
the colour factors $C(G)$ and $\tilde{C}(G)$ are as in eq.~(\ref{BCJ_form}).
Note that the symbol $G$ in $C(G)$ and $\tilde{C}(G)$ refers to the graph $G$.
The quantity $C(G)$ gives the colour factor corresponding to the ``$a$''-indices, 
$\tilde{C}(G)$ the one corresponding to the ``$b$''-indices.

\subsection{Gravity amplitudes from colour-kinematics duality}

Let us now discuss a second alternative method for computing the $n$-graviton amplitude,
based on colour-kinematics duality \cite{Bern:2010ue,Bern:2010yg}.
We recall from section~(\ref{sect_colour_kinematics_duality})
that there exists kinematic numerators $N(G)$, such that
they satisfy anti-symmetry and Jacobi-like relations, whenever the corresponding
colour factors $C(G)$ do
\bq
 C\left(G_1\right) + C\left(G_2\right) = 0
 \hspace*{5mm}
 & \Rightarrow &
 \hspace*{5mm}
 N\left(G_1\right) + N\left(G_2\right) = 0,
 \nonumber \\
 C\left(G_1\right) + C\left(G_2\right) + C\left(G_3\right) = 0
 \hspace*{5mm}
 & \Rightarrow &
 \hspace*{5mm}
 N\left(G_1\right) + N\left(G_2\right) + N\left(G_3\right) = 0,
 \hspace*{6mm}
\eq
and the full tree amplitude in Yang-Mills theory is given by
\bq
\label{BCJ_form_2}
 {\mathcal A}_n^{(0)}\left(p,\eps\right)
 & = &
 i g^{n-2}
 \sum\limits_{G \in {\mathcal U}_n}
 \frac{C\left(G\right)N\left(G\right)}{D\left(G\right)},
\eq
where ${\mathcal U}_n$ denotes as in the previous paragraph the set of all 
of all (unordered) tree diagrams with trivalent vertices.
Suppose we have a set of kinematic numerators $N(G)$, then
we may compute the {\bf $n$-graviton amplitude by squaring the kinematic numerators}:
\bq
\label{double_copy_numerators}
 {\mathcal M}_n^{(0)}\left(p, \eps, \tilde{\eps} \right)
 & = &
 i
 \left(-1\right)^{n-3}
 \left( \frac{\kappa}{4} \right)^{n-2}
 \sum\limits_{G \in {\mathcal U}_n}
 \frac{N\left(G\right)\tilde{N}\left(G\right)}{D\left(G\right)}.
\eq
For pure graviton amplitudes we take $N(G)=\tilde{N}(G)$.
Again there is a generalisation towards a theory consisting of Einstein gravity plus dilaton plus anti-symmetric
tensor field.
In this theory, the kinematic numerators $N(G)$ correspond to the set of polarisation vectors $\eps$, while
the kinematic numerators $\tilde{N}(G)$ correspond to $\tilde{\eps}$.
It is worth pointing out that the kinematic numerators are not unique, they may be changed by
a generalised gauge transformation.
However it should be stressed, that eq.~(\ref{double_copy_numerators}) is invariant under generalised gauge transformation.

In going from eq.~(\ref{BCJ_form_2}) to eq.~(\ref{double_copy_numerators})
we replaced the colour numerator $C(G)$ by another copy of the kinematic numerator $N(G)$.
Again we may ask what happens if we do the reverse, i.e. replace the kinematic numerator $N(G)$
by a second copy of a colour numerator $\tilde{C}(G)$?
From eq.~(\ref{full_bi_adjoint_amplitude}) we already know the answer.
This is nothing than the full amplitude ${\mathcal m}_{n}^{(0)}(p)$ in the bi-adjoint scalar theory with trivalent vertices:
\bq
{\mathcal m}_{n}^{(0)}\left(p\right) 
 & = &
 i \left(-1\right)^{n-3}
 \lambda^{n-2} 
 \sum\limits_{G \in {\mathcal U}_n} 
 \frac{C\left(G\right) \tilde{C}\left(G\right)}{D\left(G\right)}.
\eq

\subsection{Gravity amplitudes from the KLT relations}

As a third alternative method to compute the $n$-graviton amplitude we
consider Kawai-Lewellen-Tye (KLT) relations \cite{Kawai:1985xq}.
These relations express the tree-level $n$-graviton amplitude as products of tree-level 
primitive gauge amplitudes.
Originally the KLT relations were derived from relations between open and closed strings.
However, there is a modern formulation of these relations, which does not rely on string theory
and which we present here.

Let us recall from section~(\ref{sect:colour}) that there are $(n-3)!$ independent primitive tree-level
amplitudes in Yang-Mills theory.
Using cyclic-invariance, the Kleiss-Kuijf relations and the BCJ relations we may fix three external
particles at specified positions.
A basis of the independent cyclic orders is then for example given by
\bq
\label{def_B_basis}
 B & = & \left\{ \; \sigma = (\sigma_1,...,\sigma_n) \in S_n \; | \; \sigma_1 = 1, \sigma_2 = 2, \sigma_n = n \; \right\}.
\eq
Clearly,
\bq
 \left| B \right| & = & \left(n-3\right)!.
\eq
Let us now define a $(n-3)! \times (n-3)!$-dimensional matrix $m_{\sigma\tilde{\sigma}}$,
indexed by permutations $\sigma$ and $\tilde{\sigma}$ from $B$.
We set
\bq
 m_{\sigma\tilde{\sigma}} 
 & = & 
  m_{n}^{(0)}\left( \sigma, \tilde{\sigma}, p \right).
\eq
The entries of the matrix $m_{\sigma\tilde{\sigma}}$ are the double-ordered primitive
amplitudes for the bi-adjoint scalar theory with trivalent vertices encountered in the previous paragraphs.
The matrix $m_{\sigma\tilde{\sigma}}$ is invertible
and we set
\bq
 S_{\sigma\tilde{\sigma}}
 & = & \left( m^{-1} \right)_{\sigma\tilde{\sigma}}.
\eq
The {\bf KLT relations} read \cite{Kawai:1985xq,Bern:1999bx,BjerrumBohr:2004wh,BjerrumBohr:2010ta,Feng:2010br,Damgaard:2012fb,Cachazo:2013gna,delaCruz:2015raa,delaCruz:2016wbr}
% m_n^0 defined with factor i
% S absorbs an extra factor (-i)
\bq
\label{KLT_relation}
 M_n^{(0)}\left(p,\eps,\tilde{\eps}\right) & = &
 \sum\limits_{\sigma,\tilde{\sigma}\in B}
 A_n^{(0)}\left(\sigma,p,\eps\right) \; S_{\sigma \tilde{\sigma}} \; A_n^{(0)}\left(\tilde{\sigma},p,\tilde{\eps}\right),
\eq
where the sum runs over a basis of cyclic orders. 

Let us look at an example: For $n=4$ particles there is only one independent cyclic order, 
\begin{figure}
\begin{center}
\includegraphics[scale=0.8]{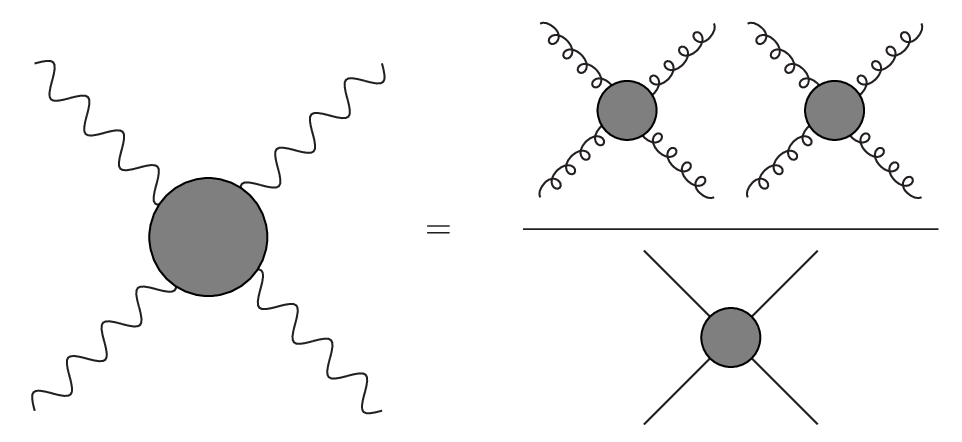}
\caption{\label{fig_KLT} Sketch of the KLT relation:
The four-graviton amplitude is the square of the four-gluon amplitude divided by a scalar amplitude.}
\end{center}
\end{figure}
which we may take as $\sigma=(1,2,3,4)$. Hence, the set $B$ has only one element $B= \{ \sigma \}$.
The matrix $m_{\sigma \sigma}$ is a $1 \times 1$-matrix, given by
\bq
 m_{\sigma \sigma} 
 \;\; = \;\;
 m_4^{(0)}\left(\sigma,\sigma,p\right)
 \;\; = \;\;
 - i \left( \frac{1}{s} + \frac{1}{t} \right)
 \;\; = \;\;
 i \frac{u}{s t}.
\eq
The $1 \times 1$-matrix $S_{\sigma \sigma}$, being the inverse of $m_{\sigma \sigma}$,  is then
\bq
 S_{\sigma \sigma} & = & - i \frac{s t}{u}.
\eq
We therefore have
\bq
 M_4^{(0)}\left( p_1^{\lambda_1 \lambda_1}, p_2^{\lambda_2 \lambda_2}, p_3^{\lambda_3 \lambda_3}, p_4^{\lambda_4 \lambda_4} \right)
 & = &
 - i \frac{s t}{u} \left[ A_4^{(0)}\left( p_1^{\lambda_1}, p_2^{\lambda_2}, p_3^{\lambda_3}, p_4^{\lambda_4} \right) \right]^2.
\eq
This is visualised in fig.~(\ref{fig_KLT}).
Note that the amplitude $A_4^{(0)}(p_1,p_2,p_3,p_4)$ has poles in $s$ and $t$, but not in $u$.
The prefactor $(s t)$ cancels the double poles in $s$ and $t$.
Therefore $M_4^{(0)}$ has only single poles in $s$, $t$ and $u$, as it should.
For four particles the four-graviton amplitude is given as the square
of the four-gluon amplitude divided by a scalar amplitude.
For $n>4$ the basis of independent cyclic orders of the primitive gluon amplitudes has more than
one element and the KLT relations turn into matrix equations.
\\
\\
\bs
{\it {\bf Exercise \theexercise}: 
Consider the case $n=5$ and determine the matrix $S_{\sigma \tilde{\sigma}}$ for the basis $B$ given 
in eq.~(\ref{def_B_basis}).
\stepcounter{exercise}
}
\es

% ----------------------------------------------------------------------------------
\newpage
\section{Outlook}
\label{sect:outlook}

In this report we considered primitive tree-level amplitudes in Yang-Mills theory.
These amplitudes may be computed with the textbook method of Feynman diagrams, involving only
simple algebraic operations like addition, multiplication, contraction of indices, etc..
One may ask, why we did not stop with this review shortly after the introduction?
The answer is two-fold.
First of all, the method based on Feynman diagrams is highly inefficient as we increase the number
of external particles.
We would like to have efficient methods to compute these scattering amplitudes.
If we opt for a numerical approach, one of the best methods is based on colour-decomposition, 
the spinor-helicity method and off-shell recurrence relations.
If on the other hand we are interested in analytical results, one of the best methods in this case
is based on on-shell recursion relations.

The second part to the answer of the question raised above is as follows:
We would have had no impetus to reveal hidden structures in quantum field theory.
We learned about colour-kinematics duality,
twistor methods, Grassmannian geometry, the scattering equations and global residues.
These mathematical developments culminate in relations between scattering amplitudes
in Yang-Mills theory and gravity.
This is a very interesting aspect, relating the three forces described by the Standard Model of particle
physics to the one force not covered by the Standard Model.

In this review we focused on primitive tree-level amplitudes in Yang-Mills theory,
presenting new developments in quantum field theory in this setting.
Although many of the themes discussed in this report provide foundational background to current research 
directions \cite{BjerrumBohr:2012mg,Monteiro:2011pc,Mafra:2011kj,Naculich:2014naa,Naculich:2015zha,Naculich:2015coa,Kalousios:2013eca,Tolotti:2013caa,Weinzierl:2014vwa,Lam:2014tga,Monteiro:2013rya,Cachazo:2015nwa,Baadsgaard:2015voa,Baadsgaard:2015hia,Baadsgaard:2015ifa,Huang:2015yka,Cardona:2016gon,Gomez:2016bmv,Cardona:2016bpi,Cardona:2016wcr,Bjerrum-Bohr:2014qwa,Mason:2013sva,Berkovits:2013xba,Gomez:2013wza,Adamo:2013tsa,Geyer:2014fka,Casali:2014hfa,Geyer:2015bja,Geyer:2016wjx,Schwab:2014xua,Afkhami-Jeddi:2014fia,Zlotnikov:2014sva,Kalousios:2014uva,White:2014qia,Cardona:2015ouc,Dolan:2015iln,delaCruz:2016wbr,Brown:2016mrh,Brown:2016hck,Cachazo:2016sdc,Cachazo:2016ror,Baadsgaard:2015twa,Bjerrum-Bohr:2016juj,Bjerrum-Bohr:2016axv,Elvang:2014fja,Hodges:2010kq}, 
it is on the other hand unavoidable that the focus on tree-level amplitudes in Yang-Mills theory
leaves some important topics uncovered.

The theory of the strong interactions, quantum chromodynamics (QCD),
is based on the (un-broken) non-abelian gauge group $\mathrm{SU}(3)$
and all results for gauge amplitudes discussed in this report directly apply to gluon amplitudes.
However, QCD does not only involve gluons, but also quarks.
Although we did not discuss fermions in detail in this review, many of the methods
described in this report extend to fermions \cite{Schwinn:2007ee,Dixon:2010ik,Melia:2013bta,Melia:2013epa,Melia:2015ika,Weinzierl:2014ava,Johansson:2015oia,delaCruz:2015dpa,delaCruz:2015raa}.

A second important topic not covered at all in this report are loop amplitudes.
These are needed for precision calculations in particle physics.
Again, many of the techniques discussed in this review have an extension towards loop level.
This is a wide field of research and it is impossible to give a complete list of references.
Therefore we limit us here to a few suggestions for further 
reading related to multiple polylogarithms \cite{Goncharov_no_note,Borwein,Vermaseren:1998uu,Remiddi:1999ew,Kotikov:1990kg,Kotikov:1991pm,Remiddi:1997ny,Gehrmann:1999as,Goncharov:2001,Moch:2001zr,Brown:2006,Brown:2008,Panzer:2014caa,Argeri:2007up,MullerStach:2012mp,Henn:2013pwa,Ablinger:2015tua}
and recent elliptic generalisations \cite{MullerStach:2011ru,Adams:2013nia,Bloch:2013tra,Adams:2014vja,Adams:2015gva,Adams:2015ydq,Remiddi:2013joa,Bloch:2016izu,Remiddi:2016gno,Adams:2016xah,Bonciani:2016qxi,Broedel:2014vla}.

There are a few mathematical topics, which we did not discuss or only mentioned briefly.
The first topic is centred around supersymmetric extensions of Yang-Mills theory, and
${\mathcal N}=4$ supersymmetric Yang-Mills theory in particular \cite{Korchemsky:2009jv,Alday:2010zy,Bern:2010tq,Kosower:2010yk,Dixon:2011ng,Bern:2012uc,Eden:2012fe,Bargheer:2014mxa,Ferro:2014gca,Dixon:2014voa,Caron-Huot:2016owq,Chicherin:2015bza,Mafra:2010jq}.
One may argue that ${\mathcal N}=4$ supersymmetric Yang-Mills theory is simpler than 
normal (non-supersymmetric) Yang-Mills theory \cite{ArkaniHamed:2008gz}.

We had no time to discuss Yangian symmetries \cite{MacKay:2004tc,Drummond:2007cf,Drummond:2008vq,Drummond:2009fd,Drummond:2010uq,Drummond:2010qh,Beisert:2010gn,Ferro:2011ph}
or cluster algebras \cite{Keller:2008,Goncharov:2010jf,Golden:2013xva,Golden:2014xqa,Drummond:2014ffa}.

The last section on the relation between gauge theories and gravity merely touched the tip of an iceberg,
there is certainly more which can be said and even more waits to be 
revealed \cite{Chen:2010ct,Bern:2011ia,Bern:2013uka,Monteiro:2014cda,Cachazo:2014nsa,Cachazo:2014xea,Johansson:2014zca,Chiodaroli:2015wal,Chiodaroli:2013upa,Bjerrum-Bohr:2014zsa,Bjerrum-Bohr:2016hpa,Nandan:2016pya,delaCruz:2016gnm,Stieberger:2014cea,Stieberger:2015qja,Stieberger:2015kia,Stieberger:2015vya,Stieberger:2016lng,Chiodaroli:2014xia,Chiodaroli:2015rdg,Chiodaroli:2016jqw}.

It is the hope that this report covers the foundations on which the readers may start their own research,
interesting research directions are not missing.

\subsection*{Acknowledgements}

This report grew out of lectures given at the Saalburg summer school.
I would like to thank the organisers for organising this summer school.
My thanks go also to the highly motivated students of this school.

% ----------------------------------------------------------------------------------

\begin{appendix}

% ----------------------------------------------------------------------------------
\newpage
\section{Solutions to the exercises}
\label{appendix:solutions}

\setcounter{exercise}{1}

\bs
{\it {\bf Exercise \theexercise}: 
Derive eq.~(\ref{relation_pull_backs}) from eq.~(\ref{relation_sections}).
\\
\\
{\bf Solution}: Let $\gamma : [a,b] \rightarrow M$ be a curve in $M$ with $\gamma(0)=x$.
A tangent vector at $x$ is given by
\bq
 X & = & \left. \frac{d}{dt} \gamma(t) \right|_{t=0}
\eq
In order to keep the notation simple we will suppress maps between a manifold and an appropriate
coordinate chart.
We have to show
\bq
 A_2\left(X\right)
 & = &
 \left( U A_1 U^\dagger + U d U^\dagger \right)\left(X\right),
\eq
where $A_1$ and $A_2$ are defined by
\bq
 A_1 \;\; = \;\; \sigma_1^\ast \omega,
 & &
 A_2 \;\; = \;\; \sigma_2^\ast \omega,
\eq
and the sections $\sigma_1$ and $\sigma_2$ are related by
\bq
 \sigma_2 & = & \sigma_1 U^\dagger.
\eq
Let us choose a local trivialisation $(x,g)$ of $P(M,G)$ and work out $\sigma_{2 \ast} X$.
With $U_0=U(\gamma(0))$ we have
\bq
 \sigma_{2 \ast} X 
 & = &
 \sigma_{2 \ast} \left( \left. \frac{d}{dt} \gamma(t) \right|_{t=0} \right)
 \;\; = \;\; 
 \left. \frac{d}{dt} \left( \gamma(t), \sigma_2\left(\gamma(t) \right) \right) \right|_{t=0}
 \;\; = \;\; 
 \left. \frac{d}{dt} \left( \gamma(t), \sigma_1\left(\gamma(t) \right) U\left(\gamma(t) \right)^\dagger \right) \right|_{t=0}
 \nonumber \\
 & = &
 \left( X, \left. \frac{d}{dt} \sigma_1\left(\gamma(t) \right) \right|_{t=0} U_0^\dagger
        + \sigma_1\left(\gamma(0) \right) \left. \frac{d}{dt} U\left(\gamma(t) \right)^\dagger \right|_{t=0} \right)
 \nonumber \\
 & = &
 R_{U_0^\dagger \ast} \left( \sigma_{1 \ast} X \right)
 + \left( 0, \sigma_2\left(\gamma(0) \right) U_0 \left. \frac{d}{dt} U\left(\gamma(t) \right)^\dagger \right|_{t=0} \right).
\eq
We then have, using eq.~(\ref{connection_projection}) and eq.~(\ref{connection_adjoint}),
\bq
 A_2\left(X\right)
 & = &
 \omega\left( \sigma_{2 \ast} X \right)
 \;\; = \;\;
 \omega_{(x,\sigma_2(\gamma(0)))}\left( R_{U_0^\dagger \ast} \left( \sigma_{1 \ast} X \right) \right)
 + \omega_{(x,\sigma_2(\gamma(0)))}\left( U_0 \left. \frac{d}{dt} U\left(\gamma(t) \right)^\dagger \right|_{t=0} \right)
 \nonumber \\
 & = &
 U_0 \left( \omega_{(x,\sigma_1(\gamma(0)))}\left( \sigma_{1 \ast} X \right) \right) U_0^\dagger
 + \left( U_0 d U^\dagger \right)\left(X\right)
 \nonumber \\
 & = & 
 \left( U A_1 U^\dagger + U d U^\dagger \right)\left(X\right).
\eq
\stepcounter{exercise}
}
\es
\\
\\
\bs
{\it {\bf Exercise \theexercise}: 
Under a gauge transformation the pull-back $A$ of the connection one-form transforms as
\bq
 A \rightarrow & U A U^\dagger + U d U^\dagger.
\eq
Show that the curvature two-form transforms as
\bq
 F & \rightarrow &
 U F U^\dagger.
\eq
{\bf Solution}: $F$ is given in terms of $A$ by
\bq
 F & = & d A + A \wedge A.
\eq
Let
\bq
 A' & = &  U A U^\dagger + U d U^\dagger
\eq
be the gauge-transformed gauge field. 
We have
\bq
 F'
 & = &
 d A' + A' \wedge A'
 \nonumber \\
 & = &
 d \left( U A U^\dagger + U d U^\dagger \right) 
 + \left( U A U^\dagger + U d U^\dagger \right) \wedge \left( U A U^\dagger + U d U^\dagger \right)
 \nonumber \\
 & = &
 \left(d U \right) \wedge A U^\dagger
 + U \left(dA\right) U^\dagger
 - U A \wedge dU^\dagger
 + \left(d U \right) \wedge \left(d U^\dagger\right)
 + U A \wedge A U^\dagger
 + U A \wedge d U^\dagger
 \nonumber \\
 & &
 + U \left( d U^\dagger \right) \wedge U A U^\dagger
 + U \left( d U^\dagger \right) \wedge U d U^\dagger.
\eq
From $U U^\dagger = {\bf 1}$ we have $d(U U^\dagger) = 0$ and hence
\bq
 U d U^\dagger & = & - \left(dU\right) U^\dagger.
\eq
Using this relation we obtain
\bq
 F'
 & = &
 U \left(dA\right) U^\dagger
 + U A \wedge A U^\dagger
 \;\; = \;\;
 U F U^\dagger.
\eq
\stepcounter{exercise}
}
\es
\\
\\
\bs
{\it {\bf Exercise \theexercise}: 
Let $\vec{\alpha}=(\alpha_1,...,\alpha_n)$ be a $n$-dimensional vector and let 
\bq
 g_i & = & g_i(\alpha_1,..,\alpha_n), \;\;\; i=1,...,n,
\eq
be $n$ functions of the $n$ varaibles $\alpha_j$. Show that
\bq
 \int \left( \prod\limits_{j=1}^n d\alpha_j \right)
  \left( \prod\limits_{i=1}^n \delta\left(g_i(\alpha_1,...,\alpha_n)\right) \right)
 \det\left( \frac{\partial g_i}{\partial \alpha_j} \right)
 & = & 1.
\eq
{\bf Solution}: 
We change the variables from $\alpha_j$ to
\bq
 \beta_i & = & g_i(\alpha_1,..,\alpha_n).
\eq
Then
\bq
 \int \left( \prod\limits_{j=1}^n d\alpha_j \right)
  \left( \prod\limits_{i=1}^n \delta\left(g_i(\alpha_1,...,\alpha_n)\right) \right)
 \det\left( \frac{\partial g_i}{\partial \alpha_j} \right)
 & = & 
 \int \left( \prod\limits_{j=1}^n d\beta_j \right)
  \left( \prod\limits_{i=1}^n \delta\left(\beta_i\right) \right)
 = 1.
\eq
\stepcounter{exercise}
}
\es
\\
\\
\bs
{\it {\bf Exercise \theexercise}: 
Derive eq.~(\ref{gluon_propagator}) from eq.~(\ref{gluon_prop_inverse_x_space}) 
and eq.~(\ref{gluon_prop_Fourier_trafo}).
\\
\\
{\bf Solution}: 
With
\bq
 P^{\mu\nu\;ab}(x) & = & \partial_\rho \partial^\rho g^{\mu\nu} \delta^{ab}
                  - \left( 1 - \frac{1}{\xi} \right) \partial^\mu \partial^\nu \delta^{ab}
\eq
and
\bq
 \left( P^{-1} \right)_{\mu\nu}^{ab}(x) & = & 
  \int \frac{d^4 p}{(2 \pi)^4} e^{-i p \cdot x} \left( \tilde{P}^{-1} \right)_{\mu\nu}^{ab}(p).
\eq
we have
\bq
 P^{\mu\sigma\;ac}(x) \left( P^{-1} \right)_{\sigma\nu}^{cb}(x-y) 
 & = & 
  \int \frac{d^4 p}{(2 \pi)^4} e^{-i p \cdot \left(x-y\right)}
  p^2 \left[ - g^{\mu\sigma} 
                  + \left( 1 - \frac{1}{\xi} \right) \frac{p^\mu p^\sigma}{p^2} 
 \right] \delta^{ac}
 \; \left( \tilde{P}^{-1} \right)_{\sigma\nu}^{cb}(p).
 \nonumber
\eq
This should be equal to
\bq
 g^\mu_{\;\;\nu} \delta^{ab} \delta^4(x-y)
 & = &
  \int \frac{d^4 p}{(2 \pi)^4} e^{-i p \cdot \left(x-y\right)} g^\mu_{\;\;\nu} \delta^{ab}.
\eq
We have for
\bq
 M^{\mu\nu} = - g^{\mu\nu} + \left( 1 - \frac{1}{\xi} \right) \frac{p^\mu p^\nu}{p^2}
 & \mbox{and} &
 N_{\mu\nu} = - g_{\mu\nu} + \left( 1 -\xi \right) \frac{p_\mu p_\nu}{p^2}
\eq
the relation
\bq
 M^{\mu \sigma} N_{\sigma \nu} & = & g^\mu_{\;\;\nu}.
\eq
Therefore
\bq
 \begin{picture}(100,20)(0,5)
 \Gluon(20,10)(70,10){-5}{5}
 \Text(15,12)[r]{\footnotesize $\mu, a$}
 \Text(75,12)[l]{\footnotesize $\nu, b$}
\end{picture} 
 & = & 
  \frac{i}{p^2} \left( - g_{\mu\nu} + \left( 1 -\xi \right) \frac{p_\mu p_\nu}{p^2} \right) \delta^{ab}.
\eq
\stepcounter{exercise}
}
\es
\\
\\
\bs
{\it {\bf Exercise \theexercise}: 
Compute the amplitude ${\mathcal A}_4^{(0)}$ from the four diagrams shown in fig.~(\ref{fig1}).
Assume that all momenta are outgoing.
Derive the Mandelstam relation
\bq 
 s + t + u & = & 0.
\eq
{\bf Solution}: Let us start with the Mandelstam relation.
Using momentum conservation and the on-shell relations we have
\bq
 0 & = &
 p_4^2
 \;\; = \;\; 
 \left( p_1 + p_2 + p_3 \right)^2
 \;\; = \;\; 
 2 p_1 \cdot p_2 + 2 p_2 \cdot p_3 + 2 p_1 \cdot p_3
 \nonumber \\
 & = &
 \left( p_1 + p_2 \right)^2 + \left( p_2 + p_3 \right)^2 + \left( p_1 + p_3 \right)^2
 \;\; = \;\; 
 s + t + u.
\eq
We then turn to the computation of the amplitude.
Let us first examine the colour factors.
The first diagrams has a colour factor
$C_s=(i f^{a_1 a_2 b}) (i f^{b a_3 a_4})$.
The second diagram we may equally well draw with legs $1$ and $4$ exchanged.
The colour factor is then given by $C_t=(i f^{a_2 a_3 b}) (i f^{b a_1 a_4})$.
(If we read off the colour factor directly from diagram $2$, we find 
$(i f^{a_2 a_3 b}) (i f^{b a_4 a_1}) = -(i f^{a_2 a_3 b}) (i f^{b a_1 a_4})$.
The minus sign cancels with another minus sign from the kinematic part.)
The third diagram has the colour factor $C_u=(i f^{a_3 a_1 b}) (i f^{b a_2 a_4})$.
The fourth diagram with the four-gluon vertex gives three terms, one contributing to each colour structure.
We may therefore write the amplitude as
\bq
 {\mathcal A}_4^{(0)}
 & = &
 i g^2
 \left[
  \frac{(i f^{a_1 a_2 b}) (i f^{b a_3 a_4}) \; N_s}{s}
  +
  \frac{(i f^{a_2 a_3 b}) (i f^{b a_1 a_4}) \; N_t}{t}
  +
  \frac{(i f^{a_3 a_1 b}) (i f^{b a_2 a_4}) \; N_u}{u}
 \right].
\eq
$N_s$ is given by
\bq
 N_s 
 & = &
 \left\{
 \left[ g^{\mu_1\mu_2} \left( p_1^{\nu} - p_2^{\nu} \right)
         +g^{\mu_2\nu} \left( p_2^{\mu_1} - p_{34}^{\mu_1} \right)
         +g^{\nu\mu_1} \left( p_{34}^{\mu_2} - p_1^{\mu_2} \right)
   \right]
 g_{\nu\rho}
 \left[ g^{\mu_3\mu_4} \left( p_3^{\rho} - p_4^{\rho} \right)
         +g^{\mu_4\rho} \left( p_4^{\mu_3} - p_{12}^{\mu_3} \right)
 \right. \right.
 \nonumber \\
 & & \left. \left.
         +g^{\rho\mu_3} \left( p_{12}^{\mu_4} - p_3^{\mu_4} \right)
   \right]
 +
 2 p_1 \cdot p_2 \left( g^{\mu_1\mu_3} g^{\mu_2\mu_4} - g^{\mu_2\mu_3} g^{\mu_1\mu_4} \right)
 \right\}
 \eps_1^{\mu_1} \eps_2^{\mu_2} \eps_3^{\mu_3} \eps_4^{\mu_4},
\eq
where we used the notation $p_{ij} = p_i + p_j$.
Using momentum conservation and $p_j \cdot \eps_j=0$ we may simplify this expression to
\bq
\label{to_be_contracted}
 N_s 
 & = &
 \left\{
 \left[ g^{\mu_1\mu_2} \left( p_1^{\nu} - p_2^{\nu} \right)
         + 2 g^{\mu_2\nu} p_2^{\mu_1} 
         - 2 g^{\nu\mu_1} p_1^{\mu_2}
   \right]
 g_{\nu\rho}
 \left[ g^{\mu_3\mu_4} \left( p_3^{\rho} - p_4^{\rho} \right)
         + 2 g^{\mu_4\rho} p_4^{\mu_3}
         - 2 g^{\rho\mu_3} p_3^{\mu_4} 
   \right]
 \right.
 \nonumber \\
 & & \left.
 +
 2 p_1 \cdot p_2 \left( g^{\mu_1\mu_3} g^{\mu_2\mu_4} - g^{\mu_2\mu_3} g^{\mu_1\mu_4} \right)
 \right\}
 \eps_1^{\mu_1} \eps_2^{\mu_2} \eps_3^{\mu_3} \eps_4^{\mu_4}.
\eq
The contraction of indices for long expressions is best done with the help
of a computer algebra program.
Here is a short FORM program, which performs the contractions in eq.~(\ref{to_be_contracted}):
{\footnotesize
\begin{verbatim}
* Example program for FORM

V p1,p2,p3,p4, e1,e2,e3,e4;
I mu1,mu2,mu3,mu4,nu,rho;

L  Ns = ((d_(mu1,mu2)*(p1(nu)-p2(nu)) + 2*d_(mu2,nu)*p2(mu1) - 2*d_(nu,mu1)*p1(mu2))*
         d_(nu,rho)*
         (d_(mu3,mu4)*(p3(rho)-p4(rho)) + 2*d_(mu4,rho)*p4(mu3) - 2*d_(rho,mu3)*p3(mu4))
         +2*p1(nu)*p2(nu)*(d_(mu1,mu3)*d_(mu2,mu4)-d_(mu2,mu3)*d_(mu1,mu4)))*
        e1(mu1)*e2(mu2)*e3(mu3)*e4(mu4);

print;

.end
\end{verbatim}
}
\noindent 
The same can be done in C++, using the GiNaC library:
{\footnotesize
\begin{verbatim}
// Example in C++ with GiNaC

#include <iostream>
#include <string>
#include <sstream>
#include <ginac/ginac.h>
using namespace std;
using namespace GiNaC;

string itos(int arg)
{
  ostringstream buffer;
  buffer << arg; 
  return buffer.str(); 
}

int main()
{
  varidx mu1(symbol("mu1"),4), mu2(symbol("mu2"),4),
         mu3(symbol("mu3"),4), mu4(symbol("mu4"),4), 
         nu(symbol("nu"),4), rho(symbol("rho"),4);

  symbol p1("p1"), p2("p2"), p3("p3"), p4("p4"),
         e1("e1"), e2("e2"), e3("e3"), e4("e4");

  vector<ex> p_vec = { p1, p2, p3, p4 };
  vector<ex> e_vec = { e1, e2, e3, e4 };

  scalar_products sp;
  for (int i=0; i<4; i++)
    {
      for (int j=i+1; j<4; j++)
        {
          sp.add(p_vec[i],p_vec[j],symbol( string("p")+itos(i+1)+string("p")+itos(j+1) ));
          sp.add(e_vec[i],e_vec[j],symbol( string("e")+itos(i+1)+string("e")+itos(j+1) ));
          sp.add(p_vec[i],e_vec[j],symbol( string("p")+itos(i+1)+string("e")+itos(j+1) ));
          sp.add(e_vec[i],p_vec[j],symbol( string("p")+itos(j+1)+string("e")+itos(i+1) ));
        }
    }

  ex Ns = ((lorentz_g(mu1,mu2)*(indexed(p1,nu)-indexed(p2,nu)) 
	    + 2*lorentz_g(mu2,nu)*indexed(p2,mu1) - 2*lorentz_g(nu,mu1)*indexed(p1,mu2))
	   * lorentz_g(nu.toggle_variance(),rho.toggle_variance())
	   * (lorentz_g(mu3,mu4)*(indexed(p3,rho)-indexed(p4,rho))
	      + 2*lorentz_g(mu4,rho)*indexed(p4,mu3) - 2*lorentz_g(rho,mu3)*indexed(p3,mu4))
           + 2*indexed(p1,nu)*indexed(p2,nu.toggle_variance())
	   *(lorentz_g(mu1,mu3)*lorentz_g(mu2,mu4) - lorentz_g(mu2,mu3)*lorentz_g(mu1,mu4)))
    *indexed(e1,mu1.toggle_variance())*indexed(e2,mu2.toggle_variance())
    *indexed(e3,mu3.toggle_variance())*indexed(e4,mu4.toggle_variance());

  Ns = Ns.expand();
  Ns = Ns.simplify_indexed(sp);

  cout << Ns << endl;
}
\end{verbatim}
}
\noindent
After a few additional simplifications one finds
\bq
 N_s & = &
% + 4 *e1e3 *p1e2 *p3e4
 4 \left( p_1 \cdot \eps_2 \right) \left( p_3 \cdot \eps_4 \right) \left( \eps_1 \cdot \eps_3 \right)
% - 4 *p4e3 *p1e2 *e1e4 
 - 4 \left( p_1 \cdot \eps_2 \right) \left( p_4 \cdot \eps_3 \right) \left( \eps_1 \cdot \eps_4 \right)
% + 4 *p4e3 *e2e4 *p2e1 
 + 4 \left( p_2 \cdot \eps_1 \right) \left( p_4 \cdot \eps_3 \right) \left( \eps_2 \cdot \eps_4 \right)
 \nonumber \\
 & &
% - 4 *p3e4 *p2e1 *e2e3
 - 4 \left( p_2 \cdot \eps_1 \right) \left( p_3 \cdot \eps_4 \right) \left( \eps_2 \cdot \eps_3 \right)
% + (-2 *p4e3 *p2e4 - 2 *p1e3 *p3e4 + 2 *p2e3 *p3e4 + 2 *p4e3 *p1e4) *e1e2
 + 4 \left[ \left( p_1 \cdot \eps_3 \right) \left( p_2 \cdot \eps_4 \right) 
          - \left( p_1 \cdot \eps_4 \right) \left( p_2 \cdot \eps_3 \right) \right] \left( \eps_1 \cdot \eps_2 \right)
 \nonumber \\
 & &
% + (-2 *p1e2 *p3e1 + 2 *p4e1 *p1e2 + 2 *p3e2 *p2e1 - 2 *p2e1 *p4e2) *e3e4 
 + 4 \left[ \left( p_3 \cdot \eps_1 \right) \left( p_4 \cdot \eps_2 \right) 
          - \left( p_3 \cdot \eps_2 \right) \left( p_4 \cdot \eps_1 \right) \right] \left( \eps_3 \cdot \eps_4 \right)
% + 2 *e2e4 *e1e3 *p1p2
 + 2 \left( p_1 \cdot p_2 \right) \left( \eps_1 \cdot \eps_3 \right) \left( \eps_2 \cdot \eps_4 \right)
 \nonumber \\
 & &
% - 2 *e1e4 *p1p2 *e2e3 
 - 2 \left( p_1 \cdot p_2 \right) \left( \eps_1 \cdot \eps_4 \right) \left( \eps_2 \cdot \eps_3 \right)
% (-p2p3 - p1p4 + p1p3 + p2p4)*e3e4*e1e2 
 - 2 \left( p_2 \cdot p_3 - p_1 \cdot p_3 \right) \left( \eps_1 \cdot \eps_2 \right) \left( \eps_3 \cdot \eps_4 \right).
\eq
The numberator $N_t$ is obtained from the numerator $N_s$ by the substitution $(1,2,3) \rightarrow (2,3,1)$,
the numerator $N_u$ is obtained from the numerator $N_s$ by the substitution $(1,2,3) \rightarrow (3,1,2)$.

In view of section~(\ref{sect_colour_kinematics_duality}), where we discuss colour kinematics duality let us add the following remark:
The colour factors satisfy (obviously) the Jacobi identity
\bq
 C_s + C_t + C_u & = & 0.
\eq
It is an easy exercise to check, that the numerators $N_s$, $N_t$ and $N_u$,
as determined above, 
satisfy the Jacobi-like identity
\bq
 N_s + N_t + N_u & = & 0.
\eq
\stepcounter{exercise}
}
\es
\\
\\
\bs
{\it {\bf Exercise \theexercise}: 
Compute the traces $\mathrm{Tr}\left( T^a T^b T^b T^a \right)$ and $\mathrm{Tr}\left( T^a T^b T^a T^b \right)$
for $\mathrm{U}(N)$ and $\mathrm{SU}(N)$.
\\
\\
{\bf Solution}: Using the Fierz identities for $\mathrm{SU}(N)$ one finds
\bq
 \mathrm{Tr}\left( T^a T^b T^b T^a \right)
 & = &
 \frac{1}{2}  \left[ \mathrm{Tr}\left( T^a T^a \right) \mathrm{Tr}\left( {\bf 1} \right)  - \frac{1}{N} \mathrm{Tr}\left( T^a T^a \right) \right]
 \;\; = \;\;
 \frac{N^2-1}{2N} \mathrm{Tr}\left( T^a T^a \right)
 \;\; = \;\; 
 \frac{\left(N^2-1\right)^2}{4N},
 \nonumber \\
 \mathrm{Tr}\left( T^a T^b T^a T^b \right)
 & = &
 - \frac{1}{2N} \mathrm{Tr}\left( T^a T^a \right) 
 \;\; = \;\;
 - \frac{N^2-1}{4N}.
\eq
For $\mathrm{U}(N)$ one obtains
\bq
 \mathrm{Tr}\left( T^a T^b T^b T^a \right)
 & = &
 \frac{1}{2}  \mathrm{Tr}\left( T^a T^a \right) \mathrm{Tr}\left( {\bf 1} \right)
 \;\; = \;\;
 \frac{N}{4}  \mathrm{Tr}\left( {\bf 1} \right) \mathrm{Tr}\left( {\bf 1} \right)
 \;\; = \;\;
 \frac{N^3}{4},
 \nonumber \\
 \mathrm{Tr}\left( T^a T^b T^a T^b \right)
 & = &
 \frac{1}{2} \mathrm{Tr}\left( T^a \right) \mathrm{Tr}\left( T^a \right)
 \;\; = \;\;
 \frac{1}{4} \mathrm{Tr}\left( {\bf 1}  \right)
 \;\; = \;\;
 \frac{N}{4}.
\eq
\stepcounter{exercise}
}
\es
\\
\\
\bs
{\it {\bf Exercise \theexercise}: 
Consider the path integral
\bq
\int {\mathcal D}\phi \; \exp \; i \int d^4x \; \mathrm{Tr}\left( \frac{1}{2} \phi P \phi + \phi K \right).
\eq
Assume that $P$ is a pseudo-differential operator of even degree and independent of the other fields.
$K$ on the other hand may depend on some other fields.
Show that under the substitution
\bq
 \phi & \rightarrow & \phi+P^{-1}K,
\eq
where $P^{-1}$ denotes the inverse pseudo-differential operator, one obtains
\bq
\int {\mathcal D} \phi \; \exp \; i \int d^4x \; 
 \mathrm{Tr}\left( \frac{1}{2} \phi P \phi - \frac{1}{2} K P^{-1} K 
 \right).
\eq
\\
\\
{\bf Solution}: Using the shift $\phi=\phi'-P^{-1}K$ one obtains
\bq
\lefteqn{
\int {\mathcal D}\phi' \; \exp \; i \int d^4x \; \mathrm{Tr}\left( \frac{1}{2} \left(\phi'-P^{-1}K\right) P \left(\phi'-P^{-1}K\right) + \left(\phi'-P^{-1}K \right) K \right)
 = }
 \\
 & = &
\int {\mathcal D}\phi' \; \exp \; i \int d^4x \; 
 \mathrm{Tr}\left( \frac{1}{2} \phi' P \phi' - \frac{1}{2} \left(P^{-1}K\right) P \phi' 
 + \frac{1}{2} \left(P^{-1}K\right) K 
 + \frac{1}{2} \phi' K 
 - \left(P^{-1}K \right) K
 \right).
 \nonumber
\eq
Using partial integration for $P^{-1}$ one obtains the result
\bq
\int {\mathcal D}\phi' \; \exp \; i \int d^4x \; 
 \mathrm{Tr}\left( \frac{1}{2} \phi' P \phi' 
 - \frac{1}{2} K P^{-1} K 
 \right).
\eq
\stepcounter{exercise}
}
\es
\\
\\
\bs
{\it {\bf Exercise \theexercise}: 
Show that at the level of cyclic-ordered Feynman rules we have
\begin{center}
\includegraphics[scale=0.8]{fig6}
\end{center}
{\bf Solution}: We calculate the left-hand side with the cyclic-ordered Feynman rules for the auxiliary
tensor field $B_{[\rho\sigma]}$:
\bq
\mathrm{l.h.s} & = &
 \frac{i}{\sqrt{2}} \left( g^{\mu_1\rho_1} g^{\mu_2 \sigma_1} - g^{\mu_1\sigma_1} g^{\mu_2 \rho_1} \right)
 \left(-\frac{i}{2} \right) \left( g_{\rho_1\rho_2} g_{\sigma_1\sigma_2} - g_{\rho_1\sigma_2} g_{\sigma_1\rho_2} \right)
 \frac{i}{\sqrt{2}} \left( g^{\mu_3\rho_2} g^{\mu_4 \sigma_2} - g^{\mu_3\sigma_2} g^{\mu_4 \rho_2} \right)
 \nonumber \\
 & &
 +
 \frac{i}{\sqrt{2}} \left( g^{\mu_2\rho_1} g^{\mu_3 \sigma_1} - g^{\mu_2\sigma_1} g^{\mu_3 \rho_1} \right)
 \left(-\frac{i}{2} \right) \left( g_{\rho_1\rho_2} g_{\sigma_1\sigma_2} - g_{\rho_1\sigma_2} g_{\sigma_1\rho_2} \right)
 \frac{i}{\sqrt{2}} \left( g^{\mu_4\rho_2} g^{\mu_1 \sigma_2} - g^{\mu_4\sigma_2} g^{\mu_1 \rho_2} \right)
 \nonumber \\
 & = &
 i
 \left( g^{\mu_1\mu_3} g^{\mu_2\mu_4} - g^{\mu_1\mu_4} g^{\mu_2\mu_3} \right)
 +
 i \left( g^{\mu_2\mu_4} g^{\mu_3 \mu_1} - g^{\mu_2\mu_1} g^{\mu_3 \mu_4} \right)
 \nonumber \\
 & = &
 i
 \left( 2 g^{\mu_1\mu_3} g^{\mu_2\mu_4} - g^{\mu_1\mu_2} g^{\mu_3 \mu_4} - g^{\mu_1\mu_4} g^{\mu_2\mu_3} \right),
\eq
which is the cyclic-ordered Feynman rule for the four-gluon vertex.
\stepcounter{exercise}
}
\es
\\
\\
\bs
{\it {\bf Exercise \theexercise}: 
Show that $[x, x]=0$ implies the anti-symmetry of the Lie bracket $[x,y]=-[y,x]$.
Show further that also the converse is true, provided $\mathrm{char}\; F \neq 2$.
Explain, why the argument does not work for $\mathrm{char}\; F = 2$.
\\
\\
{\bf Solution}: Starting from $[x, x]=0$ for all $x$ we have
\bq
 0 
 \;\; = \;\; 
 \left[ x+y, x+y \right]
 \;\; = \;\; 
 \left[ x, y \right] + \left[ y, x \right]
\eq
and therefore $[x,y]=-[y,x]$.
Now let us consider the other direction. Assuming $[x,y]=-[y,x]$ we have for $x=y$
the relation $[x,x]=-[x,x]$ or equivalently
\bq
\label{char_F_relation}
 2 \left[ x,x \right] & = & 0.
\eq
For $\mathrm{char}\; F \neq 2$ it follows that $[x,x]=0$.
For $\mathrm{char}\; F = 2$ we have $2=0 \mod 2$ and eq.~(\ref{char_F_relation}) does not give any constraint
on $[x,x]$.
\stepcounter{exercise}
}
\es
\\
\\
\bs
{\it {\bf Exercise \theexercise}: 
The helicity operator $h$ for particles is defined by
\bq
 h & = & \frac{\vec{p} \cdot \vec{S}}{\left|\vec{p}\right|}.
\eq
Assume that $p^\mu$ is a real four-vector with positive energy ($p^0>0$).
Show that the spinors $| p+ \rangle = p_A$ and $\langle p+ | = p_{\dot{A}}$ have helicity $h=+1/2$, 
while the spinors $| p- \rangle = p^{\dot{A}}$ and $\langle p- | = p^A$ have helicity $h=-1/2$.
\\
\\
{\bf Solution}: 
In the Weyl representation we have
\bq
 h & = &
 \frac{1}{2 \left|\vec{p}\right|}
 \left( \begin{array}{cc}
 \vec{p} \cdot \vec{\sigma} & 0 \\
 0 & \vec{p} \cdot \vec{\sigma} \\
 \end{array} \right).
\eq
We further have
\bq
 p_\mu \bar{\sigma}^\mu = p^0 \cdot 1 - \vec{p} \cdot \vec{\sigma},
 & &
 p_\mu \sigma^\mu = p^0 \cdot 1 + \vec{p} \cdot \vec{\sigma}.
\eq
The Weyl spinors $\left| p \pm \right\rangle$ satisfy
\bq
 p_\mu \bar{\sigma}^\mu \left| p+ \right\rangle = 0,
 & &
 p_\mu \sigma^\mu \left| p- \right\rangle = 0,
\eq
and hence 
\bq
 h \left| p+ \right\rangle = \frac{1}{2} \frac{p^0}{\left|\vec{p}\right|} \left| p+ \right\rangle,
 & &
 h \left| p- \right\rangle = - \frac{1}{2} \frac{p^0}{\left|\vec{p}\right|} \left| p- \right\rangle.
\eq
For $p^0>0$ we thus have
\bq
 h \left| p+ \right\rangle = \frac{1}{2} \left| p+ \right\rangle,
 & &
 h \left| p- \right\rangle = - \frac{1}{2} \left| p- \right\rangle.
\eq
For the bra-spinors it follows from
\bq
 \left\langle p+ \right| p_\mu \bar{\sigma}^\mu  = 0,
 & &
 \left\langle p- \right| p_\mu \sigma^\mu  = 0,
\eq
that
\bq
 \left\langle p+ \right| h = \frac{1}{2} \frac{p^0}{\left|\vec{p}\right|} \left\langle p+ \right|,
 & &
 \left\langle p- \right| h = - \frac{1}{2} \frac{p^0}{\left|\vec{p}\right|} \left\langle p- \right|,
\eq
and for $p^0>0$
\bq
 \left\langle p+ \right| h = \frac{1}{2} \left\langle p+ \right|,
 & &
 \left\langle p- \right| h = - \frac{1}{2} \left\langle p- \right|.
\eq
\stepcounter{exercise}
}
\es
\\
\\
\bs
{\it {\bf Exercise \theexercise}: The field strength in spinor notation:
For a rank-$2$ tensor $F_{\mu\nu}$ we define the spinor representation by
\bq
 F_{A B \dot{A} \dot{B}} & = & F_{\mu\nu} \sigma^\mu_{A \dot{A}} \sigma^\nu_{B \dot{B}}.
\eq
On the other hand we may decompose $F_{\mu\nu}$ into a self dual part and an anti-self dual part,
\bq
 F_{\mu\nu} & = & F_{\mu\nu}^{\mathrm{self \; dual}} + F_{\mu\nu}^{\mathrm{anti-self \; dual}},
\eq
with
\bq
 F_{\mu\nu}^{\mathrm{self \; dual}}
 \;\; = \;\;
 \frac{1}{2} \left( F_{\mu\nu} + i \tilde{F}_{\mu\nu} \right),
 & &
 F_{\mu\nu}^{\mathrm{anti-self \; dual}}
 \;\; = \;\;
 \frac{1}{2} \left( F_{\mu\nu} - i \tilde{F}_{\mu\nu} \right).
\eq
The dual field strength $\tilde{F}_{\mu\nu}$ was defined in eq.~(\ref{def_dual_field_strength}).
Show that the spinor representations of the self dual part / anti-self dual part may be written as
\bq
 F_{A B \dot{A} \dot{B}}^{\mathrm{self \; dual}}
 \;\; = \;\;
 \eps_{A B} \bar{\phi}_{\dot{A}\dot{B}},
 & &
 F_{A B \dot{A} \dot{B}}^{\mathrm{anti-self \; dual}}
 \;\; = \;\;
 \phi_{AB} \eps_{\dot{A} \dot{B}},
\eq
where $\phi_{AB}$ and $\bar{\phi}_{\dot{A}\dot{B}}$ satisfy
\bq
 \phi_{AB} \;\; = \;\; \phi_{BA},
 & &
 \bar{\phi}_{\dot{A}\dot{B}} \;\; = \;\; \bar{\phi}_{\dot{B}\dot{A}}.
\eq
{\bf Solution}: 
From the anti-symmetry $F_{\mu\nu}=-F_{\nu\mu}$ it follows
\bq
 F_{A B \dot{A} \dot{B}} & = & - F_{B A \dot{B} \dot{A}}.
\eq
We can therefore write
\bq
 F_{A B \dot{A} \dot{B}} 
 & = & 
 \frac{1}{2} \left( F_{A B \dot{A} \dot{B}} - F_{A B \dot{B} \dot{A}} \right)
 +
 \frac{1}{2} \left( F_{A B \dot{B} \dot{A}} - F_{B A \dot{B} \dot{A}} \right).
\eq
We set
\bq
 F_{A B \left[ \dot{A}, \dot{B} \right]}
 \;\; = \;\; 
 \frac{1}{2} \left( F_{A B \dot{A} \dot{B}} - F_{A B \dot{B} \dot{A}} \right),
 & &
 F_{\left[ A, B \right] \dot{B} \dot{A}} 
 \;\; = \;\; 
 \frac{1}{2} \left( F_{A B \dot{B} \dot{A}} - F_{B A \dot{B} \dot{A}} \right).
\eq
The Schouten identity
\bq
 \eps_{A B} \eps_{C D}
 +
 \eps_{B C} \eps_{A D}
 +
 \eps_{C A} \eps_{B D}
 & = &
 0.
\eq
can also be written as
\bq
 \eps_{AB} \eps^{CD} & = & \eps_A^{\;\;\;C} \eps_B^{\;\;\;D} - \eps_A^{\;\;\;D} \eps_B^{\;\;\;C}.
\eq
Contraction with $\chi_{CD}$ yields
\bq
 \eps_{AB} \chi_C^{\;\;\;C}
 & = &
 \chi_{AB} - \chi_{BA}.
\eq
Therefore we have
\bq
 F_{A B \left[ \dot{A}, \dot{B} \right]}
 \;\; = \;\; 
 \frac{1}{2} \; \eps_{\dot{A} \dot{B}} \; F_{A B \dot{C}}^{\;\;\;\;\;\;\;\; \dot{C}},
 & &
 F_{\left[ A, B \right] \dot{B} \dot{A}} 
 \;\; = \;\; 
 \frac{1}{2} \; \eps_{A B} \; F_{C \;\;\;\; \dot{B} \dot{A}}^{\;\;\; C}.
\eq
We set
\bq
 \phi_{AB} \;\; = \;\; \frac{1}{2} \; \; F_{A B \dot{C}}^{\;\;\;\;\;\;\;\; \dot{C}},
 & &
 \bar{\phi}_{\dot{B}\dot{A}} \;\; = \;\; \frac{1}{2} \; F_{C \;\;\;\; \dot{B} \dot{A}}^{\;\;\; C}.
\eq
Note that we have
\bq
 F_{\left[ A, B \right] \dot{A} \dot{B}}
 \;\; = \;\;
 F_{\left[ A, B \right] \dot{B} \dot{A}},
 & &
 F_{A B \left[ \dot{A}, \dot{B} \right]}
 \;\; = \;\;
 F_{B A \left[ \dot{A}, \dot{B} \right]},
\eq
and therefore
\bq
 \phi_{AB} \;\; = \;\; \phi_{BA},
 & &
 \bar{\phi}_{\dot{A}\dot{B}} \;\; = \;\; \bar{\phi}_{\dot{B}\dot{A}}.
\eq
We thus arrive at
\bq
 F_{A B \dot{A} \dot{B}} 
 & = & 
 \phi_{AB} \eps_{\dot{A} \dot{B}} + \eps_{A B} \bar{\phi}_{\dot{A}\dot{B}}.
\eq
Let us now consider the dual field strength
\bq
 \tilde{F}_{\mu\nu} & = & \frac{1}{2} \eps_{\mu\nu\rho\sigma} F^{\rho\sigma}
\eq
in the spinor representation
\bq
 \tilde{F}_{A B \dot{A} \dot{B}} & = & \tilde{F}_{\mu\nu} \sigma^\mu_{A \dot{A}} \sigma^\nu_{B \dot{B}}.
\eq
Using $2 g^{\mu\nu} = \sigma^\mu_{A\dot{B}} \bar{\sigma}^{\nu \dot{B} A}$ we have
\bq
 F^{\rho\sigma}
 & = & 
 \frac{1}{4} \bar{\sigma}^{\rho \dot{C} C} \bar{\sigma}^{\sigma \dot{D} D}
 F_{C D \dot{C} \dot{D}},
 \nonumber \\
 \tilde{F}_{A B \dot{A} \dot{B}} 
 & = & 
 \frac{1}{8} 
 \sigma^\mu_{A\dot{A}} \sigma^\nu_{B\dot{B}}
 \bar{\sigma}^{\rho \dot{C} C} \bar{\sigma}^{\sigma \dot{D} D}
 \eps_{\mu\nu\rho\sigma}
 F_{C D \dot{C} \dot{D}}.
\eq
Using in addition
\bq
 4 i \eps_{\mu\nu\rho\sigma}
 & = &
 \mathrm{Tr}\left( P_+ \gamma_\mu \gamma_\nu \gamma_\rho \gamma_\sigma \right)
 -
 \mathrm{Tr}\left( P_- \gamma_\mu \gamma_\nu \gamma_\rho \gamma_\sigma \right),
 \nonumber \\
 \eps_{\mu\nu\rho\sigma}
 & = &
 - \frac{i}{4}
 \left(
 \sigma_{\mu E \dot{F}} \bar{\sigma}_\nu^{\dot{F}G} \sigma_{\rho G \dot{H}} \bar{\sigma}_\sigma^{\dot{H}E}
 -
 \bar{\sigma}_\mu^{\dot{E}F} \sigma_{\nu F \dot{G}} \bar{\sigma}_\rho^{\dot{G}H} \sigma_{\sigma H \dot{E}} 
 \right),
\eq
and the Fierz identities one finds
\bq
 \tilde{F}_{A B \dot{A} \dot{B}} 
 & = & 
 \frac{i}{2}
 \left( 
   \delta_A^D \delta_B^C \eps_{\dot{A}\dot{B}} \eps^{\dot{C}\dot{D}}
 -
 \eps_{AB} \eps^{CD} \delta_{\dot{A}}^{\dot{D}} \delta_{\dot{B}}^{\dot{C}}
 \right) F_{C D \dot{C} \dot{D}}.
\eq
Finally one obtains
\bq
 \tilde{F}_{A B \dot{A} \dot{B}} 
 & = & 
 i \left( \phi_{AB} \eps_{\dot{A} \dot{B}} - \eps_{A B} \bar{\phi}_{\dot{A}\dot{B}} \right).
\eq
Therefore $\eps_{A B} \bar{\phi}_{\dot{A}\dot{B}}$ is self dual, while $\phi_{AB} \eps_{\dot{A} \dot{B}}$ is anti-self dual.
\stepcounter{exercise}
}
\es
\\
\\
\bs
{\it {\bf Exercise \theexercise}: 
Calculate the primitive helicity amplitude $A_4^{(0)}(1^-,2^+,3^+,4^-)$.
\\
\\
{\bf Solution}: As reference momenta for the polarisation vectors we choose
\bq
 q_1 \;\; = \;\; p_2,
 \;\;\;
 q_2 \;\; = \;\; p_1,
 \;\;\;
 q_3 \;\; = \;\; p_1,
 \;\;\;
 q_4 \;\; = \;\; p_2.
\eq
The polarisation vectors are then
\bq
 \eps_1^\mu \;\; = \;\; \frac{\left\langle 1- \left| \gamma^\mu \right| 2- \right\rangle}{\sqrt{2} \left[ 1 2 \right]},
 \;\;\;
 \eps_2^\mu \;\; = \;\; - \frac{\left\langle 2+ \left| \gamma^\mu \right| 1+ \right\rangle}{\sqrt{2} \left\langle 2 1 \right\rangle},
 \;\;\;
 \eps_3^\mu \;\; = \;\; - \frac{\left\langle 3+ \left| \gamma^\mu \right| 1+ \right\rangle}{\sqrt{2} \left\langle 3 1 \right\rangle},
 \;\;\;
 \eps_4^\mu \;\; = \;\; \frac{\left\langle 4- \left| \gamma^\mu \right| 2- \right\rangle}{\sqrt{2} \left[ 4 2 \right]}.
\eq
From the Fierz identity we have
\bq
 \eps_1 \cdot \eps_2
 \;\; = \;\;
 \eps_1 \cdot \eps_3
 \;\; = \;\;
 \eps_1 \cdot \eps_4
 \;\; = \;\;
 \eps_2 \cdot \eps_3
 \;\; = \;\;
 \eps_2 \cdot \eps_4
 \;\; = \;\; 0,
\eq
and only $\eps_3 \cdot \eps_4$ is non-zero.
We have to calculate the three diagrams shown in fig.~(\ref{fig2}). In the diagram with the four-gluon vertex all polarisation vectors are contracted with each other.
In particular $\eps_1$ is contracted into some other polarisation vector, giving zero.
Therefore the four-gluon diagrams does not contribute.
The $s$-channel diagram gives
\bq
 D_s & = &
 i \frac{g_{\rho\sigma}}{s} 
 \eps_{1,\mu_1} \eps_{2,\mu_2} \eps_{3,\mu_3} \eps_{4,\mu_4} 
 \left[ g^{\mu_1\mu_2} \left( p_1^{\rho} - p_2^{\rho} \right)
         +g^{\mu_2\rho} \left( p_2^{\mu_1} - p_{34}^{\mu_1} \right)
         +g^{\rho\mu_1} \left( p_{34}^{\mu_2} - p_1^{\mu_2} \right)
   \right]
 \nonumber \\
 & &
 \times 
 \left[ g^{\mu_3\mu_4} \left( p_3^{\sigma} - p_4^{\sigma} \right)
         +g^{\mu_4\sigma} \left( p_4^{\mu_3} - p_{12}^{\mu_3} \right)
         +g^{\sigma\mu_3} \left( p_{12}^{\mu_4} - p_3^{\mu_4} \right)
   \right]
 \nonumber \\
 & = &
 \frac{i}{s} 
 \eps_{1,\mu_1} \eps_{2,\mu_2} \left( \eps_{3} \cdot \eps_{4} \right) 
 \left[ 
          \left( p_2^{\mu_1} - p_{34}^{\mu_1} \right) \left( p_3^{\mu_2} - p_4^{\mu_2} \right)
         +\left( p_3^{\mu_1} - p_4^{\mu_1} \right) \left( p_{34}^{\mu_2} - p_1^{\mu_2} \right)
   \right].
\eq
Now by momentum conservation and our choice of reference momenta
\bq
 \eps_{1,\mu_1} \left( p_2^{\mu_1} - p_{34}^{\mu_1} \right)
 & = & 2 \eps_1 \cdot p_2 \;\; = \;\; 0,
 \nonumber \\
 \eps_{2,\mu_2} \left( p_{34}^{\mu_2} - p_1^{\mu_2} \right)
 & = & - 2 \eps_2 \cdot p_1 \;\; = \;\; 0,
\eq
and therefore this diagram does not contribute either.
It remains to compute the $t$-channel diagram
\bq
 D_t & = &
 i \frac{g_{\rho\sigma}}{t} 
 \eps_{1,\mu_1} \eps_{2,\mu_2} \eps_{3,\mu_3} \eps_{4,\mu_4} 
 \left[ g^{\mu_2\mu_3} \left( p_2^{\rho} - p_3^{\rho} \right)
         +g^{\mu_3\rho} \left( p_3^{\mu_2} - p_{14}^{\mu_2} \right)
         +g^{\rho\mu_2} \left( p_{14}^{\mu_3} - p_2^{\mu_3} \right)
   \right]
 \nonumber \\
 & &
 \times 
 \left[ g^{\mu_4\mu_1} \left( p_4^{\sigma} - p_1^{\sigma} \right)
         +g^{\mu_1\sigma} \left( p_1^{\mu_4} - p_{23}^{\mu_4} \right)
         +g^{\sigma\mu_4} \left( p_{23}^{\mu_1} - p_4^{\mu_1} \right)
   \right]
 \nonumber \\
 & = &
 \frac{i}{t} 
 \eps_{1,\mu_1} \eps_{2,\mu_2} \left( \eps_{3} \cdot \eps_{4} \right) 
         \left( p_{23}^{\mu_1} - p_4^{\mu_1} \right) \left( p_3^{\mu_2} - p_{14}^{\mu_2} \right) 
 \nonumber \\
 & = &
 -4 \frac{i}{t} 
 \left( \eps_{1} \cdot p_4 \right) \left( \eps_{2} \cdot p_3 \right) \left( \eps_{3} \cdot \eps_{4} \right).
\eq
This is the only non-vanishing contribution and we obtain
\bq
 A_4^{(0)}\left(1^-,2^+,3^+,4^-\right)
 & = &
 - \frac{2 i}{t} 
 \left( \eps_{1} \cdot p_4 \right) \left( \eps_{2} \cdot p_3 \right) \left( \eps_{3} \cdot \eps_{4} \right)
 \nonumber \\
 & = &
 - \frac{4 i}{\left\langle 2 3  \right\rangle \left[ 3 2 \right]} 
 \frac{\left[ 2 4 \right] \left\langle 4 1 \right\rangle}{\left[ 2 1 \right]}
 \frac{\left\langle 1 3 \right\rangle \left[ 3 2 \right] }{\left\langle 1 2 \right\rangle}
 \frac{\left\langle 1 4 \right\rangle \left[ 2 3 \right]}{\left\langle 1 3 \right\rangle \left[ 2 4 \right]}
 \nonumber \\
 & = &
 4 i
 \frac{\left\langle 1 4 \right\rangle^2 \left[ 2 3 \right]}{\left\langle 1 2 \right\rangle \left\langle 2 3  \right\rangle \left[ 2 1 \right]}.
\eq
We may still simplify this expression a little bit. 
Using $[23] \langle 34 \rangle = - [21] \langle 14 \rangle$ one arrives
at the Parke-Taylor formula for $n=4$:
\bq
 A_4^{(0)}\left(1^-,2^+,3^+,4^-\right)
 & = &
 4 i
 \frac{\left\langle 1 4 \right\rangle^4}{\left\langle 1 2 \right\rangle \left\langle 2 3  \right\rangle \left\langle 3 4  \right\rangle \left\langle 4 1 \right\rangle}.
\eq
\stepcounter{exercise}
}
\es
\\
\\
\bs
{\it {\bf Exercise \theexercise}: 
Show that the spinors defined in eq.~(\ref{def_u_spinor}) and in eq.~(\ref{def_ubar_spinor})
satisfy the Dirac equations of eq.~(\ref{Dirac_equations}),
the orthogonality relations of eq.~(\ref{spinor_orthogonality})
and the completeness relations of eq.~(\ref{spinor_completeness_relations}).
\\
\\
{\bf Solution}: 
We start with the Dirac equation.
For the spinor $u(p,+)$ we have
\bq
 \left( p\!\!\!/ - m \right) u(p,+)
 & = &
 \frac{1}{\left[ p^\flat q \right]} \left( p\!\!\!/ - m \right) \left( p\!\!\!/ + m \right) | q - \rangle
 \;\; = \;\;
 \frac{1}{\left[ p^\flat q \right]} \left( p^2 - m^2 \right) | q - \rangle
 \;\; = \;\; 0,
\eq
due to $p^2=m^2$. The argument is similar for all other spinors.
Let us now look at the orthogonality relations.
We have
\bq
 \bar{u}(p,+) u(p,+) 
 & = & 
 \frac{1}{\left\langle q p^\flat \right\rangle \left[ p^\flat q \right]} 
 \langle q - | \left( p\!\!\!/ + m \right) \left( p\!\!\!/ + m \right) | q - \rangle
 \;\; = \;\;
 \frac{2m}{\left\langle q p^\flat \right\rangle \left[ p^\flat q \right]} 
 \langle q - | p\!\!\!/ | q - \rangle
 \nonumber \\
 & = & 2 m \frac{2p \cdot q}{2 p^\flat \cdot q} 
 \;\; = \;\;
 2 m,
 \nonumber \\
 \bar{u}(p,+) u(p,-) 
 & = & 
 \frac{1}{\left\langle q p^\flat \right\rangle \left\langle p^\flat q \right\rangle} 
 \langle q - | \left( p\!\!\!/ + m \right) \left( p\!\!\!/ + m \right) | q + \rangle
 \;\; = \;\;
 - \frac{2m^2}{\left\langle q p^\flat \right\rangle^2} 
 \langle q q \rangle
 \;\; = \;\; 0.
\eq
The argumentation for the other combinations is similar.
Finally, let us consider the completeness relation:
\bq
\sum\limits_{\lambda} u(p,\lambda) \bar{u}(p,\lambda) 
 & = &
 u(p,+) \bar{u}(p,+) + u(p,-) \bar{u}(p,-)
 \nonumber \\
 & = &
 \frac{1}{\left[ p^\flat q \right] \left\langle q p^\flat \right\rangle} 
  \left[ \left( p\!\!\!/ + m \right) | q - \rangle \langle q - | \left( p\!\!\!/ + m \right)
 +
         \left( p\!\!\!/ + m \right) | q + \rangle \langle q + | \left( p\!\!\!/ + m \right)
 \right]
 \nonumber \\
 & = &
 \frac{1}{2 p \cdot q} 
  \left( p\!\!\!/ + m \right) q\!\!\!/ \left( p\!\!\!/ + m \right)
 \;\; = \;\;
 \frac{p\!\!\!/ q\!\!\!/ p\!\!\!/  + m \left(p\!\!\!/ q\!\!\!/ + q\!\!\!/ p\!\!\!/ \right)+ m^2 q\!\!\!/ }{2 p \cdot q}. 
\eq
With
\bq
 p\!\!\!/ q\!\!\!/ p\!\!\!/ \;\; = \;\; \left(2 p \cdot q\right) p\!\!\!/ \;\; - p^2 q\!\!\!/ \;\;
 & \mbox{and} &
 p\!\!\!/ q\!\!\!/ + q\!\!\!/ p\!\!\!/  \;\; = \;\; 2 p \cdot q
\eq
we obtain
\bq
\sum\limits_{\lambda} u(p,\lambda) \bar{u}(p,\lambda) 
 & = &
 p\!\!\!/ + m.
\eq
The proof of the completeness relation for the $v$-spinors is obtained by the substitution $m \rightarrow -m$.
\stepcounter{exercise}
}
\es
\\
\\
\bs
{\it {\bf Exercise \theexercise}: 
Show that the momenta defined by eq.~(\ref{bra_ket_spinors_from_W_twistor}) satisfy momentum conservation.
\\
\\
{\bf Solution}: We have
\bq
 \sum\limits_{i=1}^n p_i^\mu
 & = &
 - \frac{1}{2}
 \left(
 \frac{\left[ p_{i+1} p_i \right]}{ \left[p_{i+1} p_i \right] \left[ p_{i} p_{i-1} \right]}
   \left\langle p_i + \left| \gamma^\mu \right| \mu_{i-1} + \right\rangle
 + \frac{\left[ p_{i-1} p_{i+1} \right]}{\left[ p_{i+1} p_i \right] \left[ p_{i} p_{i-1} \right]}
   \left\langle p_i + \left| \gamma^\mu \right| \mu_{i} + \right\rangle
 \right. \nonumber \\
 & & \left.
 + \frac{\left[ p_{i} p_{i-1} \right]}{\left[ p_{i+1} p_i \right] \left[ p_{i} p_{i-1} \right]}
   \left\langle p_i + \left| \gamma^\mu \right| \mu_{i+1} + \right\rangle
 \right),
\eq
where all indices are understood $\mathrm{mod}\; n$. Let's reorder the sum:
\bq
 \sum\limits_{i=1}^n p_i^\mu
 & = &
 - \frac{1}{2}
 \left(
 \frac{\left[ p_{i+2} p_{i+1} \right]}{ \left[p_{i+2} p_{i+1} \right] \left[ p_{i+1} p_{i} \right]}
   \left\langle p_{i+1} + \left| \gamma^\mu \right| \mu_{i} + \right\rangle
 + \frac{\left[ p_{i-1} p_{i+1} \right]}{\left[ p_{i+1} p_i \right] \left[ p_{i} p_{i-1} \right]}
   \left\langle p_i + \left| \gamma^\mu \right| \mu_{i} + \right\rangle
 \right. \nonumber \\
 & & \left.
 + \frac{\left[ p_{i-1} p_{i-2} \right]}{\left[ p_{i} p_{i-1} \right] \left[ p_{i-1} p_{i-2} \right]}
   \left\langle p_{i-1} + \left| \gamma^\mu \right| \mu_{i} + \right\rangle
 \right).
\eq
Now let's use the Schouten identity for the second term,
\bq
 \left[ p_{i-1} p_{i+1} \right] \left\langle p_i + \right|
 & = &
 -
 \left[ p_{i+1} p_{i} \right] \left\langle p_{i-1} + \right|
 -
 \left[ p_{i} p_{i-1} \right] \left\langle p_{i+1} + \right|,
\eq
and the result follows:
\bq
 \sum\limits_{i=1}^n p_i^\mu
 & = & 0.
\eq
\stepcounter{exercise}
}
\es
\\
\\
\bs
{\it {\bf Exercise \theexercise}: 
Count the number of Feynman diagrams contributing to the cyclic-ordered 
amplitude $A_n^{(0)}$.
\\
\\
{\bf Solution}: Let us denote by $f(n)$ the number of Feynman diagrams contributing to the
$A_{n+1}^{(0)}$. For $f(n)$ we have the recurrence relation 
\bq
 f(n) 
 & = &
 \sum\limits_{k=1}^{j-1} f(k) f(n-k)
 + 
 \sum\limits_{k=1}^{j-2} \sum\limits_{l=k+1}^{j-1} f(k) f(l-k) f(n-l).
\eq
Compare this equation with eq.~(\ref{Berends_Giele_recursion}), $f(n)$ counts the number 
of Feynman diagrams contributing to the off-shell current $J_\mu$ with $n$ on-shell legs and
one off-shell leg.
We start with $f(1)=1$.
The first few values are easily computed
\bq
 f(2) & = & f(1) f(1) \;\; = \;\; 1,
 \nonumber \\
 f(3) & = & f(1) f(2) + f(2) f(1) + f(1) f(1) f(1) \;\; = \;\; 3,
 \nonumber \\
 f(4) & = & f(1) f(3) + f(2) f(2) + f(3) f(1) 
  + f(1) f(1) f(2) + f(1) f(2) f(1) + f(2) f(1) f(1)
 \nonumber \\
 & = & 3 + 1 + 3 + 1 + 1 + 1 \;\; = \;\; 10.
\eq
At each step we have to consider all ordered partitions of $n$ into two or three positive integers.
For $n=5$ we have the ordered partitions
\bq
 (1,4), \; (2,3), \; (3,2), \; (4,1), \; (1,1,3), \; (1,3,1), \; (3,1,1), \;
 (1,2,2), \; (2,1,2), \; (2,2,1),
\eq
giving us 
\bq
 f(5) & = & 10 + 3 + 3 + 10 + 3 + 3 + 3 + 1 + 1 + 1 \;\; = \;\; 38.
\eq
For $n=6$ we have the partitions into two numbers
\bq
 (1,5), \; (2,4), \; (3,3), \; (4,2), \; (5,1)
\eq
contributing
\bq
 38 + 10 + 9 + 10 + 38 \;\; = \;\; 105
\eq
diagrams. For the partitions into three numbers we have
the three permutations of $(1,1,4)$, six permutations of $(1,2,3)$ and one permutation of $(2,2,2)$, giving
\bq
 3 \cdot 10 + 6 \cdot 3 + 1 \cdot 1 \;\; = \;\; 49.
\eq
In total we obtain $f(5) = 105 + 49 = 154$ diagrams.
\stepcounter{exercise}
}
\es
\\
\\
\bs
{\it {\bf Exercise \theexercise}: 
Prove eq.~(\ref{off_shell_current_all_plus}) by induction.
\\
\\
{\bf Solution}: 
We start with eq.~(\ref{off_shell_current_all_plus}).
For $n=1$ we have
\bq
 J_\mu\left(p_1^+\right) 
 & = &
 \frac{\langle q- | \gamma_\mu | 1- \rangle}{\sqrt{2} \langle q 1\rangle}
 \;\; = \;\; 
 \frac{\langle q- | \gamma_\mu p\!\!\!/_{1}|q+ \rangle}{\sqrt{2} \langle q 1 \rangle \langle 1 q \rangle},
 \;
\eq
which agrees with eq.~(\ref{off_shell_current_all_plus}).
Let us now assume that eq.~(\ref{off_shell_current_all_plus}) is correct for off-shell currents 
with less than $n$ on-shell legs.
The contraction of two sub-currents vanishes due to the common choice $q$ as reference momentum.
Therefore the four-gluon vertex in the recurrence relation does not give a contribution and 
the recurrence relation reduces to
\bq
 J_\mu\left(p_1^+,...,p_n^+\right) 
 & = &
 \frac{g_{\mu\mu_3}}{p_{1,n}^2}
 \sum\limits_{j=1}^{n-1}
 \left( 2 g^{\mu_2\mu_3} p_{j+1,n}^{\mu_1} - 2 g^{\mu_3\mu_1} p_{1,j}^{\mu_2} \right) 
 J_{\mu_1}\left(p_1^+,...,p_j^+\right)
 J_{\mu_2}\left(p_{j+1}^+,...,p_n^+\right).
\eq
Using the induction hypothesis one obtains
\bq
 J_\mu\left(p_1^+,...,p_n^+\right) 
 & = &
 \left( \sqrt{2} \right)^{n-2}
 \frac{1}
      {\langle q1 \rangle \langle 1 2 \rangle ... \langle (n-1) n \rangle \langle n q \rangle}
 \frac{1}{p_{1,n}^2}
 \sum\limits_{j=1}^{n-1}
 \frac{\langle j (j+1) \rangle}{\langle j q \rangle \langle q (j+1) \rangle}
 \\
 & &
 \left[
 \langle q- | \gamma_\mu p\!\!\!/_{j+1,n}|q+ \rangle \langle q- | p\!\!\!/_{j+1,n} p\!\!\!/_{1,j}|q+ \rangle
 -
 \langle q- | \gamma_\mu p\!\!\!/_{1,j}|q+ \rangle \langle q- | p\!\!\!/_{1,j} p\!\!\!/_{j+1,n}|q+ \rangle
 \right]
 \nonumber \\
 & = &
 \left( \sqrt{2} \right)^{n-2}
 \frac{\langle q- | \gamma_\mu p\!\!\!/_{1,n}|q+ \rangle }
      {\langle q1 \rangle \langle 1 2 \rangle ... \langle (n-1) n \rangle \langle n q \rangle}
 \frac{1}{p_{1,n}^2}
 \sum\limits_{j=1}^{n-1}
 \frac{\langle j (j+1) \rangle}{\langle j q \rangle \langle q (j+1) \rangle}
 \langle q- | p\!\!\!/_{j+1,n} p\!\!\!/_{1,j}|q+ \rangle.
 \nonumber 
\eq
It remains to show that
\bq
\label{remaining_identity}
 \frac{1}{p_{1,n}^2}
 \sum\limits_{j=1}^{n-1}
 \frac{\langle j (j+1) \rangle}{\langle j q \rangle \langle q (j+1) \rangle}
 \langle q- | p\!\!\!/_{j+1,n} p\!\!\!/_{1,j}|q+ \rangle
 & = & 1.
\eq
In order to prove eq.~(\ref{remaining_identity})
we first show that
\bq
\label{eikonal_identity}
 \sum\limits_{j=k}^{l-1}
 \frac{\langle j (j+1) \rangle}{\langle j q \rangle \langle q (j+1) \rangle}
 & = & 
 \frac{\langle k l \rangle}{\langle k q \rangle \langle q l \rangle}.
\eq
The eikonal identity in eq.~(\ref{eikonal_identity}) is trivially true for $l-k=1$.
Assuming that the identity is true for sums with $(l-k-1)$ summands, we show that it is also true for $(l-k)$ summands:
\bq
 \sum\limits_{j=k}^{l-1}
 \frac{\langle j (j+1) \rangle}{\langle j q \rangle \langle q (j+1) \rangle}
 & = & 
 \sum\limits_{j=k}^{l-2}
 \frac{\langle j (j+1) \rangle}{\langle j q \rangle \langle q (j+1) \rangle}
 +
 \frac{\langle (l-1) l \rangle}{\langle (l-1) q \rangle \langle q l \rangle}
 \;\; = \;\;
 \frac{\langle k (l-1) \rangle}{\langle k q \rangle \langle q (l-1) \rangle}
 +
 \frac{\langle (l-1) l \rangle}{\langle (l-1) q \rangle \langle q l \rangle}
 \nonumber \\
 & = &
 \frac{\langle k (l-1) \rangle \langle q l \rangle - \langle k q \rangle \langle (l-1) l \rangle}{\langle k q \rangle \langle q (l-1) \rangle \langle q l \rangle}
 \;\; = \;\;
 \frac{\langle k l\rangle}{\langle k q \rangle \langle q l \rangle}.
\eq
The last step used the Schouten identity.
Let us now consider
\bq
 \sum\limits_{j=1}^{n-1}
 \frac{\langle j (j+1) \rangle}{\langle j q \rangle \langle q (j+1) \rangle}
 \langle q- | p\!\!\!/_{j+1,n} 
 & = &
 \sum\limits_{j=1}^{n-1}
 \sum\limits_{k=j+1}^{n}
 \frac{\langle j (j+1) \rangle}{\langle j q \rangle \langle q (j+1) \rangle}
 \langle q- | p\!\!\!/_k 
 \;\; = \;\;
 \sum\limits_{k=2}^{n}
 \sum\limits_{j=1}^{k-1}
 \frac{\langle j (j+1) \rangle}{\langle j q \rangle \langle q (j+1) \rangle}
 \langle q- | p\!\!\!/_k 
 \nonumber \\
 & = &
 \sum\limits_{k=2}^{n}
 \frac{\langle 1 k \rangle}{\langle 1 q \rangle \langle q k \rangle}
 \langle q- | p\!\!\!/_k 
 \;\; = \;\;
 \frac{\langle 1- | p\!\!\!/_{1,n}}{\langle 1 q \rangle}.
\eq
We may now prove eq.~(\ref{remaining_identity}):
\bq
\lefteqn{
 \frac{1}{p_{1,n}^2}
 \sum\limits_{j=1}^{n-1}
 \frac{\langle j (j+1) \rangle}{\langle j q \rangle \langle q (j+1) \rangle}
 \langle q- | p\!\!\!/_{j+1,n} p\!\!\!/_{1,j}|q+ \rangle
 = } & & \\
 & = &
 \frac{1}{p_{1,n}^2}
 \sum\limits_{j=1}^{n-1}
 \frac{\langle j (j+1) \rangle}{\langle j q \rangle \langle q (j+1) \rangle}
 \langle q- | p\!\!\!/_{j+1,n} p\!\!\!/_{1,n}|q+ \rangle
 \;\; = \;\;
 \frac{1}{p_{1,n}^2}
 \frac{\langle 1- | p\!\!\!/_{1,n} p\!\!\!/_{1,n}|q+ \rangle}{\langle 1 q \rangle}
 \;\; = \;\;
 \frac{p_{1,n}^2}{p_{1,n}^2}
 \frac{\langle 1 q \rangle}{\langle 1 q \rangle}
 \;\; = \;\;
 1. \nonumber
\eq
\stepcounter{exercise}
}
\es
\\
\\
\bs
{\it {\bf Exercise \theexercise}: 
Derive eq.~(\ref{solution_canonical_trafo}) together with eq.~(\ref{coefficient_functions}) from eq.~(\ref{eq_canonical_trafo}).
\\
\\
{\bf Solution}: 
We would like to solve the integro-differential equation, given in eq.~(\ref{eq_canonical_trafo}):
\bq
\label{eq_canonical_trafo_exercise}
 \omega A_{\bot}^a(\vec{x}) - g f^{abc} \left( \zeta A_{\bot}^b(\vec{x}) \right) A_{\bot}^c(\vec{x}) 
 =  
 \int d^3y \frac{\delta A_{\bot}^a(\vec{x})}{\delta \tilde{A}_{\bot}^b(\vec{y})} 
 \omega_y \tilde{A}_{\bot}^b(\vec{y}).
\eq
We make the ansatz
\bq
 A_{\bot}^a\left(\vec{x}\right) & = &
 \tilde{A}_{\bot}^a\left(\vec{x}\right)
 +
 \sum\limits_{n=2}^\infty
 2 \; \mathrm{Tr}\left( T^a T^{a_1} ... T^{a_n} \right)
 \int d^3x_1 ... d^3x_n 
 \Upsilon\left( \vec{x}, \vec{x_1}, ..., \vec{x}_n \right)
 \tilde{A}_{\bot}^{a_1}\left(\vec{x}_1\right)
 ...
 \tilde{A}_{\bot}^{a_n}\left(\vec{x}_n\right).
 \nonumber
\eq
We introduce the Fourier transforms
\bq
 \tilde{A}_{\bot}^{a}(\vec{x})
 & = & 
 \int \frac{d^3p}{(2\pi)^3} e^{-i \vec{p} \cdot \vec{x}}  \tilde{A}_{\bot}^{a}(\vec{p}),
 \\
 \Upsilon(\vec{x}, \vec{x}_1, ..., \vec{x}_n)
 & = &
 \int \frac{d^3p_1}{(2\pi)^3} ... \frac{d^3p_n}{(2\pi)^3}
 e^{-i \vec{p}_1 \cdot (\vec{x}-\vec{x}_1) - ... -i \vec{p}_n \cdot (\vec{x}-\vec{x}_n)}
 \Upsilon(\vec{p}_1, ..., \vec{p}_n).
 \nonumber
\eq
Expressed in terms of the Fourier transforms we obtain:
\bq
 A_{\bot}^a\left(\vec{x}\right)
 & = &
 \sum\limits_{n=1}^\infty 
 2 \; \mathrm{Tr} \left( T^a T^{a_1} ... T^{a_n} \right)
 \int \frac{d^3p_1}{(2\pi)^3} ... \frac{d^3p_n}{(2\pi)^3}
 e^{-i (\vec{p}_1 + ... + \vec{p}_n ) \cdot \vec{x}} 
 \nonumber \\
 & &
 \Upsilon\left(\vec{p}_1,...,\vec{p}_n\right)
 \tilde{A}_{\bot}^{a_1}\left(\vec{p}_1\right) ... \tilde{A}_{\bot}^{a_n}\left(\vec{p}_n\right),
\eq
with $\Upsilon(\vec{p})=1$.
The functional derivative is calculated to
\bq
 \frac{\delta A_{\bot}^a(\vec{x})}{\delta \tilde{A}_{\bot}^b(\vec{y})}
 & = &
 \sum\limits_{n=1}^\infty 
 \sum\limits_{r=1}^n
 2 \; \mathrm{Tr} \left( T^a T^{a_1} ... T^{a_{r-1}} T^b T^{a_{r+1}} ... T^{a_n} \right)
 \int \frac{d^3p_1}{(2\pi)^3} ... \frac{d^3p_n}{(2\pi)^3}
 e^{-i (\vec{p}_1 + ... + \vec{p}_n ) \cdot \vec{x} + i \vec{p}_r \cdot \vec{y}} 
 \nonumber \\
 & &
 \Upsilon\left(\vec{p}_1,...,\vec{p}_n\right)
 \tilde{A}_{\bot}^{a_1}\left(\vec{p}_1\right) 
 ... 
 \tilde{A}_{\bot}^{a_{r-1}}\left(\vec{p}_{r-1}\right) 
 \tilde{A}_{\bot}^{a_{r+1}}\left(\vec{p}_{r+1}\right) 
 ...
 \tilde{A}_{\bot}^{a_n}\left(\vec{p}_n\right).
\eq
We then plug these expressions into eq.~(\ref{eq_canonical_trafo_exercise}).
The coefficient of each trace $\mathrm{Tr} \left( T^a T^{a_1} ... T^{a_n} \right)$ has to vanish separately.
This leads to the following equation
\bq
\label{eq_coeff_Upsilon}
\lefteqn{
 \left( \omega_{p_1} + ... + \omega_{p_n} \right) \Upsilon\left(\vec{p}_1,...,\vec{p}_n\right)
 = } & &
 \\
 & &
 \omega_{p_1+...+p_n} \Upsilon\left(\vec{p}_1,...,\vec{p}_n\right)
 + 
 i g \sum\limits_{r=1}^{n-1} \left( \zeta_{p_1+...+p_r} - \zeta_{p_{r+1}+...+p_n} \right)
 \Upsilon\left(\vec{p}_1,...,\vec{p}_r\right) \Upsilon\left(\vec{p}_{r+1},...,\vec{p}_n\right).
 \nonumber
\eq
In order to simplify the notation we have set
\bq
 \omega_{p} =  
 e^{i \vec{p} \vec{x} } \omega e^{-i \vec{p} \vec{x} }
 = - i \frac{p^{\bot\ast} p^\bot}{p^-},
 & &
 \zeta_{p} = 
 e^{i \vec{p} \vec{x} } \zeta e^{-i \vec{p} \vec{x} }
 = \frac{p^\bot}{p^-}.
\eq
Eq.~(\ref{eq_coeff_Upsilon})
is a recursion relation for the coefficient functions $\Upsilon\left(\vec{p}_1,...,\vec{p}_n\right)$:
\bq
 \Upsilon\left(\vec{p}_1,...,\vec{p}_n\right) 
 & = &
 i g
 \sum\limits_{r=1}^{n-1} 
 \frac{\zeta_{p_1+...+p_r} - \zeta_{p_{r+1}+...+p_n}}{\omega_{p_1} + ... + \omega_{p_n} - \omega_{p_1+...+p_n}}
 \Upsilon\left(\vec{p}_1,...,\vec{p}_r\right) \Upsilon\left(\vec{p}_{r+1},...,\vec{p}_n\right),
\eq
with $\Upsilon(\vec{p})=1$.
This recursion relation has the solution:
\bq
 \Upsilon\left(\vec{p}_1,...,\vec{p}_n\right) 
 & = &
 \frac{\left(  \sqrt{2} g \right)^{n-1}}{\langle p_1 p_2 \rangle ... \langle p_{n-1} p_n \rangle} \frac{p_1^-+...+p_n^-}{\sqrt{p_1^- p_n^-}},
\eq
which is easily verified by inserting the solution into the recursion relation.
We note that the coefficients $\Upsilon$ satisfy a decoupling identity:
\bq
 \Upsilon\left(\vec{p}_a,\vec{p}_1,...,\vec{p}_n\right)
 + 
 \Upsilon\left(\vec{p}_1,\vec{p}_a,\vec{p}_2,...,\vec{p}_n\right)
 +
 ...
 +
 \Upsilon\left(\vec{p}_1,...,\vec{p}_n,\vec{p}_a\right)
 & = & 0.
\eq
In addition we have to express the ``old'' conjugated fields $A^a_{\bot\ast}\left(\vec{x}\right)$
in terms of the ``new'' fields.
The relevant equation to be solved is given in eq.~(\ref{new_momenta}):
\bq
\label{new_momenta_exercise}
 \partial_- \tilde{A}_{\bot\ast}^a(\vec{x})
 & = & \int d^3y \; \frac{\delta A_{\bot}^b(\vec{y})}{\delta \tilde{A}_{\bot}^a(\vec{x})} 
                 \; \partial_- A_{\bot\ast}^b(\vec{y}).
\eq
We make the ansatz
\bq
 A_{\bot\ast}^a\left(\vec{x}\right) 
 & = &
 \sum\limits_{n=1}^\infty
 \sum\limits_{r=1}^n
 2 \; \mathrm{Tr}\left( T^a T^{a_1} ... T^{a_n} \right)
 \int d^3x_1 ... d^3x_n 
 \\
 & &
 \Xi_r\left( \vec{x}, \vec{x_1}, ..., \vec{x}_n \right)
 \tilde{A}_{\bot}^{a_1}\left(\vec{x}_1\right)
 ...
 \tilde{A}_{\bot}^{a_{r-1}}\left(\vec{x}_{r-1}\right)
 \tilde{A}_{\bot\ast}^{a_r}\left(\vec{x}_r\right)
 \tilde{A}_{\bot}^{a_{r+1}}\left(\vec{x}_{r+1}\right)
 ...
 \tilde{A}_{\bot}^{a_n}\left(\vec{x}_n\right).
\eq
In Fourier space we have
\bq
 A_{\bot\ast}^a\left(\vec{x}\right)
 & = & 
 \sum\limits_{n=1}^\infty 
 \sum\limits_{r=1}^n
 2 \; \mathrm{Tr} \left( T^a T^{a_1} ... T^{a_n} \right)
 \int \frac{d^3p_1}{(2\pi)^3} ... \frac{d^3p_n}{(2\pi)^3}
 e^{-i (\vec{p}_1 + ... + \vec{p}_n ) \cdot \vec{x}} 
 \\
 & &
 \Xi_r\left(\vec{p}_1,...,\vec{p}_n\right)
 \tilde{A}_{\bot}^{a_1}\left(\vec{p}_1\right) 
 ... 
 \tilde{A}_{\bot}^{a_{r-1}}\left(\vec{p}_{r-1}\right) 
 \tilde{A}_{\bot\ast}^{a_r}\left(\vec{p}_r\right) 
 \tilde{A}_{\bot}^{a_{r+1}}\left(\vec{p}_{r+1}\right) 
 ... 
 \tilde{A}_{\bot}^{a_n}\left(\vec{p}_n\right),
\eq
with $\Xi_1\left(\vec{p}\right)=1$.
We then insert this ansatz into eq~(\ref{new_momenta_exercise}).
For $n>1$ we obtain the equation
\bq
\lefteqn{ 
 0 =
 } & & \\
 & &
 \sum\limits_{i_1=1}^r \sum\limits_{i_2=r+1}^n
 \left( p_{i_1}^- + ... + p_{i_2-1}^- \right)
 \Xi_{r-i_1+1}\left(\vec{p}_{i_1},...,\vec{p}_{i_2-1}\right)
 \Upsilon\left(\vec{p}_{i_2},...\vec{p}_{n-1},-\sum\limits_{i=1}^{n-1}\vec{p}_i,\vec{p}_1,...,\vec{p}_{i_1-1}\right).
 \nonumber 
\eq
This is again a recursion relation for the coefficients $\Xi_r\left(\vec{p}_1,...,\vec{p}_n\right)$. We can rewrite
this equation as
\bq
\lefteqn{
 \Xi_r\left(\vec{p}_1,...,\vec{p}_n\right)
 = } & &
 \nonumber \\
 & & 
 - \sum\limits_{i_1=2}^r 
   \sum\limits_{i_2=r+1}^{n+1}
    \frac{p_{i_1}^-+...+p_{i_2-1}^-}{p_1^-+...+p_n^-}
    \Xi_{r-i_1+1}\left(\vec{p}_{i_1},...,\vec{p}_{i_2-1}\right)
    \Upsilon\left(\vec{p}_{i_2},...\vec{p}_{n},-\sum\limits_{i=1}^{n}\vec{p}_i,\vec{p}_1,...,\vec{p}_{i_1-1}\right)
 \nonumber \\
 & &
 - \sum\limits_{i_2=r+1}^{n}
    \frac{p_{1}^-+...+p_{i_2-1}^-}{p_1^-+...+p_n^-}
    \Xi_{r}\left(\vec{p}_{1},...,\vec{p}_{i_2-1}\right)
    \Upsilon\left(\vec{p}_{i_2},...\vec{p}_{n},-\sum\limits_{i=1}^{n}\vec{p}_i\right).
\eq
The solution is given by
\bq
 \Xi_r\left(\vec{p}_1,...,\vec{p}_n\right) 
 & = & \left( \frac{p_r^-}{p_1^-+...+p_n^-} \right)^2 \Upsilon\left(\vec{p}_1,...,\vec{p}_n\right).
\eq
\stepcounter{exercise}
}
\es
\\
\\
\bs
{\it {\bf Exercise \theexercise}: 
Consider the MHV amplitude $A_n^{(0)}(...,\hat{p}_i^+,...,\hat{p}_j^-,...,p_k^-,...)$, where the
particles $j$ and $k$ have negative helicity, while all other particles have positive helicity.
Consider the shift of eq.~(\ref{shift_12}), which deforms the momenta of the particles $i$ and $j$.
What is the large-$z$ behaviour of the amplitude?
\\
\\
{\bf Solution}: 
The shift of eq.~(\ref{shift_12}) transforms the spinors
\bq
 \hat{p}_A^i = p_A^i  - z p_A^j,
 & &
 \hat{p}_{\dot{B}}^j = p_{\dot{B}}^j + z p_{\dot{B}}^i,
\eq
all other spinors are left untouched.
The MHV amplitude is given by
\bq
 A_n^{(0)}\left(...,\hat{p}_i^+,...,\hat{p}_j^-,...,p_k^-,...\right)
 & = &
 i \left( \sqrt{2} \right)^{n-2} 
 \frac{\langle j k \rangle^4}{\langle 1 2 \rangle ... \langle (i-1) \hat{i} \rangle \langle \hat{i} (i+1) \rangle ... \langle n 1 \rangle},
\eq
and involves only the holomorphic spinors. Therefore only the shift $\hat{p}_A^i = p_A^i  - z p_A^j$ is relevant. This spinor appears twice in the denominator.
We now have to distinguish two cases. 
The first case is given when the particles $i$ and $j$ are adjacent, i.e. $j=i-1$ or $j=i+1$.
Let's assume $j=i-1$. Then the spinor product
\bq
 \langle (i-1) \hat{i} \rangle
 & = &
 \langle j i \rangle - z \langle j j \rangle
 \;\; = \;\;
 \langle j i \rangle,
\eq
is $z$-independent, while the other one growths linearly with $z$:
\bq
 \langle \hat{i} (i+1) \rangle
 & = & 
 \langle i (i+1) \rangle - z \langle j (i+1) \rangle.
\eq
Therefore the amplitude scales like $1/z$ in this case.
\\
\\
If the particles $i$ and $j$ are not adjacent, the amplitude behaves as $1/z^2$ for large $z$.
\stepcounter{exercise}
}
\es
\\
\\
\bs
{\it {\bf Exercise \theexercise}: 
Show that a term of the form
\bq
\label{three_point_ansatz}
 \left\langle p_1 p_2 \right\rangle^{\nu_3}
 \left\langle p_2 p_3 \right\rangle^{\nu_1}
 \left\langle p_3 p_1 \right\rangle^{\nu_2}
\eq
and contributing to ${\mathcal M}_3^{(0)}(1^{h_1},2^{h_2},3^{h_3})$ must have the exponents as in eq.~(\ref{general_holomorphic_three_point}).
\\
\\
{\bf Solution}: 
Let us consider little group scaling. If we scale $p_A^1 \rightarrow \lambda p_A^1$, the function in eq.~(\ref{three_point_ansatz}) scales as $\lambda^{\nu_2+\nu_3}$.
The amplitude and every term in the amplitude scales as $\lambda^{-2h_1}$.
This gives us the equation
\bq 
 \nu_2 + \nu_3 & = & - 2 h_1.
\eq
Repeating the argument for the other particles, we obtain the system of equations
\begin{alignat}{4}
       &&  \nu_2 && +\nu_3 && = -2h_1,
 \nonumber \\
 \nu_1 &&        && +\nu_3 && = -2h_2,
 \nonumber \\
 \nu_1 && +\nu_2 &&        && = -2h_3.
\end{alignat}
Solving this linear system of equation yields
\bq
 \nu_1 = h_1 - h_2 - h_3,
 \;\;\;\;
 \nu_2 = h_2 - h_3 - h_1,
 \;\;\;\;
 \nu_3 = h_3 - h_1 - h_2.
\eq
\stepcounter{exercise}
}
\es
\\
\\
\bs
{\it {\bf Exercise \theexercise}: 
Prove the Parke-Taylor formula for two non-adjacent negative helicities from the formula of the adjacent case and
by using on-shell recursion relations.
\\
\\
{\bf Solution}: 
Let us denote the two negative helicity legs $1$ and $j$. 
The adjacent case is given by $j=n$ or $j=2$ and we may assume that the Parke-Taylor formula is already proven in this case.
We will prove the general non-adjacent case by induction in the variable $n'=n-j$.
The induction start is given by the adjacent case $n'=0$.
Let us now consider the induction step. 
We may assume that the Parke-Taylor formula is correct for $n'-1$ and we have to show that it holds for $n'$.
We consider
\bq
 A_n^{(0)}\left(1^-,2^+,...,j^-,...,n^+\right).
\eq
For the helicity configuration $(1^-,n^+)$ we use the shift
\bq
 \hat{p}_{\dot{B}}^1 = p_{\dot{B}}^1 - z p_{\dot{B}}^n,
 & &
 \hat{p}_A^n = p_A^n + z p_A^1,
\eq
In the on-shell recursion relation we obtain sub-amplitudes with one negative helicity leg.
These vanish except for the case with three external legs.
This simplifies the recursion relation significantly.
We distinguish the cases $j=n-1$ and $j<n-1$.
In the former case we find from the on-shell recursion relations
\bq
\label{case_j_eq_n_1}
 A_n^{(0)}\left(1^-,2^+,...,(n-1)^-,n^+\right)
 & = &
 A_n^{(0)}\left(\hat{1}^-,2^+,...,(n-2)^+,(-\hat{p}_{1,n-2})^- | \hat{p}_{1,n-2}^+,(n-1)^-,\hat{n}^+\right)
 \nonumber \\
 & &
 +
 A_n^{(0)}\left(\hat{1}^-,2^+, (-\hat{p}_{1,2})^+ | \hat{p}_{1,2}^-, 3^+,...,(n-1)^-,\hat{n}^+\right),
\eq
where we used the notation of eq.~(\ref{notation_factorisation_channels}).
In the case $j<n-1$ the on-shell recursion relations read
\bq
\label{case_j_neq_n_1}
 A_n^{(0)}\left(1^-,2^+,...,j^-,...,n^+\right)
 & = &
 A_n^{(0)}\left(\hat{1}^-,...,j^-,...,(-\hat{p}_{1,n-2})^+ | \hat{p}_{1,n-2}^-,(n-1)^+,\hat{n}^+\right)
 \nonumber \\
 & &
 +
 A_n^{(0)}\left(\hat{1}^-,2^+, (-\hat{p}_{1,2})^+ | \hat{p}_{1,2}^-,...,j^-,...,\hat{n}^+\right)
\eq
In both case the $z$-values of the 
two factorisation channels $p_{1,n-2}^2=2 p_{n-1} p_n$ and $p_{1,2}^2=2p_1 p_2$ are
\bq 
\label{def_z_shift_values}
 z_{1,n-2} 
 \;\; = \;\;
 -\frac{2 p_{n-1} p_n}{\left\langle n+ \left| p\!\!\!/_{n-1} \right| 1+ \right\rangle}
 \;\; = \;\;
 \frac{\left\langle(n-1) n\right\rangle}{\left\langle1 (n-1)\right\rangle},
 & &
 z_{1,2} 
 \;\; = \;\;
 \frac{2 p_{1} p_2}{\left\langle n+ \left| p\!\!\!/_{2} \right| 1+ \right\rangle}
 \;\; = \;\;
 - \frac{\left[ 1 2 \right]}{\left[ 2 n \right]}.
\eq
From eq.~(\ref{case_j_eq_n_1}) we obtain
\bq
\label{case_j_eq_n_1_v1}
\lefteqn{
 A_n^{(0)}\left(1^-,2^+,...,(n-1)^-,n^+\right)
 = } & & 
 \nonumber \\
 & &
 \left.
 i \left( \sqrt{2} \right)^{n-2} 
 \frac{\langle \hat{1} \hat{p}_{1,n-2} \rangle^4}{\langle \hat{1} 2 \rangle ... \langle (n-2) \hat{p}_{1,n-2}\rangle \langle \hat{p}_{1,n-2} \hat{1} \rangle}
 \frac{1}{2 p_{n-1} p_n}
 \frac{[\hat{p}_{1,n-2} \hat{n}]^4}{[\hat{n} (n-1)] [ (n-1) \hat{p}_{1,n-2}][\hat{p}_{1,n-2}  \hat{n}]}
 \right|_{z=z_{1,n-2}}
 \nonumber \\
 & &
 +
 \left.
 i \left( \sqrt{2} \right)^{n-2} 
 \frac{[\hat{p}_{1,2} 2]^4}{[\hat{p}_{1,2} 2] [2 \hat{1}] [\hat{1} \hat{p}_{1,2}] }
 \frac{1}{2 p_1 p_2}
 \frac{\langle \hat{p}_{1,2} (n-1) \rangle^4}{\langle \hat{p}_{1,2} 3 \rangle ... \langle (n-1) \hat{n} \rangle \langle \hat{n} \hat{p}_{1,2} \rangle}
 \right|_{z=z_{1,2}}.
\eq
From eq.~(\ref{case_j_neq_n_1}) we obtain
\bq
\label{case_j_neq_n_1_v1}
\lefteqn{
 A_n^{(0)}\left(1^-,2^+,...,j^-,...,n^+\right)
 = } & & 
 \nonumber \\
 & &
 \left.
 i \left( \sqrt{2} \right)^{n-2} 
 \frac{\langle \hat{1} j \rangle^4}{\langle \hat{1} 2 \rangle ... \langle (n-2) \hat{p}_{1,n-2}\rangle \langle \hat{p}_{1,n-2} \hat{1} \rangle}
 \frac{1}{2 p_{n-1} p_n}
 \frac{[\hat{n} (n-1)]^4}{[\hat{n} (n-1)] [ (n-1) \hat{p}_{1,n-2}][\hat{p}_{1,n-2}  \hat{n}]}
 \right|_{z=z_{1,n-2}}
 \nonumber \\
 & &
 +
 \left.
 i \left( \sqrt{2} \right)^{n-2} 
 \frac{[\hat{p}_{1,2} 2]^4}{[\hat{p}_{1,2} 2] [2 \hat{1}] [\hat{1} \hat{p}_{1,2}] }
 \frac{1}{2 p_1 p_2}
 \frac{\langle \hat{p}_{1,2} j \rangle^4}{\langle \hat{p}_{1,2} 3 \rangle ... \langle (n-1) \hat{n} \rangle \langle \hat{n} \hat{p}_{1,2} \rangle}
 \right|_{z=z_{1,2}}.
\eq
It is easy to see that eq.~(\ref{case_j_neq_n_1_v1}) reduces for $j=n-1$ to eq.~(\ref{case_j_eq_n_1_v1}),
therefore it is sufficient to consider in the following only eq.~(\ref{case_j_neq_n_1_v1}).
Let us look at the second term of eq.~(\ref{case_j_neq_n_1_v1}).
With $z_{1,2}$ given in eq.~(\ref{def_z_shift_values}) we find
\bq
 \left. [\hat{p}_{1,2} 2] \right|_{z=z_{1,2}}
 \;\; = \;\; 
 \left. [2 \hat{1}] \right|_{z=z_{1,2}}
 \;\; = \;\; 
 \left. [\hat{1} \hat{p}_{1,2}] \right|_{z=z_{1,2}}
 \;\; = \;\; 0,
\eq
and therefore the three-point amplitude $A_3^{(0)}(\hat{1}^-,2^+, (-\hat{p}_{1,2})^+)$ contained
in this term vanishes.
Thus the second term of eq.~(\ref{case_j_neq_n_1_v1}) does not give a contribution.
Let us now look at the first term of eq.~(\ref{case_j_neq_n_1_v1}).
We have $|\hat{1}+\rangle=|1+\rangle$ and $\langle \hat{n}+ |=\langle n+ |$.
Therefore we obtain
\bq
\lefteqn{
 A_n^{(0)}\left(1^-,2^+,...,j^-,...,n^+\right)
 = } & &
 \nonumber \\
 & &
 i \left( \sqrt{2} \right)^{n-2} \frac{\langle 1 j \rangle^4}{\langle 1 2 \rangle ... \langle n 1 \rangle}
 \left(
 \left.
 \frac{\langle (n-2) (n-1) \rangle \langle n 1 \rangle }{\langle (n-2) \hat{p}_{1,n-2}\rangle \langle \hat{p}_{1,n-2} 1 \rangle}
 \frac{[n (n-1)]^2}{[ (n-1) \hat{p}_{1,n-2}][\hat{p}_{1,n-2} n]}
 \right|_{z=z_{1,n-2}} 
 \right).
\eq
It remains to show that the term in the bracket equals one:
\bq
\lefteqn{
 \left.
 \frac{\langle (n-2) (n-1) \rangle \langle n 1 \rangle }{\langle (n-2) \hat{p}_{1,n-2}\rangle \langle \hat{p}_{1,n-2} 1 \rangle}
 \frac{[n (n-1)]^2}{[ (n-1) \hat{p}_{1,n-2}][\hat{p}_{1,n-2} n]}
 \right|_{z=z_{1,n-2}} 
 = } & &
 \nonumber \\
 & = &
 \left.
 \frac{\langle (n-2) (n-1) \rangle \langle n 1 \rangle [n (n-1)]^2}
      {\langle (n-2) - | \hat{p}\!\!\!/_{1,n-2} | n- \rangle  
       \langle 1- | \hat{p}\!\!\!/_{1,n-2} | (n-1) - \rangle}
 \right|_{z=z_{1,n-2}} 
 \nonumber \\
 & = &
 \frac{\langle (n-2) (n-1) \rangle \langle n 1 \rangle [n (n-1)]^2}
      {\langle (n-2) - | p\!\!\!/_{1,n-2} | n- \rangle  
       \langle 1- | p\!\!\!/_{1,n-2} | (n-1) - \rangle}
 \;\; = \;\;
 \frac{\langle (n-2) (n-1) \rangle \langle n 1 \rangle [n (n-1)]^2}
      {\langle (n-2) - | p\!\!\!/_{n-1} | n- \rangle  
       \langle 1- | p\!\!\!/_{n} | (n-1) - \rangle}
 \nonumber \\
 & = & 1. 
\eq
\stepcounter{exercise}
}
\es
\\
\\
\bs
{\it {\bf Exercise \theexercise}: 
Re-derive the expression for the MHV amplitude $A_3^{(0)}\left(1^-,2^-,3^+\right)$
and the anti-MHV amplitude $A_3^{(0)}\left(1^+,2^+,3^-\right)$ given in eq.~(\ref{three_point_MHV_and_MHVbar})
from eq.~(\ref{three_point_SYM_MHV}) and eq.~(\ref{three_point_SYM_MHVbar}), respectively.
\\
\\
{\bf Solution}: 
Let us start with the anti-MHV amplitude $A_3^{(0)}\left(1^+,2^+,3^-\right)$.
In order to project out this helicity configuation from the super-Yang-Mills amplitude
\bq
 A_3^{(0)}\left(1,2,3\right)
 & = &
 i \sqrt{2}
 \frac{\delta^4\left( [ 3 2 ] \eta_I^1 + [ 1 3 ] \eta_I^2 + [ 2 1 ] \eta_I^3 \right) }{[3 2] [2 1] [1 3]},
\eq
we set $\eta_I^1=\eta_I^2=0$ and apply the differential operator
\bq
 \frac{\partial}{\partial \eta_1^3}
 \frac{\partial}{\partial \eta_2^3}
 \frac{\partial}{\partial \eta_3^3}
 \frac{\partial}{\partial \eta_4^3}.
\eq
We have
\bq
 \frac{\partial}{\partial \eta_1^3}
 \frac{\partial}{\partial \eta_2^3}
 \frac{\partial}{\partial \eta_3^3}
 \frac{\partial}{\partial \eta_4^3}
 \delta\left( [ 2 1 ] \eta_1^3 \right)
 \delta\left( [ 2 1 ] \eta_2^3 \right)
 \delta\left( [ 2 1 ] \eta_3^3 \right)
 \delta\left( [ 2 1 ] \eta_4^3 \right)
 & = &
 [ 2 1 ]^4,
\eq
and therefore
\bq
 A_3^{(0)}\left(1^+,2^+,3^-\right)
 & = &
 i \sqrt{2}
 \frac{[ 2 1 ]^4}{[3 2] [2 1] [1 3]}.
\eq
Let us now consider the MHV amplitude $A_3^{(0)}\left(1^-,2^-,3^+\right)$.
The super-Yang-Mills amplitude is 
\bq
 A_3^{(0)}\left(1,2,3\right)
 & = &
 i \sqrt{2}
 \frac{\delta^8\left( p_A^1 \eta_I^1 + p_A^2 \eta_I^2 + p_A^3 \eta_I^3 \right)}{\langle 1 2 \rangle \langle 2 3  \rangle \langle 3 1  \rangle},
\eq
and in order to project out the helicity configuration $(1^-,2^-,3^+)$ we set $\eta_I^3=0$ and we work out
\bq
\lefteqn{
 \frac{\partial}{\partial \eta_1^1}
 \frac{\partial}{\partial \eta_2^1}
 \frac{\partial}{\partial \eta_3^1}
 \frac{\partial}{\partial \eta_4^1}
 \frac{\partial}{\partial \eta_1^2}
 \frac{\partial}{\partial \eta_2^2}
 \frac{\partial}{\partial \eta_3^2}
 \frac{\partial}{\partial \eta_4^2}
 \delta^8\left( p_A^1 \eta_I^1 + p_A^2 \eta_I^2 \right)
 = } & & \nonumber \\
 & = &
 \left[ 
 \frac{\partial}{\partial \eta_1^1}
 \frac{\partial}{\partial \eta_1^2}
 \delta\left( p_1^1 \eta_1^1 + p_1^2 \eta_1^2 \right)
 \delta\left( p_2^1 \eta_1^1 + p_2^2 \eta_1^2 \right)
 \right]^4
 \;\; = \;\;
 \left( p_2^1 p_1^2 - p_1^1 p_2^2 \right)^4
 \;\; = \;\;
 \left\langle 1 2 \right\rangle^4.
\eq
Therefore
\bq
 A_3^{(0)}\left(1^-,2^-,3^+\right)
 & = &
 i \sqrt{2}
 \frac{\left\langle 1 2 \right\rangle^4}{\langle 1 2 \rangle \langle 2 3  \rangle \langle 3 1  \rangle}.
\eq
\stepcounter{exercise}
}
\es
\\
\\
\bs
{\it {\bf Exercise \theexercise}: 
Show that for $k=1$ the form $\omega$ given in eq.~(\ref{def_omega_general_k}) reduces to the expression
given in eq.~(\ref{def_omega_k_eq_1}).
\\
\\
{\bf Solution}: 
For  $k=1$ we write for an element $C \in \mathrm{Gr}_{1,n}({\mathbb R}) = {\mathbb R} {\mathbb P}^{n-1}$
\bq
 C & = &
 \left( \begin{array}{ccc}
 C_{1 1} & ... & C_{1 n} \\
 \end{array} 
 \right)
 \;\; = \;\;
 \left( \begin{array}{ccc}
 x_{1} & ... & x_{n} \\
 \end{array} 
 \right).
\eq
The form $\omega$, as defined by eq.~(\ref{def_omega_general_k}), reads
\bq
 \omega 
 & = &
 \left\langle C^1, \left( dC^1 \right)^{(n-1)} \right\rangle
 \;\; = \;\;
 \eps^{a_1 a_2 ... a_n}
 C_{1 a_1} \; dC_{1,a_2} \wedge ... \wedge dC_{1,a_n}
 \nonumber \\
 & = &
 \eps^{a_1 a_2 ... a_n}
 x_{a_1} \; dx_{a_2} \wedge ... \wedge dx_{a_n}.
\eq
Consider now the case $a_1=1$. The sum over the $(n-1)!$ permutations of the remaining indices
$(2,...,n)$ gives
\bq
 \eps^{1 a_2 ... a_n}
 x_{1} dx_{a_2} \wedge ... \wedge dx_{a_n}.
 & = &
 \left(n-1\right)! \; x_1 \; dx_2 \wedge ... \wedge dx_n,
\eq
the minus signs for odd permutations from the totally anti-symmetric tensor $\eps^{1 a_2 ... a_n}$ compensate
exactly the minus signs needed to bring the differentials $dx_{a_2} \wedge ... \wedge dx_{a_n}$ into the order
$dx_2 \wedge ... \wedge dx_n$.
For $a_2=2$ we obtain
\bq
 \eps^{2 a_2 ... a_n}
 x_{2} dx_{a_2} \wedge ... \wedge dx_{a_n}.
 & = &
 - \left(n-1\right)! \; x_2 \; dx_1 \wedge dx_3 \wedge ... \wedge dx_n,
\eq
where the additional minus sign comes from
\bq
 \eps^{2 1 3 ... n} & = & - \eps^{1 2 3 ... n}.
\eq
In general we find for $a_1=j$
\bq
 \eps^{j a_2 ... a_n}
 x_{j} dx_{a_2} \wedge ... \wedge dx_{a_n}.
 & = &
 \left(-1\right)^{j-1} \left(n-1\right)! \; x_j \; dx_1 \wedge ... \wedge \widehat{dx_j} \wedge ... \wedge dx_n.
\eq
Thus we have
\bq
 \omega & = & 
 \left(n-1\right)!
 \sum\limits_{j=1}^n (-1)^{j-1}
  \; x_j \; dx_1 \wedge ... \wedge \widehat{dx_j} \wedge ... \wedge dx_n,
\eq
which is the expression given in eq.~(\ref{def_omega_k_eq_1}).
\stepcounter{exercise}
}
\es
\\
\\
\bs
{\it {\bf Exercise \theexercise}: 
Derive the expression in eq.~(\ref{def_five_bracket_2}) from eq.~(\ref{def_five_bracket}).
\\
\\
{\bf Solution}: 
We start from eq.~(\ref{def_five_bracket}):
\bq
 \left[ Z^1, Z^2, Z^3, Z^4, Z^5 \right]
 =
 \frac{1}{4!}
 \int d^4\phi
 \frac{\left\langle 1, 2, 3, 4, 5 \right\rangle^4}
      {
       \left\langle 0, 1, 2, 3, 4 \right\rangle
       \left\langle 0, 2, 3, 4, 5 \right\rangle
       \left\langle 0, 3, 4, 5, 1 \right\rangle
       \left\langle 0, 4, 5, 1, 2 \right\rangle
       \left\langle 0, 5, 1, 2, 3 \right\rangle
      }
 \nonumber
\eq
Let us look at the denominator first: 
Since $I_\alpha^{\; 0}=0$ for $\alpha=1,...,4$ and $I_5^{\;0}=1$, 
the five-brackets in the denominator
reduce to four-brackets:
\bq
 \left\langle 0, i, j, k, l \right\rangle
 & = & 
 \left\langle i, j, k, l \right\rangle.
\eq
Let us now look at the five-bracket $\langle 1,2,3,4,5 \rangle$ appearing in the numerator.
Expanding the determinant by Laplace's formula yields
\bq
 \left\langle 1,2,3,4,5 \right\rangle
 & = &
 \left\langle 1,2,3,4 \right\rangle \phi_I \eta_I^5
 +
 \left\langle 2,3,4,5 \right\rangle \phi_I \eta_I^1
 +
 \left\langle 3,4,5,1 \right\rangle \phi_I \eta_I^2
 +
 \left\langle 4,5,1,2 \right\rangle \phi_I \eta_I^3
 +
 \left\langle 5,1,2,3 \right\rangle \phi_I \eta_I^4.
 \nonumber 
\eq
We have four factors of this expression in the numerator.
Now let us perform the integration over $\phi_1$. This will produce a factor
\bq
\lefteqn{
 \left\langle 1,2,3,4 \right\rangle \eta_1^5
 +
 \left\langle 2,3,4,5 \right\rangle \eta_1^1
 +
 \left\langle 3,4,5,1 \right\rangle \eta_1^2
 +
 \left\langle 4,5,1,2 \right\rangle \eta_1^3
 +
 \left\langle 5,1,2,3 \right\rangle \eta_1^4
 = } & & \nonumber \\
 & = &
 \delta\left(
 \left\langle 1,2,3,4 \right\rangle \eta_1^5
 +
 \left\langle 2,3,4,5 \right\rangle \eta_1^1
 +
 \left\langle 3,4,5,1 \right\rangle \eta_1^2
 +
 \left\langle 4,5,1,2 \right\rangle \eta_1^3
 +
 \left\langle 5,1,2,3 \right\rangle \eta_1^4
 \right).
\eq
We have four possibilities to produce this factor.
Now let us proceed to the integration over $\phi_2$. 
This will give us a Grassmann delta-function similar to the one above, with
$\eta_1^j$ replaced by $\eta_2^j$.
We now have three possibilities to produce this delta-function.
Proceeding in this way with the integration over $\phi_3$ and $\phi_4$ we see that the prefactor
$1/4!$ gets cancelled by the total number of possibilities and we obtain
\bq
 \left[ Z^1, Z^2, Z^3, Z^4, Z^5 \right]
 & = &
 \frac{\delta^4\left( \left\langle 1, 2, 3, 4 \right\rangle \eta_I^5 + \mbox{cyclic} \right)}
      {
       \left\langle 1, 2, 3, 4 \right\rangle
       \left\langle 2, 3, 4, 5 \right\rangle
       \left\langle 3, 4, 5, 1 \right\rangle
       \left\langle 4, 5, 1, 2 \right\rangle
       \left\langle 5, 1, 2, 3 \right\rangle
      }.
\eq
\stepcounter{exercise}
}
\es
\\
\\
\bs
{\it {\bf Exercise \theexercise}: 
Let $g \in \mathrm{PSL}(2,{\mathbb C})$ and let $(z_1,z_2, ..., z_n)$ be a solution
of the scattering equations.
Show that $(z_1',z_2', ..., z_n') = g \cdot (z_1,z_2, ..., z_n)$ is also a solution
of the scattering equations.
\\
\\
{\bf Solution}: 
We have
\bq
\label{GL2_trafo}
 \frac{1}{z_i'-z_j'}  
 & = &
 \frac{A}{z_i-z_j} + B,
\eq
where
\bq
 A = \frac{\left(c z_i + d \right)^2}{ad-bc} = \left(c z_i + d \right)^2,
 & &
 B = -\frac{c\left(c z_i +d \right)}{ad-bc} = -c\left(c z_i +d \right).
\eq
$A$ and $B$ are independent of $z_j$. When eq.~(\ref{GL2_trafo}) is inserted into the scattering equations,
the terms proportional to $A$ vanish due to our assumption that the unprimed variables
$(z_1,z_2, ..., z_n)$ are a solution of the scattering equations
\bq
 \sum\limits_{j=1, j \neq i}^n 2 p_i \cdot p_j \frac{A}{z_i - z_j}
 \;\; = \;\;
 A \sum\limits_{j=1, j \neq i}^n \frac{2 p_i \cdot p_j }{z_i - z_j}
 \;\; = \;\;
 0.
\eq
The terms proportional to $B$ vanish due to momentum conservation and the on-shell condition $p_i^2=0$:
\bq
 \sum\limits_{j=1, j \neq i}^n \left( 2 p_i \cdot p_j \right) B
 \;\; = \;\;
 2 p_i \cdot \left(\sum\limits_{j=1, j \neq i}^n p_j \right) B
 \;\; = \;\;
 - 2 p_i^2 B
 \;\; = \;\;
 0.
\eq
\stepcounter{exercise}
}
\es
\\
\\
\bs
{\it {\bf Exercise \theexercise}: Prove the relations in eq.~(\ref{relations_scattering_equations}).
\\
\\
{\bf Solution}: 
We start with the first relation:
\bq
 \sum\limits_{i=1}^n f_i\left(z,p\right) 
 \;\; = \;\;
 \sum\limits_{i=1}^n
 \;\;
 \sum\limits_{j=1, j \neq i}^n 
 \;\;
 \frac{2 p_i \cdot p_j}{z_i-z_j} & = & 0,
\eq
due to the antisymmetry of the denominators $1/(z_i-z_j)$.
Let us now turn to the second relation:
We use $z_i=(z_i-z_j)+z_j$, momentum conservation together with the on-shell conditions and 
obtain
\bq
 \sum\limits_{i=1}^n z_i f_i\left(z,p\right) 
 & = &
 \sum\limits_{i=1}^n
 \;\;
 z_i
 \sum\limits_{j=1, j \neq i}^n 
 \;\;
 \frac{2 p_i \cdot p_j}{z_i-z_j} 
 \;\; = \;\;
 \sum\limits_{i=1}^n
 \sum\limits_{j=1, j \neq i}^n 
 2 p_i \cdot p_j
 +
 \sum\limits_{i=1}^n
 \sum\limits_{j=1, j \neq i}^n 
 \; z_j \;
 \frac{2 p_i \cdot p_j}{z_i-z_j} 
 \nonumber \\
 & = &
 -
 \sum\limits_{i=1}^n
 \;\;
 z_i
 \sum\limits_{j=1, j \neq i}^n 
 \;\;
 \frac{2 p_i \cdot p_j}{z_i-z_j} 
 \;\; = \;\;
 -
 \sum\limits_{i=1}^n z_i f_i\left(z,p\right).
\eq
In the second line we re-labeled the summation indices $i \leftrightarrow j$. 
It follows
\bq
 \sum\limits_{i=1}^n z_i f_i\left(z,p\right) 
 & = &
 0.
\eq
The third relation is proven as follows:
\bq
 \sum\limits_{i=1}^n z_i^2 f_i\left(z,p\right) 
 & = &
 \sum\limits_{i=1}^n
 \;\;
 z_i^2
 \sum\limits_{j=1, j \neq i}^n 
 \;
 \frac{2 p_i \cdot p_j}{z_i-z_j} 
 \;\; = \;\;
 \frac{1}{2}
 \sum\limits_{i=1}^n
 \sum\limits_{j=1, j \neq i}^n 
 \;
 \left( z_i^2 - z_j^2 \right)
 \frac{2 p_i \cdot p_j}{z_i-z_j} 
 \nonumber \\
 & = &
 \frac{1}{2}
 \sum\limits_{i=1}^n
 \sum\limits_{j=1, j \neq i}^n 
 2 p_i \cdot p_j
 \left( z_i + z_j \right)
 \;\; = \;\;
 \sum\limits_{i=1}^n
 z_i
 \sum\limits_{j=1, j \neq i}^n 
 2 p_i \cdot p_j
 \;\; = \;\;
 0.
\eq
\stepcounter{exercise}
}
\es
\\
\\
\bs
{\it {\bf Exercise \theexercise}: 
Consider the Koba-Nielsen function
\bq
 U\left(z,p\right)
 & = &
 \prod\limits_{i<j} \left( z_i - z_j \right)^{2 p_i \cdot p_j }.
\eq
Show
\bq
 U^{-1} \frac{\partial}{\partial z_i} U
 & = &
 f_i\left(z,p\right).
\eq
\\
\\
{\bf Solution}: 
We have
\bq
 \frac{\partial U}{\partial z_i}
 & = &
 \frac{\partial}{\partial z_i}
 \prod\limits_{j=1}^{n-1} \prod\limits_{k=j+1}^n
 \left( z_j - z_k \right)^{2 p_j \cdot p_k}
 \;\; = \;\;
 \sum\limits_{k=i+1}^n
 \frac{2 p_i \cdot p_k}{z_i-z_k} \; U
 -
 \sum\limits_{j=1}^{i-1} 
 \frac{2 p_j \cdot p_i}{z_j-z_i} \; U
 \nonumber \\
 & = &
 \sum\limits_{j \neq i} 
 \frac{2 p_i \cdot p_j}{z_i-z_j} \; U
 \;\; = \;\;
 f_i\left(z,p\right) U.
\eq
Hence
\bq
 U^{-1} \frac{\partial}{\partial z_i} U
 & = &
 f_i\left(z,p\right).
\eq
\stepcounter{exercise}
}
\es
\\
\\
\bs
{\it {\bf Exercise \theexercise}: 
Consider the case $n=5$ and determine the matrix $S_{\sigma \tilde{\sigma}}$ for the basis $B$ given 
in eq.~(\ref{def_B_basis}).
\\
\\
{\bf Solution}: 
We take all momenta as outgoing and set $p=(p_1,p_2,p_3,p_4,p_5)$.
For $n=5$ the basis $B$ consists of $(n-3)!=2$ elements, which with the choice as in eq.~(\ref{def_B_basis})
are given by
\bq
 B & = & 
 \left\{ \left(1,2,3,4,5\right), \left(1,2,4,3,5\right) \right\}.
\eq
We set $\rho=(1,2,3,4,5)$ and $\tau=(1,2,4,3,5)$.
The matrix $m_{\sigma \tilde{\sigma}}$ (with indices $\sigma,\tilde{\sigma} \in \{\rho,\tau\}$) is given by
\bq
 m & = &
 \left( \begin{array}{cc}
 m_5^{(0)}\left(\rho,\rho,p\right) & m_5^{(0)}\left(\rho,\tau,p\right) \\
 m_5^{(0)}\left(\tau,\rho,p\right) & m_5^{(0)}\left(\tau,\tau,p\right) \\
 \end{array} \right).
\eq
The entries are given by
\bq
 m_5^{(0)}\left(\rho,\rho,p\right)
 & = &
 i \left( \frac{1}{s_{12}s_{34}} + \frac{1}{s_{23}s_{45}} + \frac{1}{s_{15}s_{34}} + \frac{1}{s_{45}s_{12}} + \frac{1}{s_{23}s_{15}} \right),
 \nonumber \\
 m_5^{(0)}\left(\rho,\tau,p\right)
 & = &
 m_5^{(0)}\left(\tau,\rho,p\right)
 \;\; = \;\;
 i \left( - \frac{1}{s_{12}s_{34}} - \frac{1}{s_{15}s_{34}} \right),
 \nonumber \\
 m_5^{(0)}\left(\tau,\tau,p\right)
 & = &
 i \left( \frac{1}{s_{12}s_{34}} + \frac{1}{s_{24}s_{35}} + \frac{1}{s_{15}s_{34}} + \frac{1}{s_{35}s_{12}} + \frac{1}{s_{24}s_{15}} \right),
\eq
where we used the notation $s_{ij}=(p_i+p_j)^2=2p_i\cdot p_j$.
We calculate $\det m$ and find
\bq
 \det m & = &
 \frac{ s_{13} s_{14} s_{25}}{s_{12} s_{15} s_{23} s_{24} s_{34} s_{35} s_{45}}.
\eq
Thus
\bq
 S & = &
 \frac{s_{12} s_{15} s_{23} s_{24} s_{34} s_{35} s_{45}}{ s_{13} s_{14} s_{25}}
 \left( \begin{array}{rr}
 m_5^{(0)}\left(\tau,\tau,p\right) & -m_5^{(0)}\left(\rho,\tau,p\right) \\
 -m_5^{(0)}\left(\tau,\rho,p\right) & m_5^{(0)}\left(\rho,\rho,p\right) \\
 \end{array} \right).
\eq
\stepcounter{exercise}
}
\es

\end{appendix}

% ----------------------------------------------------------------------------------
% references
\newpage
\bs
\bibliography{/home/stefanw/notes/biblio}

\begin{thebibliography}{100}

\bibitem{Mangano:1990by}
M.~L. Mangano and S.~J. Parke,
\newblock Phys. Rept. {\bf 200}, 301 (1991), arXiv:hep-th/0509223.
%%CITATION = HEP-TH/0509223;%%

\bibitem{Dixon:1996wi}
L.~J. Dixon,
\newblock {Calculating scattering amplitudes efficiently},
\newblock in {\em {QCD and beyond. Proceedings, Theoretical Advanced Study
  Institute in Elementary Particle Physics, TASI-95, Boulder, USA, June 4-30,
  1995}}, pp. 539--584, 1996, arXiv:hep-ph/9601359.
%%CITATION = HEP-PH/9601359;%%

\bibitem{Elvang:2013cua}
H.~Elvang and Y.-t. Huang,
\newblock {\em {Scattering Amplitudes}} (Cambridge University Press, 2015),
  arXiv:1308.1697.
%%CITATION = ARXIV:1308.1697;%%

\bibitem{Henn:2014yza}
J.~M. Henn and J.~C. Plefka,
\newblock {\em {Scattering Amplitudes in Gauge Theories}} (Springer, Lecture
  Notes in Physics 883, 2014).
%%CITATION = LNPHA,883,1;%%

\bibitem{Peskin}
M.~E. Peskin and D.~V. Schroeder,
\newblock {\em An Introduction to Quantum Field Theory} (Perseus Books, 1995).

\bibitem{Srednicki:2007qs}
M.~Srednicki,
\newblock {\em {Quantum field theory}} (Cambridge University Press, 2007).
%%CITATION = INSPIRE-752478;%%

\bibitem{Schwartz}
M.~D. Schwartz,
\newblock {\em Quantum Field Theory and the Standard Model} (Cambridge
  University Press, 2014).

\bibitem{Yang:1954ek}
C.-N. Yang and R.~L. Mills,
\newblock Phys. Rev. {\bf 96}, 191 (1954).
%%CITATION = PHRVA,96,191;%%

\bibitem{Nakahara}
M.~Nakahara,
\newblock {\em Geometry, Topology and Physics} (Institute of Physics
  Publishing, 2003).

\bibitem{Isham:1999qu}
C.~J. Isham,
\newblock {\em {Modern differential geometry for physicists}} (World
  Scientific, 1999).
%%CITATION = INSPIRE-508872;%%

\bibitem{Gribov:1977wm}
V.~N. Gribov,
\newblock Nucl. Phys. {\bf B139}, 1 (1978).
%%CITATION = NUPHA,B139,1;%%

\bibitem{Faddeev:1967fc}
L.~D. Faddeev and V.~N. Popov,
\newblock Phys. Lett. {\bf B25}, 29 (1967).
%%CITATION = PHLTA,B25,29;%%

\bibitem{Smirnov:2006ry}
V.~A. Smirnov,
\newblock {\em Feynman integral calculus} (Springer, Berlin, 2006).

\bibitem{Weinzierl:2006qs}
S.~Weinzierl,
\newblock Fields Inst. Commun. {\bf 50}, 345 (2006), hep-ph/0604068.
%%CITATION = HEP-PH 0604068;%%

\bibitem{Weinzierl:2010ps}
S.~Weinzierl,
\newblock {Introduction to Feynman Integrals},
\newblock in {\em {Geometric and Topological Methods for Quantum Field Theory:
  Proceedings, 6th Summer School, Villa de Leyva, Colombia, 6-23 Jul 2009}},
  pp. 144--187, 2013, arXiv:1005.1855.
%%CITATION = ARXIV:1005.1855;%%

\bibitem{Duhr:2014woa}
C.~Duhr,
\newblock {Mathematical aspects of scattering amplitudes},
\newblock in {\em {Theoretical Advanced Study Institute in Elementary Particle
  Physics: Journeys Through the Precision Frontier: Amplitudes for Colliders
  (TASI 2014) Boulder, Colorado, June 2-27, 2014}}, 2014, arXiv:1411.7538.
%%CITATION = ARXIV:1411.7538;%%

\bibitem{Henn:2014qga}
J.~M. Henn,
\newblock J. Phys. {\bf A48}, 153001 (2015), arXiv:1412.2296.
%%CITATION = ARXIV:1412.2296;%%

\bibitem{Vermaseren:2000nd}
J.~A.~M. Vermaseren,
\newblock (2000), math-ph/0010025.
%%CITATION = MATH-PH 0010025;%%

\bibitem{Bauer:2000cp}
C.~Bauer, A.~Frink, and R.~Kreckel,
\newblock J. Symbolic Computation {\bf 33}, 1 (2002), cs.sc/0004015.
%%CITATION = CS.SC 0004015;%%

\bibitem{Bassetto:1984ik}
A.~Bassetto, M.~Ciafaloni, and G.~Marchesini,
\newblock Phys. Rept. {\bf 100}, 201 (1983).
%%CITATION = PRPLC,100,201;%%

\bibitem{Catani:1997vz}
S.~Catani and M.~H. Seymour,
\newblock Nucl. Phys. {\bf B485}, 291 (1997), hep-ph/9605323.
%%CITATION = NUPHA,B485,291;%%

\bibitem{'tHooft:1972fi}
G.~'t~Hooft and M.~J.~G. Veltman,
\newblock Nucl. Phys. {\bf B44}, 189 (1972).
%%CITATION = NUPHA,B44,189;%%

\bibitem{Bollini:1972ui}
C.~G. Bollini and J.~J. Giambiagi,
\newblock Nuovo Cim. {\bf B12}, 20 (1972).
%%CITATION = NUCIA,B12,20;%%

\bibitem{Cicuta:1972jf}
G.~M. Cicuta and E.~Montaldi,
\newblock Nuovo Cim. Lett. {\bf 4}, 329 (1972).
%%CITATION = NCLTA,4,329;%%

\bibitem{Kinoshita:1962ur}
T.~Kinoshita,
\newblock J. Math. Phys. {\bf 3}, 650 (1962).
%%CITATION = JMAPA,3,650;%%

\bibitem{Lee:1964is}
T.~D. Lee and M.~Nauenberg,
\newblock Phys. Rev. {\bf 133}, B1549 (1964).
%%CITATION = PHRVA,133,B1549;%%

\bibitem{Cvitanovic:1980bu}
P.~Cvitanovic, P.~G. Lauwers, and P.~N. Scharbach,
\newblock Nucl. Phys. {\bf B186}, 165 (1981).
%%CITATION = NUPHA,B186,165;%%

\bibitem{Berends:1987cv}
F.~A. Berends and W.~Giele,
\newblock Nucl. Phys. {\bf B294}, 700 (1987).
%%CITATION = NUPHA,B294,700;%%

\bibitem{Mangano:1987xk}
M.~L. Mangano, S.~J. Parke, and Z.~Xu,
\newblock Nucl. Phys. {\bf B298}, 653 (1988).
%%CITATION = NUPHA,B298,653;%%

\bibitem{Kosower:1987ic}
D.~Kosower, B.-H. Lee, and V.~P. Nair,
\newblock Phys. Lett. {\bf B201}, 85 (1988).
%%CITATION = PHLTA,B201,85;%%

\bibitem{Bern:1990ux}
Z.~Bern and D.~A. Kosower,
\newblock Nucl. Phys. {\bf B362}, 389 (1991).
%%CITATION = NUPHA,B362,389;%%

\bibitem{DelDuca:1999rs}
V.~Del~Duca, L.~J. Dixon, and F.~Maltoni,
\newblock Nucl. Phys. {\bf B571}, 51 (2000), hep-ph/9910563.
%%CITATION = HEP-PH 9910563;%%

\bibitem{Maltoni:2002mq}
F.~Maltoni, K.~Paul, T.~Stelzer, and S.~Willenbrock,
\newblock Phys. Rev. {\bf D67}, 014026 (2003), hep-ph/0209271.
%%CITATION = HEP-PH 0209271;%%

\bibitem{Bern:1994fz}
Z.~Bern, L.~J. Dixon, and D.~A. Kosower,
\newblock Nucl. Phys. {\bf B437}, 259 (1995), hep-ph/9409393.
%%CITATION = HEP-PH 9409393;%%

\bibitem{Ellis:2008qc}
R.~K. Ellis, W.~Giele, Z.~Kunszt, K.~Melnikov, and G.~Zanderighi,
\newblock JHEP {\bf 0901}, 012 (2009), arXiv:0810.2762.
%%CITATION = ARXIV:0810.2762;%%

\bibitem{Ellis:2011cr}
R.~K. Ellis, Z.~Kunszt, K.~Melnikov, and G.~Zanderighi,
\newblock Phys.Rept. {\bf 518}, 141 (2012), arXiv:1105.4319.
%%CITATION = ARXIV:1105.4319;%%

\bibitem{Ita:2011ar}
H.~Ita and K.~Ozeren,
\newblock JHEP {\bf 1202}, 118 (2012), arXiv:1111.4193.
%%CITATION = ARXIV:1111.4193;%%

\bibitem{Badger:2012pg}
S.~Badger, B.~Biedermann, P.~Uwer, and V.~Yundin,
\newblock Comput.Phys.Commun. {\bf 184}, 1981 (2013), arXiv:1209.0100.
%%CITATION = ARXIV:1209.0100;%%

\bibitem{Reuschle:2013qna}
C.~Reuschle and S.~Weinzierl,
\newblock Phys.Rev. {\bf D88}, 105020 (2013), arXiv:1310.0413.
%%CITATION = ARXIV:1310.0413;%%

\bibitem{Schuster:2013aya}
T.~Schuster,
\newblock Phys. Rev. {\bf D89}, 105022 (2014), arXiv:1311.6296.
%%CITATION = ARXIV:1311.6296;%%

\bibitem{Kleiss:1988ne}
R.~Kleiss and H.~Kuijf,
\newblock Nucl. Phys. {\bf B312}, 616 (1989).
%%CITATION = NUPHA,B312,616;%%

\bibitem{Draggiotis:1998gr}
P.~Draggiotis, R.~H.~P. Kleiss, and C.~G. Papadopoulos,
\newblock Phys. Lett. {\bf B439}, 157 (1998), hep-ph/9807207.
%%CITATION = HEP-PH 9807207;%%

\bibitem{Duhr:2006iq}
C.~Duhr, S.~Hoche, and F.~Maltoni,
\newblock JHEP {\bf 08}, 062 (2006), hep-ph/0607057.
%%CITATION = HEP-PH 0607057;%%

\bibitem{Bern:2008qj}
Z.~Bern, J.~J.~M. Carrasco, and H.~Johansson,
\newblock Phys. Rev. {\bf D78}, 085011 (2008), arXiv:0805.3993.
%%CITATION = 0805.3993;%%

\bibitem{BjerrumBohr:2009rd}
N.~Bjerrum-Bohr, P.~H. Damgaard, and P.~Vanhove,
\newblock Phys.Rev.Lett. {\bf 103}, 161602 (2009), arXiv:0907.1425.
%%CITATION = ARXIV:0907.1425;%%

\bibitem{Stieberger:2009hq}
S.~Stieberger,
\newblock (2009), arXiv:0907.2211.
%%CITATION = ARXIV:0907.2211;%%

\bibitem{Feng:2010my}
B.~Feng, R.~Huang, and Y.~Jia,
\newblock Phys.Lett. {\bf B695}, 350 (2011), arXiv:1004.3417.
%%CITATION = ARXIV:1004.3417;%%

\bibitem{Bern:2010ue}
Z.~Bern, J.~J.~M. Carrasco, and H.~Johansson,
\newblock Phys.Rev.Lett. {\bf 105}, 061602 (2010), arXiv:1004.0476.
%%CITATION = ARXIV:1004.0476;%%

\bibitem{Gotz:2012zz}
D.~G{\"o}tz, C.~Schwan, and S.~Weinzierl,
\newblock Phys.Rev. {\bf D85}, 116011 (2012).
%%CITATION = PHRVA,D85,116011;%%

\bibitem{Czakon:2009ss}
M.~Czakon, C.~G. Papadopoulos, and M.~Worek,
\newblock JHEP {\bf 08}, 085 (2009), arXiv:0905.0883.
%%CITATION = 0905.0883;%%

\bibitem{Dittmaier:2008md}
S.~Dittmaier, A.~Kabelschacht, and T.~Kasprzik,
\newblock Nucl. Phys. {\bf B800}, 146 (2008), arXiv:0802.1405.
%%CITATION = 0802.1405;%%

\bibitem{Berends:1981rb}
F.~A. Berends, R.~Kleiss, P.~De~Causmaecker, R.~Gastmans, and T.~T. Wu,
\newblock Phys. Lett. {\bf B103}, 124 (1981).
%%CITATION = PHLTA,B103,124;%%

\bibitem{DeCausmaecker:1982bg}
P.~De~Causmaecker, R.~Gastmans, W.~Troost, and T.~T. Wu,
\newblock Nucl. Phys. {\bf B206}, 53 (1982).
%%CITATION = NUPHA,B206,53;%%

\bibitem{Gunion:1985vc}
J.~F. Gunion and Z.~Kunszt,
\newblock Phys. Lett. {\bf B161}, 333 (1985).
%%CITATION = PHLTA,B161,333;%%

\bibitem{Kleiss:1986qc}
R.~Kleiss and W.~J. Stirling,
\newblock Phys. Lett. {\bf B179}, 159 (1986).
%%CITATION = PHLTA,B179,159;%%

\bibitem{Xu:1987xb}
Z.~Xu, D.-H. Zhang, and L.~Chang,
\newblock Nucl. Phys. {\bf B291}, 392 (1987).
%%CITATION = NUPHA,B291,392;%%

\bibitem{Gastmans_book}
R.~Gastmans and T.~T. Wu,
\newblock {\em The Ubiquitous photon: Helicity method for QED and QCD}
  (Clarendon Press, 1990).

\bibitem{Schwinn:2005pi}
C.~Schwinn and S.~Weinzierl,
\newblock JHEP {\bf 05}, 006 (2005), hep-th/0503015.
%%CITATION = HEP-TH 0503015;%%

\bibitem{Rodrigo:2005eu}
G.~Rodrigo,
\newblock JHEP {\bf 09}, 079 (2005), hep-ph/0508138.
%%CITATION = HEP-PH 0508138;%%

\bibitem{Hodges:2005bf}
A.~P. Hodges,
\newblock (2005), arXiv:hep-th/0503060.
%%CITATION = HEP-TH/0503060;%%

\bibitem{Hodges:2005aj}
A.~P. Hodges,
\newblock (2005), arXiv:hep-th/0512336.
%%CITATION = HEP-TH/0512336;%%

\bibitem{Hodges:2009hk}
A.~Hodges,
\newblock JHEP {\bf 05}, 135 (2013), arXiv:0905.1473.
%%CITATION = ARXIV:0905.1473;%%

\bibitem{Mason:2009sa}
L.~J. Mason and D.~Skinner,
\newblock JHEP {\bf 01}, 064 (2010), arXiv:0903.2083.
%%CITATION = ARXIV:0903.2083;%%

\bibitem{Eastwood:1981jy}
M.~G. Eastwood, R.~Penrose, and R.~Wells,
\newblock Commun.Math.Phys. {\bf 78}, 305 (1981).
%%CITATION = CMPHA,78,305;%%

\bibitem{Penrose:1972ia}
R.~Penrose and M.~A.~H. MacCallum,
\newblock Phys. Rept. {\bf 6}, 241 (1972).
%%CITATION = PRPLC,6,241;%%

\bibitem{Byckling:1971vca}
E.~Byckling and K.~Kajantie,
\newblock {\em {Particle Kinematics}} (John Wiley \& Sons, 1973).

\bibitem{Kleiss:1986gy}
R.~Kleiss, W.~J. Stirling, and S.~D. Ellis,
\newblock Comput. Phys. Commun. {\bf 40}, 359 (1986).
%%CITATION = CPHCB,40,359;%%

\bibitem{vanHameren:2002tc}
A.~van Hameren and C.~G. Papadopoulos,
\newblock Eur. Phys. J. {\bf C25}, 563 (2002), hep-ph/0204055.
%%CITATION = HEP-PH/0204055;%%

\bibitem{vanHameren:2010gg}
A.~van Hameren,
\newblock (2010), arXiv:1003.4953.
%%CITATION = ARXIV:1003.4953;%%

\bibitem{Berends:1987me}
F.~A. Berends and W.~T. Giele,
\newblock Nucl. Phys. {\bf B306}, 759 (1988).
%%CITATION = NUPHA,B306,759;%%

\bibitem{Parke:1986gb}
S.~J. Parke and T.~R. Taylor,
\newblock Phys. Rev. Lett. {\bf 56}, 2459 (1986).
%%CITATION = PRLTA,56,2459;%%

\bibitem{Cachazo:2004kj}
F.~Cachazo, P.~Svrcek, and E.~Witten,
\newblock JHEP {\bf 09}, 006 (2004), hep-th/0403047.
%%CITATION = HEP-TH 0403047;%%

\bibitem{Kosower:2004yz}
D.~A. Kosower,
\newblock Phys. Rev. {\bf D71}, 045007 (2005), hep-th/0406175.
%%CITATION = HEP-TH 0406175;%%

\bibitem{Bena:2004ry}
I.~Bena, Z.~Bern, and D.~A. Kosower,
\newblock Phys. Rev. {\bf D71}, 045008 (2005), hep-th/0406133.
%%CITATION = HEP-TH 0406133;%%

\bibitem{Gorsky:2005sf}
A.~Gorsky and A.~Rosly,
\newblock JHEP {\bf 01}, 101 (2006), hep-th/0510111.
%%CITATION = HEP-TH/0510111;%%

\bibitem{Mansfield:2005yd}
P.~Mansfield,
\newblock JHEP {\bf 03}, 037 (2006), hep-th/0511264.
%%CITATION = HEP-TH/0511264;%%

\bibitem{Ettle:2006bw}
J.~H. Ettle and T.~R. Morris,
\newblock JHEP {\bf 08}, 003 (2006), hep-th/0605121.
%%CITATION = HEP-TH/0605121;%%

\bibitem{Ettle:2007qc}
J.~H. Ettle, C.-H. Fu, J.~P. Fudger, P.~R.~W. Mansfield, and T.~R. Morris,
\newblock JHEP {\bf 05}, 011 (2007), hep-th/0703286.
%%CITATION = HEP-TH/0703286;%%

\bibitem{Ettle:2008ey}
J.~H. Ettle, T.~R. Morris, and Z.~Xiao,
\newblock JHEP {\bf 08}, 103 (2008), arXiv:0805.0239.
%%CITATION = 0805.0239;%%

\bibitem{Buchta:2010qr}
S.~Buchta and S.~Weinzierl,
\newblock JHEP {\bf 09}, 071 (2010), arXiv:1007.2742.
%%CITATION = 1007.2742;%%

\bibitem{Mason:2005kn}
L.~J. Mason and D.~Skinner,
\newblock Phys. Lett. {\bf B636}, 60 (2006), hep-th/0510262.
%%CITATION = HEP-TH/0510262;%%

\bibitem{Mason:2005zm}
L.~J. Mason,
\newblock JHEP {\bf 10}, 009 (2005), hep-th/0507269.
%%CITATION = HEP-TH/0507269;%%

\bibitem{Boels:2007qn}
R.~Boels, L.~Mason, and D.~Skinner,
\newblock Phys. Lett. {\bf B648}, 90 (2007), hep-th/0702035.
%%CITATION = HEP-TH/0702035;%%

\bibitem{Boels:2006ir}
R.~Boels, L.~Mason, and D.~Skinner,
\newblock JHEP {\bf 02}, 014 (2007), hep-th/0604040.
%%CITATION = HEP-TH/0604040;%%

\bibitem{Boels:2007gv}
R.~Boels,
\newblock Phys. Rev. {\bf D76}, 105027 (2007), arXiv:hep-th/0703080.
%%CITATION = HEP-TH/0703080;%%

\bibitem{Boels:2008fc}
R.~Boels, K.~J. Larsen, N.~A. Obers, and M.~Vonk,
\newblock JHEP {\bf 11}, 015 (2008), arXiv:0808.2598.
%%CITATION = 0808.2598;%%

\bibitem{Risager:2005vk}
K.~Risager,
\newblock JHEP {\bf 12}, 003 (2005), hep-th/0508206.
%%CITATION = HEP-TH 0508206;%%

\bibitem{Britto:2004ap}
R.~Britto, F.~Cachazo, and B.~Feng,
\newblock Nucl. Phys. {\bf B715}, 499 (2005), hep-th/0412308.
%%CITATION = HEP-TH 0412308;%%

\bibitem{Britto:2005fq}
R.~Britto, F.~Cachazo, B.~Feng, and E.~Witten,
\newblock Phys. Rev. Lett. {\bf 94}, 181602 (2005), hep-th/0501052.
%%CITATION = HEP-TH 0501052;%%

\bibitem{ArkaniHamed:2008yf}
N.~Arkani-Hamed and J.~Kaplan,
\newblock JHEP {\bf 0804}, 076 (2008), arXiv:0801.2385.
%%CITATION = ARXIV:0801.2385;%%

\bibitem{Cheung:2015cba}
C.~Cheung, C.-H. Shen, and J.~Trnka,
\newblock JHEP {\bf 06}, 118 (2015), arXiv:1502.05057.
%%CITATION = ARXIV:1502.05057;%%

\bibitem{Dinsdale:2006sq}
M.~Dinsdale, M.~Ternick, and S.~Weinzierl,
\newblock JHEP {\bf 03}, 056 (2006), hep-ph/0602204.
%%CITATION = HEP-PH 0602204;%%

\bibitem{Badger:2012uz}
S.~Badger {\em et~al.},
\newblock Phys. Rev. {\bf D87}, 034011 (2013), arXiv:1206.2381.
%%CITATION = ARXIV:1206.2381;%%

\bibitem{Benincasa:2007xk}
P.~Benincasa and F.~Cachazo,
\newblock (2007), arXiv:0705.4305.
%%CITATION = ARXIV:0705.4305;%%

\bibitem{Jia:2010nz}
Y.~Jia, R.~Huang, and C.-Y. Liu,
\newblock Phys.Rev. {\bf D82}, 065001 (2010), arXiv:1005.1821.
%%CITATION = ARXIV:1005.1821;%%

\bibitem{Chen:2011jxa}
Y.-X. Chen, Y.-J. Du, and B.~Feng,
\newblock JHEP {\bf 1102}, 112 (2011), arXiv:1101.0009.
%%CITATION = ARXIV:1101.0009;%%

\bibitem{delaCruz:2015dpa}
L.~de~la Cruz, A.~Kniss, and S.~Weinzierl,
\newblock JHEP {\bf 09}, 197 (2015), arXiv:1508.01432.
%%CITATION = ARXIV:1508.01432;%%

\bibitem{Postnikov:2013}
A.~Postnikov,
\newblock (2013),
\newblock Lecture notes.

\bibitem{Brion:2004}
M.~{Brion},
\newblock (2004), math/0410240.

\bibitem{ArkaniHamed:2009si}
N.~Arkani-Hamed, F.~Cachazo, C.~Cheung, and J.~Kaplan,
\newblock JHEP {\bf 03}, 110 (2010), arXiv:0903.2110.
%%CITATION = ARXIV:0903.2110;%%

\bibitem{ArkaniHamed:2009dn}
N.~Arkani-Hamed, F.~Cachazo, C.~Cheung, and J.~Kaplan,
\newblock JHEP {\bf 03}, 020 (2010), arXiv:0907.5418.
%%CITATION = ARXIV:0907.5418;%%

\bibitem{ArkaniHamed:2009sx}
N.~Arkani-Hamed, J.~Bourjaily, F.~Cachazo, and J.~Trnka,
\newblock JHEP {\bf 01}, 108 (2011), arXiv:0912.3249.
%%CITATION = ARXIV:0912.3249;%%

\bibitem{ArkaniHamed:2009dg}
N.~Arkani-Hamed, J.~Bourjaily, F.~Cachazo, and J.~Trnka,
\newblock JHEP {\bf 01}, 049 (2011), arXiv:0912.4912.
%%CITATION = ARXIV:0912.4912;%%

\bibitem{ArkaniHamed:2012nw}
N.~Arkani-Hamed {\em et~al.},
\newblock {\em {Scattering Amplitudes and the Positive Grassmannian}}
  (Cambridge University Press, 2012), arXiv:1212.5605.
%%CITATION = ARXIV:1212.5605;%%

\bibitem{Bourjaily:2010kw}
J.~L. Bourjaily, J.~Trnka, A.~Volovich, and C.~Wen,
\newblock JHEP {\bf 01}, 038 (2011), arXiv:1006.1899.
%%CITATION = ARXIV:1006.1899;%%

\bibitem{Nair:1988bq}
V.~P. Nair,
\newblock Phys. Lett. {\bf B214}, 215 (1988).
%%CITATION = PHLTA,B214,215;%%

\bibitem{ArkaniHamed:2008gz}
N.~Arkani-Hamed, F.~Cachazo, and J.~Kaplan,
\newblock JHEP {\bf 1009}, 016 (2010), arXiv:0808.1446.
%%CITATION = ARXIV:0808.1446;%%

\bibitem{Wess:1992cp}
J.~Wess and J.~Bagger,
\newblock {\em {Supersymmetry and supergravity}} (Princeton University Press,
  1992).
%%CITATION = INSPIRE-350988;%%

\bibitem{Arkani-Hamed:2013jha}
N.~Arkani-Hamed and J.~Trnka,
\newblock JHEP {\bf 10}, 030 (2014), arXiv:1312.2007.
%%CITATION = ARXIV:1312.2007;%%

\bibitem{Arkani-Hamed:2013kca}
N.~Arkani-Hamed and J.~Trnka,
\newblock JHEP {\bf 12}, 182 (2014), arXiv:1312.7878.
%%CITATION = ARXIV:1312.7878;%%

\bibitem{Arkani-Hamed:2014dca}
N.~Arkani-Hamed, A.~Hodges, and J.~Trnka,
\newblock JHEP {\bf 08}, 030 (2015), arXiv:1412.8478.
%%CITATION = ARXIV:1412.8478;%%

\bibitem{Bai:2014cna}
Y.~Bai and S.~He,
\newblock JHEP {\bf 02}, 065 (2015), arXiv:1408.2459.
%%CITATION = ARXIV:1408.2459;%%

\bibitem{Franco:2014csa}
S.~Franco, D.~Galloni, A.~Mariotti, and J.~Trnka,
\newblock JHEP {\bf 03}, 128 (2015), arXiv:1408.3410.
%%CITATION = ARXIV:1408.3410;%%

\bibitem{Bern:2015ple}
Z.~Bern, E.~Herrmann, S.~Litsey, J.~Stankowicz, and J.~Trnka,
\newblock JHEP {\bf 06}, 098 (2016), arXiv:1512.08591.
%%CITATION = ARXIV:1512.08591;%%

\bibitem{Cachazo:2013gna}
F.~Cachazo, S.~He, and E.~Y. Yuan,
\newblock Phys.Rev. {\bf D90}, 065001 (2014), arXiv:1306.6575.
%%CITATION = ARXIV:1306.6575;%%

\bibitem{Cachazo:2013hca}
F.~Cachazo, S.~He, and E.~Y. Yuan,
\newblock Phys.Rev.Lett. {\bf 113}, 171601 (2014), arXiv:1307.2199.
%%CITATION = ARXIV:1307.2199;%%

\bibitem{Cachazo:2013iea}
F.~Cachazo, S.~He, and E.~Y. Yuan,
\newblock JHEP {\bf 1407}, 033 (2014), arXiv:1309.0885.
%%CITATION = ARXIV:1309.0885;%%

\bibitem{Dolan:2014ega}
L.~Dolan and P.~Goddard,
\newblock JHEP {\bf 1407}, 029 (2014), arXiv:1402.7374.
%%CITATION = ARXIV:1402.7374;%%

\bibitem{He:2014wua}
Y.-H. He, C.~Matti, and C.~Sun,
\newblock JHEP {\bf 1410}, 135 (2014), arXiv:1403.6833.
%%CITATION = ARXIV:1403.6833;%%

\bibitem{Griffiths:book}
P.~Griffiths and J.~Harris,
\newblock {\em Principles of Algebraic Geometry} (John Wiley \& Sons, New York,
  1994).

\bibitem{Dolan:2013isa}
L.~Dolan and P.~Goddard,
\newblock JHEP {\bf 1405}, 010 (2014), arXiv:1311.5200.
%%CITATION = ARXIV:1311.5200;%%

\bibitem{Sogaard:2015dba}
M.~Søgaard and Y.~Zhang,
\newblock Phys. Rev. {\bf D93}, 105009 (2016), arXiv:1509.08897.
%%CITATION = ARXIV:1509.08897;%%

\bibitem{Bosma:2016ttj}
J.~Bosma, M.~Søgaard, and Y.~Zhang,
\newblock Phys. Rev. {\bf D94}, 041701 (2016), arXiv:1605.08431.
%%CITATION = ARXIV:1605.08431;%%

\bibitem{DeWitt:1967yk}
B.~S. DeWitt,
\newblock Phys. Rev. {\bf 160}, 1113 (1967).
%%CITATION = PHRVA,160,1113;%%

\bibitem{DeWitt:1967ub}
B.~S. DeWitt,
\newblock Phys. Rev. {\bf 162}, 1195 (1967).
%%CITATION = PHRVA,162,1195;%%

\bibitem{DeWitt:1967uc}
B.~S. DeWitt,
\newblock Phys. Rev. {\bf 162}, 1239 (1967).
%%CITATION = PHRVA,162,1239;%%

\bibitem{Veltman:1975vx}
M.~J.~G. Veltman,
\newblock {Quantum Theory of Gravitation},
\newblock in {\em {Methods in Field Theory: Proceedings, 28th Les Houches
  Summer School, France, July 28-September 6, 1975}}, pp. 265--327, 1975.
%%CITATION = CONFP,C7507281,265;%%

\bibitem{White:2016jzc}
C.~D. White,
\newblock Phys. Lett. {\bf B763}, 365 (2016), arXiv:1606.04724.
%%CITATION = ARXIV:1606.04724;%%

\bibitem{Bern:2010yg}
Z.~Bern, T.~Dennen, Y.-t. Huang, and M.~Kiermaier,
\newblock Phys.Rev. {\bf D82}, 065003 (2010), arXiv:1004.0693.
%%CITATION = ARXIV:1004.0693;%%

\bibitem{Kawai:1985xq}
H.~Kawai, D.~Lewellen, and S.~Tye,
\newblock Nucl.Phys. {\bf B269}, 1 (1986).
%%CITATION = NUPHA,B269,1;%%

\bibitem{Bern:1999bx}
Z.~Bern, A.~De~Freitas, and H.~L. Wong,
\newblock Phys. Rev. Lett. {\bf 84}, 3531 (2000), arXiv:hep-th/9912033.
%%CITATION = HEP-TH/9912033;%%

\bibitem{BjerrumBohr:2004wh}
N.~E.~J. Bjerrum-Bohr and K.~Risager,
\newblock Phys. Rev. {\bf D70}, 086011 (2004), arXiv:hep-th/0407085.
%%CITATION = HEP-TH/0407085;%%

\bibitem{BjerrumBohr:2010ta}
N.~Bjerrum-Bohr, P.~H. Damgaard, B.~Feng, and T.~Sondergaard,
\newblock Phys.Rev. {\bf D82}, 107702 (2010), arXiv:1005.4367.
%%CITATION = ARXIV:1005.4367;%%

\bibitem{Feng:2010br}
B.~Feng and S.~He,
\newblock JHEP {\bf 09}, 043 (2010), arXiv:1007.0055.
%%CITATION = ARXIV:1007.0055;%%

\bibitem{Damgaard:2012fb}
P.~H. Damgaard, R.~Huang, T.~Sondergaard, and Y.~Zhang,
\newblock JHEP {\bf 08}, 101 (2012), arXiv:1206.1577.
%%CITATION = ARXIV:1206.1577;%%

\bibitem{delaCruz:2015raa}
L.~de~la Cruz, A.~Kniss, and S.~Weinzierl,
\newblock JHEP {\bf 11}, 217 (2015), arXiv:1508.06557.
%%CITATION = ARXIV:1508.06557;%%

\bibitem{delaCruz:2016wbr}
L.~de~la Cruz, A.~Kniss, and S.~Weinzierl,
\newblock Phys. Rev. Lett. {\bf 116}, 201601 (2016), arXiv:1601.04523.
%%CITATION = ARXIV:1601.04523;%%

\bibitem{BjerrumBohr:2012mg}
N.~Bjerrum-Bohr, P.~H. Damgaard, R.~Monteiro, and D.~O'Connell,
\newblock JHEP {\bf 1206}, 061 (2012), arXiv:1203.0944.
%%CITATION = ARXIV:1203.0944;%%

\bibitem{Monteiro:2011pc}
R.~Monteiro and D.~O'Connell,
\newblock JHEP {\bf 1107}, 007 (2011), arXiv:1105.2565.
%%CITATION = ARXIV:1105.2565;%%

\bibitem{Mafra:2011kj}
C.~R. Mafra, O.~Schlotterer, and S.~Stieberger,
\newblock JHEP {\bf 1107}, 092 (2011), arXiv:1104.5224.
%%CITATION = ARXIV:1104.5224;%%

\bibitem{Naculich:2014naa}
S.~G. Naculich,
\newblock JHEP {\bf 1409}, 029 (2014), arXiv:1407.7836.
%%CITATION = ARXIV:1407.7836;%%

\bibitem{Naculich:2015zha}
S.~G. Naculich,
\newblock JHEP {\bf 05}, 050 (2015), arXiv:1501.03500.
%%CITATION = ARXIV:1501.03500;%%

\bibitem{Naculich:2015coa}
S.~G. Naculich,
\newblock JHEP {\bf 09}, 122 (2015), arXiv:1506.06134.
%%CITATION = ARXIV:1506.06134;%%

\bibitem{Kalousios:2013eca}
C.~Kalousios,
\newblock J.Phys. {\bf A47}, 215402 (2014), arXiv:1312.7743.
%%CITATION = ARXIV:1312.7743;%%

\bibitem{Tolotti:2013caa}
M.~Tolotti and S.~Weinzierl,
\newblock JHEP {\bf 07}, 111 (2013), arXiv:1306.2975.
%%CITATION = ARXIV:1306.2975;%%

\bibitem{Weinzierl:2014vwa}
S.~Weinzierl,
\newblock JHEP {\bf 1404}, 092 (2014), arXiv:1402.2516.
%%CITATION = ARXIV:1402.2516;%%

\bibitem{Lam:2014tga}
C.~S. Lam,
\newblock Phys. Rev. {\bf D91}, 045019 (2015), arXiv:1410.8184.
%%CITATION = ARXIV:1410.8184;%%

\bibitem{Monteiro:2013rya}
R.~Monteiro and D.~O'Connell,
\newblock JHEP {\bf 1403}, 110 (2014), arXiv:1311.1151.
%%CITATION = ARXIV:1311.1151;%%

\bibitem{Cachazo:2015nwa}
F.~Cachazo and H.~Gomez,
\newblock JHEP {\bf 04}, 108 (2016), arXiv:1505.03571.
%%CITATION = ARXIV:1505.03571;%%

\bibitem{Baadsgaard:2015voa}
C.~Baadsgaard, N.~E.~J. Bjerrum-Bohr, J.~L. Bourjaily, and P.~H. Damgaard,
\newblock JHEP {\bf 09}, 129 (2015), arXiv:1506.06137.
%%CITATION = ARXIV:1506.06137;%%

\bibitem{Baadsgaard:2015hia}
C.~Baadsgaard, N.~E.~J. Bjerrum-Bohr, J.~L. Bourjaily, P.~H. Damgaard, and
  B.~Feng,
\newblock JHEP {\bf 11}, 080 (2015), arXiv:1508.03627.
%%CITATION = ARXIV:1508.03627;%%

\bibitem{Baadsgaard:2015ifa}
C.~Baadsgaard, N.~E.~J. Bjerrum-Bohr, J.~L. Bourjaily, and P.~H. Damgaard,
\newblock JHEP {\bf 09}, 136 (2015), arXiv:1507.00997.
%%CITATION = ARXIV:1507.00997;%%

\bibitem{Huang:2015yka}
R.~Huang, J.~Rao, B.~Feng, and Y.-H. He,
\newblock JHEP {\bf 12}, 056 (2015), arXiv:1509.04483.
%%CITATION = ARXIV:1509.04483;%%

\bibitem{Cardona:2016gon}
C.~Cardona, B.~Feng, H.~Gomez, and R.~Huang,
\newblock JHEP {\bf 09}, 133 (2016), arXiv:1606.00670.
%%CITATION = ARXIV:1606.00670;%%

\bibitem{Gomez:2016bmv}
H.~Gomez,
\newblock JHEP {\bf 06}, 101 (2016), arXiv:1604.05373.
%%CITATION = ARXIV:1604.05373;%%

\bibitem{Cardona:2016bpi}
C.~Cardona and H.~Gomez,
\newblock JHEP {\bf 06}, 094 (2016), arXiv:1605.01446.
%%CITATION = ARXIV:1605.01446;%%

\bibitem{Cardona:2016wcr}
C.~Cardona and H.~Gomez,
\newblock JHEP {\bf 10}, 116 (2016), arXiv:1607.01871.
%%CITATION = ARXIV:1607.01871;%%

\bibitem{Bjerrum-Bohr:2014qwa}
N.~E.~J. Bjerrum-Bohr, P.~H. Damgaard, P.~Tourkine, and P.~Vanhove,
\newblock Phys.Rev. {\bf D90}, 106002 (2014), arXiv:1403.4553.
%%CITATION = ARXIV:1403.4553;%%

\bibitem{Mason:2013sva}
L.~Mason and D.~Skinner,
\newblock JHEP {\bf 1407}, 048 (2014), arXiv:1311.2564.
%%CITATION = ARXIV:1311.2564;%%

\bibitem{Berkovits:2013xba}
N.~Berkovits,
\newblock JHEP {\bf 1403}, 017 (2014), arXiv:1311.4156.
%%CITATION = ARXIV:1311.4156;%%

\bibitem{Gomez:2013wza}
H.~Gomez and E.~Y. Yuan,
\newblock JHEP {\bf 1404}, 046 (2014), arXiv:1312.5485.
%%CITATION = ARXIV:1312.5485;%%

\bibitem{Adamo:2013tsa}
T.~Adamo, E.~Casali, and D.~Skinner,
\newblock JHEP {\bf 1404}, 104 (2014), arXiv:1312.3828.
%%CITATION = ARXIV:1312.3828;%%

\bibitem{Geyer:2014fka}
Y.~Geyer, A.~E. Lipstein, and L.~J. Mason,
\newblock Phys.Rev.Lett. {\bf 113}, 081602 (2014), arXiv:1404.6219.
%%CITATION = ARXIV:1404.6219;%%

\bibitem{Casali:2014hfa}
E.~Casali and P.~Tourkine,
\newblock JHEP {\bf 04}, 013 (2015), arXiv:1412.3787.
%%CITATION = ARXIV:1412.3787;%%

\bibitem{Geyer:2015bja}
Y.~Geyer, L.~Mason, R.~Monteiro, and P.~Tourkine,
\newblock Phys. Rev. Lett. {\bf 115}, 121603 (2015), arXiv:1507.00321.
%%CITATION = ARXIV:1507.00321;%%

\bibitem{Geyer:2016wjx}
Y.~Geyer, L.~Mason, R.~Monteiro, and P.~Tourkine,
\newblock Phys. Rev. {\bf D94}, 125029 (2016), arXiv:1607.08887.
%%CITATION = ARXIV:1607.08887;%%

\bibitem{Schwab:2014xua}
B.~U.~W. Schwab and A.~Volovich,
\newblock Phys.Rev.Lett. {\bf 113}, 101601 (2014), arXiv:1404.7749.
%%CITATION = ARXIV:1404.7749;%%

\bibitem{Afkhami-Jeddi:2014fia}
N.~Afkhami-Jeddi,
\newblock (2014), arXiv:1405.3533.
%%CITATION = ARXIV:1405.3533;%%

\bibitem{Zlotnikov:2014sva}
M.~Zlotnikov,
\newblock JHEP {\bf 1410}, 148 (2014), arXiv:1407.5936.
%%CITATION = ARXIV:1407.5936;%%

\bibitem{Kalousios:2014uva}
C.~Kalousios and F.~Rojas,
\newblock JHEP {\bf 01}, 107 (2015), arXiv:1407.5982.
%%CITATION = ARXIV:1407.5982;%%

\bibitem{White:2014qia}
C.~White,
\newblock Phys.Lett. {\bf B737}, 216 (2014), arXiv:1406.7184.
%%CITATION = ARXIV:1406.7184;%%

\bibitem{Cardona:2015ouc}
C.~Cardona and C.~Kalousios,
\newblock Phys. Lett. {\bf B756}, 180 (2016), arXiv:1511.05915.
%%CITATION = ARXIV:1511.05915;%%

\bibitem{Dolan:2015iln}
L.~Dolan and P.~Goddard,
\newblock JHEP {\bf 10}, 149 (2016), arXiv:1511.09441.
%%CITATION = ARXIV:1511.09441;%%

\bibitem{Brown:2016mrh}
R.~W. Brown and S.~G. Naculich,
\newblock JHEP {\bf 10}, 130 (2016), arXiv:1608.04387.
%%CITATION = ARXIV:1608.04387;%%

\bibitem{Brown:2016hck}
R.~W. Brown and S.~G. Naculich,
\newblock JHEP {\bf 11}, 060 (2016), arXiv:1608.05291.
%%CITATION = ARXIV:1608.05291;%%

\bibitem{Cachazo:2016sdc}
F.~Cachazo and G.~Zhang,
\newblock (2016), arXiv:1601.06305.
%%CITATION = ARXIV:1601.06305;%%

\bibitem{Cachazo:2016ror}
F.~Cachazo, S.~Mizera, and G.~Zhang,
\newblock (2016), arXiv:1609.00008.
%%CITATION = ARXIV:1609.00008;%%

\bibitem{Baadsgaard:2015twa}
C.~Baadsgaard {\em et~al.},
\newblock Phys. Rev. Lett. {\bf 116}, 061601 (2016), arXiv:1509.02169.
%%CITATION = ARXIV:1509.02169;%%

\bibitem{Bjerrum-Bohr:2016juj}
N.~E.~J. Bjerrum-Bohr, J.~L. Bourjaily, P.~H. Damgaard, and B.~Feng,
\newblock Nucl. Phys. {\bf B913}, 964 (2016), arXiv:1605.06501.
%%CITATION = ARXIV:1605.06501;%%

\bibitem{Bjerrum-Bohr:2016axv}
N.~E.~J. Bjerrum-Bohr, J.~L. Bourjaily, P.~H. Damgaard, and B.~Feng,
\newblock JHEP {\bf 09}, 094 (2016), arXiv:1608.00006.
%%CITATION = ARXIV:1608.00006;%%

\bibitem{Elvang:2014fja}
H.~Elvang {\em et~al.},
\newblock JHEP {\bf 12}, 181 (2014), arXiv:1410.0621.
%%CITATION = ARXIV:1410.0621;%%

\bibitem{Hodges:2010kq}
A.~Hodges,
\newblock JHEP {\bf 08}, 051 (2013), arXiv:1004.3323.
%%CITATION = ARXIV:1004.3323;%%

\bibitem{Schwinn:2007ee}
C.~Schwinn and S.~Weinzierl,
\newblock JHEP {\bf 04}, 072 (2007), hep-ph/0703021.
%%CITATION = HEP-PH/0703021;%%

\bibitem{Dixon:2010ik}
L.~J. Dixon, J.~M. Henn, J.~Plefka, and T.~Schuster,
\newblock JHEP {\bf 01}, 035 (2011), arXiv:1010.3991.
%%CITATION = ARXIV:1010.3991;%%

\bibitem{Melia:2013bta}
T.~Melia,
\newblock Phys.Rev. {\bf D88}, 014020 (2013), arXiv:1304.7809.
%%CITATION = ARXIV:1304.7809;%%

\bibitem{Melia:2013epa}
T.~Melia,
\newblock Phys.Rev. {\bf D89}, 074012 (2014), arXiv:1312.0599.
%%CITATION = ARXIV:1312.0599;%%

\bibitem{Melia:2015ika}
T.~Melia,
\newblock JHEP {\bf 12}, 107 (2015), arXiv:1509.03297.
%%CITATION = ARXIV:1509.03297;%%

\bibitem{Weinzierl:2014ava}
S.~Weinzierl,
\newblock JHEP {\bf 1503}, 141 (2015), arXiv:1412.5993.
%%CITATION = ARXIV:1412.5993;%%

\bibitem{Johansson:2015oia}
H.~Johansson and A.~Ochirov,
\newblock JHEP {\bf 01}, 170 (2016), arXiv:1507.00332.
%%CITATION = ARXIV:1507.00332;%%

\bibitem{Goncharov_no_note}
A.~B. Goncharov,
\newblock Math. Res. Lett. {\bf 5}, 497 (1998).

\bibitem{Borwein}
J.~M. Borwein, D.~M. Bradley, D.~J. Broadhurst, and P.~Lisonek,
\newblock Trans. Amer. Math. Soc. {\bf 353:3}, 907 (2001), math.CA/9910045.

\bibitem{Vermaseren:1998uu}
J.~A.~M. Vermaseren,
\newblock Int. J. Mod. Phys. {\bf A14}, 2037 (1999), hep-ph/9806280.
%%CITATION = IMPAE,A14,2037;%%

\bibitem{Remiddi:1999ew}
E.~Remiddi and J.~A.~M. Vermaseren,
\newblock Int. J. Mod. Phys. {\bf A15}, 725 (2000), hep-ph/9905237.
%%CITATION = IMPAE,A15,725;%%

\bibitem{Kotikov:1990kg}
A.~V. Kotikov,
\newblock Phys. Lett. {\bf B254}, 158 (1991).
%%CITATION = PHLTA,B254,158;%%

\bibitem{Kotikov:1991pm}
A.~V. Kotikov,
\newblock Phys. Lett. {\bf B267}, 123 (1991).
%%CITATION = PHLTA,B267,123;%%

\bibitem{Remiddi:1997ny}
E.~Remiddi,
\newblock Nuovo Cim. {\bf A110}, 1435 (1997), hep-th/9711188.
%%CITATION = HEP-TH 9711188;%%

\bibitem{Gehrmann:1999as}
T.~Gehrmann and E.~Remiddi,
\newblock Nucl. Phys. {\bf B580}, 485 (2000), hep-ph/9912329.
%%CITATION = NUPHA,B580,485;%%

\bibitem{Goncharov:2001}
A.~B. Goncharov,
\newblock (2001), math.AG/0103059.

\bibitem{Moch:2001zr}
S.~Moch, P.~Uwer, and S.~Weinzierl,
\newblock J. Math. Phys. {\bf 43}, 3363 (2002), hep-ph/0110083.
%%CITATION = HEP-PH 0110083;%%

\bibitem{Brown:2006}
F.~Brown,
\newblock C. R. Acad. Sci. Paris {\bf 342}, 949 (2006).

\bibitem{Brown:2008}
F.~Brown,
\newblock Commun. Math. Phys. {\bf 287}, 925 (2008), arXiv:0804.1660.

\bibitem{Panzer:2014caa}
E.~Panzer,
\newblock Comput. Phys. Commun. {\bf 188}, 148 (2014), arXiv:1403.3385.
%%CITATION = ARXIV:1403.3385;%%

\bibitem{Argeri:2007up}
M.~Argeri and P.~Mastrolia,
\newblock Int. J. Mod. Phys. {\bf A22}, 4375 (2007), arXiv:0707.4037.
%%CITATION = 0707.4037;%%

\bibitem{MullerStach:2012mp}
S.~M{\"u}ller-Stach, S.~Weinzierl, and R.~Zayadeh,
\newblock Commun.Math.Phys. {\bf 326}, 237 (2014), arXiv:1212.4389.
%%CITATION = ARXIV:1212.4389;%%

\bibitem{Henn:2013pwa}
J.~M. Henn,
\newblock Phys. Rev. Lett. {\bf 110}, 251601 (2013), arXiv:1304.1806.
%%CITATION = ARXIV:1304.1806;%%

\bibitem{Ablinger:2015tua}
J.~Ablinger {\em et~al.},
\newblock Comput. Phys. Commun. {\bf 202}, 33 (2016), arXiv:1509.08324.
%%CITATION = ARXIV:1509.08324;%%

\bibitem{MullerStach:2011ru}
S.~M{\"u}ller-Stach, S.~Weinzierl, and R.~Zayadeh,
\newblock Commun. Num. Theor. Phys. {\bf 6}, 203 (2012), arXiv:1112.4360.
%%CITATION = 1112.4360;%%

\bibitem{Adams:2013nia}
L.~Adams, C.~Bogner, and S.~Weinzierl,
\newblock J. Math. Phys. {\bf 54}, 052303 (2013), arXiv:1302.7004.
%%CITATION = ARXIV:1302.7004;%%

\bibitem{Bloch:2013tra}
S.~Bloch and P.~Vanhove,
\newblock J. Numb. Theor. {\bf 148}, 328 (2015), arXiv:1309.5865.
%%CITATION = ARXIV:1309.5865;%%

\bibitem{Adams:2014vja}
L.~Adams, C.~Bogner, and S.~Weinzierl,
\newblock J. Math. Phys. {\bf 55}, 102301 (2014), arXiv:1405.5640.
%%CITATION = ARXIV:1405.5640;%%

\bibitem{Adams:2015gva}
L.~Adams, C.~Bogner, and S.~Weinzierl,
\newblock J. Math. Phys. {\bf 56}, 072303 (2015), arXiv:1504.03255.
%%CITATION = ARXIV:1504.03255;%%

\bibitem{Adams:2015ydq}
L.~Adams, C.~Bogner, and S.~Weinzierl,
\newblock J. Math. Phys. {\bf 57}, 032304 (2016), arXiv:1512.05630.
%%CITATION = ARXIV:1512.05630;%%

\bibitem{Remiddi:2013joa}
E.~Remiddi and L.~Tancredi,
\newblock Nucl.Phys. {\bf B880}, 343 (2014), arXiv:1311.3342.
%%CITATION = ARXIV:1311.3342;%%

\bibitem{Bloch:2016izu}
S.~Bloch, M.~Kerr, and P.~Vanhove,
\newblock (2016), arXiv:1601.08181.
%%CITATION = ARXIV:1601.08181;%%

\bibitem{Remiddi:2016gno}
E.~Remiddi and L.~Tancredi,
\newblock Nucl. Phys. {\bf B907}, 400 (2016), arXiv:1602.01481.
%%CITATION = ARXIV:1602.01481;%%

\bibitem{Adams:2016xah}
L.~Adams, C.~Bogner, A.~Schweitzer, and S.~Weinzierl,
\newblock J. Math. Phys. {\bf 57}, 122302 (2016), arXiv:1607.01571.
%%CITATION = ARXIV:1607.01571;%%

\bibitem{Bonciani:2016qxi}
R.~Bonciani {\em et~al.},
\newblock JHEP {\bf 12}, 096 (2016), arXiv:1609.06685.
%%CITATION = ARXIV:1609.06685;%%

\bibitem{Broedel:2014vla}
J.~Broedel, C.~R. Mafra, N.~Matthes, and O.~Schlotterer,
\newblock JHEP {\bf 07}, 112 (2015), arXiv:1412.5535.
%%CITATION = ARXIV:1412.5535;%%

\bibitem{Korchemsky:2009jv}
G.~P. Korchemsky and E.~Sokatchev,
\newblock Nucl. Phys. {\bf B829}, 478 (2010), arXiv:0907.4107.
%%CITATION = ARXIV:0907.4107;%%

\bibitem{Alday:2010zy}
L.~F. Alday, B.~Eden, G.~P. Korchemsky, J.~Maldacena, and E.~Sokatchev,
\newblock JHEP {\bf 09}, 123 (2011), arXiv:1007.3243.
%%CITATION = ARXIV:1007.3243;%%

\bibitem{Bern:2010tq}
Z.~Bern, J.~J.~M. Carrasco, L.~J. Dixon, H.~Johansson, and R.~Roiban,
\newblock Phys. Rev. {\bf D82}, 125040 (2010), arXiv:1008.3327.
%%CITATION = ARXIV:1008.3327;%%

\bibitem{Kosower:2010yk}
D.~A. Kosower, R.~Roiban, and C.~Vergu,
\newblock Phys. Rev. {\bf D83}, 065018 (2011), arXiv:1009.1376.
%%CITATION = ARXIV:1009.1376;%%

\bibitem{Dixon:2011ng}
L.~J. Dixon, J.~M. Drummond, and J.~M. Henn,
\newblock JHEP {\bf 06}, 100 (2011), arXiv:1104.2787.
%%CITATION = ARXIV:1104.2787;%%

\bibitem{Bern:2012uc}
Z.~Bern, J.~J.~M. Carrasco, H.~Johansson, and R.~Roiban,
\newblock Phys. Rev. Lett. {\bf 109}, 241602 (2012), arXiv:1207.6666.
%%CITATION = ARXIV:1207.6666;%%

\bibitem{Eden:2012fe}
B.~Eden, P.~Heslop, G.~P. Korchemsky, V.~A. Smirnov, and E.~Sokatchev,
\newblock Nucl. Phys. {\bf B862}, 123 (2012), arXiv:1202.5733.
%%CITATION = ARXIV:1202.5733;%%

\bibitem{Bargheer:2014mxa}
T.~Bargheer, Y.-t. Huang, F.~Loebbert, and M.~Yamazaki,
\newblock Phys. Rev. {\bf D91}, 026004 (2015), arXiv:1407.4449.
%%CITATION = ARXIV:1407.4449;%%

\bibitem{Ferro:2014gca}
L.~Ferro, T.~Łukowski, and M.~Staudacher,
\newblock Nucl. Phys. {\bf B889}, 192 (2014), arXiv:1407.6736.
%%CITATION = ARXIV:1407.6736;%%

\bibitem{Dixon:2014voa}
L.~J. Dixon, J.~M. Drummond, C.~Duhr, and J.~Pennington,
\newblock JHEP {\bf 06}, 116 (2014), arXiv:1402.3300.
%%CITATION = ARXIV:1402.3300;%%

\bibitem{Caron-Huot:2016owq}
S.~Caron-Huot, L.~J. Dixon, A.~McLeod, and M.~von Hippel,
\newblock Phys. Rev. Lett. {\bf 117}, 241601 (2016), arXiv:1609.00669.
%%CITATION = ARXIV:1609.00669;%%

\bibitem{Chicherin:2015bza}
D.~Chicherin {\em et~al.},
\newblock JHEP {\bf 03}, 031 (2016), arXiv:1506.04983.
%%CITATION = ARXIV:1506.04983;%%

\bibitem{Mafra:2010jq}
C.~R. Mafra, O.~Schlotterer, S.~Stieberger, and D.~Tsimpis,
\newblock Phys.Rev. {\bf D83}, 126012 (2011), arXiv:1012.3981.
%%CITATION = ARXIV:1012.3981;%%

\bibitem{MacKay:2004tc}
N.~J. MacKay,
\newblock Int. J. Mod. Phys. {\bf A20}, 7189 (2005), arXiv:hep-th/0409183.
%%CITATION = HEP-TH/0409183;%%

\bibitem{Drummond:2007cf}
J.~M. Drummond, J.~Henn, G.~P. Korchemsky, and E.~Sokatchev,
\newblock Nucl. Phys. {\bf B795}, 52 (2008), arXiv:0709.2368.
%%CITATION = ARXIV:0709.2368;%%

\bibitem{Drummond:2008vq}
J.~M. Drummond, J.~Henn, G.~P. Korchemsky, and E.~Sokatchev,
\newblock Nucl. Phys. {\bf B828}, 317 (2010), arXiv:0807.1095.
%%CITATION = ARXIV:0807.1095;%%

\bibitem{Drummond:2009fd}
J.~M. Drummond, J.~M. Henn, and J.~Plefka,
\newblock JHEP {\bf 05}, 046 (2009), arXiv:0902.2987.
%%CITATION = ARXIV:0902.2987;%%

\bibitem{Drummond:2010uq}
J.~M. Drummond and L.~Ferro,
\newblock JHEP {\bf 12}, 010 (2010), arXiv:1002.4622.
%%CITATION = ARXIV:1002.4622;%%

\bibitem{Drummond:2010qh}
J.~M. Drummond and L.~Ferro,
\newblock JHEP {\bf 07}, 027 (2010), arXiv:1001.3348.
%%CITATION = ARXIV:1001.3348;%%

\bibitem{Beisert:2010gn}
N.~Beisert, J.~Henn, T.~McLoughlin, and J.~Plefka,
\newblock JHEP {\bf 04}, 085 (2010), arXiv:1002.1733.
%%CITATION = ARXIV:1002.1733;%%

\bibitem{Ferro:2011ph}
L.~Ferro,
\newblock (2011), arXiv:1107.1776.
%%CITATION = ARXIV:1107.1776;%%

\bibitem{Keller:2008}
B.~{Keller},
\newblock (2008), arXiv:0807.1960.

\bibitem{Goncharov:2010jf}
A.~B. Goncharov, M.~Spradlin, C.~Vergu, and A.~Volovich,
\newblock Phys. Rev. Lett. {\bf 105}, 151605 (2010), arXiv:1006.5703.
%%CITATION = ARXIV:1006.5703;%%

\bibitem{Golden:2013xva}
J.~Golden, A.~B. Goncharov, M.~Spradlin, C.~Vergu, and A.~Volovich,
\newblock JHEP {\bf 01}, 091 (2014), arXiv:1305.1617.
%%CITATION = ARXIV:1305.1617;%%

\bibitem{Golden:2014xqa}
J.~Golden, M.~F. Paulos, M.~Spradlin, and A.~Volovich,
\newblock J. Phys. {\bf A47}, 474005 (2014), arXiv:1401.6446.
%%CITATION = ARXIV:1401.6446;%%

\bibitem{Drummond:2014ffa}
J.~M. Drummond, G.~Papathanasiou, and M.~Spradlin,
\newblock JHEP {\bf 03}, 072 (2015), arXiv:1412.3763.
%%CITATION = ARXIV:1412.3763;%%

\bibitem{Chen:2010ct}
Y.-X. Chen, Y.-J. Du, and B.~Feng,
\newblock JHEP {\bf 01}, 081 (2011), arXiv:1011.1953.
%%CITATION = ARXIV:1011.1953;%%

\bibitem{Bern:2011ia}
Z.~Bern and T.~Dennen,
\newblock Phys.Rev.Lett. {\bf 107}, 081601 (2011), arXiv:1103.0312.
%%CITATION = ARXIV:1103.0312;%%

\bibitem{Bern:2013uka}
Z.~Bern, S.~Davies, T.~Dennen, A.~V. Smirnov, and V.~A. Smirnov,
\newblock Phys. Rev. Lett. {\bf 111}, 231302 (2013), arXiv:1309.2498.
%%CITATION = ARXIV:1309.2498;%%

\bibitem{Monteiro:2014cda}
R.~Monteiro, D.~O'Connell, and C.~D. White,
\newblock JHEP {\bf 12}, 056 (2014), arXiv:1410.0239.
%%CITATION = ARXIV:1410.0239;%%

\bibitem{Cachazo:2014nsa}
F.~Cachazo, S.~He, and E.~Y. Yuan,
\newblock JHEP {\bf 01}, 121 (2015), arXiv:1409.8256.
%%CITATION = ARXIV:1409.8256;%%

\bibitem{Cachazo:2014xea}
F.~Cachazo, S.~He, and E.~Y. Yuan,
\newblock JHEP {\bf 07}, 149 (2015), arXiv:1412.3479.
%%CITATION = ARXIV:1412.3479;%%

\bibitem{Johansson:2014zca}
H.~Johansson and A.~Ochirov,
\newblock JHEP {\bf 11}, 046 (2015), arXiv:1407.4772.
%%CITATION = ARXIV:1407.4772;%%

\bibitem{Chiodaroli:2015wal}
M.~Chiodaroli, M.~Gunaydin, H.~Johansson, and R.~Roiban,
\newblock Phys. Rev. Lett. {\bf 117}, 011603 (2016), arXiv:1512.09130.
%%CITATION = ARXIV:1512.09130;%%

\bibitem{Chiodaroli:2013upa}
M.~Chiodaroli, Q.~Jin, and R.~Roiban,
\newblock JHEP {\bf 01}, 152 (2014), arXiv:1311.3600.
%%CITATION = ARXIV:1311.3600;%%

\bibitem{Bjerrum-Bohr:2014zsa}
N.~E.~J. Bjerrum-Bohr, J.~F. Donoghue, B.~R. Holstein, L.~Planté, and
  P.~Vanhove,
\newblock Phys. Rev. Lett. {\bf 114}, 061301 (2015), arXiv:1410.7590.
%%CITATION = ARXIV:1410.7590;%%

\bibitem{Bjerrum-Bohr:2016hpa}
N.~E.~J. Bjerrum-Bohr, J.~F. Donoghue, B.~R. Holstein, L.~Plante, and
  P.~Vanhove,
\newblock (2016), arXiv:1609.07477.
%%CITATION = ARXIV:1609.07477;%%

\bibitem{Nandan:2016pya}
D.~Nandan, J.~Plefka, O.~Schlotterer, and C.~Wen,
\newblock JHEP {\bf 10}, 070 (2016), arXiv:1607.05701.
%%CITATION = ARXIV:1607.05701;%%

\bibitem{delaCruz:2016gnm}
L.~de~la Cruz, A.~Kniss, and S.~Weinzierl,
\newblock (2016), arXiv:1607.06036.
%%CITATION = ARXIV:1607.06036;%%

\bibitem{Stieberger:2014cea}
S.~Stieberger and T.~R. Taylor,
\newblock Phys. Lett. {\bf B739}, 457 (2014), arXiv:1409.4771.
%%CITATION = ARXIV:1409.4771;%%

\bibitem{Stieberger:2015qja}
S.~Stieberger and T.~R. Taylor,
\newblock Phys. Lett. {\bf B744}, 160 (2015), arXiv:1502.00655.
%%CITATION = ARXIV:1502.00655;%%

\bibitem{Stieberger:2015kia}
S.~Stieberger and T.~R. Taylor,
\newblock Phys. Lett. {\bf B750}, 587 (2015), arXiv:1508.01116.
%%CITATION = ARXIV:1508.01116;%%

\bibitem{Stieberger:2015vya}
S.~Stieberger and T.~R. Taylor,
\newblock Nucl. Phys. {\bf B903}, 104 (2016), arXiv:1510.01774.
%%CITATION = ARXIV:1510.01774;%%

\bibitem{Stieberger:2016lng}
S.~Stieberger and T.~R. Taylor,
\newblock Nucl. Phys. {\bf B913}, 151 (2016), arXiv:1606.09616.
%%CITATION = ARXIV:1606.09616;%%

\bibitem{Chiodaroli:2014xia}
M.~Chiodaroli, M.~Günaydin, H.~Johansson, and R.~Roiban,
\newblock JHEP {\bf 01}, 081 (2015), arXiv:1408.0764.
%%CITATION = ARXIV:1408.0764;%%

\bibitem{Chiodaroli:2015rdg}
M.~Chiodaroli, M.~Gunaydin, H.~Johansson, and R.~Roiban,
\newblock (2015), arXiv:1511.01740.
%%CITATION = ARXIV:1511.01740;%%

\bibitem{Chiodaroli:2016jqw}
M.~Chiodaroli,
\newblock (2016), arXiv:1607.04129.
%%CITATION = ARXIV:1607.04129;%%

\end{thebibliography}
\bibliographystyle{/home/stefanw/latex-style/h-physrev5}
\es

\end{document}